\documentclass{article}
\usepackage[T1]{fontenc}
\usepackage[utf8]{inputenc}
\usepackage{lmodern}
\usepackage{graphicx}
\usepackage{subcaption}
\usepackage{amsmath}
\usepackage{fontawesome5}
\usepackage{csquotes}
\usepackage{comment}
\usepackage{enumitem}
\usepackage[table,xcdraw]{xcolor}
\usepackage{listings}
\usepackage{authblk}
\usepackage{lipsum}
\usepackage{minitoc}
\usepackage{booktabs}
\usepackage{longtable}
\usepackage{array}
\usepackage[justification=justified]{caption}
\usepackage{orcidlink}
\usepackage{placeins}
\usepackage{hyperref}
\usepackage{siunitx}
\usepackage{float}
\usepackage{tocloft}
\usepackage{wrapfig}
\usepackage{soul}

\usepackage{amsmath,amssymb,amsthm}
\usepackage{bm}
\usepackage{algorithm}
\usepackage{algpseudocode}
\usepackage{booktabs}
\usepackage{multirow}
\usepackage{tgadventor}

\usepackage[margin=1in]{geometry}
\usepackage[most]{tcolorbox}

\usepackage[version=3]{mhchem}

\definecolor{myblue}{rgb}{0.1, 0.2, 0.6}
\definecolor{mytheme}{RGB}{242, 246, 255}
\definecolor{mygray}{gray}{0.35}
\definecolor{mygreen}{RGB}{0,120,0}
\definecolor{codebg}{RGB}{248,248,248}
\definecolor{keywordcolor}{RGB}{0,0,180}
\definecolor{stringcolor}{RGB}{153,0,0}
\definecolor{commentcolor}{RGB}{0,128,0}

\lstdefinestyle{bandgap}{
  backgroundcolor=\color{codebg},
  basicstyle=\ttfamily\footnotesize,
  keywordstyle=\color{keywordcolor}\bfseries,
  stringstyle=\color{stringcolor},
  commentstyle=\color{commentcolor}\itshape,
  showstringspaces=false,
  columns=fullflexible,
  keepspaces=true,
  frame=single,
  rulecolor=\color{mygray},
  breaklines=true,
  postbreak=\mbox{\textcolor{mygray}{$\hookrightarrow$}\space},
  language=Python
}

\hypersetup{
    colorlinks=true,
    linkcolor=[RGB]{0,51,153},
    urlcolor=[RGB]{0,51,153},
    citecolor=[RGB]{0,51,153},
    filecolor=[RGB]{0,51,153},
    allcolors=[RGB]{0,51,153}
}

\newcommand{\github}[1]{\href{#1}{\faGithubSquare}}
\newcommand{\youtube}[1]{\href{#1}{\faYoutubeSquare}}
\newcommand{\database}[1]{\href{#1}{\faDatabase}}
\newcommand{\gitlab}[1]{\href{#1}{\faGitlab}}
\newcommand{\globe}[1]{\href{#1}{\faGlobe}}

\newcounter{teamcount}

\newcommand{\authorsblock}[1]{}
\newcommand{\affiliationsblock}[1]{}

\newcommand{\RNum}[1]{\uppercase\expandafter{\romannumeral #1\relax}}

\begin{document}





\begin{tcolorbox}[
    enhanced,
    colback=mytheme,    
    colframe=mytheme,   
    arc=3mm,            
    boxrule=0pt,        
    left=15pt,          
    right=15pt,
    top=12pt,
    bottom=12pt
]
    \begin{minipage}{\linewidth}
    \centering
    {\par
    \fontsize{20pt}{22pt}\selectfont
    \fontseries{eb}\selectfont
    From Knowledge to Action: Outcomes of the 2025 Large Language Models (LLM) Hackathon for Applications in Materials Science and Chemistry
    \par}

    \vspace{15pt}

    {\large\textbf{
    2025 LLM Hackathon for Applications in Materials Science and Chemistry Participants$^1$}}\\
    {\normalsize $^1$A detailed contributor list can be found in the appendix of this paper.}
    
\end{minipage}

\vspace{10pt}
\end{tcolorbox}

\section*{Introduction}
The integration of large language models (LLMs) into scientific workflows is rapidly reshaping how researchers perform knowledge discovery, automate routine tasks, and orchestrate end-to-end computational and experimental pipelines.
In chemistry and materials science, where progress is often constrained by heterogeneous data types, fragmented tooling, and rapidly evolving literature, LLMs have emerged as general-purpose natural language interfaces that bridge domain knowledge, code, laboratory {\small \&} computational procedures, knowledge extraction, and more \cite{miret2025enabling,ruan2024automatic,roy2026comproscanner,choudhary2023chemnlp,prein2025language}.
Early efforts have demonstrated promise across a wide spectrum of applications, from understanding the literature and extracting information,
to code synthesis and simulation setup, to agentic systems that connect planning, execution, and reporting \cite{zheng2025large,lee2023towards}.

However, the rapid pace of change in available foundation models, new tools (e.g., retrieval-augmented generation and agentic frameworks), and cloud deployment platforms make it difficult for individual research groups to continuously track capabilities, identify high-impact use cases, and validate what works in their domain of practice.
As such, community-driven, time-bounded hackathons have proven to be an effective tool for rapid prototyping and exploration in a friendly but competitive manner, including for applications in science, while fostering interdisciplinary collaborations, contextual learning, and cross-domain innovations and knowledge transfer~\cite{pe2019understanding, nolte2020support, heller2023hack,treyde2026organizing}.
We have found that such hackathons can be organized cost-effectively while maximizing both their impact and reach through the use of social media, virtual platforms, and hybrid event structures.

Our previous hackathon events in 2023~\cite{jablonka202314} and 2024~\cite{llm2024} demonstrated the potential of LLM applications in wide-ranging areas of materials science and chemistry, while also revealing remaining limitations and open challenges. Compared to those earlier events, the 2025 hackathon saw a significant shift from simpler tools towards integrated, multi-agent systems that orchestrate research workflows, reflecting the maturation of agentic frameworks, retrieval-augmented generation, tool-use capabilities, and researcher knowledge and capability.

In this paper, we discuss the results and outcomes of the third LLM Hackathon for Applications in Materials Science and Chemistry, held September 11--12, 2025, which engaged a large global community in a hybrid format.
The event produced a broad portfolio of submissions that collectively illustrate how LLMs are being used not only as multipurpose models for scientific machine-learning tasks, but also as enabling platforms for rapid prototyping of domain-specific applications.
We begin by detailing and analyzing the project submissions, organizing the results into two broad categories: \emph{Knowledge Infrastructure} (new ways to know things) and \emph{Action Systems} (new ways to do things), reflecting the complementary roles of LLMs in (i) structuring, retrieving, and synthesizing scientific knowledge and (ii) executing or coordinating scientific work across computational and experimental environments. We also categorize projects with more in-depth sub-classifications and discuss emerging themes. 

Finally, we conclude by suggesting future directions of LLM applications in materials science and chemistry, including increasingly autonomous and integrated research workflows aligned with the longer-term vision of self-driving laboratories \cite{yoshikawa2023large,duo2025autonomous,cao2024agents}. Given the breadth of this paper, we encourage readers to use \autoref{tab:hackathon_projects} as an entry point: it summarizes all 88 projects with methodology tags and contains direct links to code repositories, allowing practitioners to navigate directly to the projects most relevant to their domain or technical interest.

\begin{figure}[h!]
    \centering
    \includegraphics[width=1\linewidth]{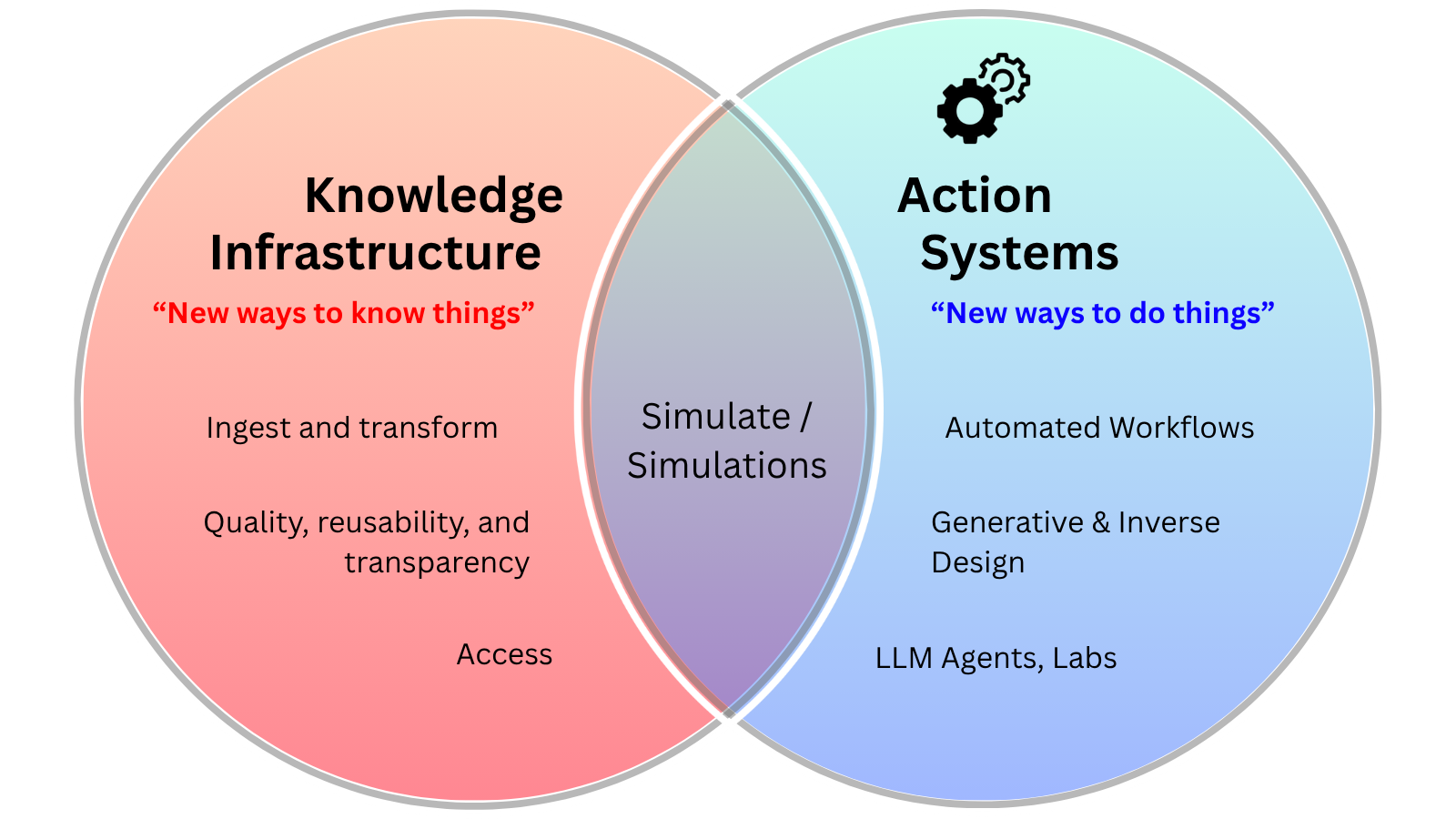}
    \caption{Venn diagram showing the classification of 88 submitted hackathon projects from the third LLM Hackathon for Applications in Materials Science and Chemistry along with their sub-classes: Knowledge Infrastructure (\textit{New ways to know things}) and Action Systems (\textit{New ways to do things}).}
    \label{fig:groups-classification}
\end{figure}

\section*{Submissions Overview}
The third LLM Hackathon for Applications in Materials Science and Chemistry received 120 submissions from more than 350 participants around the world. The paper covers 88 team submissions in detail, including accompanying links to code repositories, video demonstrations, datasets where applicable, next steps, and more. The remaining submissions were not included in this paper due to incomplete documentation submission by the teams. We categorized submissions based on their primary objectives, clustering them into two broad categories that span the materials and chemistry research lifecycle: \textit{Knowledge Infrastructure} (new ways to know things) and \textit{Action Systems }(new ways to do things)(\autoref{fig:groups-classification}):

\begin{enumerate}
    \item \textbf{Knowledge Infrastructure (New ways to know things):} Submissions focused on structuring,
    retrieving, and synthesizing scientific knowledge, including tools for data curation and analysis, reproducible artificial intelligence for science (AI4Science) workflows and trust, and educational and collaborative artificial intelligence (AI) tools.
    
    \item \textbf{Action Systems (New ways to do things):} Submissions focused on executing or coordinating
    scientific work across computational and experimental environments, including automated workflows, generative and inverse design, and LLM agents and lab-oriented systems that orchestrate
    domain tools.
\end{enumerate}


Notably, simulations occupy the intersection of these categories (\autoref{fig:groups-classification}), reflecting how LLM-based
systems increasingly couple knowledge-centric capabilities (e.g., retrieval, synthesis, and verification) with
action-centric execution (e.g., running or steering computational workflows). Collectively, this portfolio of
submissions forms a coherent planning-to-action constellation where knowledge infrastructure enables reliable context
and scientific grounding, while action systems convert intent into measurable progress via tool use and
workflow automation. Together, these systems accelerate the longer-term vision of integrated, increasingly autonomous
research workflows embodied in self-driving laboratories.

\subsection*{Emerging Themes}
Five themes stand out from the 88 projects submitted this year including 1) agentic orchestration, 2) continued importance of retrieval-augmented generation for context bridging, 3) extraction of information into persistent structured forms, 4) deeper integration of of multimodal and multilingual inputs, 5) paths towards closing the loop toward laboratory integration. Below we discuss these themes in more detail and include exemplar projects to orient readers and to connect the individual project descriptions that follow to the broader trajectory of LLM-enabled science.

\vspace{10pt}
\noindent
\textbf{Agentic orchestration as a dominant paradigm.} The most striking feature of this year's submissions is the prevalence of agentic architectures. Over 40 of the 88 projects employed LLM agents that plan, call domain tools, or iteratively refine their outputs. This represents a qualitative shift from the 2023 and 2024 hackathons, where most projects used LLMs as single-turn generators or classifiers. Projects now routinely chain multiple tools within a single workflow. For example, MixSense~(\autoref{sec:mixsense}) couples NMR spectral deconvolution with yield quantification and property prediction, while the MaterialSim AI Agent~(\autoref{sec:materialsim}) configures, executes, and post-processes molecular dynamics simulations from natural language alone. The BAKER~(\autoref{sec:baker}) framework, which autonomously spawns and configures task-specific research assistants, signals a further maturation toward scalable, modular architectures rather than monolithic models.

\vspace{10pt}
\noindent
\textbf{Retrieval-augmented generation (RAG) as connective infrastructure.} RAG appears as a component in over a third of submissions, but its role has evolved. Projects used retrieval to ground agent reasoning in curated literature (CaMEL-RAG, \autoref{sec:camel-rag}; CatalystAssistant, \autoref{sec:catalyst-assistant}), to enforce provenance and traceability in generated outputs (XAScribe, \autoref{sec:xascribe}; SKY, \autoref{sec:sky}), and to connect structured databases to natural-language interfaces (MP-LLM, \autoref{sec:mp-llm}; NOMAD RAGBOT, \autoref{sec:nomad-ragbot}). RAG now serves as infrastructure, acting as a key grounding layer that most agentic systems now leverage.

\vspace{10pt}
\noindent
\textbf{From extraction to persistent structured knowledge.} Projects sought to expand beyond one-shot information extraction toward building and maintaining structured knowledge representations. OntoKG (\autoref{sec:ontokg}) dynamically expands a Neo4j knowledge graph with LLM-inferred material-application relationships; CAMEL (\autoref{sec:camel}) constructs causal processing-structure-property graphs from multimodal literature; MOF-Genie (\autoref{sec:mof-genie}) integrates graph neural networks with a knowledge graph for hydrogen storage discovery; and ChemUnityQA (\autoref{sec:chemunity}) uses GraphRAG over a MOF knowledge graph as persistent memory for scientific agents. These projects suggest an emerging consensus that durable, machine-queryable knowledge structures --- not ephemeral context windows --- are essential for grounding autonomous scientific reasoning.

\vspace{10pt}
\noindent
\textbf{Multimodal and multilingual inputs.} A growing number of projects operate on non-textual inputs: e.g., scanning or tunnelling electron microscopy (S/TEM) micrographs (\autoref{sec:atombridge}), nuclear magnetic resonance (NMR) spectra (\autoref{sec:mixsense}), x-ray diffraction (XRD) patterns (\autoref{sec:crystalenz}), Fourier transform infrared or ultra-violet and visible light (FTIR/UV-Vis) spectra (\autoref{sec:spectrobot}), chromatography data (\autoref{sec:chromatographyminer}), nanoparticle images (\autoref{sec:ms2nano}), audio laboratory logs (\autoref{sec:ulna}), and multilingual patent text (MuMMIE, \autoref{sec:mummie}. These projects address a practical bottleneck in materials science and chemistry where experimental data is often generated as images, spectra, or instrument logs, rather than text. The ability of multimodal LLMs to operate directly on these modalities, rather than requiring manual transcription, substantially broadens the scope of LLM-enabled workflows.

\vspace{10pt}
\noindent
\textbf{Closing the loop toward laboratory integration.} Several projects explicitly target the interface between computational reasoning and physical and even autonomous experimentation. Unified Natural Language Control of Lab Modules (\autoref{sec:unlclm}) uses the Model Context Protocol (MCP) to standardize LLM interactions with hardware, including 3D printers; Natural-Language-Driven Closed-Loop Optimization (\autoref{sec:sdl-smart}) enables robotic liquid-handling systems to be steered by text instructions; and the Critical Materials Extraction project (\autoref{sec:acme}) connects RAG-driven molecular design with quantum-mechanical screening and automated benchtop electrochemistry. At the same time, platforms like LARA-HPC (\autoref{sec:lara-hpc}) and AESOP (\autoref{sec:aesop}) emphasize that human-in-the-loop validation remains essential for reliable execution, particularly for high-performance computing and safety-critical laboratory tasks. Together, these projects illustrate a community moving deliberately --- but cautiously --- towards the self-driving laboratory vision.

\vspace{10pt}
\noindent
\autoref{tab:hackathon_projects} provides a summary of all 88 projects, including condensed descriptions and standardized methodology tags. Each entry links to the corresponding detailed section and code repository.
\begin{longtable}{| m{4.5cm} | >{\centering\arraybackslash}m{1.2cm} | m{6cm} | m{3.5cm} |}
\caption{Summary of Projects from the 2025 LLM Hackathon} \label{tab:hackathon_projects} \\
\hline
\textbf{Title} & \textbf{Link} & \textbf{Description} & \textbf{Tags} \\ \hline
\endfirsthead

\multicolumn{4}{c}%
{{\bfseries \tablename\ \thetable{} -- continued from previous page}} \\
\hline
\textbf{Title} & \textbf{Link} & \textbf{Description} & \textbf{Tags} \\ \hline
\endhead

\hline \multicolumn{4}{|r|}{{Continued on next page}} \\ \hline
\endfoot

\hline \hline
\endlastfoot

1. \hyperref[sec:chemcode-bench]{ChemCodeBench: A Holistic Benchmark for Generating Efficient and Accurate PySCF Code} & \github{https://github.com/hansgundlach/ChemCodeBench} & Benchmarks LLMs on generating efficient, chemically accurate PySCF density-functional theory code. & Benchmarking, Code Generation \\ \hline
2. \hyperref[sec:umadock]{UMADock: Docking molecules in protein binding sites using Meta's UMA MLIP} & \github{https://github.com/MauricioCafiero/UMADock/tree/main} & Uses an AI agent to dock molecules into protein binding sites utilizing Meta's UMA MLIP as a scoring function. & Agentic Workflows, Simulations \\ \hline
3. \hyperref[sec:best_team]{Machine learning supported by LLMs for advanced functional materials} & \github{https://github.com/Arthurns16/hacktoon2025} & Extracts high-entropy oxide data from unstructured literature to train predictive ML phase classification models. & RAG, Data Extraction, Property Prediction \\ \hline
4. \hyperref[sec:CuPirates]{LLM-Engineered Prompt-Driven Discovery of Cuprate Superconductors} & \github{https://github.com/kinzani/Inzani-Group/tree/main/LLM\_Hackathon\_2025} & Fine-tunes an LLM to generate targeted structural prompts for text-guided diffusion models proposing new superconductors. & Generative Design \\ \hline
5. \hyperref[sec:NeuroSymbolic]{A Neurosymbolic System for Unified Autonomous Scientific Hypothesis Generation} & \github{https://github.com/dreamboat26/hackathon} & Unifies LLMs with spiking neural networks and symbolic logic to autonomously generate verifiable materials hypotheses. & Agentic Workflows, Knowledge Graphs \\ \hline
6. \hyperref[sec:aria]{Autonomous Reasoning Intelligence for Atomics} & \github{https://github.com/yicao-elina/LLM4Chem-Explainable-synthesis.git} & Addresses contextual tunneling by selectively using causal knowledge graphs for inverse synthesis design. & RAG, Knowledge Graphs, Generative Design \\ \hline
7. \hyperref[sec:lara-hpc]{LARA-HPC: A Language Model Powered Research Assistant for HPC} & \github{https://github.com/BigDFT-group/llm-hackathon-2025}  & An agent that safely synthesizes, validates, and submits scientific workflows for execution on HPC clusters. & Agentic Workflows, Code Generation, Simulations \\ \hline
8. \hyperref[sec:mixsense]{Autonomous LLM Agent Workflow for Automated NMR Mixture Analysis} & \github{https://github.com/jdsanc/MixSense.git} & An autonomous agent that integrates NMR spectral deconvolution, yield quantification, and chemosensory property prediction. & Agentic Workflows, Property Prediction \\ \hline
9. \hyperref[sec:agentlearn]{Agent Learn: Accelerating Active Learning with LLM Agents} & \github{https://github.com/tonylifepix/AgentLearn} & Embeds an LLM agent within an active learning loop to autonomously propose informative molecules and expand datasets. & Agentic Workflows, Generative Design \\ \hline
10. \hyperref[sec:fullerene_factory]{The Fullerene Factory: A Multi-Agent Workflow} & \github{https://github.com/aritraroy24/the-fullerene-factory} & A multi-agent system that optimizes functionalized fullerene isomers from natural language using ML interatomic potentials. & Agentic Workflows, Simulations \\ \hline
11. \hyperref[sec:fordham]{Reasoning Model Interpretation and Knowledge Extraction} & \github{https://github.com/BaosenZ/mach-interp-LLM-hackathon-2025.git} & Analyzes LLM reasoning traces as directed graphs to understand how models organize and apply chemical knowledge. & Benchmarking \\ \hline
12. \hyperref[sec:chembot]{A Context-Aware Chemistry Research Assistant} & \github{https://github.com/aishahasim/ChemBot---AI-Assisted-Chemistry-Assistant} & A context-aware chemistry assistant leveraging BioGPT and RAG to retrieve literature-supported explanations. & RAG, Education \\ \hline
13. \hyperref[sec:ontokg]{Dynamic Materials Ontology Expansion with LLM-KG Integration} & \github{https://github.com/deepaksaipendyala/OntoKG} & Dynamically expands a Neo4j knowledge graph using an LLM to infer and validate new material-application relationships. & Knowledge Graphs \\ \hline
14. \hyperref[sec:mint_llm]{MINT LLM: Nature Language Interface for Automated MD Analysis} & \github{https://github.com/ncsu-llm-hackathon-materials-2025/MINT-LLM.git} & A natural language interface to automatically plot, compute observables, and analyze molecular dynamics trajectories. & Simulations \\ \hline
15. \hyperref[sec:ontomapper]{Ontology-Aware Data Mapping with LLM-Driven Tree Search} & \github{https://github.com/materialdigital/ontomapper} & Couples LLM reasoning with structured tree search and node embeddings to map materials data to complex ontologies. & Knowledge Graphs \\ \hline
16. \hyperref[sec:atombridge]{Extracting crystallographic information from S/TEM literature} & \github{https://github.com/dpalmer-anl/AtomBridge.git} & Converts S/TEM literature micrographs into simulation-ready CIFs using RAG, vision models, and physics validation. & RAG, Data Extraction \\ \hline
17. \hyperref[sec:gains]{GAINS: Generative AI for No-PAIN Structures} & \github{https://github.com/napoles-uach/GAINS} & Proposes minimal, scaffold-preserving molecular edits via LLMs to remove assay-interfering PAINS motifs. & Generative Design, Property Prediction \\ \hline
18. \hyperref[sec:chemgraph_IR]{Computation and Visualization of Infrared Spectroscopy} & \github{https://github.com/argonne-lcf/ChemGraph} & Orchestrates end-to-end computational workflows to optimize geometry and predict IR spectra directly from text prompts. & Agentic Workflows, Simulations \\ \hline
19. \hyperref[sec:CLUE]{Crystal Learning for Understandable Explanations (CLUE)} & \github{https://github.com/epatyukova/llm2025-hackathon-CLUE} & Generates counterfactual structure-property explanations for crystalline materials to interpret black-box ML predictions. & Property Prediction \\ \hline
20. \hyperref[sec:NEDD]{Next Experiment Data Driven (NEDD)} & \github{https://github.com/ViktoriiaBaib/NEDD} & Recommends the next best experiment using active learning and allows natural-language querying of datasets. & Lab Automation \\ \hline
21. \hyperref[sec:T2_relax]{Text to text (T2) crystal relaxation} & \github{https://github.com/anky3733/Traj} & Predicts DFT-relaxed crystal structures directly in CIF format using a fine-tuned T5-Transformer LLM. & Simulations \\ \hline
22. \hyperref[sec:materialsim]{MaterialSim AI Agent} & \github{https://github.com/Awwal41/MaterialSim.git} & Automates molecular dynamics configurations, execution, and property extraction using natural language instructions. & Agentic Workflows, Simulations \\ \hline
23. \hyperref[sec:SCARA]{SCARA: Steel Corrosion Agent for Risk Assessment} & \github{https://github.com/mo-alkubaish/SCARA} & Predicts corrosion risk and recommends alloy alternatives using an LLM trained on industry standards and literature. & Agentic Workflows, Property Prediction \\ \hline
24. \hyperref[sec:chromatographyminer]{Chromatography Miner} & \github{https://github.com/msehabibur/gcxgc\_peakcards} & An interactive AI platform that rapidly identifies compounds from GC-MS data using vendor-neutral spectral matching. & Data Extraction \\ \hline
25. \hyperref[sec:guillemot]{guillemot: Automated Rietveld Refinement Using Agents} & \github{https://github.com/datalab-org/guillemot} & Automates PXRD pattern fitting by allowing a multimodal LLM agent to interpret visual data and configure refinement software. & Agentic Workflows \\ \hline
26. \hyperref[sec:xascribe]{XAScribe: Automated X-ray Absorption Spectroscopy Analysis} & \github{https://github.com/Oscuro-Phoenix/xascribe} & Automates X-ray absorption spectroscopy analysis and manuscript generation using predictive ML models and RAG. & RAG, Data Extraction \\ \hline
27. \hyperref[sec:baker]{BAKER: Automated Spawning of Specialized Research Assistants} & \github{https://github.com/mattiasutancykeln/Baker} & An automated framework that autonomously spawns, configures, and tests specialized multi-agent research assistants. & Agentic Workflows \\ \hline
28. \hyperref[sec:polypredictor]{PolyPredictor: Multimodal Representation of Polymers} & \github{https://github.com/554181320angela-lang/Poly-predictor} & Utilizes multimodal LLM embeddings to capture polymer microstructural features for improved property prediction. & Property Prediction \\ \hline
29. \hyperref[sec:sdl-smart]{Natural-Language-Driven Closed-Loop Optimization} & \gitlab{https://gitlab.com/heingroup/llm-hackathon-2025} & Enables closed-loop optimization for robotic liquid-handling systems via natural language instructions. & Lab Automation, Agentic Workflows \\ \hline
30. \hyperref[sec:unlclm]{Unified Natural Language Control of Lab Modules} & \github{https://github.com/ivoryzh/MCP4SDL} & Employs the Model Context Protocol (MCP) to standardize LLM interactions with diverse lab hardware like 3D printers. & Lab Automation \\ \hline
31. \hyperref[sec:FADE]{F.A.D.E: A Fully Agentic Drug Engine} & \github{https://github.com/Naveen-R-M/F.A.D.E} & Autonomously orchestrates target identification, ligand generation, and binding affinity prediction from text queries. & Agentic Workflows, Generative Design \\ \hline
32. \hyperref[sec:MatFOMGen]{MatFOMGen: Domain-Specific Figures-of-Merit} & \github{https://github.com/gopal-iyer/MatFOMGen} & Uses human-in-the-loop LLM agents to extract and mathematically formulate domain-specific figures-of-merit. & Agentic Workflows, Data Extraction \\ \hline
33. \hyperref[sec:DFTPilot]{DFTPilot: Automation and Previewing of DFT Calculation Setups} & \github{https://github.com/chiku-parida/DFTPilot} & Leverages RAG and crystal graph neural networks to help non-experts set up and preview VASP DFT calculations. & RAG, Simulations \\ \hline
34. \hyperref[sec:Parse_Patrol]{Parse Patrol: Dual-Mode Scientific Parsing Infrastructure} & \github{https://github.com/ndaelman-hu/parse-patrol} & A dual-mode framework automating scientific parser infrastructure testing and deployment using MCP servers. & RAG, Data Extraction \\ \hline
35. \hyperref[sec:catalyst-assistant]{Catalyst Assistant} & \href{https://chatgpt.com/g/g-68c09896f11c81918e86b7ddcbad47e3-catalyst-assistant}{GPT Store} & An evidence-traced agent that proposes ammonia decomposition catalysts and automates characterization plots. & Agentic Workflows, RAG \\ \hline
36. \hyperref[sec:ThinFilmAI]{Data-Driven Prediction of Thin Film Properties} & \github{https://github.com/ram123-debug/mat-chem-llm-hackathon.git} & A data-driven project predicting thin film qualities in physical and chemical deposition methods. & Property Prediction, Data Extraction \\ \hline
37. \hyperref[sec:SCALE]{Scaffold Conscious Agent for Learning \& Exploration} & \github{https://github.com/schandy2211/scale} & Guides scaffold-preserving molecular optimization loops bounded by cheminformatics guardrails and physics-aware scoring. & Agentic Workflows, Generative Design \\ \hline
38. \hyperref[sec:aerogel]{LLM as Aerogel Research Assistant (L.A.R.A)} & \github{https://github.com/sugannathan/L.A.R.A-LLMs-as-Aerogel-Research-Assistants/tree/main} & A modular framework deploying a fine-tuned LLM as a specialized research assistant for aerogel materials. & RAG, Agentic Workflows \\ \hline
39. \hyperref[sec:odeforge]{Agentic AI for Differential Equation Models} & \github{https://github.com/souvikta/ODEForge} & A multi-agent framework that translates natural language into runnable Python ODE simulations with automatic debugging. & Agentic Workflows, Code Generation, Simulations \\ \hline
40. \hyperref[sec:mp-llm]{MP-LLM: Materials Project Query Interface} & \github{https://github.com/killiansheriff/MP-LLM} & A tool-augmented interface translating conversational queries into structured API calls for the Materials Project database. & RAG, Data Extraction \\ \hline
41. \hyperref[sec:PALS]{PALS: Property Analogies with LLMs} & \github{https://github.com/ahaibel/mp-property-analogies} & Evaluates the ability of LLMs to predict molecular and materials properties through analogical reasoning. & Property Prediction \\ \hline
42. \hyperref[sec:camel-rag]{CaMEL-RAG: Retrieval-Augmented Catalytic Screening} & \github{https://github.com/ashikiut/CaMEL-RAG} & Translates natural-language queries into predictive heterogeneous catalysis insights for sustainable energy applications. & RAG, Property Prediction \\ \hline
43. \hyperref[sec:SuperconLLM]{SuperconLLM: Automating Superconductivity Knowledge Extraction} & \github{https://github.com/fpriante/SuperconLLM-Multi-Agent-Framework-for-Automating-Superconductivity-Knowledge-Extraction} & An end-to-end multi-agent pipeline converting raw arXiv PDFs into a structured superconductivity JSON database. & Agentic Workflows, Data Extraction \\ \hline
44. \hyperref[sec:Catalyze]{Catalyze: Multi-Agent Chemistry Assistant} & \github{https://github.com/srustisain/mit-catalyze} & An AI-powered assistant focusing on automated protocol generation, dual-platform lab automation, and comprehensive safety analysis. & Agentic Workflows, Lab Automation \\ \hline
45. \hyperref[sec:camel]{CAMEL: Causal Analysis of Materials Extracted from Literature} & \github{https://github.com/gourav-k/LLM4Causal.git} & Employs multimodal agents to extract causal processing-structure-property relationships into a unified knowledge graph. & Knowledge Graphs, Data Extraction \\ \hline
46. \hyperref[sec:ZeroMAT]{ZeroMAT: Zero-training MATerial Autonomous Analysis} & \github{https://github.com/Ahri111/ZEROMAT} & Achieves zero-shot materials property prediction by combining TabPFN's architecture with structural RAG embeddings. & RAG, Property Prediction \\ \hline
47. \hyperref[sec:ulna]{ULNA: Unstructured Lab Notebook Assistant} & \github{https://github.com/iarretche/Unstructured-Lab-Notebook-Assistant-ULNA-} & Converts unstructured audio logs of polymer experiments into structured data to reveal human-level procedural trends. & Data Extraction \\ \hline
48. \hyperref[sec:mummie]{Multilingual Multimodal Materials Information Extraction (MuMMIE)} & \github{https://github.com/zakidotai/MuMMIE} & A benchmark and pipeline evaluating the extraction of multilingual materials composition and properties from patents. & Benchmarking, Data Extraction \\ \hline
49. \hyperref[sec:electrolyte_ml]{Smarter Electrolyte Design via Machine Learning} & \github{https://github.com/jaydenlee97/2025-LLM-Hackathon-for-Applications-in-Materials-and-Chemistry/tree/main/code} & Uses offline reinforcement learning and high-throughput screening data to autonomously optimize solid electrolyte mixtures. & Property Prediction, Lab Automation \\ \hline
50. \hyperref[sec:benzene-boyz]{Once Upon a Time in Chromatography} & \github{https://github.com/SubramanyamSahoo/LLM-Hackathon-for-Applications-in-Materials-and-Chemistry-2025} & A platform focused on adaptive denoising, peak discovery, and compound identification in chromatography data. & Data Extraction \\ \hline
51. \hyperref[sec:sol-agent]{Sol-Agent} & \github{https://github.com/rafaelespinosacastaneda/Hackaton-for-Applications-in-Materials-Science-and-Chemistry} & Extracts sol-gel synthesis procedures from the literature to construct a queryable, FAISS-indexed knowledge base. & RAG, Data Extraction \\ \hline
52. \hyperref[sec:RedoxFlow]{RedoxFlow} & \github{https://github.com/CGruich/RedoxFlow} & Autonomously prepares NWChem DFT inputs for high-throughput Nernstian redox potential screening of generated SMILES. & Agentic Workflows, Simulations \\ \hline
53. \hyperref[sec:AutoFeatSci]{AutoFeatSci: Automated Feature Engineering} & \github{https://github.com/guannant/LLM\_auto\_featurization} & Automates feature engineering via a multi-agent loop that interprets literature to construct and evaluate physical descriptors. & Agentic Workflows, Property Prediction \\ \hline
54. \hyperref[sec:MAGE]{MAGE: Materials Agent for Generative and Evaluative Design} & \github{https://github.com/truptimohanty/MAGE} & Integrates property prediction and inverse generative design of structures into a unified natural language interface. & Agentic Workflows, Generative Design \\ \hline
55. \hyperref[sec:BASIS]{BASIS: Bulk and Surface Interface Simulations} & \github{https://github.com/abir0/dft-agent} & Automates surface chemistry setups and Quantum ESPRESSO DFT modeling utilizing a multi-agent system. & Agentic Workflows, Simulations \\ \hline
56. \hyperref[sec:Titanarium]{Titanarium: A Digital Terrarium} & \github{https://github.com/JSR-ISM-Smart-Chemistry-Lab/Titanarium} & Creates a digital arena where LLM personas of historical scientists autonomously debate chemical concepts. & Education, Agentic Workflows \\ \hline
57. \hyperref[sec:LLM4ConProp]{Predicting Concrete Materials Properties} &  & Evaluates LLM capabilities in quantitative prediction of concrete materials properties. & Property Prediction \\ \hline
58. \hyperref[sec:ms2nano]{m2snano: Nanoparticles Image Analysis} & \github{https://github.com/icn2-ai/m2snano-llm} & Automates nanoparticle image analysis to extract morphological descriptors and generates insights via an integrated LLM. & Data Extraction \\ \hline
59. \hyperref[sec:DynaAgent]{DynaAgent} & \github{https://github.com/schwallergroup/MDAgent} & A modular multi-agent framework orchestrating autonomous protein-ligand molecular dynamics simulations. & Agentic Workflows, Simulations \\ \hline
60. \hyperref[sec:CrysTalk]{CrysTalk} & \github{https://github.com/MatAgentHub/crystalk} & An agent system that edits and optimizes crystal structures driven purely by natural language instructions. & Generative Design, Agentic Workflows \\ \hline
61. \hyperref[sec:spectrobot]{SpectroBot: One-Click FTIR/UV-Vis Analysis} & \github{https://github.com/MUzair20/spectrobot} & Analyzes FTIR/UV-Vis spectra directly and grounds its peak identification in locally retrieved, literature-supported interpretations. & RAG, Data Extraction \\ \hline
62. \hyperref[sec:MindMesh]{Collaboration Learning AI-Agents (MindMesh)} & \github{https://github.com/estefaniavazquez/MindMesh} & Generates personalized AI learning agents tailored to multidisciplinary researchers' backgrounds and learning styles. & Education, Agentic Workflows \\ \hline
63. \hyperref[sec:SyntheSeek]{SyntheSeek} & \github{https://github.com/alexchen5/syntheseek} & Retrieves synthesis procedures from literature and generates customized, constraint-aware experimental plans via agents. & Agentic Workflows, RAG \\ \hline
64. \hyperref[sec:V-RAPIDS]{Validated Rapid Adsorption Probe Interaction Discovery System} & \github{https://github.com/ruiding-uchicago/auto\_CPT\_uma\_simul} & Delivers rapid, qualitative dry-lab simulations of probe-target interactions using an agentic mixture-of-experts validation loop. & Simulations, Agentic Workflows \\ \hline
65. \hyperref[sec:nomad-ragbot]{NOMAD RAGBOT} & \github{https://github.com/FAIRmat-NFDI/nomad-bot-rag-docs-discord} & Navigates distributed community documentation using structure-aware semantic retrieval and citation-grounded generation. & RAG \\ \hline
66. \hyperref[sec:MIDAS]{MIDAS: Language Controlled Molecular Design} & \github{https://github.com/pagel-s/MIDAS.git} & Guides 3D structure-based molecular diffusion models using natural language semantic constraints and tool calling. & Generative Design, Agentic Workflows \\ \hline
67. \hyperref[sec:AdsKRK]{AdsKRK: Agentic atomistic simulation framework} & \github{https://github.com/schwallergroup/llm\_adsorbate} & Provides an iterative agentic loop that attempts, learns from, and optimizes stable adsorbate configurations on catalytic surfaces. & Agentic Workflows, Simulations \\ \hline
68. \hyperref[sec:AssemblAI]{AssemblAI} & \github{https://github.com/ndharms/peptide-agent} & Automates the generation of peptide self-assembly protocols by retrieving experimental examples from a curated vector store. & Agentic Workflows, RAG \\ \hline
69. \hyperref[sec:MaterialMind]{MaterialMind} & \github{https://github.com/AzizahAlq/MaterialMind} & Recommends materials based on constraints using RAG reasoning coupled with an independent-weight multi-criteria scoring model. & RAG, Property Prediction \\ \hline
70. \hyperref[sec:chemtutor-ai]{ChemTutor-AI} & \href{https://huggingface.co/Abbasaabdul/AI\_ChemTutor/tree/main}{Hugging Face} & Dynamically generates tailored chemistry problems and JSON-structured solutions based on user knowledge profiles. & Education, RAG \\ \hline
71. \hyperref[sec:crystalenz]{CrystaLenz: Agentic XRD Analysis Pipeline} & \github{https://github.com/MJRaei/CrystaLenz} & An agentic pipeline that performs baseline correction, peak picking, and automated phase identification on XRD patterns. & Agentic Workflows, Data Extraction \\ \hline
72. \hyperref[sec:acme]{Prototyping Autonomous Critical Materials Extraction} & \github{https://github.com/HassanHarb92/ACME} & A closed-loop system connecting RAG-driven molecular design, QM screening, and automated benchtop electrochemistry testing. & Lab Automation, Generative Design \\ \hline
73. \hyperref[sec:HEAQuery]{HEAQuery} & \github{https://github.com/staradutt/HEAquery/tree/main} & An intelligent search engine synthesizing responses from a harmonized database and literature corpus of high-entropy alloys. & RAG, Data Extraction \\ \hline
74. \hyperref[sec:PackSynth]{PackSynth} & \github{https://github.com/devanshu-777/PackSynth} & An agent system automating 3D structure generation, LAMMPS molecular dynamics setup, and RMSD simulation analysis. & Agentic Workflows, Simulations \\ \hline
75. \hyperref[sec:ExpAlign]{Standardizing Material Property Data} & \github{https://github.com/sptiwari/ExpAlign\_paper} & Automates the extraction and harmonization of CO2 polymer permeability data into structured, ML-ready datasets. & Data Extraction \\ \hline
76. \hyperref[sec:QSPHAgents]{QSPHAgents} & \github{https://github.com/sbanik2/QSPHAgents} & Predicts qualitative DOS features from structural descriptors using a multi-agent generator-critic reasoning loop. & Agentic Workflows, Property Prediction \\ \hline
77. \hyperref[sec:CrysGen]{Generative Modeling of Stable Inorganic Crystals} & \github{https://github.com/RishikeshMagar/GPT-OSS-MAT} & Utilizes fine-tuned GPT-OSS for the generative modeling and text-conditional generation of stable inorganic crystals. & Generative Design \\ \hline
78. \hyperref[sec:yolo-vs-mllm]{YOLO vs. Multimodal LLMs} & \github{https://github.com/abdulazizashy5/Alhaytham\_Eye} & Benchmarks 2D plot digitization accuracy and speed, comparing lightweight YOLO detectors against multimodal LLM reasoning. & Benchmarking, Data Extraction \\ \hline
79. \hyperref[sec:VERA]{VERA: AI-Powered Compliance Co-Pilot} & \github{https://github.com/puppalasaisrikar/VERA} & An AI-powered compliance co-pilot streamlining materials testing standards and validation workflows. & Agentic Workflows \\ \hline
80. \hyperref[sec:MaterEase]{MaterEase} & \github{https://github.com/Areeb2735/MaterEase-Your-AI-Powered-Materials-Science-Assistant} & Synthesizes early-stage materials literature and generates interactive visualization distributions using dual LLM reasoning. & RAG, Data Extraction \\ \hline
81. \hyperref[sec:MatSciAgent]{MatSciAgent} &  & A domain-aware coding agent integrating targeted literature/code retrieval to synthesize customized scientific Python solutions. & Agentic Workflows, Code Generation \\ \hline
82. \hyperref[sec:chemunity]{Implementing Knowledge Graphs as Long-term Memory} & \github{https://github.com/AI4ChemS/ChemUnityQA} & Explores using knowledge graphs as the persistent memory layer for autonomous scientific agents. & Knowledge Graphs, Agentic Workflows \\ \hline
83. \hyperref[sec:aesop]{AESOP: Accelerated Expert-in-loop Scientific Output Protocol} & \github{https://github.com/HallucinatingStrikeTeam/AESOP} & Uses a two-LLM expert-in-the-loop protocol to suggest methodologies and generate runnable computational chemistry scripts. & Agentic Workflows, Code Generation \\ \hline
84. \hyperref[sec:MatGen]{MATLAB-Integrated Generative Plugin} &  & Allows experimentalists to execute RDKit cheminformatics functions and property predictions via text directly inside MATLAB. & Agentic Workflows, Simulations \\ \hline
85. \hyperref[sec:explain-automation]{Explain that Automation} & \github{https://github.com/darianSmalley/explain-that-automation} & Uses multimodal RAG to interpret raw automated instrument logs and provide conversational laboratory session summaries. & Lab Automation, RAG \\ \hline
86. \hyperref[sec:AIssistant]{AIssistant} & \github{https://github.com/DIYANAPV/llm-hackathon-MS-and-Chem-2025/tree/main} & A human-AI collaborative framework intended to accelerate scientific discovery in atomic layer deposition. & Agentic Workflows \\ \hline
87. \hyperref[sec:mof-genie]{MOF-Genie} & \github{https://github.com/AntoRoyan/MOF-Genie} & Integrates graph neural networks and a Neo4j knowledge graph with a natural language interface for MOF hydrogen storage discovery. & Knowledge Graphs, Property Prediction \\ \hline
88. \hyperref[sec:sky]{SKY: An Agent for Materials Synthesis Planning} & \github{https://github.com/hspark1212/sky} & Automates synthesis planning by identifying structural analogs and grounding proposals in recursive nearest-neighbor retrieval. & Agentic Workflows, RAG \\ \hline
\end{longtable}

\section{ChemCodeBench: A Holistic Benchmark for Generating Efficient and Accurate PySCF Code}
\label{sec:chemcode-bench}



AI systems now exhibit strong code-generation abilities, and many problems in chemistry and materials science could be accelerated by AI-generated scientific code. However, most existing benchmarks emphasize passing synthetic unit tests rather than evaluating multidimensional outcomes such as code efficiency and chemical accuracy \cite{paul2024benchmarks}\cite{mirza2024large}. 

We therefore introduce a benchmark of PySCF density-functional theory (DFT) tasks that evaluates models on \emph{accuracy}, \emph{speed}, and \emph{correctness}. Writing efficient PySCF code is both a technical and a chemical challenge: models must select appropriate exchange–correlation functionals, basis sets, numerical grids, and other parameters to meet accuracy and runtime targets. A given accuracy threshold is set and a model is prompted to produce the fastest-running method that attains the set target.

The scope of this work focuses on total electronic ground-state energies for small molecules. As reference data, we use tabulated energies from NIST Computational Chemistry Comparison and Benchmark Database (CCCBDB); unless noted, our reference level corresponds to CCSD(T) with a cc-pVDZ basis.

\subsection*{Results}

\begin{figure}[h!]
    \centering
    \includegraphics[width=0.6\linewidth]{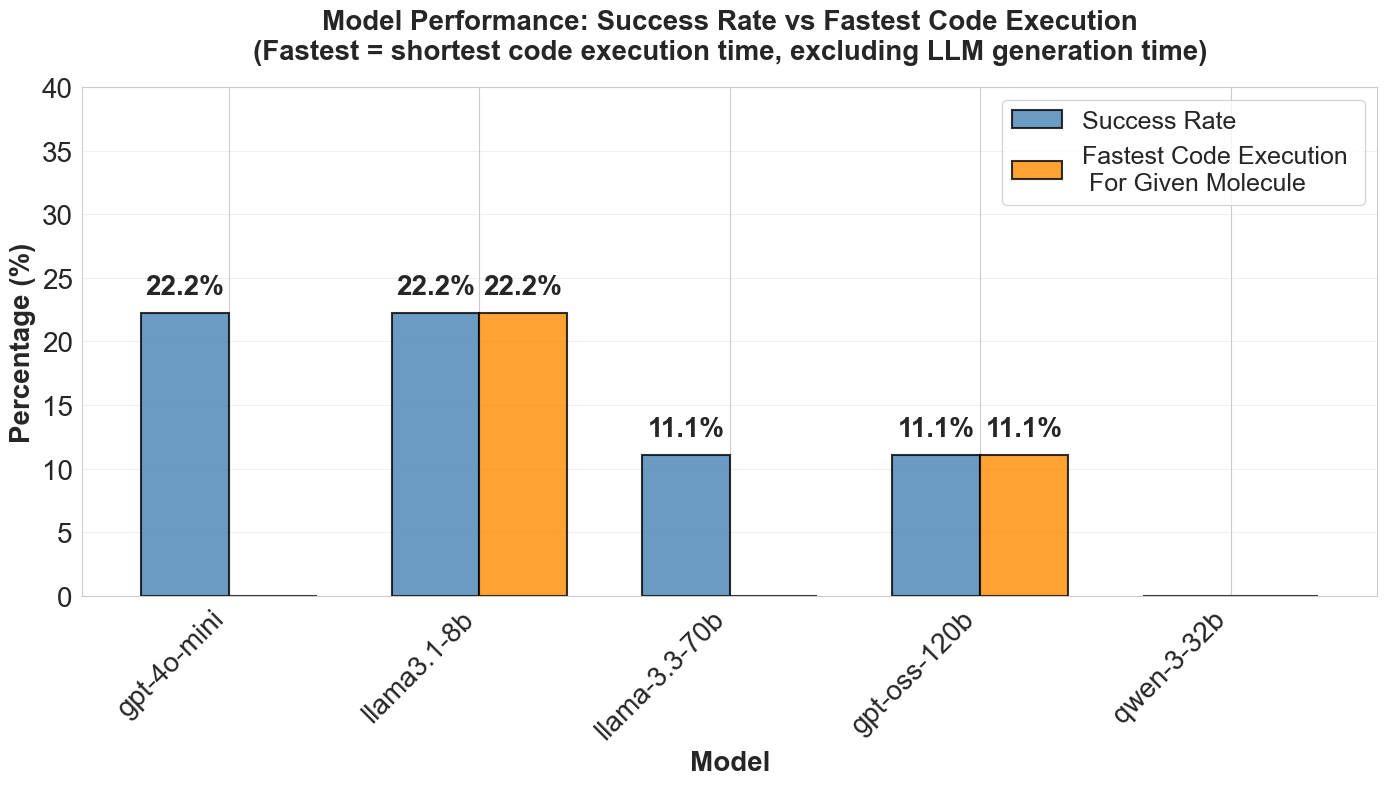}
    \caption{The blue bars represent the percentage of molecules for which the given AI model is able to generate PySCF code meeting the set accuracy standard. Orange represents the percentage of molecules for which the model generates the fastest-running code out of all models that meet set criteria (excluding LLM generation time).}
    \label{fig:time-to-threshold}
\end{figure}


Across the models evaluated, few were able to generate PySCF code that both runs and meets the accuracy criterion of $10^{-2}$ hartree. Llama 3.1-8B, rather than larger models such as Llama 3.3-70B, performed best overall at producing executable code and selecting DFT settings that satisfy the accuracy target while also being faster than other models in most cases. One possible explanation is that larger models tend to generate higher-fidelity (and often more computationally expensive) setups than this task requires. GPT-4o mini achieved the lowest mean execution time among successful runs while matching Llama 3.1-8B’s success rate under our criterion. Qwen-3-32B performed worst, failing to solve any problems despite its general-purpose coding ability. 
Overall performance is low given the elementary nature of the task: at best, models meet our standard on only $22\%$ of the molecules tested. These results suggest a need for chemistry-specific code training, as well as more holistic scientific coding benchmarks.



\subsection*{Future Work}
Next steps include broadening molecule diversity, including additional electronic-structure methods beyond DFT, and evaluating more models (including longer reasoning budgets). In addition, a worthwhile exploration would include expanding the benchmark to other problems in scientific computing like molecular dynamics.
\subsection*{Open-source Materials}
\textbf{Code and data:} \github{https://github.com/hansgundlach/ChemCodeBench}\, \textbf{Demo video:} \youtube{https://youtu.be/cLGeocfd8VM}



\section{UMADock: Docking molecules in protein binding sites using Meta's UMA MLIP}\label{sec:umadock}
The CafChem team wrote original docking code to dock molecules in protein binding sites using Meta's UMA MLIP as the energy scoring function. The code creates conformations of the input molecule with RDKit, then docks them in a pruned version of the protein crystal structures (defined as the location of a previously known binder and all residues and metals within 4 \AA. The interaction energy of each pose is evaluated with the UMA MLIP. The best pose for each conformer is then optimized with the MLIP, and the explicit desolvation and strain energies of the ligand are calculated. These additional terms are combined with the interaction energy for an overall electronic binding energy. The code then chooses the best overall, and performs rudimentary dynamics to examine stability. The code developed can be run from an AI agent, built using LangGraph. The Phi-4-mini-instruct model was used to drive the agent. 

\section*{Results}
\begin{figure}[h]
    \centering
    \includegraphics[width=0.6\linewidth]{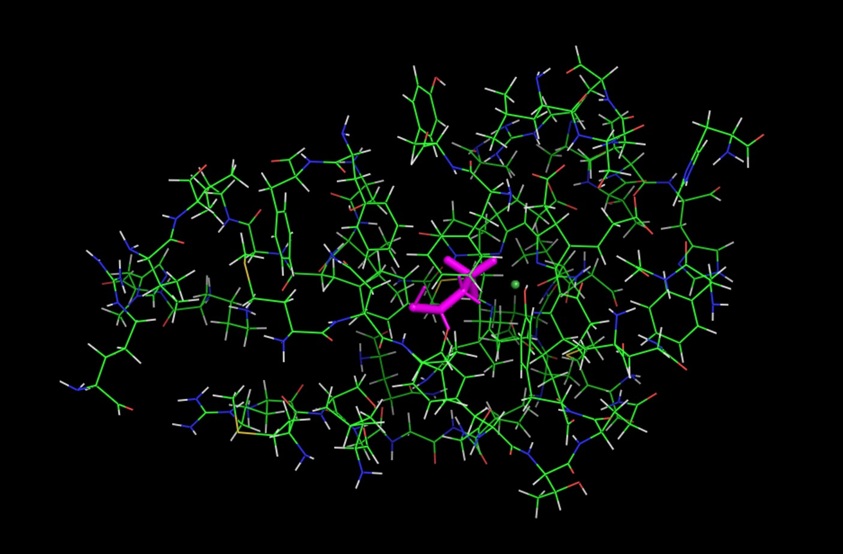}
    \caption{Ligand in the MR binding site; pose found using UMADock.}
    \label{fig:CC_UMADock}
\end{figure}

Docking poses and electronic binding energies were calculated for several ligands with the DRD2, MCR(see~\ref{fig:CC_UMADock}) and HMCGR proteins. These tests were conducted using 20 conformations of each ligand and 50-100 random poses for each conformation.  In all cases, UMADock was able to place ligands in realistic poses in the protein binding sites. The overall electronic binding energies calculated were in the range of ~-20 kcal/mol to ~+10 kcal/mol, roughly in line with values produced by conventional docking codes. 

\section*{Future Work}
In the future, exploration of expanded dynamics is of interest. This code takes the best overall pose and performs dynamics using the UMA MLIP to test for stability. The Agent may be expanded to perform other drug design tasks commonly performed before and after docking calculations. 

\subsection*{Open-source Materials}
Code available on GitHub: \github{https://github.com/MauricioCafiero/UMADock/tree/main}\,; Prototypes of the Agent can be found on \href{https://huggingface.co/cafierom}{HuggingFace}\,; Demo video: \youtube{https://www.linkedin.com/posts/mauricio-cafiero-5481259b_llm-uma-activity-7372341201200369664-mx6p?utm_source=share&utm_medium=member_ios&rcm=ACoAABUygsgB1ocjqc_UIBRFQJEZ3QMcuNRWSas}


\section{Machine learning supported by large language models for advanced functional materials}\label{sec:best_team}
\begin{figure}[t]
    \centering
    \captionsetup{justification=justified, singlelinecheck=false}
    \includegraphics[width=1\linewidth]{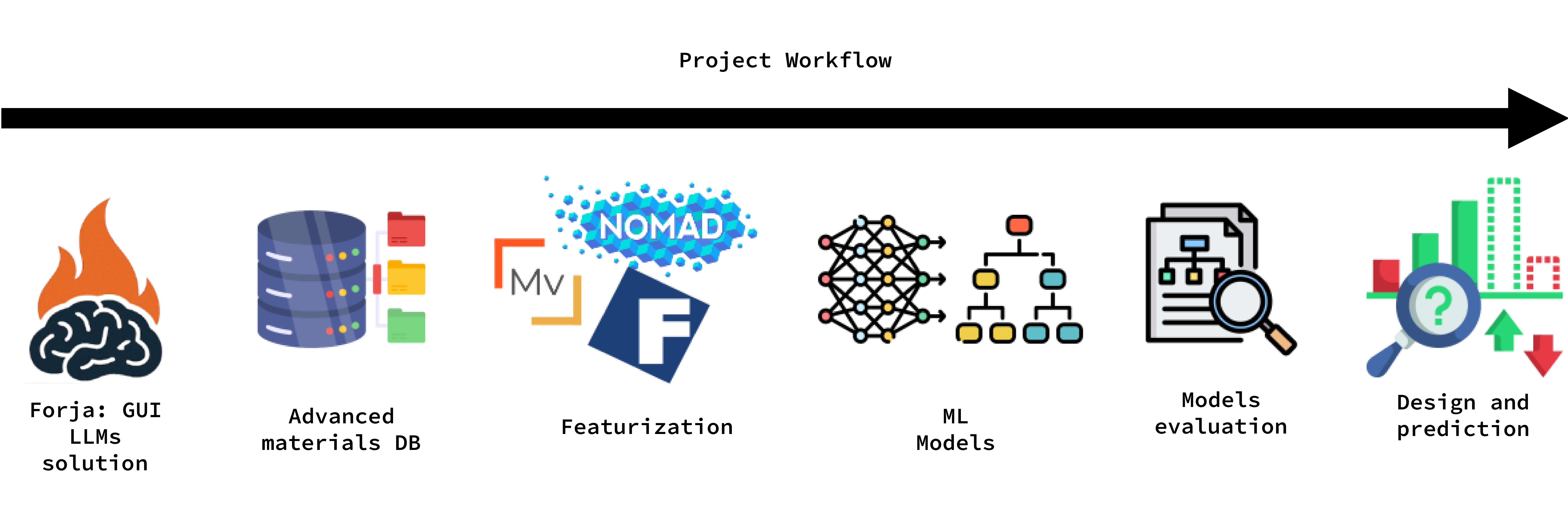}
    \caption{A schematic representation of the complete workflow from database generation using Forja de Mat\'{e}ria AI and ML Validation.}
    \label{fig:BT_Workflow}
\end{figure}

The development of AIML frameworks promises to accelerate discoveries in materials science, but progress is still limited by the lack of large, high-quality, and diverse datasets. This challenge is especially acute in high-entropy materials, where the design space is vast and available data are minimal~\cite{Zhang2023}. Retrieval-Augmented Generation (RAG)-based AI agents help overcome this bottleneck by using LLMs to extract and structure information from unstructured scientific literature. Following this approach, the team (Best Team) developed Forja de Mat\'{e}ria AI, a GUI tool for LLM-driven dataset generation and a workflow for training and generating machine-learning models for advanced materials research. A schematic of the workflow, starting from database generation using Forja de Mat\'{e}ria AI and its subsequent utilization and validation using Machine learning frameworks, is shown in Figure~\ref{fig:BT_Workflow}.
\subsection*{Material Database Generation and Initial Model Validation}
The robustness of Forja de Mat\'{e}ria AI was tested by generating a database on high-entropy oxides. 1{,}773 scientific article abstracts were retrieved from the \textit{Web of Science}~\cite{webofscience} using advanced search queries specifically designed to filter and collect studies focused on high-entropy oxides. Subsequently, two distinct models (both integrable into Forja de Mat\'{e}ria AI for small-scale executions) --- \textsf{mistral-large:123b-instruct-2407-q4\_0}~\cite{mistral_large_123b_instruct_2407_q4_0} and \textsf{gpt-oss:120b}~\cite{gpt_oss_120b} --- were employed to process the abstracts and extract pairs of compositions and crystal structures of high-entropy oxides. \\

\noindent In addition to high-entropy oxides, the same LLM-driven pipeline was extended to carbon-based materials, particularly those studied in adsorption and gas-separation experiments. Using RAG-enabled agents within Forja de Mat\'{e}ria AI, the models were prompted to extract adsorption capacities, textural properties (BET surface area, pore volume, pore-size distributions), synthesis routes, and adsorption conditions from unstructured literature. 
This also demonstrated the model's capability to generate structured datasets for multiple research areas.
Next, feature calculation was performed and the task was formulated as a classification problem, with the phase as the target variable and the features computed as composition-based vectors.
The Python library Mendeleev~\cite{mendeleev_python} and resources from the NOMAD Lab~\cite{nomad_lab} were used to derive atomic-level features, while FactSage~\cite{factsage} was employed to compute thermodynamic descriptors. The dataset generated using Mistral contained 757 samples, while the one produced with GPT comprised 369 samples. However, the data accuracy was higher in the GPT-generated dataset. A manual verification of 25 randomly selected samples from each dataset revealed accuracies of 96\% for GPT and 76\% for Mistral.
\subsection*{Feature Engineering and Machine Learning for Phase Prediction}
The generated dataset included diverse numerical descriptors of atomic and material properties. Feature engineering refined it by highlighting property variability, boundary conditions, and synthesis–property interactions. Three statistical metrics: range (max–min), coefficient of variation (std/mean), and min/max ratio captured the data distribution, scale-normalized variability, and homogeneity. Two binary indicators identified cases of zero standard deviation and zero minimum. Domain knowledge added nine synthesis–property interaction features linking specific synthesis methods to relevant thermodynamic, atomic-electronic, and physical-mechanical properties. The resulting \textit{engineered} dataset combined these additions with the remaining original features to create a more informative input for modeling.\\

\noindent Both the generated and engineered datasets were used to train Random Forest, XGBoost, CatBoost, Logistic Regression, SVM (RBF), ANN and CNN models. Data were split into train, validation, and test sets, preprocessed with imputation, encoding, and standard scaling. For efficiency, the evaluation of each ML model was established through a baseline training on the training set, followed by refitting on the combined train+validation data together in order to select optimal parameters for each model. Test performance showed macro accuracy between 88.7–94.4\% and F1-macro 49.4–61.5\% (up to 52.7\% for engineered data), emphasizing the imbalance between classes in the datasets. Among the ML models, Logistic Regression best described the generated dataset, while XGBoost favored the engineered one.

\subsection*{Future Work}
Future work of interest involves training of models more methodologically, increasing training computational resources, along with the interpretation of the best-performing models to study phase formation in high-entropy oxides and carbon-based materials for adsorption applications. The Forja de Mat\'{e}ria AI can also be extended to assist in the training and interpretation phase by extracting key parameters and explanations directly from the output. 

\subsection*{Open-source Materials}
All datasets, models and codes are available on GitHub: \github{https://github.com/Arthurns16/hacktoon2025}\,; Demo video: \youtube{https://youtu.be/wlbnUKJ7t1s}




\section{LLM-Engineered Prompt-Driven Discovery of Cuprate Superconductors}\label{sec:CuPirates}
Superconductivity at room temperature remains one of the major unsolved challenges in condensed matter physics. While cuprate materials exhibit superconductivity up to \SI{138}{\kelvin} at ambient pressure~\cite{dai_synthesis_1995}, the underlying mechanisms driving their high critical temperatures ($T_c$) remain poorly understood, limiting progress toward higher-$T_c$ systems. Notably, certain chemical and structural motifs are common to superconducting cuprates, and certain parameters may even correlate with superconducting behavior. \\

\noindent To explore the chemical and structural landscape of complex materials, crystal-structure generation tools can be integrated with machine-learning models trained on large structural databases. However, the relatively small number of known superconducting cuprates is in a data-scarce regime, limiting this direct approach. On the other hand, decades of intense interest in this material class have produced a vast scientific literature that richly documents the key structural motifs thought to be relevant for high-$T_c$ behavior. The CuPirates team explored whether LLMs can identify and extract these motifs from scientific literature on cuprate superconductors and further be used to generate new candidate materials via an existing text-guided diffusion model. 

\subsection*{Results}
\begin{figure}[H]
    \centering
    \includegraphics[width=5.5in]{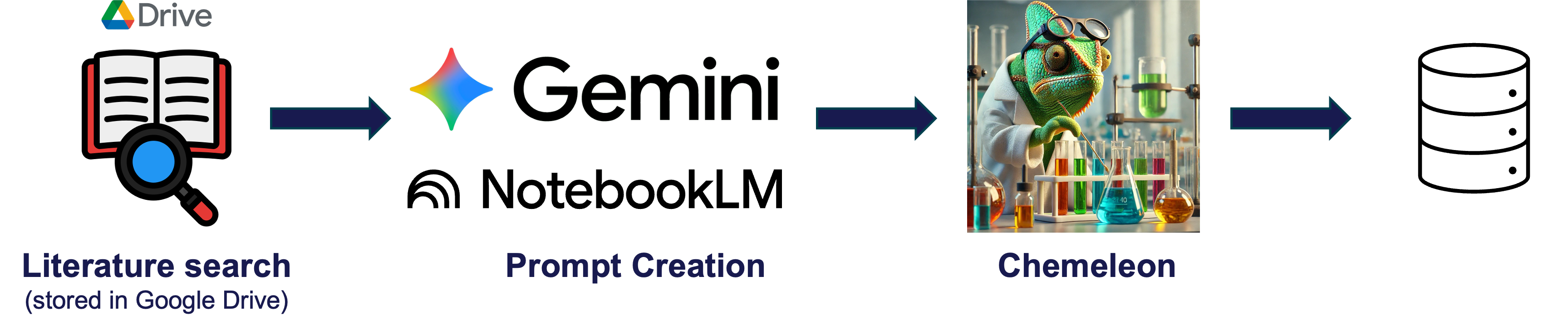}
    \caption{Schematic of the workflow used for the LLM-driven generation of new cuprate materials.}
    \label{fig:CuPirates_Workflow}
\end{figure}

Figure~\ref{fig:CuPirates_Workflow} summarises the workflow in four simple steps. First, a database of scientific literature (31 sources) describing cuprate superconducting structures was curated and used as the basis for this work. Utilizing the NotebookLM AI-powered research and writing tool, this database was then used to fine-tune an LLM, namely the Gemini~\cite{Gemini} multimodal model, to create targeted, chemist-informed prompts describing key structural features. The generated prompts were then input into a text-guided diffusion model for crystal structure generation, i.e. Chemeleon~\cite{ParkNC25_Chemeleon}, to propose new potential high-$T_c$ cuprate superconductors. The generated materials were finally screened manually for chemical feasibility.

To evaluate the differences between text prompts generated using the base and fine-tuned models, LLMs ChatGPT and Gemini were used to compare the outputs. These indicated that the non-fine-tuned model produces text outputs that are rich in formal crystallographic details, such as precise space groups and layer chemistries; they tend to be systematic, repetitive, and database-like. In contrast, the fine-tuned Gemini model generates more diverse and chemically insightful descriptions, emphasizing structural motifs, coordination environments, and prototype families, demonstrating greater flexibility that may be valuable for materials design. \\

\begin{figure}[H]
    \centering
    \includegraphics[width=0.7\linewidth]{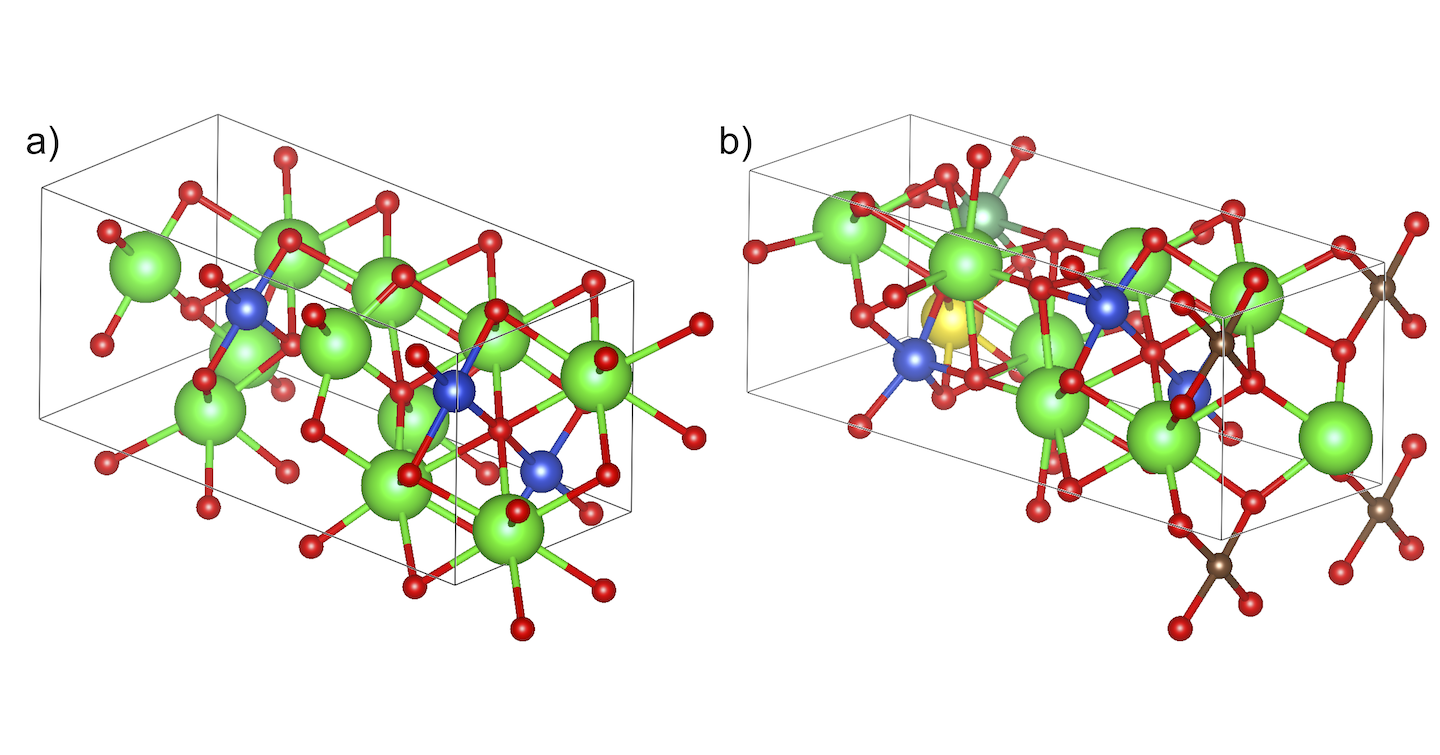}
    \caption{Crystal structures of two candidate cuprate superconductors generated by Chemeleon using the prompt ``Construct a monolayered cuprate structure featuring fourfold square-planar oxygen-coordinated copper atoms, intergrown with rock-salt \ce{SrO} and fluorite-type rare-earth oxide layers to prevent apical oxygen formation (\texttt{-{}-n-atoms 30 -{}-n-samples 100})''. Shown are (a) \ce{Sr_{11}Cu_{3}O_{16}} with square-planar \ce{Cu^{2+}} and (b) \ce{Sr_{8}PrNbCu_{3}CO_{16}} with mixed square-planar and square-pyramidal \ce{Cu^{2+}} coordination. Cu, O, Sr, C, Pr, and Nb ions are blue, red, green, brown, yellow, and teal, respectively.}
    \label{fig:CuPirates_crystalstruct}
\end{figure}

Two candidate materials, generated by Chemeleon using prompts produced by the fine-tuned LLM, are presented in Figure~\ref{fig:CuPirates_crystalstruct}. The copper--oxygen planes in both materials are arranged in distinctive sublattices, with squares of \ce{O^2-} anions and a \ce{Cu^2+} cation at the center of each square. This checkerboard pattern is typical of cuprate superconductors and highlights the promising nature of our results. While still early-stage, this approach demonstrates how LLMs can bridge scientific knowledge and computational materials design, opening new avenues for exploring complex quantum materials.

\subsection*{Future Work}
Future work will focus on automating the screening of generated structures, including computational checks of whether outputs satisfy the prompt-defined cuprate motifs and basic chemical feasibility. Scaling the workflow to larger sets of generated structures will enable systematic cross-checking against existing crystallographic databases to verify reproduction of known superconductors and identify any novel candidates. Promising materials identified through this pipeline can then be studied using advanced electronic structure and many-body methods to assess their stability and potential for superconducting behavior.

\subsection*{Open-source Materials}
Resources used are available on GitHub: \github{https://github.com/kinzani/Inzani-Group/tree/main/LLM_Hackathon_2025}\,; Demo video: \youtube{https://www.linkedin.com/feed/update/urn:li:activity:7372299525052448768/}





\section{A Neurosymbolic System for Unified Autonomous Scientific Hypothesis Generation}\label{sec:NeuroSymbolic}

Scientific progress depends on iterative hypothesis generation, experimental validation, and theory refinement, yet the exponential growth of published research has exceeded human capacity to integrate knowledge and identify new directions \cite{Fortunato2018}. Artificial intelligence, particularly large language models (LLMs), offers a potential solution by rapidly synthesizing vast scientific corpora \cite{brown2020language}. However, LLMs lack causal reasoning, physical constraint awareness, and logical consistency, often producing ``hallucinations''—plausible but unfounded statements—that limit their reliability in fields demanding falsifiable reasoning, such as materials science \cite{Bender2021}. Neurosymbolic AI, which combines neural learning with symbolic reasoning, has been proposed to address these issues \cite{Garcez2022}, yet current systems remain loosely coupled, separating inference from constraint logic.

To overcome these limitations, the NeuroSymbolic team introduces the \textit{Neurosymbolic Hypothesis Engine (NSHE)}, a fully integrated framework that combines LLMs, spiking neural networks, energy-based models, and symbolic knowledge retrieval within a constraint-sensitive architecture for hypothesis generation. Implemented in Python using PyTorch and Hugging Face Transformers, NSHE maintains symbolic grounding throughout all processing stages and is tailored for materials science applications, enabling robust, interpretable, and scientifically grounded hypothesis generation \cite{Butler2018}.

\subsection*{Results}

The NeuroSymbolic team demonstrates that the Neurosymbolic Hypothesis Engine (NSHE) unifies large language models (LLMs), spiking neural networks (SNNs), energy-based transformers (EBTs), and symbolic reasoning into a constraint-sensitive framework for scientific hypothesis generation. Built on Qwen2.5-7B, NSHE extracts structured concepts, models temporal relationships, and scores hypotheses by plausibility, novelty, and testability while maintaining symbolic grounding. When applied to materials science tasks, NSHE produced literature-grounded, experimentally testable hypotheses and outperformed six benchmark systems (quality: $0.75 \pm 0.04$, $+4$--$27\%$) with high confidence ($0.78$) and sub-second generation time. Expert reviewers rated outputs highly for plausibility ($4.2$), novelty ($3.9$), and testability ($4.0$).

Experimental validation of an NSHE-predicted Li--Si--O interlayer confirmed improved battery performance, showing an 80\% reduction in interfacial resistance and a $2\times$ increase in stability. Retrospective analysis further demonstrated NSHE’s predictive capability through rediscovery of later advances in perovskite materials. The Quantum-Evolutionary-Chaotic Hypothesis Synthesis (QE-CHS) module enhanced creative exploration, yielding higher novelty scores ($0.83 \pm 0.06$). Overall, NSHE demonstrates an integrated and interpretable AI-driven reasoning system capable of generating verifiable, cross-domain scientific insights.

\begin{figure}[h]
    \centering
    \includegraphics[width=0.35\linewidth]{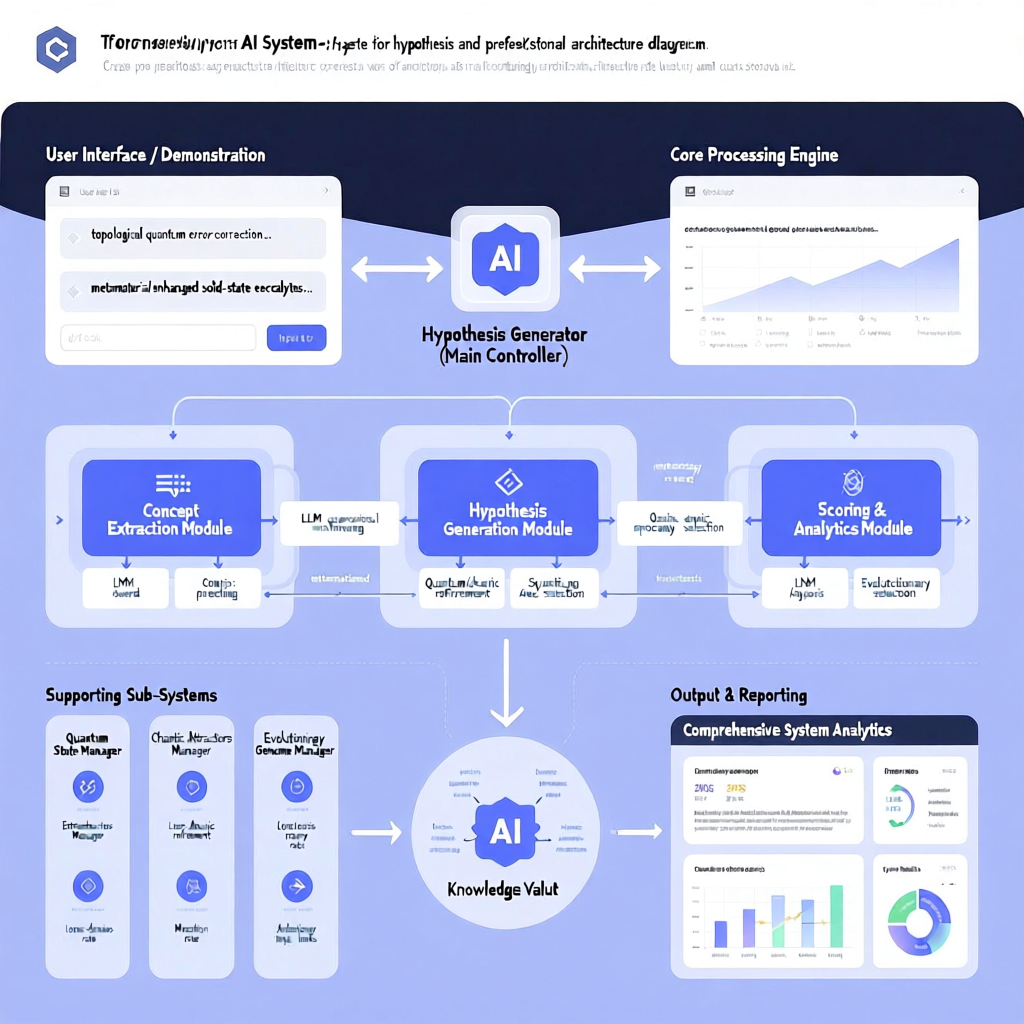}
    \caption{Workflow of the NeuroSymbolic hypothesis generation pipeline.}
    \label{fig:nshe}
\end{figure}

\begin{figure}[h]
    \centering
    \includegraphics[width=0.65\linewidth]{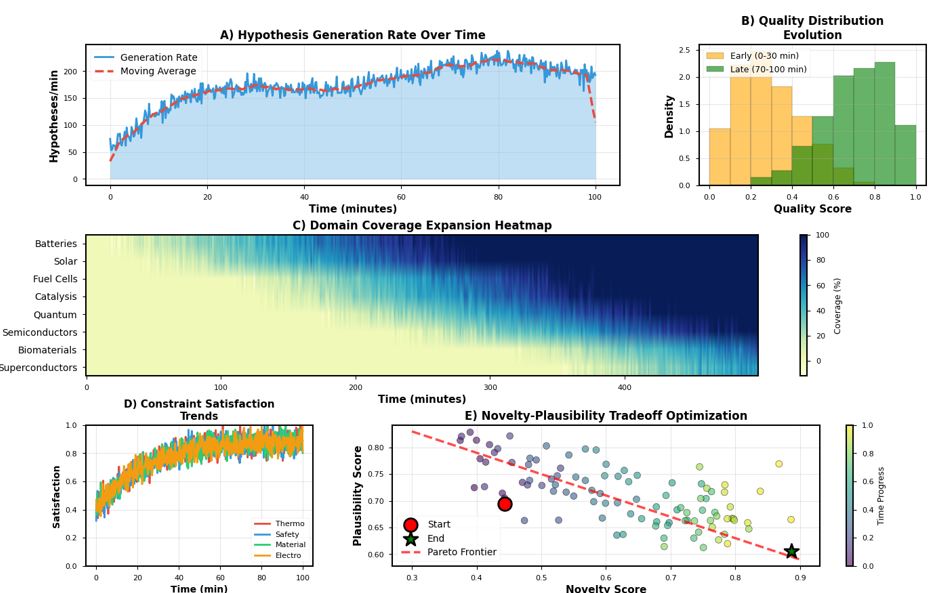}
    \caption{Real-time system analytics dashboard for NSHE.}
    \label{fig:nshe1}
\end{figure}

\subsection*{Future Work}

Future work will focus on expanding NSHE’s scientific utility along several key directions. An \textbf{experimental validation pipeline} will systematically test and refine generated hypotheses in collaboration with experimental laboratories. The \textbf{QE-CHS framework} will be extended with stronger quantitative grounding and theoretical rigor. Continued efforts toward \textbf{simplification and explainability} will provide intuitive interfaces and transparent reasoning tools for domain scientists.

In addition, \textbf{cross-domain adaptation} and \textbf{multi-modal integration} will extend NSHE to additional scientific fields and data types, linking simulations, experimental data, and literature. \textbf{Dynamic ontology learning} will enable continual knowledge updates as scientific understanding evolves, while coupling NSHE with \textbf{automated experimentation} will support closed-loop discovery workflows. Finally, \textbf{long-term impact assessment} will track NSHE’s real-world influence on scientific productivity and discovery. Together, these directions position NSHE as a rigorous, interpretable, and collaborative AI system for accelerating scientific research.

\subsection*{Open-source Materials}
Code and pretrained weights are available on GitHub: \github{https://github.com/dreamboat26/hackathon}\,; Video demo: \youtube{https://www.youtube.com/watch?v=lKxwF3fFfkM}





\section{Autonomous Reasoning Intelligence for Atomics: A Causal-Reasoning Framework for Materials Discovery}\label{sec:aria}

While large language models (LLMs) have demonstrated strong reasoning capabilities, their knowledge remains constrained by training data cutoffs and finite parametric capacity \cite{brown2020language}, often leading to factual inaccuracies and hallucinations that undermine reliability in rigorous scientific inquiry \cite{li-etal-2024-dawn}. Retrieval-augmented generation with knowledge graphs has emerged as a common solution, grounding LLMs in structured, domain-specific facts \cite{amayuelas2025groundingllmreasoningknowledge}. However, recent studies challenge the assumption that external augmentation consistently improves reasoning, showing that incomplete or poorly matched retrieval can degrade performance rather than enhance it \cite{yoran2024making}.

The ARIA team introduces \textbf{ARIA} (Autonomous Reasoning Intelligence for Atomics), a causal-reasoning framework designed to address the critical failure mode of \emph{contextual tunneling}, in which AI systems become constrained by narrow retrieved evidence. This limitation is particularly acute in materials discovery, where progress depends on multi-step causal reasoning across processing--structure--property relationships \cite{Butler2018}. ARIA explicitly models causality and selectively integrates external knowledge to preserve reasoning flexibility while maintaining scientific grounding.

\subsection*{Results}

\begin{figure}[h!]
\centering
\includegraphics[width=0.95\linewidth]{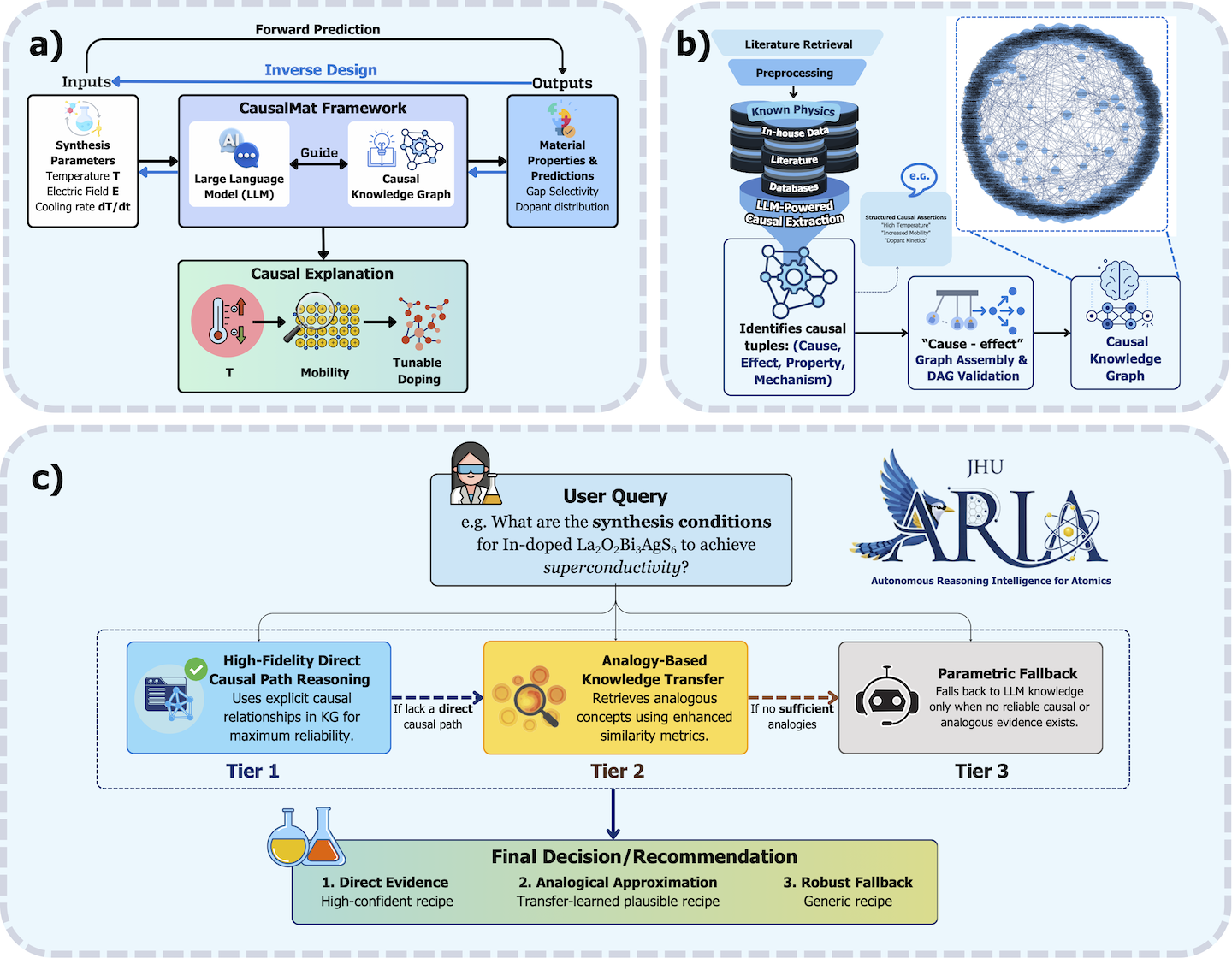}
\caption{\textbf{The ARIA framework for bidirectional materials reasoning.} (a) Bidirectional materials discovery enables both forward property prediction from synthesis parameters and inverse design of synthesis protocols from target properties, with complete causal traceability through the knowledge graph. (b) Automated knowledge graph construction extracts causal relationships from scientific literature using LLM-powered information extraction, producing structured synthesis--property repositories with provenance. (c) A three-tiered reasoning architecture: Tier~1 applies direct graph-constrained reasoning for established causal paths, Tier~2 performs analogy-based reasoning for novel materials, and Tier~3 provides parametric fallback when external evidence is unavailable.}
\end{figure}

The ARIA team designed ARIA around a three-tiered reasoning cascade that selectively leverages external evidence while preserving model adaptability. The system automatically constructs causal knowledge graphs from scientific literature using LLM-powered information extraction, generating structured repositories of synthesis--property relationships with full provenance traceability. Tier~1 performs direct graph-constrained reasoning for queries with well-established causal paths, Tier~2 enables analogy-based reasoning by extrapolating from related materials systems, and Tier~3 provides a parametric fallback when external evidence is insufficient.

Evaluation across 149 synthesis--property relationships spanning 85 distinct materials systems shows that naive knowledge integration degrades baseline LLM performance by 20--35\%. In contrast, ARIA not only recovers this loss but also delivers complete mechanistic transparency. For inverse design tasks involving complex semiconductors such as In-doped \ce{La2O2Bi3AgS6}, ARIA generated comprehensive synthesis protocols specifying temperature ranges (800--1200$^\circ$C), controlled atmospheres, and mechanistic justification for each design choice. These results demonstrate ARIA’s ability to support both forward property prediction and inverse synthesis design within the processing--structure--property paradigm.

\subsection*{Future Work}

Future work will expand ARIA’s knowledge graph to encompass broader classes of materials and integrate real-time experimental feedback loops to enable continual learning. The ARIA team also plans to develop domain-specific reasoning modules tailored to applications such as energy storage and catalysis, further strengthening causal interpretability and practical utility.

\subsection*{Open-Source Materials}

Code and pretrained weights are publicly available on GitHub: \github{https://github.com/yicao-elina/LLM4Chem-Explainable-synthesis.git}

\section{LARA-HPC: A Language Model Powered Research Assistant for High Performance Computing}\label{sec:lara-hpc}

Preparing and validating scientific workflows for execution on high-performance computing (HPC) resources is a time-consuming task that requires deep technical expertise. While large language models (LLMs) offer a promising interface for automating such workflows, scientific computing environments are dominated by bespoke resources and laboratory-specific Python libraries that are often sparsely documented, rapidly evolving, or partially confidential. Because these libraries cannot be embedded directly into LLM pre-training corpora, naïve prompting risks the \textbf{catastrophic wasting of computational resources} through incorrect or unsafe code execution.

The LARA-HPC team introduces \textbf{LARA-HPC}, a language-model-powered research assistant designed to safely bridge LLMs with laboratory-level software stacks and HPC execution environments. LARA-HPC combines local retrieval-augmented generation (RAG) with rigorous validation and human oversight to expose local knowledge while enforcing correctness prior to execution. By integrating documentation-aware code generation with automated verification and controlled HPC submission, LARA-HPC improves accessibility, reliability, and enables new research workflows in which LLMs collaborate safely with specialized scientific codebases.

\subsection*{Results}

\begin{figure}[h!]
\centering
\includegraphics[width=0.73\linewidth]{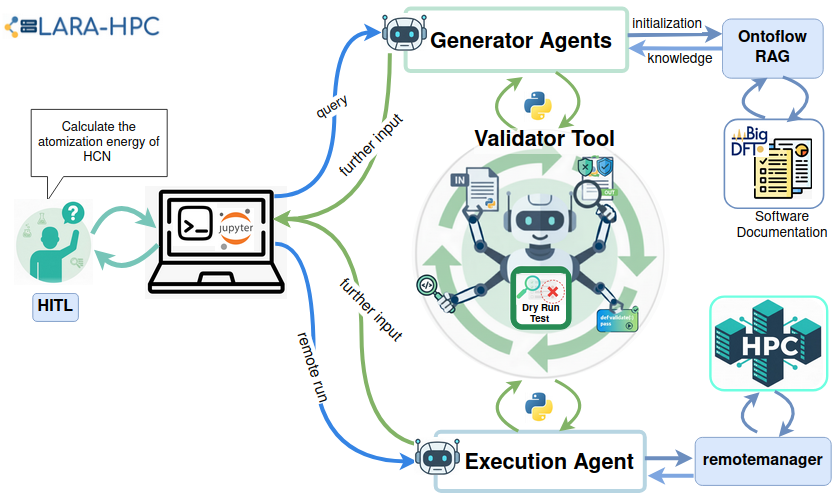}
\caption{
\textbf{The LARA-HPC architecture.}
Workflows are orchestrated by a set of autonomous agents and incorporate a \textbf{human-in-the-loop (HITL)} component to ensure scientific validity and computational safety. A \textbf{Generator Agent} receives a natural language scientific query (e.g., ``Calculate the atomization energy of HCN'') and, through the \textbf{Ontoflow RAG} pipeline, retrieves domain knowledge from software documentation such as \textbf{PyBigDFT} to synthesize a candidate Python script. The \textbf{Validator Tool} performs multi-layer verification, including syntactic checks, scientific consistency, and execution feasibility, supported by a low-cost \textbf{dry-run test} in a local simulation environment. When required, HITL intervention guides correction and interpretation. Upon successful validation and explicit human approval, the workflow is submitted via \textbf{remotemanager} to the target HPC system for execution and monitoring. Failed validations trigger a feedback loop back to the Generator Agent, forming a closed, self-correcting workflow.}
\label{fig:lara_hpc_architecture}
\end{figure}

The LARA-HPC team demonstrates a unified framework that integrates domain-specific scientific libraries and automated reasoning tools into a single assistant. Figure~\ref{fig:lara_hpc_architecture} illustrates the system in the context of electronic structure calculations, a domain that places particularly high demands on HPC resources \cite{Gavini2023}. The framework leverages two key components developed within the team: \textbf{PyBigDFT} \cite{ratcliff2020, pybigdft} for electronic structure simulations and \textbf{remotemanager} \cite{Dawson2024, remotemanager} for job submission, execution, and monitoring on HPC systems.

LARA-HPC employs a retrieval-augmented generation pipeline that dynamically queries PyBigDFT documentation to synthesize executable Python functions from natural language requests. Generated code is passed to a dedicated validation tool that performs multi-layer checks of syntactic correctness, scientific consistency, and execution feasibility before remote submission via remotemanager.

As a representative use case, the LARA-HPC team tasked the system with computing the atomization energy of a molecule. The LLM-generated workflow was iteratively corrected, validated, and executed on a remote ``virtual'' HPC system implemented as a locally networked container. Successful execution was confirmed through CPU monitoring and job logs. This case study highlights how \textbf{documentation quality directly impacts the reliability of LLM-driven HPC workflows} and provides concrete guidance for making scientific software more ``LLM-friendly.'' Crucially, LARA-HPC supports laboratory-specific resources—such as PyBigDFT tutorials—that are absent from public LLM training data but essential for accurate electronic-structure workflows on HPC architectures.

The broader impact of this work is significant. By lowering the barrier to entry for researchers who lack extensive scripting or HPC expertise, LARA-HPC demonstrates a practical path toward democratizing access to large-scale computational resources. Automated workflow generation and validation substantially reduce the time and effort required to design and execute complex simulations, accelerating the pace of computational materials research.

\subsection*{Future Work}

Future work will focus on making LARA-HPC fully LLM-agnostic and exploring alternative multi-agent orchestration strategies. The LARA-HPC team also plans systematic benchmarking across diverse documentation styles and scientific software ecosystems. These efforts aim to establish concrete guidelines for designing AI-interpretable scientific codes, documentation, and tutorials, ultimately improving the reliability, safety, and accessibility of LLM-assisted HPC workflows.

\section{An Autonomous LLM Agent Workflow for Automated NMR Mixture Analysis and Property Prediction}\label{sec:mixsense}

Automating reaction analysis remains a central challenge in chemical research due to the time, cost, and product losses associated with sample preparation and manual spectral interpretation. The MixSense team introduces \emph{MixSense}, an autonomous large language model (LLM) agent that integrates product prediction, spectral deconvolution, quantitative analysis, and property estimation into a single end-to-end workflow. Given only a \ce{^1H}-NMR spectrum, MixSense predicts plausible reaction products, deconvolves mixture spectra using simulated and reference libraries, and quantifies yields and purity of the resulting components.

Extending beyond structural analysis, the framework incorporates SMILES-based prediction models for chemosensory properties, including compound-level flavor and odor profiles. This enables simultaneous analysis of chemical composition and estimation of functional properties directly from raw spectral data. By eliminating extensive purification and manual interpretation, MixSense aims to accelerate reaction characterization while reducing material losses and experimental overhead. Preliminary results demonstrate the feasibility of LLM-driven, end-to-end analysis of chemical mixtures, highlighting the potential of autonomous agents to transform analytical chemistry and chemical discovery workflows.

\subsection*{Results}

The MixSense team designed MixSense as an LLM-orchestrated pipeline integrating three core capabilities. First, given user-specified reactants and an NMR spectrum, a forward-reaction module proposes plausible products and side products present in the measured mixture. To accomplish this, the agent retrieves or simulates pure-component \ce{^1H}-NMR references from public repositories or predictive models and matches them to experimentally observed peaks. Second, users may supply multiple NMR spectra for time-resolved analysis, enabling quantification of reactant and product evolution over time. For each hypothesized product set, the agent deconvolves crude spectra by fitting a non-negative linear combination of candidate spectra with alignment and baseline correction, using integrated peak areas to track temporal trends. Third, MixSense estimates chemosensory descriptors with associated confidence metrics directly from SMILES representations using a lightweight property predictor.

This workflow builds on prior chemoinformatic tools adapted to operate as interactive agents \cite{2025zimmermann, domzal2024magnetstein, kuhn2024twenty, wang2025nmrextractor, sagawa2025reactiont5}, enabling seamless integration of structure elucidation, mixture quantification, and property estimation.

\begin{figure}[H]
    \centering
    \includegraphics[width=0.7\linewidth]{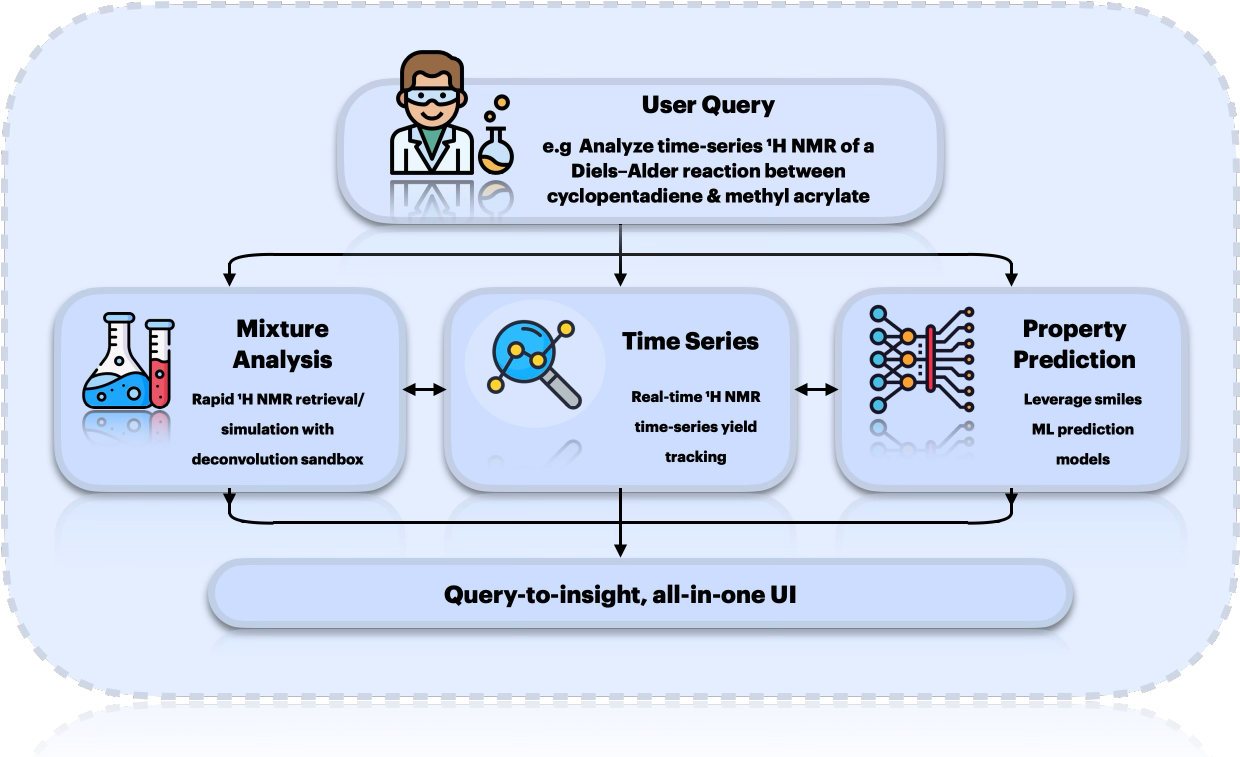}
    \caption{\textbf{Overview of \emph{MixSense}.} The system ingests \ce{^1H}-NMR spectra and routes them to task-specific modules for product hypothesis generation, spectral deconvolution, time-series quantification, and SMILES-based chemosensory property prediction.}
    \label{fig:mixsense_workflow}
\end{figure}

\noindent\textbf{Demo and Interface.} The MixSense team demonstrated all three capabilities through an interactive Gradio interface \cite{gradio}: mixture identification and quantification, time-series analysis, and property prediction. A DeepSeek model (V3:together) \cite{deepseek} was used for query parsing and natural-language output generation. For mixture identification, the interface was prompted with a query such as ``Analyze my anisole reaction,'' along with a \ce{^1H}-NMR spectrum from the chlorination of anisole. The system predicted plausible components, estimated mole fractions, and retrieved reference spectra for individual species from a spectral database.

For time-series analysis, the query ``Track the conversion of anisole during conversion to p-bromoanisole'' was paired with multiple CSV files acquired during bromination. MixSense quantified component evolution over time and produced an accompanying statistical summary. To demonstrate property prediction, SMILES strings for ethanol and sucrose were provided as input. The system returned predicted flavor descriptors with confidence scores and associated probabilities across multiple sensory classes, including bitter, sour, sweet, umami, and undefined.

\subsection*{Future Work}

Future work will expand spectral coverage to include two-dimensional NMR techniques and extend the framework to additional analytical modalities such as IR and Raman spectroscopy. These developments aim to further generalize MixSense as a unified, multimodal platform for autonomous reaction analysis.

\subsection*{Open-Source Materials}

The MixSense team provides an open-source repository containing the agent, interactive application, spectral deconvolution tools, chemosensory property predictors, NMR datasets, and example mixtures:  
\github{https://github.com/jdsanc/MixSense.git}.  
A demonstration video is available at: \youtube{https://drive.google.com/file/d/1qLt48Yh7qlWbZ0pMRnlUzLZP4Wmqsjaw/view}.






\section{AgentLearn: Accelerating Active Learning with LLM Agents}\label{sec:agentlearn}

Machine learning has become a powerful driver of scientific discovery, accelerating progress across a wide range of domains. However, training high-performing models typically requires large, accurate, and diverse datasets, which are often expensive and time-consuming to acquire \cite{8862913}. The AgentLearn team introduces \textbf{AgentLearn}, a framework that embeds a large language model (LLM) agent within an enhanced active learning loop. The system is designed to intelligently expand datasets, continuously improve a target model, and adapt its data-generation strategy based on performance feedback.

\subsection*{Results}

The AgentLearn team designed AgentLearn as a flexible framework that can be integrated with a wide range of predictive models with minimal modification. To maximize compatibility with external tools for knowledge retrieval, validation, and experiment automation, the framework adopts Model Context Protocols (MCP), providing the LLM agent with a unified and automated interface to downstream resources.

\begin{figure}[H]
    \centering
    \includegraphics[width=0.75\linewidth]{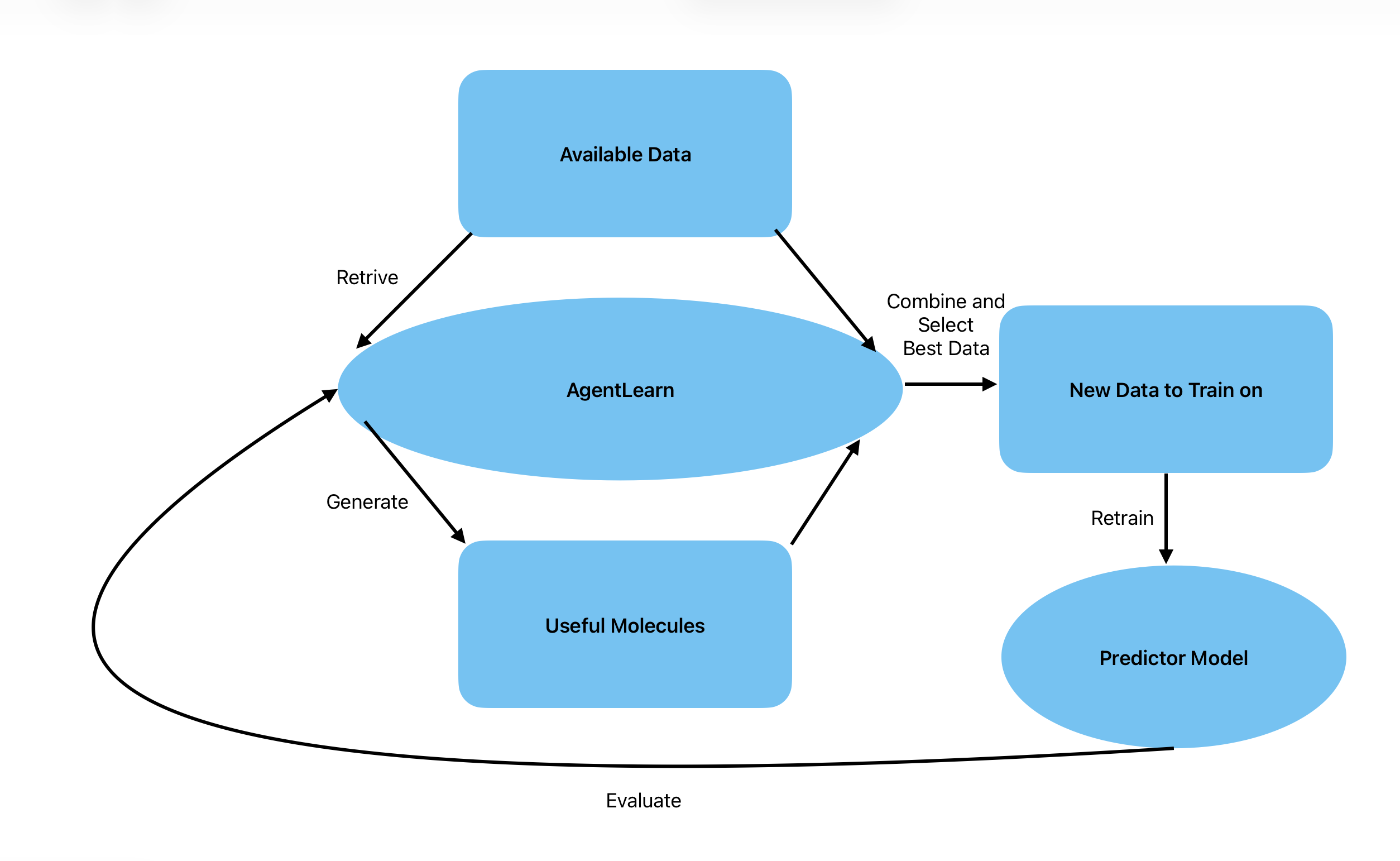}
    \caption{The AgentLearn workflow for LLM-driven active learning and iterative dataset expansion.}
    \label{fig:AL-diagram}
\end{figure}

As illustrated in Figure~\ref{fig:AL-diagram}, the AgentLearn workflow begins by retrieving knowledge from available sources, including existing datasets, structured knowledge bases, or internet search results. Using this information, the agent generates new candidate molecules deemed informative for improving the target model. These generated molecules are combined with the existing dataset to form an expanded training set. The predictor model is then retrained on this augmented data, evaluated, and the resulting performance metrics are fed back to the agent. This feedback loop guides subsequent molecule generation, enabling iterative and adaptive dataset expansion.

To demonstrate the effectiveness of the framework, the AgentLearn team developed a model for predicting molecular aromaticity and used AgentLearn to iteratively improve its performance. Experiments were conducted using both Gemini-2.5-Flash and Qwen-3-235B as foundation models. In all cases, AgentLearn successfully integrated with the training loop and autonomously proposed new molecules that contributed to continued model improvement.

\subsection*{Future Work}

Future work will integrate computational chemistry methods to calculate properties of generated molecules, enabling access to effectively unbounded data pools for active learning. The AgentLearn team also plans to explore integration with autonomous laboratory platforms, further closing the loop between model-driven hypothesis generation, data acquisition, and experimental validation.

\subsection*{Open-Source Materials}

The AgentLearn team provides open-source access to the framework at:  
\github{https://github.com/tonylifepix/AgentLearn}.  
A demonstration video is available at: \youtube{https://youtu.be/gkuUykswBHg}.





\section{The Fullerene Factory: A Multi-Agent Workflow for Functionalized Fullerene Modeling}\label{sec:fullerene_factory}

Fullerenes exhibit unique cage-like structures and electron-deficient character that make their functionalization an active area of research with promising applications in medicine and energy storage \cite{guirado2006structural, zhang2025functionalized}. However, the vast design space arising from multiple equivalent binding sites and sequential functionalization generates thousands of possible regioisomers, creating substantial experimental and computational challenges. The Fullerene Factory team introduces \textbf{The Fullerene Factory}, a multi-agent automated workflow that efficiently identifies promising functionalized fullerene isomers directly from natural language input.

\subsection*{Results}

The Fullerene Factory enables users to generate functionalized fullerene structures with specified addends using simple natural language queries. The system returns the most stable optimized geometries as \textit{.xyz} files using the UMA machine learning interatomic potential (MLIP) developed by Meta FAIR \cite{wood2025umafamilyuniversalmodels}, with optional storage as Hugging Face datasets for downstream use.

\begin{figure}[t]
    \centering
    \includegraphics[width=0.95\linewidth]{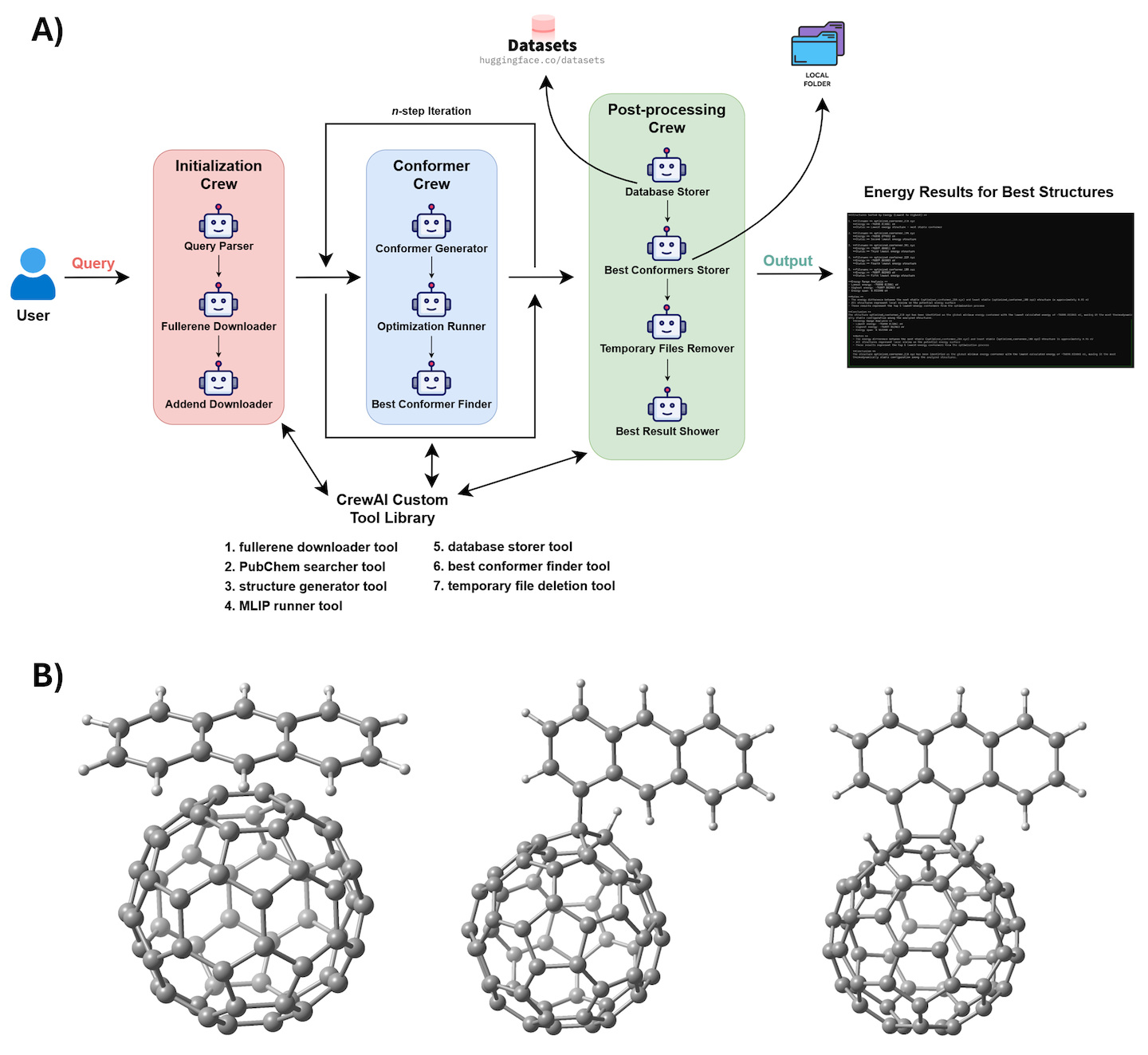}
    \caption{(A) Overall workflow of \textbf{The Fullerene Factory} multi-agent framework. (B) Examples of MLIP-optimized functionalized fullerene structures generated by the system.}
    \label{fig:fullerene_factory}
\end{figure}

The Fullerene Factory team designed the framework as a coordinated multi-agent workflow composed of three distinct CrewAI \cite{moura2023crewai} crews: the \textit{Initialization Crew} (pink zone), \textit{Conformer Crew} (blue zone), and \textit{Post-processing Crew} (green zone), as shown in Figure~\ref{fig:fullerene_factory}A. The Initialization and Conformer Crews each consist of three agents, while the Post-processing Crew contains four agents. All agents leverage a shared custom tool library that includes PubChem search utilities, structure generation modules, MLIP execution tools, and auxiliary analysis functions. The Conformer Crew operates iteratively, adapting to the number of sequential functionalization steps specified by the user.

Figure~\ref{fig:fullerene_factory}B illustrates the three most thermodynamically stable structures generated from the following example query, executed using the DeepSeek-V3.2-Exp model \cite{deepseek} across all agents:

\begin{tcolorbox}[title=\textbf{Example Query}]
Generate a C60 fullerene structure with the addend as anthracene. Get single-step addition products where the total number of angles to make conformers is 15. Also, store all the optimized structures in the database. Get the 5 best structures and return the energy report as text data.
\end{tcolorbox}

\subsection*{Future Work}

Future work will extend the framework to a broader range of reaction chains, including both controlled and uncontrolled functionalization pathways. The Fullerene Factory team also plans to integrate density functional theory (DFT) screening following initial MLIP-based optimization to improve energetic accuracy. In addition, a fine-tuned model trained on the curated dataset generated by the workflow will enable \textit{chemical vibe coding} for fullerene chemistry, allowing users to specify desired chemical outcomes through concise natural language queries.

\subsection*{Open-Source Materials}

Source code and documentation are available at:  
\github{https://github.com/aritraroy24/the-fullerene-factory}.  
The generated Hugging Face dataset is available at:  
\database{https://huggingface.co/datasets/aritraroy24/optimized_conformers_dataset}.  
A demonstration video is available at:  
\youtube{https://youtu.be/-9HRh9dXvWc}.





\section{Reasoning Model Interpretation and Knowledge Extraction for Chemistry Problems}\label{sec:fordham}

Reasoning models such as DeepSeek-R1, Gemini-2.5-Pro, and Ether0 are emerging, and have demonstrated strong problem-solving capabilities by generating step-by-step reasoning traces.~\cite{Guo2025-kn,narayanan2025trainingscientificreasoningmodel} However, their reasoning traces, especially how models organize and apply chemical knowledge, remain underexplored.

Inspired by Thought Anchors,~\cite{bogdan2025thoughtanchorsllmreasoning} the Fordham Reasoning Team began exploring reasoning-trace knowledge extraction and interpretability in LLMs for chemistry problems, which is a new direction toward understanding their inner workings. This study represents an early step toward LLM interpretability. By analyzing reasoning traces as structured knowledge graphs, the Fordham Reasoning Team aims to uncover which reasoning steps carry the greatest influence and how different models differ in reasoning structure and their inner knowledge.

\subsection*{Results}

The Fordham Reasoning Team analyzed reasoning behavior across three reasoning models including Ether0, DeepSeek-R1, and Gemini-2.5-Pro, on the dataset introduced by the Ether0 paper, which contains chemistry-related multiple-choice problems. From this dataset, the team selected 75 problems specifically related to safety, LD-50, and scent. An example problem asks:
\begin{quote}
\textbf{Example (Problem \#67).} From these molecules, identify the one most likely to possess a nutty scent:
\begin{verbatim}
C(CCCC(OCCC)=O)CC
O=C(CCCCC(=O)OCCC)OCCC
O(C(=O)CCCCCCCCCCCCCCC)CCCC
C(C(OCCC)=O)CCCC
C(CCCCC)CCC(=O)OCCCCCC
\end{verbatim}
\end{quote}

\begin{figure}[h]
    \centering
    \includegraphics[width=0.75\linewidth]{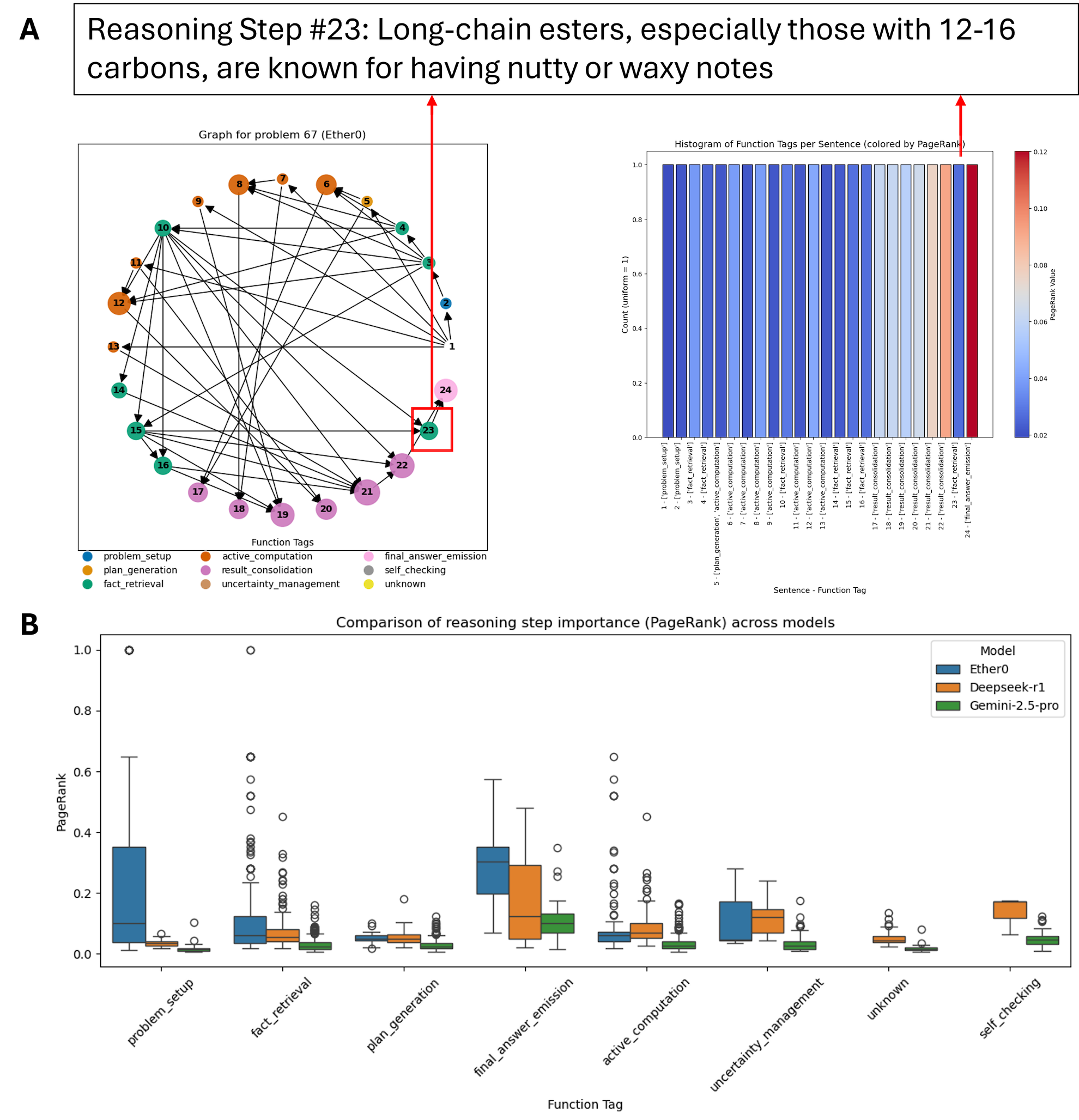}
    \caption{(A) Closeness centrality circular graph (Left) and PageRank centrality analysis (Right) of Ether0-generated reasoning step for problem \#67 in the dataset. The important fact retrieval reasoning step is extracted, as shown on top of the figure. (B) Comparison of reasoning step importance across Ether0, DeepSeek-R1, and Gemini-2.5-Pro models.}
    \label{fig:llm-reasoning-interp}
\end{figure}

Ether0 is a 24B parameter model fine-tuned with supervised finetuning (SFT) and group relative policy optimization (GRPO) from Mistral-Small-24B-Instruct-2501, pretrained on DeepSeek-R1, and further refined with Gemini-2.5-Flash. The model was deployed on NERSC’s Perlmutter HPC using four A100 GPUs. DeepSeek-R1 was accessed through OpenRouter. Gemini-2.5-Pro was evaluated via the Gemini API. Ether0 solved 25, DeepSeek-R1 24, and Gemini-2.5-Pro 12 of the 75 problems. Using GPT-5, the Fordham Reasoning Team parsed the reasoning outputs, classified sentences into functional categories, and identified dependencies among them to reveal structural relationships. Each reasoning trace was then represented as a directed graph, where nodes correspond to reasoning steps and edges capture logical dependencies. Graph-based methods were applied to visualize reasoning relationships. By combining sentence functions with closeness and PageRank centrality, key knowledge was extracted, often in the “fact retrieval” category with the highest closeness or PageRank centrality (Figure~\ref{fig:llm-reasoning-interp}A).

The Fordham Reasoning Team then performed cross-model comparison. Gemini-2.5-Pro produced the longest reasoning traces, while Ether0 generated the shortest and most concise ones. The functional category distributions revealed domain-specific reasoning: chemistry problems rely heavily on knowledge recall, unlike mathematical reasoning which emphasizes computation. Box-plot (Figure ~\ref{fig:llm-reasoning-interp}B) comparisons further showed that reasoning-step importance varied systematically among models. Ether0 and DeepSeek-R1 concentrated importance on a few key steps, while Gemini distributed it more evenly across the trace.

\subsection*{Future Work}
Future work can be focused on introducing chemistry knowledge judging and developing more efficient reasoning-trace metrics to quantify reasoning quality. Another promising direction is to locate where chemistry knowledge is represented within model structures. Such insights could guide the development of advanced reasoning LLMs.

\subsection*{Open-source Materials}
Code available on GitHub: \github{https://github.com/BaosenZ/mach-interp-LLM-hackathon-2025.git}\,; Demo video: \youtube{https://youtu.be/tW3ArrFU4_Y}\,;




\section{A Context-Aware Chemistry Research Assistant}\label{sec:chembot}

ChemBot is an AI-assisted chemistry research tool designed to help students and researchers quickly extract, retrieve, and understand information from scientific literature. The goal of the project is to build a context-aware assistant that supports chemistry learning and research by combining Retrieval-Augmented Generation (RAG) with BioGPT, enabling users to ask chemistry-related questions and receive grounded, literature-supported answers.

\subsection*{Architecture}
ChemBot is implemented using a lightweight Retrieval-Augmented Generation (RAG) pipeline designed to
provide grounded, literature-supported answers to chemistry questions. Research papers in PDF format are
first loaded and converted to raw text, which is then cleaned and split into overlapping chunks using a
recursive text splitter to preserve scientific coherence across sections.

Each chunk is transformed into a vector representation using the \texttt{sentence-transformers/all-MiniLM-L6-v2}
embedding model and stored in a FAISS vector database, enabling efficient semantic search over the
collection of scientific documents. This allows ChemBot to retrieve the most contextually relevant excerpts
for any given user query. BioGPT (\texttt{microsoft/biogpt}) then generates a domain-aware response that is grounded in the
retrieved material, improving factual accuracy and reducing hallucinations.

\begin{figure}[H]
    \centering
    \includegraphics[width=0.2\linewidth]{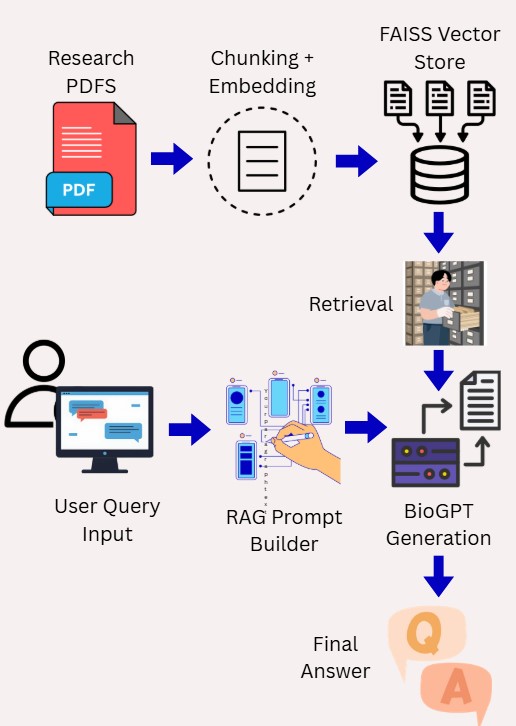}
    \caption{High-level conceptual architecture of the ChemBot system, showing the sequential stages from document processing and vector indexing to retrieval-augmented reasoning using BioGPT.}
    \label{fig:chembot-architecture}
\end{figure}

Figure~\ref{fig:chembot-architecture} illustrates the high-level conceptual flow of the ChemBot system, from document ingestion to vector retrieval and BioGPT-driven reasoning.

Prompt engineering played a critical role in reliability. Constraining prompts, adding domain-specific instructions, and iteratively refining context reduced hallucination rates and improved factual consistency across chemistry subdomains.

\subsection*{Results}
ChemBot successfully retrieves chemistry-specific information from research papers and generates context-aware answers using BioGPT. By using a sample of 50 Q\& A's, the accuracy was found to be 67\%. The RAG setup significantly reduces hallucinations compared to using BioGPT alone, and the system performs well on tasks such as summarizing reactions, explaining material properties, and clarifying scientific concepts. User testing during the hackathon showed that ChemBot provides faster and more reliable explanations than manual document search, making it useful for both students and researchers. Below are graphs displaying the accuracy of the ChemBot LLM.

\begin{figure}[H]
    \centering
    \includegraphics[width=0.6\linewidth]{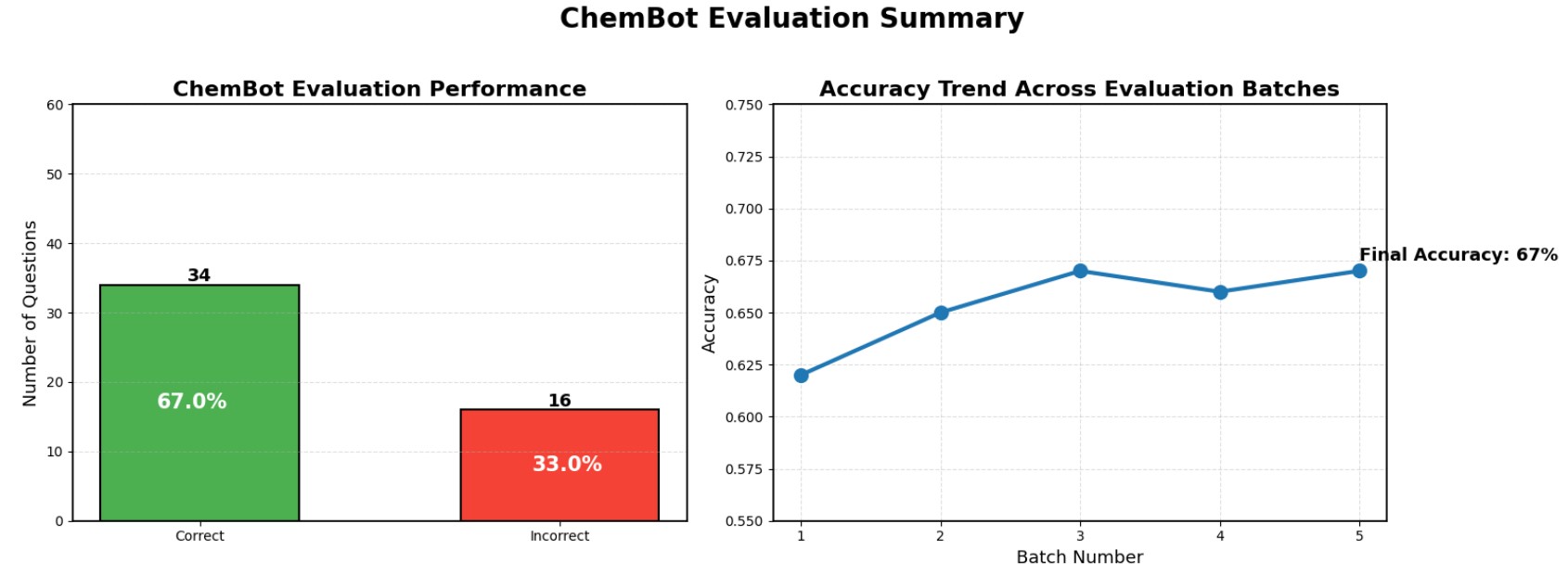}
    \caption{Graphs displaying the accuracy of the ChemBot LLM.}
    \label{fig:chembot-accuracy}
\end{figure}

\subsection*{Future Work}
Future improvements include expanding the dataset with more diverse chemistry literature, integrating additional safety filters, and improving retrieval accuracy. The ChemBot team also aims to extend ChemBot into a multimodal assistant by adding support for interpreting chemical structures, graphs, and reaction schemes through vision-language models. These enhancements would enable deeper reasoning and broaden ChemBot’s utility in chemistry and materials science research.

\subsection*{Open-source Materials}
The full implementation, including code, interface, fine-tuned weights, and evaluation resources, is provided in the following open-source repositories and notebooks:

\begin{itemize}
    \item \href{https://github.com/aishahasim/ChemBot---AI-Assisted-Chemistry-Assistant}{GitHub RAG Pipeline}
    \item \href{https://www.linkedin.com/posts/s-aishah-asim-335b191a7_llmhackathon-materialsscience-chemistry-activity-7372329143130034177-1-2D?utm_source=share&utm_medium=member_desktop&rcm=ACoAADBlw5kBLVvXchGkkN0nCe7h7fo-N2clA1k}{Video Explaining the Project}
    \item \href{https://huggingface.co/umairbinmansoor/finetuned_model/tree/main}{Fine-tuned BioGPT (gpt-oss 20B, LoRA)}
\end{itemize}





\section{Dynamic Materials Ontology Expansion with LLM--KG Integration}\label{sec:ontokg}

Accelerating materials discovery requires unifying structural, property, and application data into a form that supports reasoning, validation, and traceability. Existing databases capture isolated attributes but lack mechanisms for adaptive hypothesis generation and explainable inference. The OntoKG team presents \textbf{OntoKG}, an AI-driven framework that dynamically expands a materials ontology by coupling a Neo4j knowledge graph (KG) with a locally hosted large language model (LLM) via Ollama (Llama3/Mistral). The LLM proposes candidate relationships based on patterns in the KG, while a rule- and data-driven validator ensures only evidence-supported hypotheses are written back. This approach enables continuous ontology growth with full provenance and confidence tracking.

\subsection*{Architecture}

\begin{figure}[!ht]
    \centering
    \includegraphics[width=0.9\linewidth]{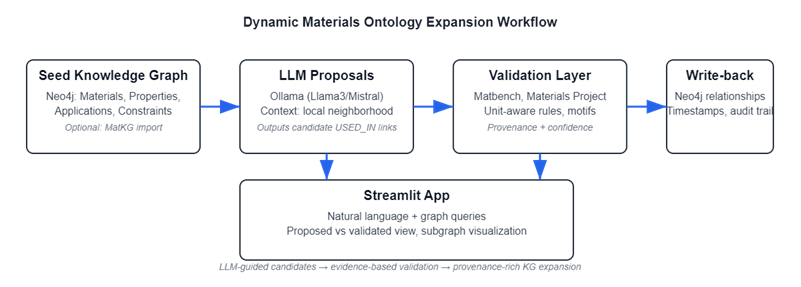}
    \caption{OntoKG pipeline: seed knowledge graph (Neo4j) $\rightarrow$ LLM-guided hypothesis generation (Ollama: Llama3/Mistral) $\rightarrow$ validation (Matbench, Materials Project, rule-based constraints) $\rightarrow$ write-back with provenance and timestamps $\rightarrow$ Streamlit-based visualization and analytics.}
    \label{fig:kg_ontology_workflow}
\end{figure}

OntoKG follows a modular, closed-loop architecture consisting of four primary components:

\begin{enumerate}
    \item \textbf{Knowledge Graph Layer:} The Neo4j graph encodes entities such as \texttt{Material}, \texttt{Property}, and \texttt{Application} with explicit relationships (\texttt{HAS\_PROPERTY}, \texttt{USED\_IN}) and schema constraints. Each edge carries metadata including validation status, confidence score, and provenance.
    \item \textbf{LLM Hypothesis Engine:} Using contextual graph queries, the Ollama-hosted LLM generates candidate material–application links. The generation process leverages few-shot templates and domain-specific prompts to reason analogically (e.g., perovskite structure similarity or band-gap alignment).
    \item \textbf{Validation and Evidence Module:} Candidate links are automatically validated using property thresholds, unit-aware consistency rules, and external data sources such as Matbench, Materials Project, and MatKG. Validation results are annotated with numeric confidence scores derived from property-based metrics and structural similarity measures.
    \item \textbf{Interface and Analytics Layer:} A Streamlit-based frontend provides natural-language querying, ontology visualization, and discovery analytics. Users can view proposed and validated relations, explore 3D subgraphs, and monitor system expansion in real time.
\end{enumerate}

This design establishes a feedback-driven loop where every validated discovery enhances both the KG and subsequent model reasoning, forming a continuously learning ontology.

\subsection*{Results}

The OntoKG team demonstrates an end-to-end workflow using curated seed data and optional MatKG integration. The base ontology encodes 15 materials, 10 properties, and 8 application domains. The LLM successfully inferred novel \texttt{Material}–\texttt{Application} hypotheses, several of which were validated against property databases with confidence exceeding 80\%. The system maintains full traceability, with every accepted relationship annotated by source, timestamp, and validation evidence. A representative workflow is shown in Figure~\ref{fig:kg_ontology_workflow}, illustrating the KG initialization, AI-guided inference, automated validation, and live visualization.

\subsection*{Future Work}

Planned enhancements include: (i) expansion to larger-scale MatKG datasets, (ii) uncertainty calibration for confidence estimation, (iii) property normalization and dimensional consistency checks, and (iv) integration of a human-in-the-loop review layer for high-impact discoveries. Longer-term directions involve applying graph neural networks for structure-property prediction and incorporating explainable AI modules for hypothesis interpretability.

\subsection*{Open-source Materials}

\noindent Code: \github{https://github.com/deepaksaipendyala/OntoKG}\,; Reference Dataset: \database{https://github.com/olivettigroup/MatKG}\,; Demo Video: \youtube{https://youtu.be/t9jvssU48KQ}\,;




\section{MINT LLM: Nature Language Interface for Automated Molecular Dynamics Analysis}\label{sec:mint_llm}

Molecular dynamics (MD) simulations use Newtonian mechanics to model the time-dependent behavior of atoms and molecules, enabling researchers to examine the structural, dynamical, and thermodynamic properties of complex systems at atomistic resolution \cite{lemkul2024introductory}. Consequently, MD simulations generate large volumes of trajectory data that require systematic analysis to extract meaningful insights. Traditional workflows, however, are often slow and labor-intensive, relying heavily on manual plotting and scripting. The Molecular INsights Toolkit (MINT LLM) addresses this challenge by providing a lightweight platform that integrates natural language interfaces with automated trajectory analysis. MINT LLM allows researchers to query simulation data conversationally and automatically compute key observables across commonly used MD simulation packages.

\subsection*{Results}

The MINT LLM team developed a fully functional Streamlit Cloud–hosted web application \cite{streamlit} for rapid analysis of molecular dynamics (MD) logs and trajectories. The platform comprises five modules: Simulation Assessment, Analysis Query, Results, and Trajectory Conversion, as illustrated in Figure~\ref{fig:MintLLM} below.

\begin{figure}[!ht]
  \centering
  \includegraphics[width=.8\linewidth]{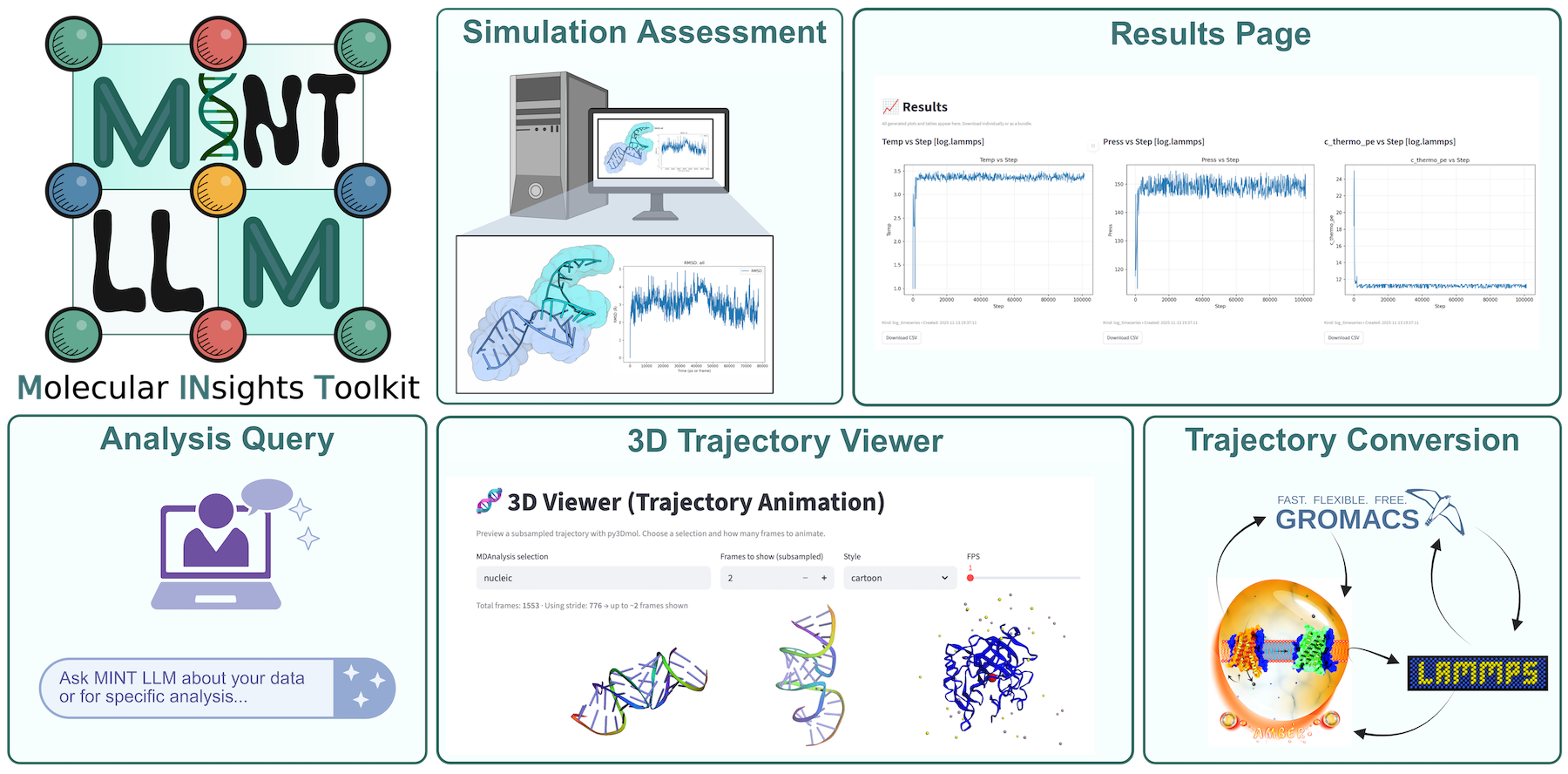}
  \caption{Overview of the MINT LLM web application, showing its five core modules: Simulation Assessment, Trajectory Upload, Analysis Query, Results, and Trajectory Conversion. [Created with Biorender.com]}
  \label{fig:MintLLM}
\end{figure}

Under the Simulation Assessment page, users can upload log files from any supported MD engine, including Large-scale Atomic/Molecular Massively Parallel Simulator (LAMMPS) \cite{thompson2022lammps}, GROningen MAchine for Chemical Simulations (GROMACS) \cite{berendsen1995gromacs}, and Assisted Model Building with Energy Refinement (AMBER) \cite{Amber2025}, and MINT LLM automatically generates time-series plots for all relevant variables. For example, uploading a LAMMPS log file produces plots such as Temperature vs.\ Step, Pressure vs.\ Step, and Potential Energy vs.\ Step. All plots generated in this module are automatically stored under the Results page, where users can download individual CSV files or a ZIP archive containing all associated datasets and images.

Users can upload topology and trajectory files through the Trajectory Upload page and initialize them by selecting “Create/Refresh Universe.” Successfully uploaded files are listed below and can be cleared or replaced as needed. The left-hand panel includes an option to upload an OpenAI API key \cite{OpenAI_API}. If a valid API key is provided, the Analysis Query tab allows users to chat with GPT-5 to ask questions about the data and request automated analysis and plotting. For example, given the query “plot root mean square deviation of the whole structure over time,” the chatbot provides general information about plotting RMSD and notifies the user when the plot has been completed. Generated plots are available in the Results tab with messages such as “Ran RMSD all and added the plot to Results.” Users can also view trajectories in the 3D viewer tab. The Trajectory Conversion tab allows users to convert uploaded trajectories to one of the other supported MD engine trajectory formats.

\subsection*{Future Work}
Future development of MINT LLM will focus on implementing complete functionality in the trajectory conversion module and refining other components of the framework. At present, MINT LLM supports trajectory and log file inputs from AMBER, GROMACS, and LAMMPS simulations. In upcoming versions, the MINT LLM team aims to expand compatibility to additional molecular dynamics engines, including oxDNA \cite{poppleton2021oxdna}, HOOMD-blue \cite{anderson2020hoomd}, and others.

\subsection*{Open-source Materials}

All source code and related materials are available on GitHub:
\github{https://github.com/ncsu-llm-hackathon-materials-2025/MINT-LLM.git}\,; Project website: \globe{https://mint-llm.streamlit.app/}\,; Video demo: \youtube{https://www.linkedin.com/feed/update/urn:li:activity:7372408867328274432/}\,;





\section{Ontology-Aware Data Mapping with LLM-Driven Tree Search}\label{sec:ontomapper}

In order to provide access to data across distributed resources with different local labels, the data need to be annotated using a semantic approach. This is realized using ontologies, which provide an explicit, formal specification of a shared conceptualization, thereby enabling semantic interoperability across heterogeneous data sources. In many materials-science collaborations, researchers lack ontology expertise, making it difficult to annotate data for cross-dataset search and reuse \cite{Bayerlein2025, Schilling2025}. Searching and editing ontologies requires in-depth knowledge of their internal structure and specialized query languages. The OntoMapper team addresses this gap by coupling large-language-model (LLM) reasoning with a structured tree search over the target ontology.

\subsection*{Results}

The workflow starts by parsing RDF/Turtle files with rdflib and constructing a NetworkX graph enriched with \texttt{rdfs:label}, \texttt{skos:altLabel}, and \texttt{skos:definition} for each node.
A bounded-oracle LLM scores candidate nodes on a 1--100 scale, while a depth penalty discourages overly generic matches, yielding a composite quality score $S(v)=q(v)-p(d(v))$.
The agent performs a beam search on this scored tree, expanding only promising branches; typical runs require $O(k\cdot d)$ LLM calls (with $k = 3$, $d = 5$) instead of exhaustive enumeration.

To locate a term or a related term within an ontology, the OntoMapper team relies on node embeddings. Similarity matching in an embedded vector space is orders of magnitude faster than performing graph traversals. It is important to note that exact matches are not the objective; therefore, methods such as SPARQL queries are not suitable for this task.

\begin{figure}[!ht]
\centering
\includegraphics[width=\textwidth]{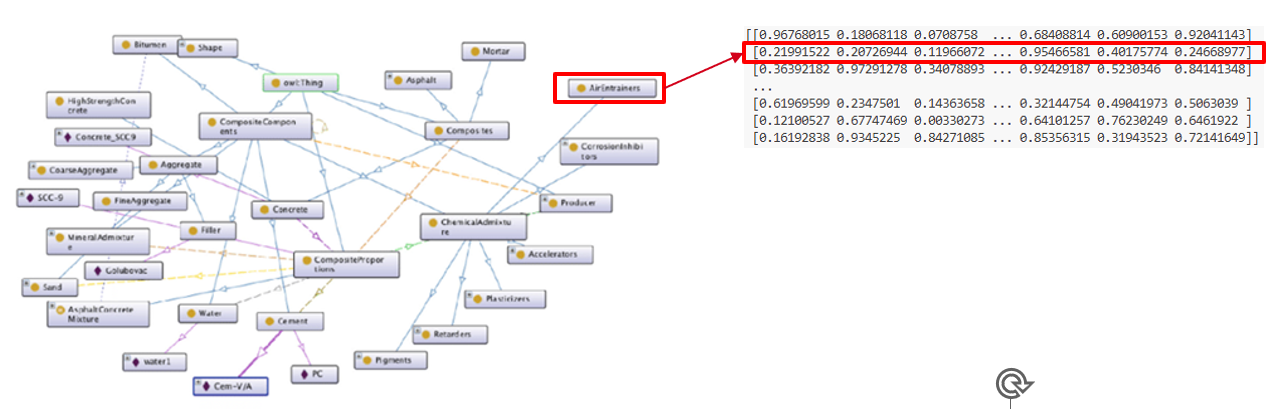}
\caption{Node embedding performed on an ontology}
\label{fig:node-embedding_schematic}
\end{figure}

Two node-embedding approaches were evaluated. The first, direct node embedding, embeds only the description of each individual node. The second, context-enhanced node embedding, incorporates both the node’s description and the descriptions and connectivity of its neighbouring nodes (Fig.~\ref{fig:node-embedding_schematic}).

\begin{figure}
\centering
\includegraphics[width=0.5\textwidth]{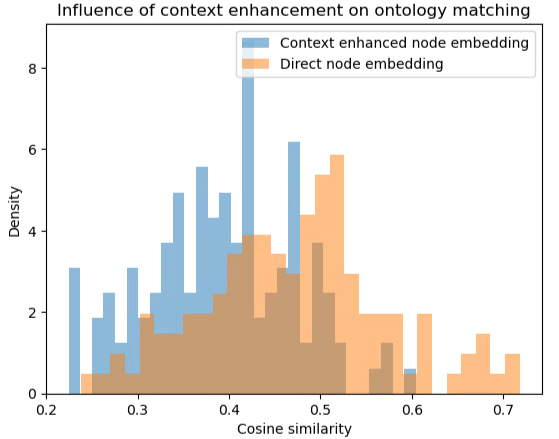}
\caption{Influence of context enhancement on ontology matching.}
\label{fig:node-embedding_result}
\end{figure}

To validate results, the OntoMapper team used the pre-annotated MuLMS text corpus \cite{schrader2023mulms} to extract domain-relevant terms mentioned in steel-related papers and then used these terms to evaluate the different approaches. The results indicated that the direct node-embedding approach performed marginally better (Fig.~\ref{fig:node-embedding_result}), attributed to additional noise likely introduced by the context-enhanced method. Context-enhanced embeddings are expected to excel when matching higher-level concepts rather than isolated keywords, which remains an avenue for future exploration.

\subsection*{Future Work}
Planned enhancements include integrating a hybrid pre-pruning stage that leverages vector-based lexical similarity to further reduce LLM query count, and evaluating the approach on larger, domain-specific ontologies such as the PMD core ontology \cite{Bayerlein2024}.

\subsection*{Open-source Materials}
All code, demonstration notebooks, and the MCP-compatible service for subtree retrieval are released under the BSD-3 license at GitHub: \github{https://github.com/materialdigital/ontomapper}\,; Video demo: \youtube{https://www.youtube.com/watch?v=NuG3IHJun4M}

\section{Extracting crystallographic information for structure generation from S/TEM literature}
\label{sec:atombridge}

Experimental electron microscopy (EM) now routinely produces atomic-resolution scanning transmission electron microscopy or transmission electron microscopy (S/TEM) images at scale, offering a direct yet underutilized route from experiments to atomistic simulations. While toolkits such as AtomAI \cite{ghosh2022bridging, ziatdinov2021atomai} and related atom-vision frameworks have enabled large-scale, automated extraction of atom positions from S/TEM images, converting these micrographs into crystallographic information files (CIFs)—especially in the absence of raw data—remains challenging and labor-intensive. \texttt{AtomBridge} addresses this gap by providing an end-to-end, reproducible workflow that integrates robust column localization, symmetry detection, and textual descriptions found in manuscripts using large language models (LLMs), along with geometry- and physics-based validation, facilitating reliable CIF generation and atomic simulation environment (ASE)-compatible exports. By automating this process, \texttt{AtomBridge} reduces manual intervention, enhances reproducibility, and accelerates the creation of simulation-ready, experimentally validated datasets for high-throughput computational materials discovery \cite{eliasson2024localization}.

\subsection*{Results}
\texttt{AtomBridge} integrates literature- and image-driven crystallography approaches into a single, auditable Streamlit interface. Its PDF path allows user uploads or selections from a curated library, with PyPDFLoader processing sectioned text and images. The application features auto-suggested prompts derived from abstracts for batch execution over targets, alongside support for ad-hoc custom queries. Following this, an ASE-aware RAG pipeline retrieves relevant ASE snippets from a ChromaDB index (all-MiniLM-L6-v2; top-$k=15$) to anchor Gemini-2.5-Pro code generation. This process constructs \texttt{ase.Atoms} objects and writes CIFs. The generated code runs in an isolated subprocess, utilizing an iterative auto-fix routine (up to three LLM-guided retries). Final outputs undergo multi-tier validation: interatomic-overlap checks ($\leq$ 0.5{\AA}; per-file distance summaries), optional M3GNET relaxation for stability estimation, and CIF-to-CIF geometry comparison against references ($\pm$ 15\% lattice, $\pm 5^\circ$ angles).

The image path in \texttt{AtomBridge} automatically identifies figures from PDFs, correlates subpanels with captions, and permits manual editing or LLM-augmented (Gemini-2.5-Flash) picking of TEM/HAADF-STEM images. After the user manually selects an area of interest and inputs the \mbox{pixel-to-\si{nm}} conversion factor, the pipeline applies CLAHE and a peak search to identify atomic columns. It then computes two non-collinear lattice vectors via DBSCAN clustering of nearest-neighbor shifts, using an FFT fallback to identify reciprocal-space peaks and transform to Crystallography CBF format $(a,b,\gamma)$. These CIF lattice parameters are injected to create generation tasks. The corresponding validation CIFs are immediately displayed and tested in conjunction with the purely text-derived structures. This end-to-end PDF-to-CIF cycle completes within minutes, having successfully reconstructed micrographs of layered oxides (e.g., LCO) \cite{LCO}, 2D crystals with point defects (e.g., WSe\textsubscript{2}) \cite{WSe2_2D}, and structures with dimensions consistent with existing literature standards. Despite successes with LCO and defects in 2D crystals, the workflow failed to reproduce reasonable .CIF files for more complicated structures, such as semiconducting nanoclusters.

\begin{figure}[H]
    \centering
    \includegraphics[width=0.8\linewidth]{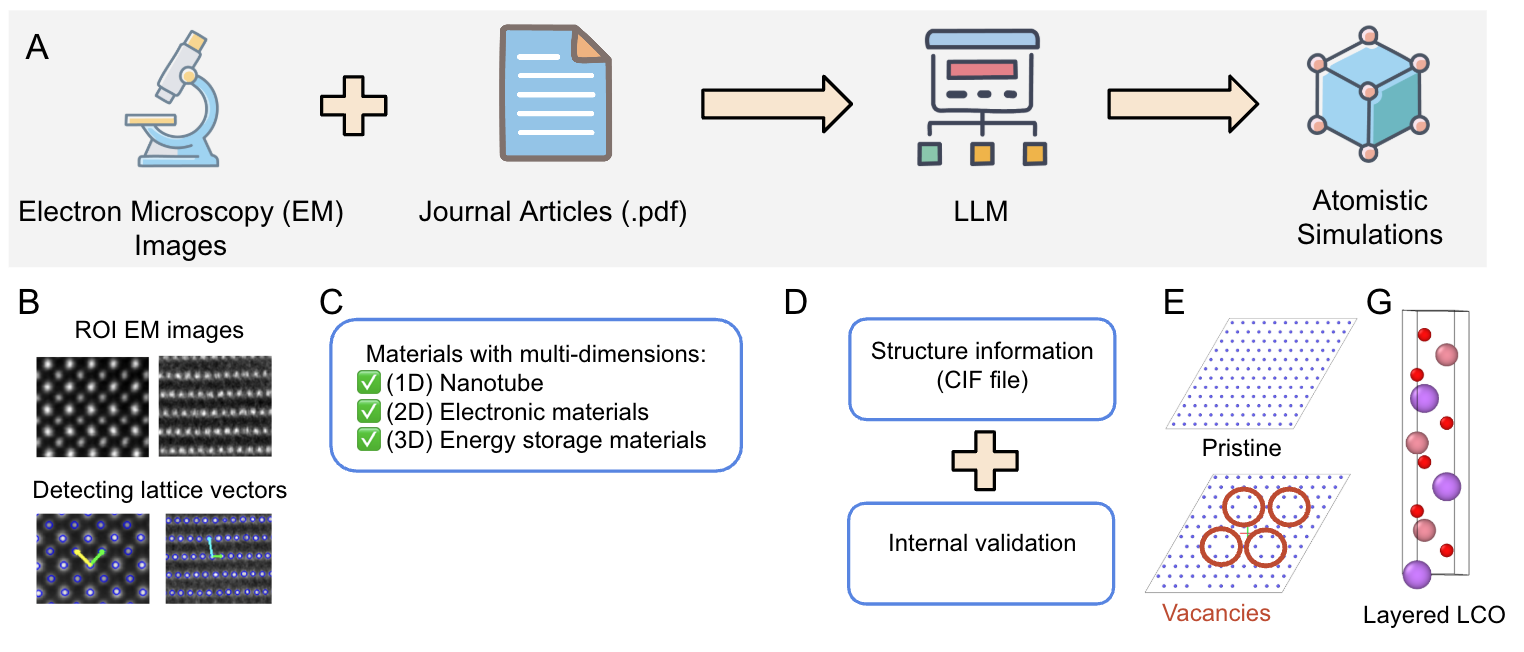}
    \caption{(A) Schematic illustration of \texttt{AtomBridge} workflow. (B) \texttt{AtomBridge} allows users to select the region of interest (ROI) in electron microscopy (EM) images (top) and detects lattice vectors from selected EM images (bottom). (C) \texttt{AtomBridge} is tested for materials with multi-dimensions. (D) \texttt{AtomBridge} extracts structural information in CIF form from journals \cite{LCO, STO_NT} and is equipped with an internal structure validation function. The resulting simulated atomic structures of (E) WSe\textsubscript{2} showing from pristine (top) to vacancies (bottom) \cite{WSe2_2D} and (G) layered LCO that are relaxed by DFT. (B) is reproduced from journals \cite{LCO, STO_NT} with permission. Copyright 2025 American Chemical Society (left), Copyright 2016 American Chemical Society (right), respectively.}
    \label{fig:AtomBridge_fig}
\end{figure}

\subsection*{Future Work}
Given the difficulties in extracting real-space information from low-quality images in manuscripts, the AtomBridge team plans to develop a more robust workflow for extracting both real-space and reciprocal-space information from images. Additionally, the AtomBridge team plans to add further stability checks to the generated structures using density functional theory, and a more robust pipeline for foundational machine learning interatomic potentials \cite{batatia2023foundation}.

\subsection*{Open-source Materials}
Code available on GitHub: \github{https://github.com/dpalmer-anl/AtomBridge.git}\,; Demo video: \youtube{https://youtu.be/PQHEWl9STFM}\,;




\section{GAINS: Generative AI for No-PAIN Structures}\label{sec:gains}

Pan-assay interference compounds (PAINS) are substructures that frequently generate spurious signals in biological assays, leading to false positives and misallocation of experimental efforts \cite{Baell2010,Baell2014}. Classical mitigation relies on rule-based filtering of SMARTS patterns and manual medicinal chemistry.
A key idea is to use LLM generative models to enable structure-aware edits subject to property constraints, offering a complementary path: \emph{repair} problematic motifs rather than discard entire scaffolds.
In this sense, the GAINS team defines \textbf{GAINS} (\emph{Generative AI for No-PAIN Structures}) as a reproducible pipeline that (i) detects PAINS motifs using RDKit filter catalogs, (ii) proposes \emph{minimal} structural edits via a large language model (LLM), and (iii) validates the resulting molecules with cheminformatics descriptors and a synthetic accessibility score (PAINS status, QED, MW, logP, TPSA, HBD/HBA, ROT, SA).

\subsection*{Results}
From an initial dataset of 2,000 ChEMBL-derived molecules (filtered to molecular weights below 400~Da), a total of 100 compounds were identified as containing PAINS-related substructures using the RDKit FilterCatalog (PAINS A/B/C). These flagged molecules served as the input set for the GAINS pipeline.

The generative stage, executed with the \texttt{gpt-5-mini-2025-08-07} model, produced 448 valid molecular candidates. Among these, 373 molecules (83.3\%) successfully removed all PAINS alerts (\texttt{False}), while the remaining 75 (16.7\%) retained partial or modified PAINS motifs (\texttt{True}). This distribution indicates that the majority of GAINS-curated molecules overcame assay-interfering structural patterns while maintaining overall scaffold integrity.

\begin{figure}[H]
    \centering
    \includegraphics[width=0.75\linewidth]{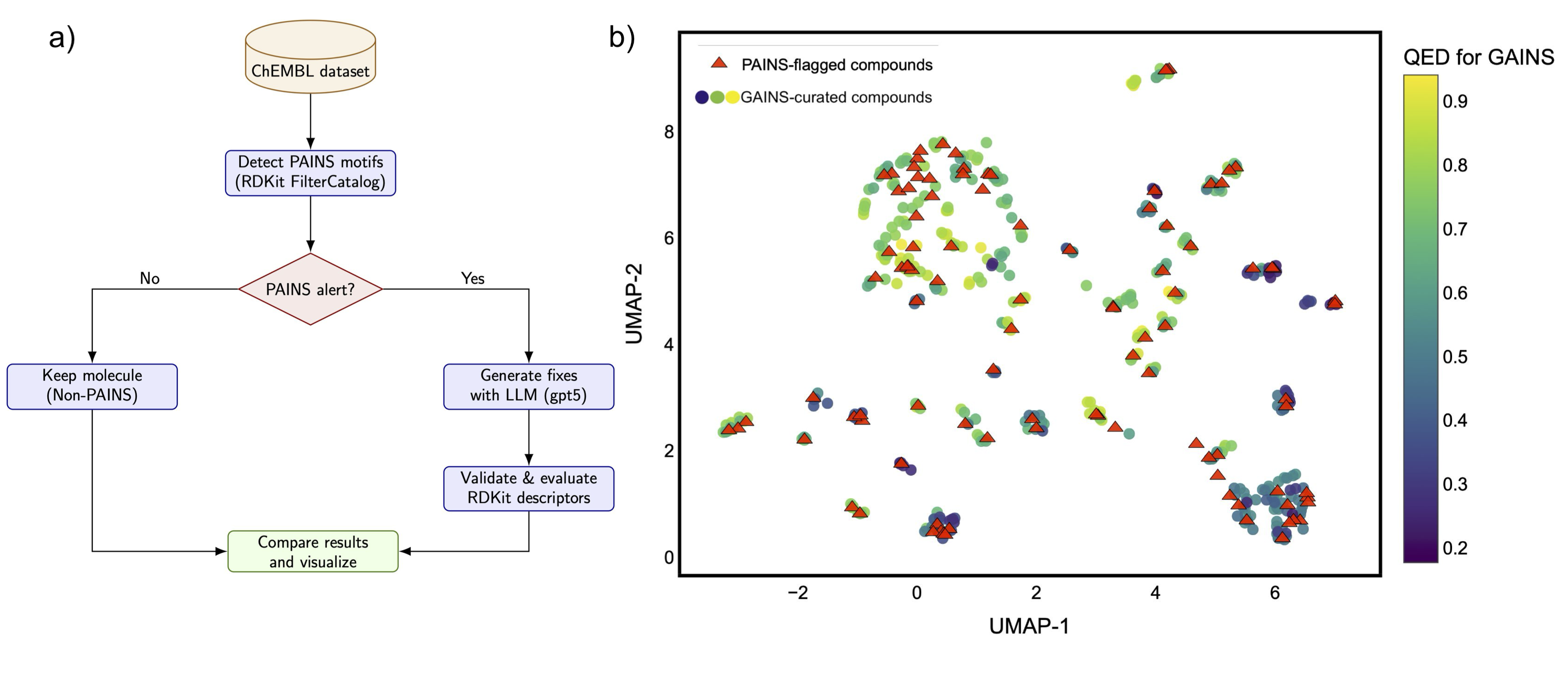}
    \caption{a) GAINS workflow. b) Two-dimensional UMAP projection of molecular fingerprints showing the distribution of original ChEMBL compounds and LLM-generated structures.
    Red triangles highlight PAINS-flagged motifs, and colored circles represent GPT-repaired candidates (GAINS) according to their quantitative estimate of drug-likeness (QED, color scale).}
    \label{fig:figGAINS}
\end{figure}

The analysis of molecular transformations proposed by the GAINS pipeline (Figure~\ref{fig:figGAINS}a) revealed a consistent tendency toward \emph{minimal structural repair} rather than full redesign. The most frequent modifications (approximately 70\%) correspond to minor atomic substitutions, in which the global molecular scaffold is preserved while local reactive moieties are subtly altered. This pattern indicates that the LLM internalized a minimal-edit chemical strategy, focusing on structural conservation and local correction of problematic motifs such as styrenes, catechols, or anilines.

Quantitatively, repaired compounds exhibited an average increase in QED of +0.10, a moderate rise in logP (+0.14), and a decrease in polar surface area ($\Delta$TPSA $\approx$ --11~\AA$^2$), suggesting an overall improvement in drug-likeness and physicochemical balance. A detailed breakdown of the most common transformations and their average effects is presented in Table~\ref{tab:transformations}, while the global distribution of original and repaired compounds in the molecular embedding space is shown in Figure~\ref{fig:figGAINS}b.

As shown in Table~\ref{tab:transformations}, minor atomic substitutions were dominant, improving both QED and logP while moderately reducing TPSA. Chain extensions or substituent additions tended to increase hydrophobicity (logP) but maintained acceptable polarity levels. In contrast, hydroxyl and ether insertions increased TPSA by approximately 10~\AA$^2$, enhancing solubility. Carbonyl additions produced the strongest polarity shifts ($\Delta$TPSA $\approx$ +20~\AA$^2$), whereas scaffold simplifications achieved the highest QED gains (+0.17) by reducing redundant or highly polar fragments.

The UMAP projection (Figure~\ref{fig:figGAINS}b) provides a visual overview of this effect: PAINS-flagged compounds (red triangles) cluster in high-density regions characterized by lower QED values, while the GAINS-repaired structures (colored circles) occupy nearby zones in the latent chemical space, yet shift toward regions associated with higher drug-likeness.

Altogether, GAINS is a chemically conservative workflow, capable of selectively editing assay-interfering structures while maintaining key pharmacophoric features. This behavior aligns with the intended goal of structural repair rather than de novo generation, providing a realistic and interpretable route for transforming PAINS-containing molecules into more drug-like analogs suitable for downstream optimization.

\vspace{0.5cm}

\begin{table}[H]
\centering
\caption{Most common structural transformations proposed by the GAINS workflow and their average effects on physicochemical properties.}
\label{tab:transformations}
\begin{tabular}{lcccc}
\toprule
\textbf{Transformation type} & \textbf{Occurrences} & \textbf{$\Delta$QED} & \textbf{$\Delta$logP} & \textbf{$\Delta$TPSA} \\
\midrule
Minor atomic substitution               & 321 & +0.098 & +0.141 & --11.105 \\
Chain extension / substituent added     & 50  & +0.077 & +0.481 & --3.848 \\
Added hydroxyl or ether group           & 23  & +0.017 & --0.304 & +10.735 \\
Simplified scaffold                     & 20  & +0.168 & --0.880 & --13.115 \\
Added carbonyl (=O)                     & 19  & +0.054 & --0.168 & +19.958 \\
\bottomrule
\end{tabular}
\end{table}

\subsection*{Future Work}
Future work will focus on leveraging pretrained molecular language models such as ChemBERTa, ChemGPT, and Tx-Gemma to enhance GAINS through domain-specific fine-tuning and property-conditioned generation \cite{Chemberta,chemgpt,txgemma}.

\subsection*{Open-source Materials}
Source code available on GitHub: \github{https://github.com/napoles-uach/GAINS}\,; Video demo: \youtube{https://youtu.be/gd6tvAtTTiQ}\,;




\section{Computation and Visualization of Infrared Spectroscopy with ChemGraph}\label{sec:chemgraph_IR}

Infrared (IR) spectroscopy is a powerful experimental technique for probing molecular structure and bonding, yet interpreting IR spectra often requires detailed computational analysis to assign vibrational modes and understand underlying atomic motions.
This analysis typically proceeds by (i) optimizing the equilibrium geometry, (ii) computing the Hessian (mass--weighted second derivatives) to obtain harmonic frequencies and normal modes, and (iii) evaluating IR intensities from the square of the dipole--moment derivative with respect to each normal coordinate \cite{wilson1980molecular}.

ChemGraph\cite{pham2025chemgraph} is an open-source agentic framework that orchestrates end-to-end computational-chemistry workflows from natural-language requests, automating everything from parameter selection to visualization of final results.
Building on this foundation, the ChemGraph-IR team provides a specialized pipeline for IR spectroscopy: given a user query and a molecular structure, it performs geometry optimization, vibrational analysis, IR intensity calculations, and generates both spectra and normal-mode animations with consistent file outputs for downstream analysis.
Together, these capabilities enable users to compute and visualize IR spectra directly from natural language queries, making spectral interpretation faster and more accessible.

Recent foundation models for atomistic simulation, particularly equivariant graph neural network force fields such as MACE and UMA, deliver first-principles-level accuracy at orders-of-magnitude lower cost, allowing routine frequency computations via finite differences of ML forces and rapid exploration of conformers \cite{batatia2023mace,wood2025uma}.
In parallel, large language models (LLMs) serve as high-level controllers in ChemGraph, translating user intent into reproducible workflows and reducing manual effort in setup, error recovery, and provenance capture.

Dipole moments (and their derivatives) can be obtained at multiple levels of theory: (i) \emph{ab initio} methods provide high fidelity for intensities and selection rules; (ii) modern semiempirical models such as GFN2-xTB offer speedy dipoles suitable for screening and solvent/ensemble studies \cite{bannwarth2019gfn2}; and (iii) machine-learning models such as MACE4IR\cite{bhatia2025mace4ir} predict dipoles directly alongside energies and forces, enabling fast IR-intensity estimation when coupled with normal modes from ML potentials.
Examples include PhysNet, which jointly learns energies, forces, dipoles, and partial charges \cite{PhysNet_JCTC_2019}, and AIMNet2, which learns physics-informed charges to reproduce dipole and quadrupole moments across diverse chemical spaces \cite{AIMNet2_PNASplus_2025}.

\subsection*{Results}
\begin{figure}[htbp]
\centering
\includegraphics[width=0.6\linewidth]{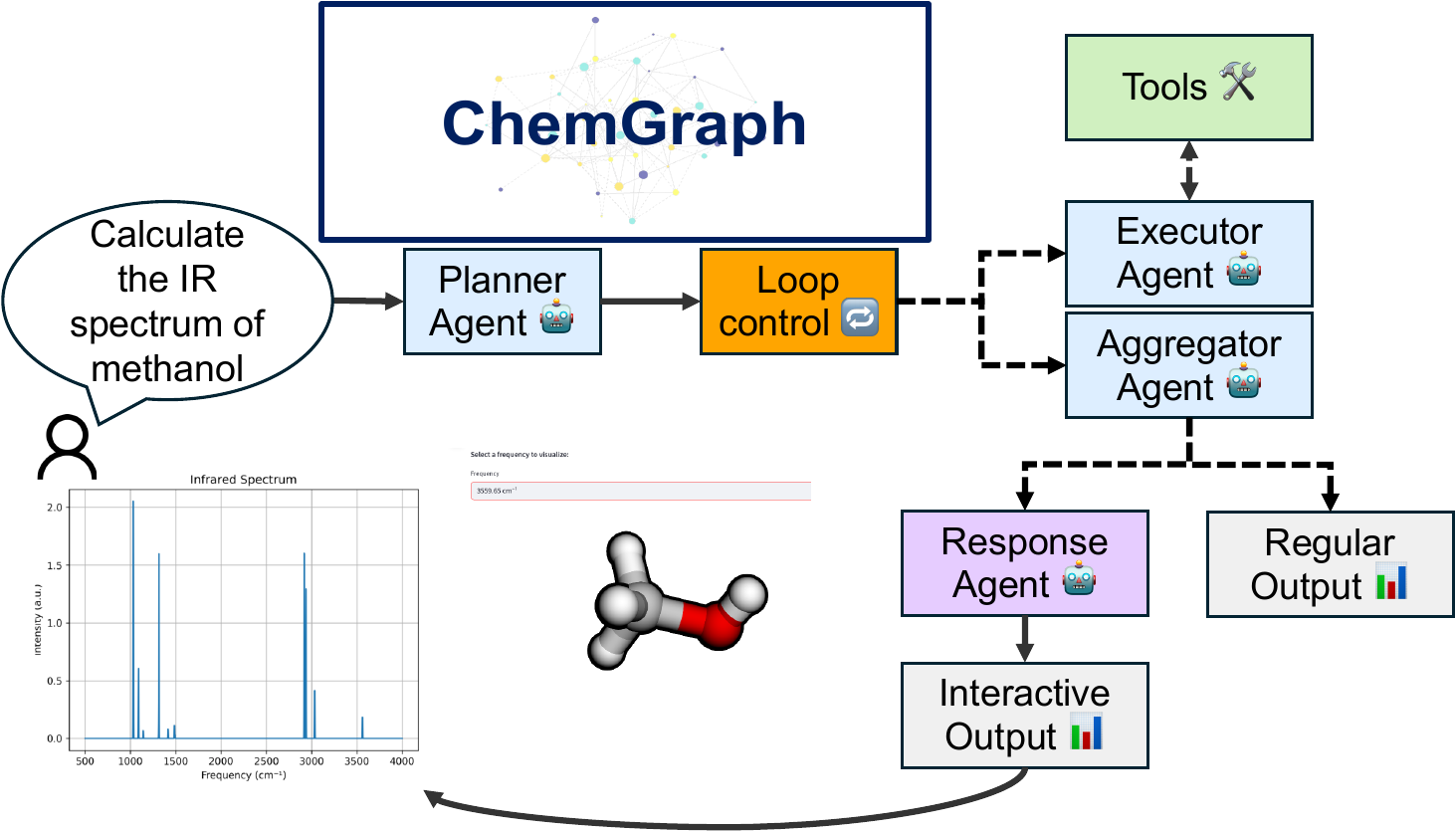}
\caption{ChemGraph-IR workflow: from a natural-language request to optimized structure, vibrational analysis, IR intensities, spectrum visualization, and normal-mode animation.}
\label{fig:chemgraph-ir}
\end{figure}

During the LLM Hackathon, the ChemGraph-IR team implemented and validated the following capabilities:
\begin{itemize}
  \item \textbf{End-to-end IR pipeline:} Automated geometry optimization, harmonic frequency analysis, and IR-intensity generation, with robust fallbacks (restarts, stale-file cleanup, and parameter retries) to improve reliability.
  \item \textbf{Calculator backends:} Integration of an ASE calculator based on AIMNet2~\cite{AIMNet2_PNASplus_2025} and dipole-moment calculations via GFN2-xTB~\cite{bannwarth2019gfn2}, enabling fast ML-accelerated runs and efficient screening.
  \item \textbf{Interactive visualization:} A user interface that renders the IR spectrum (image and peak table), supports mode selection, and previews animated normal-mode displacements in 3D to aid interpretation.
  \item \textbf{Reproducible file artifacts:} Standardized outputs for downstream analysis, including \texttt{frequencies.csv}, per-mode \texttt{.traj} files, and a saved spectrum image (\texttt{ir\_spectrum.png}) for reproducible workflows.
  \item \textbf{LLM backends:} Added Groq support end-to-end (dependencies, configuration, loader, and supported-model registry) to improve portability and provider choice.
\end{itemize}

The ChemGraph-IR team also added a Jupyter notebook that demonstrates these features end-to-end, and set up documentation to build and deploy automatically via a GitHub Actions workflow using MkDocs, ensuring guidance stays current with each repository update.

\subsection*{Future Work}
Planned enhancements include incorporating anharmonic effects through frequency-scaling schemes alongside explicit treatments based on high-dimensional representations of the potential energy surface and solution of the vibrational Schrödinger equation.
Additional plans include integrating a curated database of experimental IR spectra with solvent and temperature metadata to enable side-by-side comparison, automated peak matching, and rigorous provenance.
To quantify accuracy and cost--benefit trade-offs, ML-derived normal modes, frequencies, and intensities will be benchmarked against the experimental database.

\subsection*{Open-source Materials}
ChemGraph source code is available on GitHub: \github{https://github.com/argonne-lcf/ChemGraph}\,;
A special release for the hackathon is provided at \href{https://github.com/argonne-lcf/ChemGraph/releases/tag/v2025.09.12}{v2025.09.12}\,; Demo video: \youtube{https://youtu.be/LLz4MXTztWA}\,;




\section{Crystal Learning for Understandable Explanations}\label{sec:CLUE}

In traditional physics and chemistry, explanations and definitions form the scaffolding of scientific progress. They guide model development, enable fair comparison between approaches, and transform results into reusable knowledge. In contrast, black-box machine learning (ML) models lack inherent interpretability. Thus, developing dedicated methods to explain and interpret their decisions is crucial for building trust, enabling evaluation beyond output statistics, and guiding more purposeful model refinement.

There are roughly four groups of XAI methods used for post-hoc explanation of black-box models: (1) feature-importance methods (e.g., SHAP~\cite{lundberg2017shap}); (2) gradient-based methods~\cite{Sundararajan2017}; (3) surrogate-model methods (e.g., LIME~\cite{Ribeiro2016}); and (4) analysis of counterfactuals~\cite{Wellawatte2023, Wellawatte2025} and examples. This work focuses on counterfactual explanations.

A \textit{counterfactual} in the context of chemistry is a molecule or compound that is structurally similar to a given sample but is predicted to exhibit an opposite property. Explanations emerge by identifying and reasoning about structural differences between the sample and its counterfactuals, and relating these differences to the predicted property. Traditionally, this interpretive step relies on human expertise, as it requires flexible reasoning and domain knowledge. However, recent studies have shown that large language models (LLMs) such as ChatGPT and Claude can effectively perform this reasoning for molecular systems~\cite{Wellawatte2025}.  
The CLUE team extends these ideas to the interpretation of structure–property relationships in crystalline materials—a setting that, to the team’s knowledge, has not been explored before.

\subsection*{Results}
\begin{figure}[h]
    \centering
    \includegraphics[width=1\linewidth]{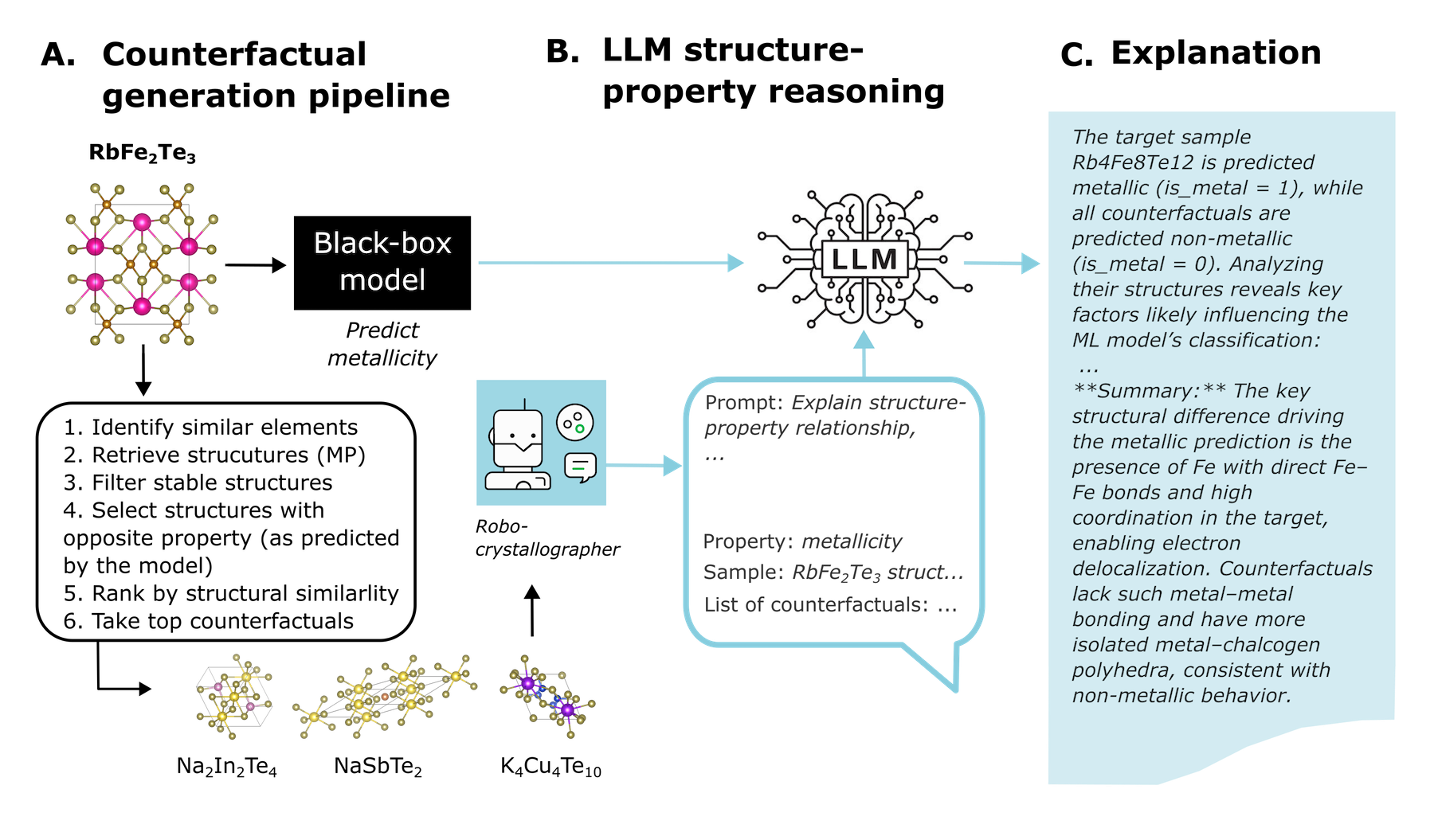}
    \caption{CLUE workflow.  
    (A) Counterfactual generation: property prediction for the sample and identification of counterfactual structures.  
    (B) LLM reasoning: prompt compilation and structural–property analysis.  
    (C) Explanation: model-generated interpretation of structure–property relationships.}
    \label{fig:CLUE-pipeline}
\end{figure}

In this hackathon project, the CLUE team developed a pipeline to generate counterfactual explanations for structure–property relationships in crystalline materials (Figure~\ref{fig:CLUE-pipeline}). The workflow comprises three stages.

\textbf{Stage 1: Property prediction and counterfactual generation.}  
As the black-box model, a random forest (RF) classifier from \texttt{scikit-learn} was used, chosen for its strong baseline performance and simplicity. The target property is metallicity. The model was trained on the JARVIS 3D-DFT dataset~\cite{Choudhary2020} (metal:nonmetal ratio 0.75:0.25), using an 80:20 train–test split. Features included \texttt{matminer}~\cite{ward2018matminer} composition and structural descriptors combined with SOAP features (computed for the lattice with atom identities removed). The trained model achieved $\mathrm{ACC}=0.92$, $\mathrm{MCC}=0.81$, and $\mathrm{F1}=0.94$. High performance is essential, as meaningful explanations can only be expected when the model captures genuine structure–property relationships.

To generate counterfactuals, a list of candidate elements was compiled by selecting the five nearest neighbours for each element in the sample on the Pettifor scale~\cite{Pettifor1984}, as chemically similar elements are more likely to substitute each other. All corresponding compounds were downloaded from the Materials Project database, filtered for thermodynamic stability, and predicted using the trained model. Compounds predicted to have the opposite property were retained. Structural similarity between the sample and each candidate was then computed, and the top ten most similar compounds were selected as counterfactuals.

The overall structural similarity metric is defined as
$S_{\text{struct}} = w_{\text{comp}} S_{\text{comp}} + w_{\text{soap}} S_{\text{soap}} + w_{\text{vol}} S_{\text{vol}}$,
where $S_{\text{comp}}$, $S_{\text{soap}}$, and $S_{\text{vol}}$ measure similarity in composition, local structure, and volume per atom, respectively.  
Compositional distance is computed via the Earth Mover’s Distance (ElMD~\cite{Hargreaves2020}) using Pettifor distances, rescaled and transformed to similarity as $S_{\text{comp}} = \exp(-\tilde{D}_{\text{comp}})$.  
Structural similarity uses SOAP descriptors~\cite{Bartok2013, himanen2020dscribe}, averaged into orbit vectors per symmetry-equivalent atom. These vectors are matched via the Hungarian algorithm with Pettifor-weighted species similarity. To correct for the insensitivity of the SOAP similarity to uniform cell expansion, a volume-based similarity term is added:
$S_{\text{vol}} = \exp\{-|v_1 - v_2|/2\}$.

\textbf{Stage 2: LLM reasoning over counterfactuals.}  
Each structure is described using the Robocrystallographer tool~\cite{Ganose2019}, which produces human-readable structural descriptions. These are combined into a prompt instructing the LLM (ChatGPT-4-mini) to analyze the structure–property relationships and evaluate their consistency with known chemical principles.

\textbf{Stage 3: Explanation synthesis.}  
The LLM generates a structured explanation linking features of the sample and counterfactuals to the predicted property. The output is stored locally for review.  
The pipeline was tested on five structurally complex compounds; the resulting explanations are available on GitHub.

\subsection*{Future Work}
Future directions include:  
(1) investigating how different similarity measures affect the selected counterfactuals and resulting explanations;  
(2) developing quantitative metrics to evaluate explanation quality~\cite{Wellawatte2025}; and  
(3) extending from single-sample (local) explanations to dataset-level analyses that assess model reliability and uncover broader structure–property trends beyond accuracy metrics.

\subsection*{Open-source Materials}
Code is available on GitHub: \github{https://github.com/epatyukova/llm2025-hackathon-CLUE}\,; Demo video: \youtube{https://x.com/ElenaPatyukova/status/1966598741105746362}\,;




\section{Next Experiment Data Driven}\label{sec:NEDD}

When experiments are conducted, the parameter space is often sampled non-uniformly, resulting in skewed data distributions and regions of ignorance—areas where no experiments have been performed. Machine learning models perform best when trained on diverse, representative datasets; if experiments are clustered in a narrow region of the parameter space, the model’s ability to generalize is severely limited, leading to biased training and inaccurate predictions. NEDD mitigates this issue by visualizing the parameter space, identifying unexplored regions, and suggesting next experiments using active learning rather than randomness~\cite{wei2025discovering}. Additionally, the integrated LLM assistant enables natural-language querying of experimental data and provides informed guidance for future experiments.

\subsection*{Results}
NEDD (Next Experiment Data-Driven) is implemented as a local server or Streamlit-based user interface, allowing scientists to interact with machine learning–driven experiment planning without writing code~\cite{khorasani2022web}. Because the system runs locally, all experimental data remain on the user’s machine, preserving confidentiality and enabling secure use in laboratory or industrial environments.

\begin{figure}[h]
    \centering
    \includegraphics[width=0.92\linewidth]{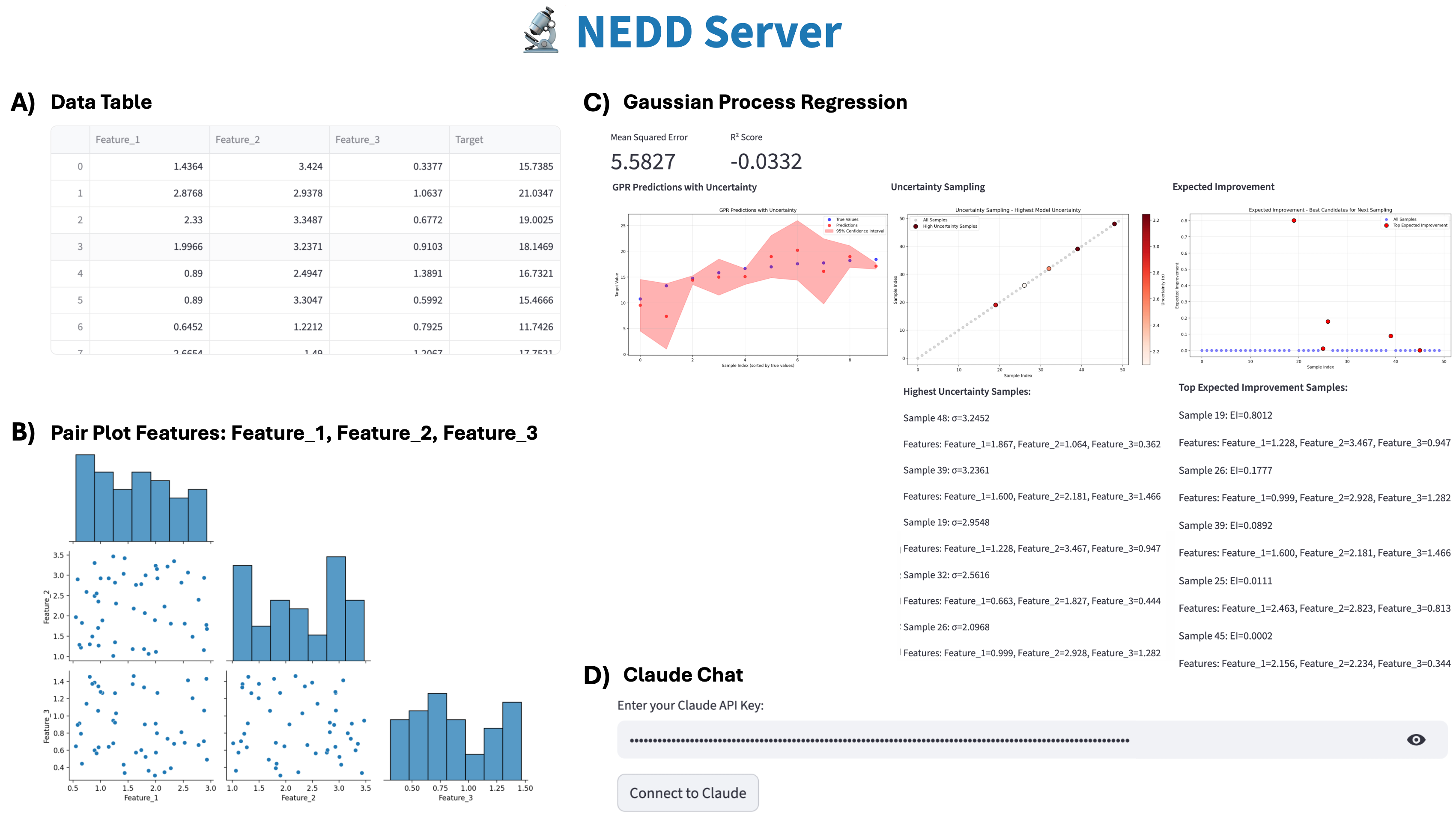}
    \caption{NEDD user interface for data-driven experiment planning. A) Data upload and tabular view where the rightmost column is interpreted as the target variable. B) Interactive pairplot visualizing parameter distributions and gaps in explored space. C) GPR output showing predictive mean, uncertainty, and top five suggested next experiments using uncertainty sampling and expected improvement. D) Integrated Claude LLM chat interface enabling natural-language queries about data trends, experimental strategies, and model outputs.}
    \label{fig:nedd}
\end{figure}

The workflow begins with users uploading a dataset in tabular format (Figure~\ref{fig:nedd}). NEDD automatically parses the table, interprets all columns except the rightmost as input features, and treats the rightmost column as the target variable. The platform supports an unlimited number of numerical features (categorical support under development), making it compatible with high-dimensional synthesis or characterization datasets. Within the user interface, the dataset is visualized through interactive pairwise parameter plots and kernel-density projections, enabling users to detect data skewness, clustering, and unvisited regions of parameter space. These visual tools are designed specifically for non-coding users, allowing experimentalists to understand dataset structure and coverage without scripting.

NEDD automatically trains a Gaussian Process Regression (GPR) model with an ARD kernel on the uploaded dataset to estimate predictive mean and uncertainty~\cite{khorasani2022web, chhajer2024rationalised}. Based on the trained surrogate model, NEDD recommends the top five next experiments using two acquisition strategies. Uncertainty sampling prioritizes regions with high model uncertainty to improve overall predictive confidence (exploration mode), while expected improvement (EI) targets conditions likely to exceed current best performance (exploitation mode). The platform provides both modes via UI toggles, enabling users to shift between knowledge expansion and performance optimization.

A key innovation of NEDD is its integration of a Claude LLM chat interface (via API), which enables natural-language interrogation of the dataset and model outputs (e.g., “How can I reduce parameter imbalance?”)~\cite{claude2024nedd}. The LLM does not access raw data externally; instead, it analyzes metadata, summaries, or computed statistics to provide guidance on trends, biases, and next steps.

Data-guided experimental approaches significantly reduce the number of iterations required to reach optimal conditions compared to random or expert-only sampling methods. These strategies leverage prior results or model predictions to select new, informative experimental conditions, resulting in faster convergence toward desired outcomes while requiring fewer trial runs and saving time and computational resources~\cite{chhajer2024rationalised}.

The NEDD team notes the use of the Cline AI coding assistant with the Claude LLM to accelerate code development and debugging during implementation~\cite{cline2024nedd}.

\subsection*{Future Work}
Planned extensions include expanded data-handling capabilities, increased robustness of active-learning models, and integration of a local LLM to enable direct, privacy-preserving interaction with experimental data.

\subsection*{Open-source Materials}
Code is available on GitHub: \github{https://github.com/ViktoriiaBaib/NEDD}\,; Demo video: \youtube{https://www.linkedin.com/posts/viktoriia-baibakova_ai-materialsscience-hackathon-activity-7372296092257800192-pssH?utm_source=share&utm_medium=member_desktop&rcm=ACoAADoFARMBcTXMnjRDWSM6OoM9_ggldCivzp8}\,;




\section{Text to text (T2) crystal relaxation}\label{sec:T2_relax}

Crystal structure relaxation is an integral part of high-throughput computational workflows, which often optimize the geometry of thousands of structures, selecting a DFT code and the corresponding computational parameters~\cite{speckhard2025big, speckhard2025workflows}. In this project, the T2-Relax team uses the LeMatTraj dataset~\cite{ramlaoui2025lemat} to train a large language model (T5-Transformer)~\cite{raffel2020exploring} to perform crystal structure relaxation. To feed the model textual input data, the ASE library~\cite{larsen2017atomic} is used to transform the crystal structure into a crystallographic input file (CIF), which is the standard in the Inorganic Crystallographic Database (ICSD)~\cite{hellenbrandt2004inorganic}. The objective is to predict the relaxed structure in CIF format via $\Delta$-learning~\cite{speckhard2025extrapolation}. This work was inspired by the MD-LLM-1 approach~\cite{murtada2025md}, which predicted relaxed protein structures using an LLM. To improve performance, the model is fine-tuned specifically on the relaxation task.

\subsection*{Results}

\begin{figure}[h]
    \centering
    \includegraphics[width=0.92\linewidth]{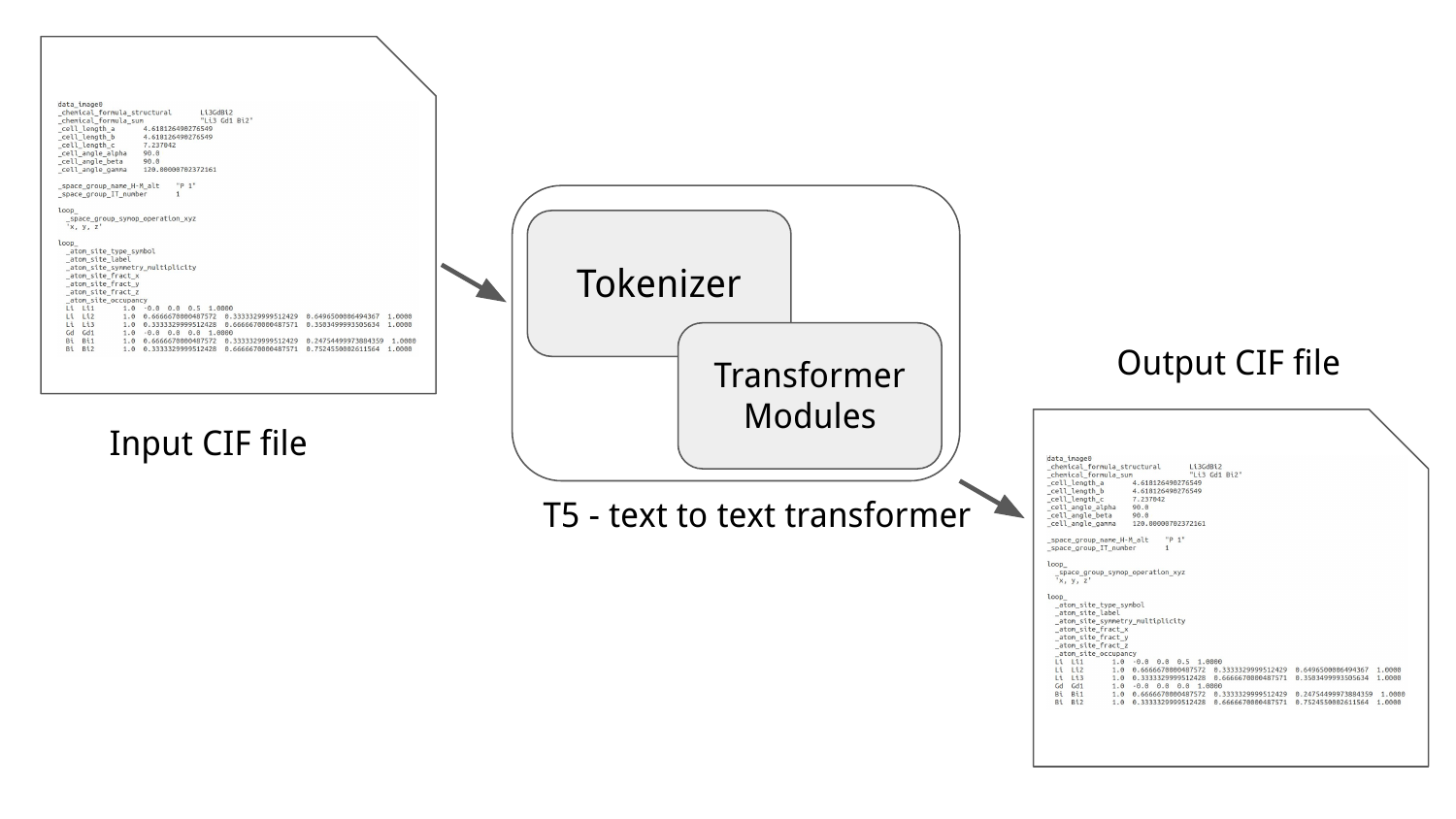}
    \caption{Overview of the relaxation training pipeline. The input structure from the LeMatTraj dataset is transformed to a CIF file using ASE. It is then fed into the T5 tokenizer and the fine-tuned transformer block layers to output a relaxed structure CIF file.}
    \label{fig:t2_relax_pipeline}
\end{figure}

The T5-Transformer is an LLM trained on the C4 text corpus for natural language processing tasks such as translation, summarization, and question answering. The model is fine-tuned to optimize text-based metrics (ROUGE~\cite{lin2004rouge} and BLEU~\cite{papineni2002bleu}) to ensure that much of the CIF file remains unchanged after relaxation, particularly the atomic elements in the structure. Using a weighted sum, the model is also fine-tuned to minimize the RMSE of the lattice parameters and atomic positions, which are given relative to the basis vectors. Since the crystal structures in the LeMatTraj dataset have a wide distribution, $\Delta$-learning~\cite{speckhard2025extrapolation} is applied to predict differences in lattice parameters (lengths and angles) from the initial to the DFT-relaxed structures, yielding a more Gaussian target distribution. The general framework of the approach is shown in Figure~\ref{fig:t2_relax_pipeline}.

The results are shown in Table~\ref{tab:t2_relax_parameter_crystal_cell_errors}. The MAE and RMSE of the predicted change in atomic positions for each structure in the dataset are shown in Table~\ref{tab:t2_relax_crystal_positions}. Overall, the observed errors are small, indicating that the LLM predicts relaxed structures with a good degree of accuracy. Without the ROUGE and BLEU constraints to preserve elemental composition between input and predicted CIF outputs, the model often removes or alters atoms. More extensive benchmarking is required to fully evaluate the approach, but the initial results presented here are promising.

\begin{table}[h!]
\centering
\begin{tabular}{|l|c|c|c|}
\hline
\textbf{Parameter} & \textbf{MAE} & \textbf{MSE} & \textbf{RMSE} \\
\hline
$\Delta a$ (\AA) & 0.053 & 0.006 & 0.080 \\
$\Delta b$ (\AA) & 0.053 & 0.006 & 0.080 \\
$\Delta c$ (\AA) & 0.074 & 0.020 & 0.140 \\
$\Delta \alpha$ ($^\circ$) & 0.086 & 0.036 & 0.189 \\
$\Delta \beta$ ($^\circ$) & 0.101 & 0.070 & 0.265 \\
$\Delta \gamma$ ($^\circ$) & 0.160 & 0.104 & 0.322 \\
\hline
\end{tabular}
\caption{Errors in predicting the lattice cell parameters.}
\label{tab:t2_relax_parameter_crystal_cell_errors}
\end{table}

\begin{table}[h!]
\centering
\begin{tabular}{|l|c|}
\hline
\textbf{Metric} & \textbf{Value} \\
\hline
MAE & 0.066 \AA \\
MSE & 0.252 \AA$^2$ \\
RMSE & 0.502 \AA \\
\hline
\end{tabular}
\caption{Errors in predicting the relaxed atom positions.}
\label{tab:t2_relax_crystal_positions}
\end{table}

\subsection*{Future Work}
Future work includes exploring larger LLMs, expanding the training data with additional datasets such as NOMAD~\cite{draxl2019nomad}, and performing systematic comparisons with state-of-the-art graph neural network methods used in materials science~\cite{rhodes2025orb, bechtel2025band}.

\subsection*{Open-source Materials}
Code and related resources are available on GitHub: \github{https://github.com/anky3733/Traj}\,; Video demo: \youtube{https://drive.google.com/file/d/1vn3kjPw_2SXSUMvP6i4DSSY9x9Igco8E/view}\,;




\section{MaterialSim AI Agent: An AI-Driven Agent for Automating Computational Materials Simulations}\label{sec:materialsim}



Designing new materials traditionally involves time-consuming, expert-driven processes to configure molecular dynamics (MD) simulations, select interatomic potentials, tune parameters, and analyze results. These manual steps hinder rapid exploration of chemical space. The MaterialSim AI Agent is an open-source, Python-based system that integrates large language models (LLMs) with computational materials science tools to automate these workflows. Using natural-language instructions, researchers can design, execute, and analyze MD simulations, extract properties such as radial distribution functions, mean squared displacements, elastic constants, and thermal conductivity, and leverage machine learning for accelerated predictions. Scalable from laptops to high-performance computing (HPC) clusters and featuring a web-based graphical interface for interactive 3D visualization, MaterialSim streamlines materials discovery by making simulation workflows more efficient, adaptive, and accessible. By reducing manual effort, the MaterialSim AI Agent team aims to accelerate innovation in materials design and scientific research.


\subsection*{Results}
The MaterialSim AI Agent is implemented as an open-source Python package that automates computational materials science tasks by integrating large language models with established simulation tools such as LAMMPS \cite{plimpton1995fast} and the Atomic Simulation Environment (ASE) \cite{larsen2017atomic}. It enables users to perform molecular dynamics simulations, compute material properties including radial distribution functions, mean squared displacements, elastic constants, and thermal conductivity, and apply machine-learning models for predictive analysis through natural-language queries. The framework interfaces with external databases such as Materials Project \cite{jain2013commentary}, NOMAD \cite{draxl2019nomad}, and the Open Catalyst Project \cite{chanussot2021open} to retrieve structures and augment data. A Streamlit-based graphical user interface provides interactive 2D and 3D visualizations, while compatibility with SLURM-managed HPC clusters ensures scalability. By translating prompts such as ``Simulate the thermal conductivity of silicon at 300 K'' into executable workflows, the MaterialSim AI Agent lowers the barrier to entry for researchers, educators, and interdisciplinary teams, thereby accelerating materials discovery.


\begin{figure}[H]
    \centering
    \includegraphics[width=0.60\linewidth]{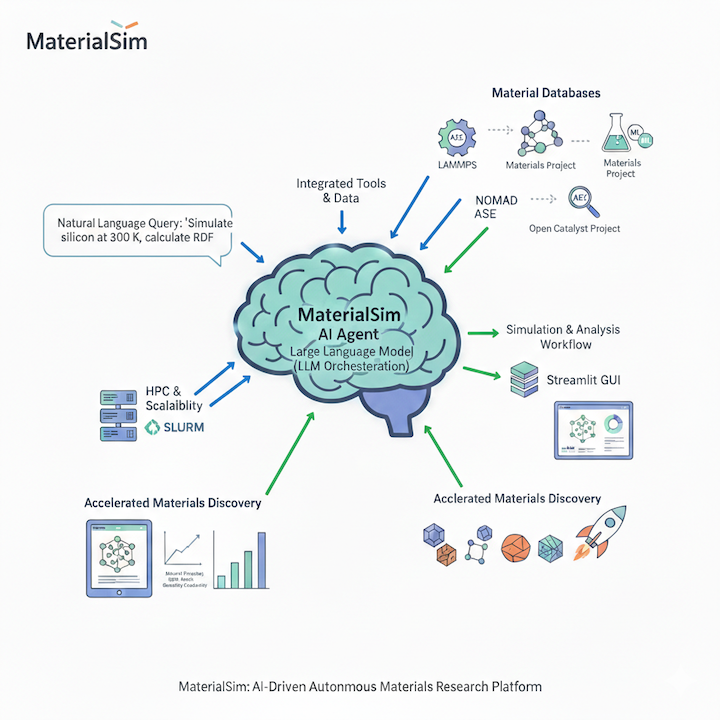}
    \caption{Workflow of the MaterialSim AI Agent.}
    \label{fig:materialSim pipeline}
\end{figure}

\subsection*{Future Work}
Future development will focus on evolving the MaterialSim AI Agent into a fully autonomous computational framework. The long-term goal is to enable researchers to initiate simulations, monitor execution, analyze results, and generate predictive insights entirely through natural-language interaction. By allowing complex modeling tasks to be conducted through conversational interfaces rather than specialized software environments, this approach has the potential to significantly broaden access to advanced computational methods and further accelerate simulation-driven scientific discovery.


\subsection*{Open-source Materials}
Code: \github{https://github.com/Awwal41/MaterialSim.git}\,; 
Demo video: \youtube{https://youtu.be/1wNyPK6Py5U}\,.





\section{SCARA: Steel Corrosion Agent for Risk Assessment}\label{sec:SCARA}



Steel is the backbone of industrial infrastructure; however, corrosion in aggressive environments such as (\ce{CO2}/\ce{H2S}/HCl) mixtures degrades structural integrity and contributes to billions of dollars in annual global losses~\cite{SCARA1}. Conventional corrosion assessment methods based on ASTM and NACE standards are slow, fragmented, and largely reactive, limiting timely decision-making~\cite{SCARA2}. Large language models (LLMs) present a transformative opportunity by integrating standards, datasets, and scientific literature to enable rapid corrosion risk prediction and informed alloy selection.

\begin{figure}[h]
    \centering
    \includegraphics[width=0.65\linewidth]{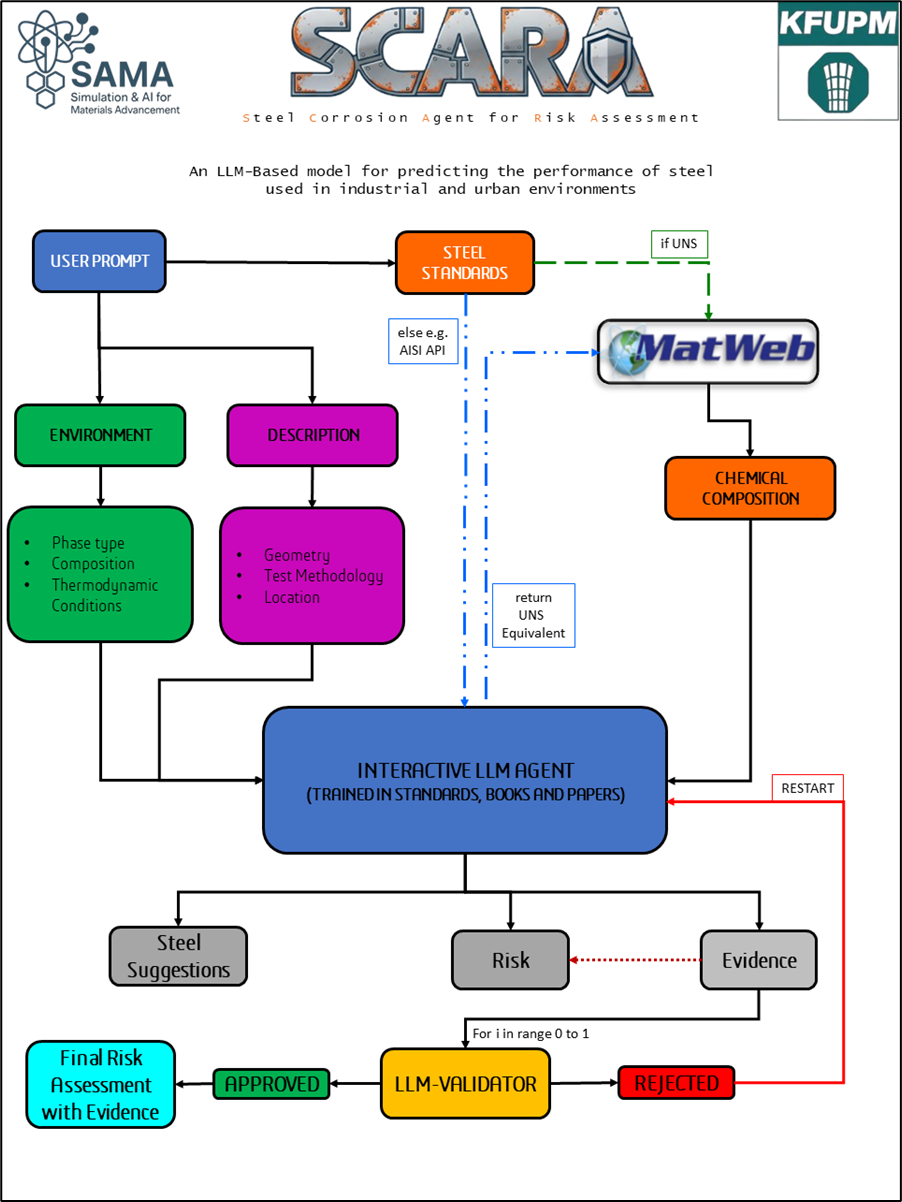}
    \caption{SCARA (Steel Corrosion Agent for Risk Assessment) workflow.}
    \label{fig:SCARA}
\end{figure}

The SCARA team introduces SCARA (Steel Corrosion Agent for Risk Assessment), an LLM-powered assistant designed to predict corrosion risk, assign standardized risk scores, and recommend safer alloy alternatives. As illustrated in Fig.~\ref{fig:SCARA}, the SCARA workflow begins with user input describing the operating environment and steel standard. The system links the specified standard to steel composition using an LLM in conjunction with the MatWeb database. The LLM agent, trained on scientific literature and industry standards, analyzes the input data to estimate corrosion risk and propose alternative steels. A validation module checks prediction consistency: validated outputs are reported with supporting evidence, while inconsistent results trigger recalculation. By integrating documentation, materials data, and LLM-based reasoning, SCARA delivers evidence-driven corrosion risk evaluations.

%

\subsection*{Results}
\begin{figure}[H]
    \centering
    \includegraphics[width=0.5\linewidth]{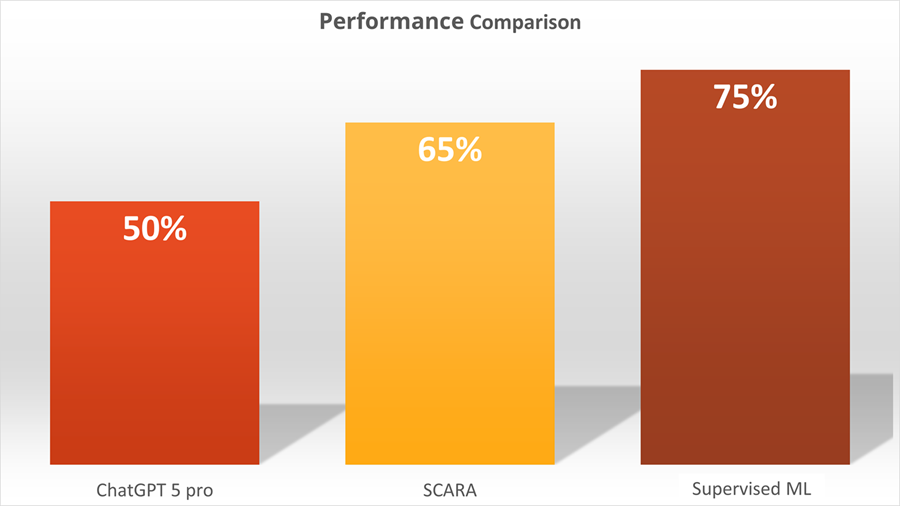}
    \captionof{figure}{Performance comparison of SCARA against a general-purpose LLM and supervised ML approaches.}
    \label{fig:SCARresults}
\end{figure}

Figure~\ref{fig:SCARresults} compares the performance of the SCARA agent against two alternative predictive approaches: a general-purpose OpenAI model prompted with identical inputs and supervised machine-learning algorithms. SCARA outperforms the general-purpose LLM by approximately 15\%, highlighting the importance of domain-aware agent design. As expected, the specialized supervised learning approach achieves higher overall accuracy, reflecting its task-specific training. These results are reasonable given that SCARA currently represents a demonstration version of a broader, general-purpose corrosion expert agent.

Overall, SCARA demonstrates the potential of LLM-based agents to evaluate corrosion risk by integrating standards, scientific literature, and materials databases into a unified decision-support platform. Unlike conventional corrosion assessment methods, which are often fragmented and scope-limited, SCARA enables rapid prediction, flexible risk evaluation, and alloy recommendation grounded in scientific evidence.

%

\subsection*{Future Work}
Although currently in a demonstration phase, SCARA already outperforms general-purpose AI models as a scalable digital corrosion expert. While specialized supervised machine-learning methods achieve higher predictive accuracy, SCARA’s primary strengths lie in its adaptability, explainability, and integration of domain knowledge. With continued refinement, SCARA has the potential to become a reliable tool for improving safety, reducing costs, and enhancing materials selection across critical industrial sectors.


\subsection*{Open-source Materials}
\textbf{Code:} \github{https://github.com/mo-alkubaish/SCARA}\,; 
\textbf{Demo video:} \youtube{https://www.linkedin.com/posts/chahd-rahyl-adjmi-47735a230_excited-to-share-our-project-from-the-llm-activity-7372417195986952192-Yvi5}\,.





\section{ChromatographyMiner: Interactive Platform for Gas Chromatography Data Analysis with AI-Assisted Interpretation}
\label{sec:chromatographyminer}



The Team ChromatographyMiner presents \texttt{ChromatographyMiner}, an interactive web-based platform designed to simplify the analysis of gas chromatography--mass spectrometry (GC--MS) and two-dimensional gas chromatography--mass spectrometry (GC$\times$GC--MS) data. The platform supports files from any instrument manufacturer and accepts multiple input types, including raw data files, two-dimensional chromatogram images, and one-dimensional mass spectra provided in spreadsheet or image formats. A key feature of the system is its flexible and robust spectral library management, which enables automatic downloading of reference spectra from the MassBank library, loading of local databases such as MassBank and MassBank of North America, and ingestion of user-uploaded spectral libraries.

To efficiently identify chemical compounds, \texttt{ChromatographyMiner} employs a fast matching algorithm that rapidly narrows candidate compounds from very large libraries. Candidate matches are ranked using a combined metric based on mass spectral similarity and retention-time deviation. In addition, an artificial intelligence module generates clear, structured explanations for top-ranked matches, helping users understand the rationale behind compound identification. Identified results can be curated and exported as downloadable PeakCards, providing complete, transparent, and traceable records of each analysis.


\begin{figure}[H]
  \centering
  \includegraphics[width=0.98\textwidth]{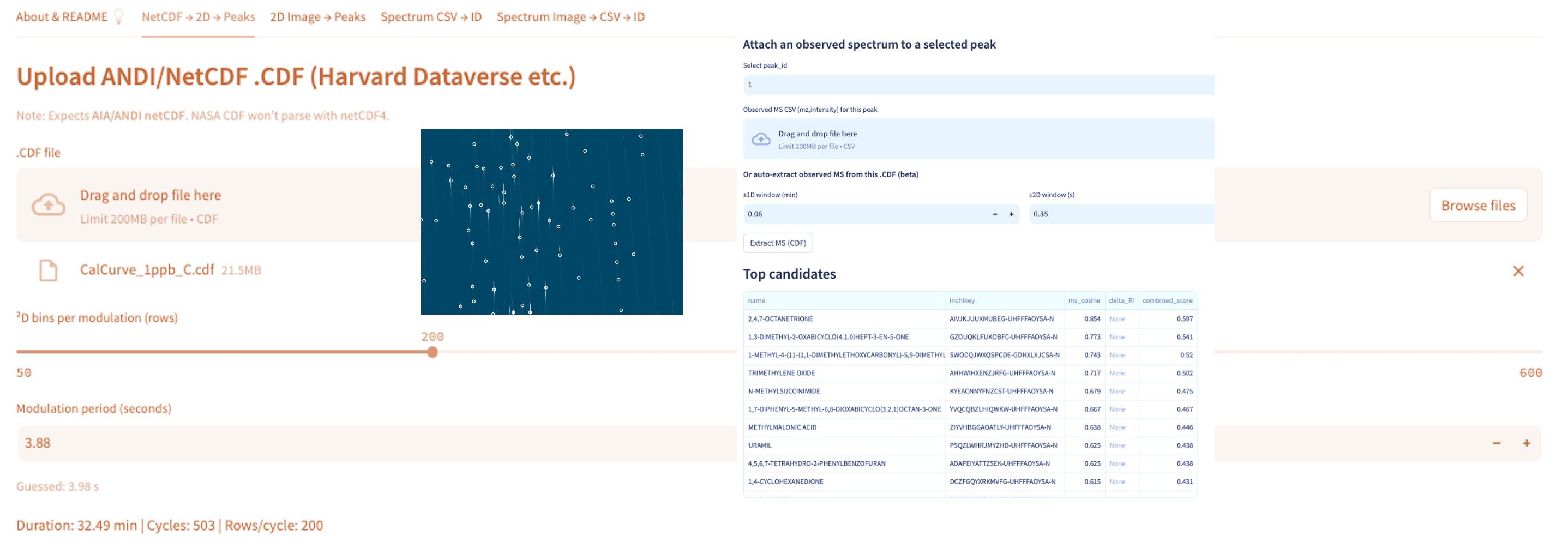}
  \caption{User interface of \texttt{ChromatographyMiner}, illustrating the workflow for uploading and analyzing two-dimensional gas chromatography--mass spectrometry (GC$\times$GC--MS) data in NetCDF format. The platform supports drag-and-drop input of .CDF files, chromatogram visualization, automatic mass spectrum extraction, and compound identification using spectral libraries such as MassBank and MassBank of North America.}
  \label{fig:ChromatographyMiner}
\end{figure}

\subsection*{Results}
The platform provides a fast, flexible, and vendor-neutral solution for identifying chemical compounds from GC--MS and GC$\times$GC--MS datasets. By combining efficient top-ion pre-filtering with large-scale public spectral libraries such as MassBank and MassBank of North America, \texttt{ChromatographyMiner} achieves rapid and high-quality tentative compound identifications, as shown in Fig.~\ref{fig:ChromatographyMiner}.

The curated output file, \texttt{saved\_refs.json}, links observed chromatographic peaks to definitive reference spectra from the spectral libraries, ensuring traceability and reproducibility across experiments. Benchmark evaluations demonstrate:
\begin{itemize}
    \item \textbf{Speed:} Up to 30$\times$ faster performance compared to conventional full-spectrum cosine-matching workflows on libraries exceeding 50{,}000 spectra.
    \item \textbf{Accuracy:} Correct compound retrieval for approximately 90\% of benchmark peaks from the MassBank dataset within the top-five ranked candidates.
    \item \textbf{Scalability:} Efficient processing of MassBank of North America archives larger than 5~GB using a streaming mass spectral parser without memory limitations.
\end{itemize}

%

\subsection*{Future Work}
Future development will focus on expanding compatibility with additional vendor-neutral formats such as mzML, implementing automated co-elution deconvolution algorithms, and integrating external chemical databases including PubChem and ChemSpider to enrich compound identification with chemical properties and metadata. Planned updates also include the release of a dedicated Python software development kit (SDK) and a web-based application programming interface (API) to support high-throughput, programmatic analysis workflows.


\subsection*{Open-source Materials}
\textbf{Code and data:} \github{https://github.com/msehabibur/gcxgc_peakcards}\,; 
\textbf{Demo video:} \youtube{https://www.youtube.com/watch?v=EUzJ4XTkCG4}\,.




\section{\texttt{guillemot}: Automated Rietveld Refinement Using Agents}\label{sec:guillemot}



The datalab team investigates whether a multimodal large language model (LLM) agent can automate a traditionally human-intensive data-fitting task: pattern fitting of powder X-ray diffraction (PXRD) data. A major bottleneck in materials discovery is the analysis of the rapidly increasing volume of characterization data generated by high-throughput experiments. Existing AI approaches typically rely on ensemble models trained on synthetic data or automated refinement pipelines that select the best fit from user-supplied phases \cite{Szymanski2021, Szymanski2023, Chang2025}. These approaches contrast with traditional Rietveld refinement workflows, where experts iteratively improve fits through visual inspection rather than relying solely on quantitative metrics.

High-quality Rietveld analysis often requires expert judgement to avoid overfitting and non-physical results. The team hypothesized that the ability of multimodal LLMs to interpret visual data while incorporating chemical knowledge could enable automated systems that emulate this expert-driven workflow. By allowing an AI agent to perform refinements and interpret both visual and textual outputs directly, a wide range of structural modeling problems—within the scope of refinement software—can be explored. This approach offers a key advantage over pre-trained models, which are constrained by the structures represented in their training data.


\subsection*{Results}

\begin{figure}[ht]
  \centering
  \includegraphics[width=0.9\textwidth]{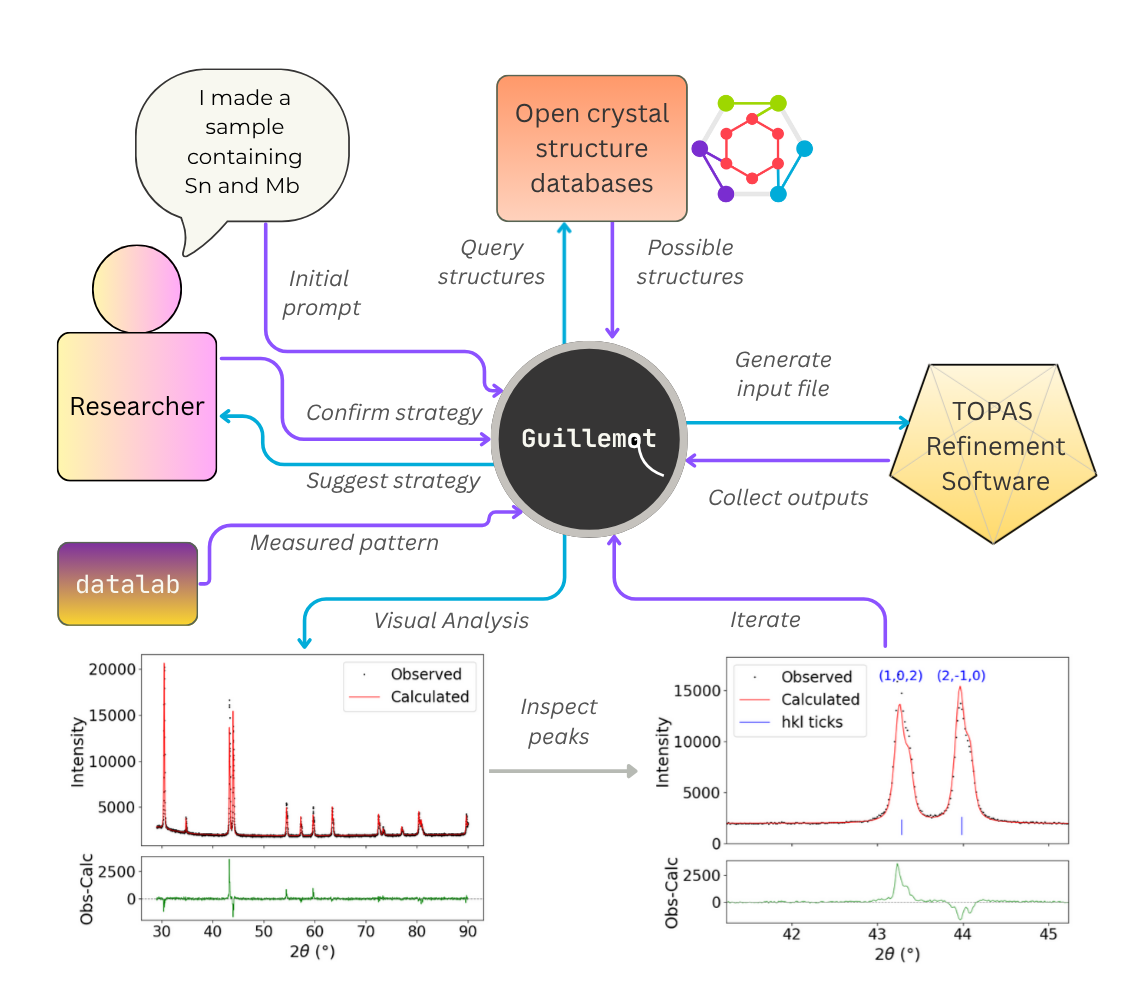}
  \caption{Workflow cycle of \texttt{guillemot}. The AI agent interacts with users, crystal structure databases, and TOPAS to automate a human-like Rietveld refinement process.}
  \label{fig:Guillemot}
\end{figure}

A provider-agnostic, LLM-powered agent named \texttt{guillemot} was developed using the Pydantic AI framework and equipped with a suite of task-specific tools. Upon receiving a diffraction pattern—optionally retrieved from a \emph{datalab} instance \cite{Evans2025a}—and a user-provided rubric (e.g., ``I have made a sample containing Mn and Sb''), the agent performs a unified query across the federation of approximately 20 open crystal structure databases within the OPTIMADE consortium \cite{Evans2024a, Evans2021}. Structures with prior experimental validation are prioritized.

Based on the retrieved candidates, \texttt{guillemot} determines an initial set of trial phases and generates a TOPAS \cite{Coelho2018} input file to execute a Rietveld refinement. The agent processes the refinement output into a graphical summary and can replot or zoom into specific regions of interest, such as dominant peaks or impurity features, mirroring expert human practice. The agent then evaluates refinement quality and autonomously decides on subsequent actions, including phase substitution, impurity modeling, strain inclusion, or other refinement strategies exposed by TOPAS.

Notably, without any model fine-tuning, frontier LLMs were able to generate valid TOPAS input files after only a small number of examples, iteratively correcting formatting or syntax errors reported by TOPAS during execution.

\subsection*{Future Work}
This study focused on a specific analysis task: Rietveld refinement using TOPAS Academic, a widely used proprietary software for advanced PXRD analysis. Future work will extend this approach to additional refinement tools, including open-source alternatives such as GSAS and Profex/BGMN, as well as software used for fitting other types of experimental data. Integration with research data management platforms such as \emph{datalab} is also planned, enabling background analysis of stored PXRD datasets.

Further improvements may be achieved through model fine-tuning or enhanced prompting using TOPAS documentation and domain-specific diffraction knowledge. Another observed limitation was occasional misinterpretation of refinement output plots. Systematic evaluation of vision–language models’ ability to parse scientific figures will be an important direction for improving robustness and designing output formats optimized for automated visual interpretation.


\subsection*{Open-source Materials}
\textbf{Code:} \github{https://github.com/datalab-org/guillemot}\,; 
\textbf{Demo video:} \youtube{https://www.youtube.com/watch?v=0hpQB-fnDRQ}\,.




\section{XAScribe: An AI-Powered Platform for Automated, Manuscript-Ready X-ray Absorption Spectroscopy Analysis}
\label{sec:xascribe}


X-ray Absorption Spectroscopy (XAS) is a powerful technique that measures the absorption of X-rays over a range of energies, enabling the determination of oxidation states and coordination environments of atoms in materials \cite{bunker2010xafs}. This capability is particularly important for nickel–cobalt–manganese oxide (NCM) battery cathodes, where the oxidation state and local coordination of Ni evolve during electrochemical cycling \cite{yan2020nmcxas}. Despite the widespread reporting of XAS measurements, thorough analysis and interpretation remain time-intensive and require significant domain expertise.  

The XAScribe team explored how this bottleneck can be addressed by combining the text-generation capabilities of modern transformer-based large language models with state-of-the-art machine learning (ML) models for quantitative spectroscopy analysis. Building on the ML framework developed by Chen \emph{et al.} \cite{chen2024cdf}, the team demonstrates an end-to-end workflow that automates the generation of manuscript-ready analysis and discussion for XAS datasets. This approach is designed to significantly reduce analysis time for both academic and industrial research teams studying coordination environments and oxidation states using XAS.

\begin{figure}[H]
    \centering
    \includegraphics[width=1\linewidth]{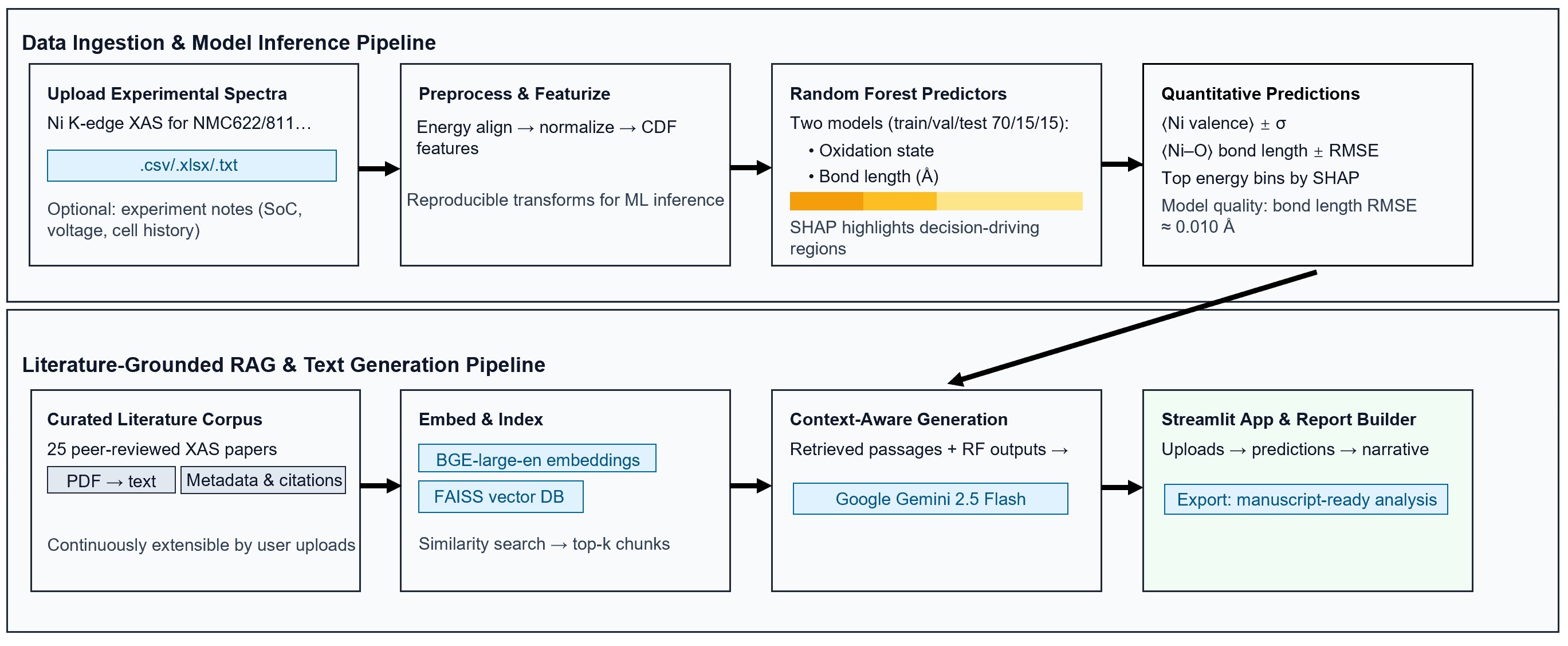}
    \caption{Workflow of XAScribe.}
    \label{fig:XAScribe workflow}
\end{figure}

\subsection*{Results}
XAScribe is an AI-assisted platform developed to automate the analysis and interpretation of Ni K-edge X-ray Absorption Spectroscopy data (Fig.~\ref{fig:XAScribe workflow}). The system illustrates how large language models (LLMs) and machine learning can accelerate spectroscopy workflows by producing manuscript-ready scientific analyses.  

\begin{figure}[H]
    \centering
    \includegraphics[width=1\linewidth]{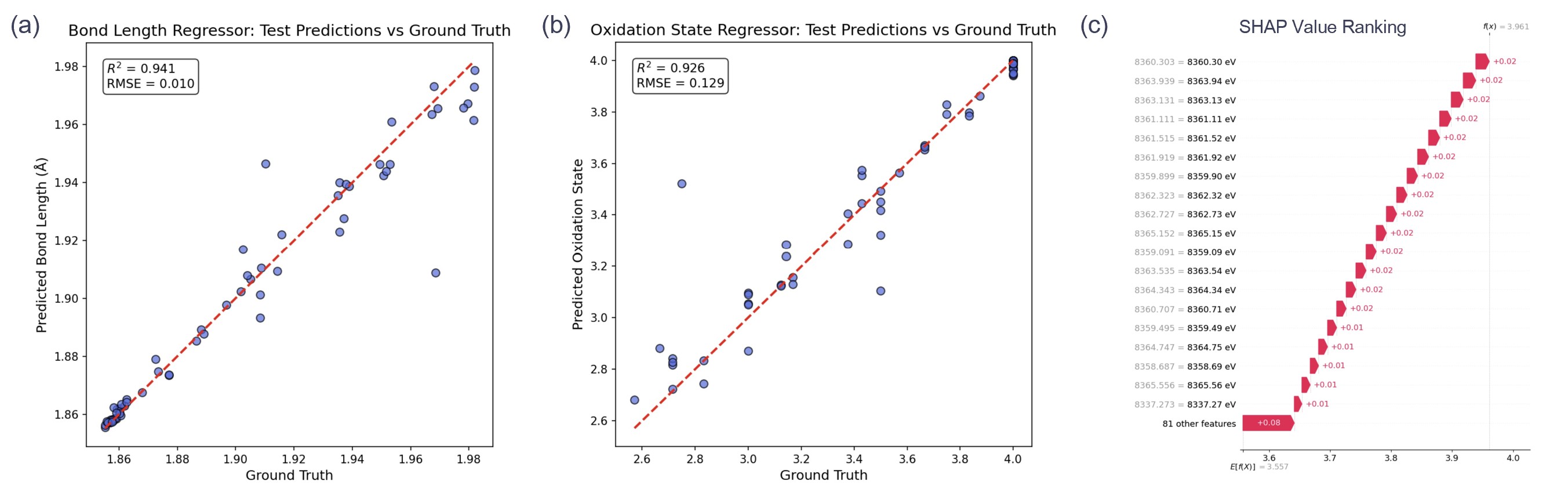}
    \caption{Results of the random forest models: (a) test predictions for Ni–O bond length, (b) test predictions for Ni oxidation state, and (c) representative SHAP value rankings highlighting energy regions that drive model decisions.}
    \label{fig:XAScribe result}
\end{figure}

The workflow integrates a Retrieval-Augmented Generation (RAG) framework using Google Gemini \cite{lewis2020retrieval}, which performs context-aware text generation from a curated corpus of 25 peer-reviewed publications containing XAS data and expert analysis. In parallel, a Random Forest Regressor trained on publicly available XAS datasets \cite{chen2024cdf} predicts oxidation states and Ni–O bond lengths. Physical interpretability is provided through Shapley Additive Explanations (SHAP) \cite{lundberg2017shap}, which identify energy regions in the spectrum that drive model predictions (Fig.~\ref{fig:XAScribe result}). Efficient retrieval and embedding are enabled by FAISS similarity search \cite{johnson2017faiss}.  

The full pipeline is deployed as an interactive Streamlit web application, allowing users to upload experimental spectra and receive quantitative predictions alongside automatically generated, manuscript-ready analysis text.

\subsection*{Future Work}
Future directions are twofold. First, the approach can be extended to additional elements and absorption edges by expanding the training datasets. Second, the overall workflow can be adapted to other experimental characterization techniques, such as X-ray diffraction (XRD), electron energy loss spectroscopy (EELS), and related modalities, broadening its applicability across materials characterization.

\subsection*{Open-source Materials}
\textbf{Code:} \github{https://github.com/Oscuro-Phoenix/xascribe}\,; \textbf{Demo video:} \youtube{https://www.youtube.com/watch?v=E2KHzhEcL8c}\,.




\section{BAKER: Automated Spawning of Specialized Research Assistants}\label{sec:baker}


The \textit{LLM Hackathon for Applications in Materials Science \& Chemistry} highlights the potential of domain-specialized, agentic research assistants to accelerate scientific workflows or to act as modular components within more general-purpose systems. As noted by Hu \textit{et al.}~\cite{hu2025automateddesignagenticsystems}, “the history of machine learning teaches us that hand-designed solutions are eventually replaced by learned solutions,” suggesting that automated generation of research agents themselves is a natural next step. In addition, recent work by the team~\cite{akke2025bayesian} shows that agents often benefit from deliberately limited toolsets and context, further motivating the automated development of minimal, task-specific assistants.

\subsection*{Results}
\begin{figure}[h!]
    \centering
    \includegraphics[width=0.7\linewidth,trim=0 250 0 300,clip]{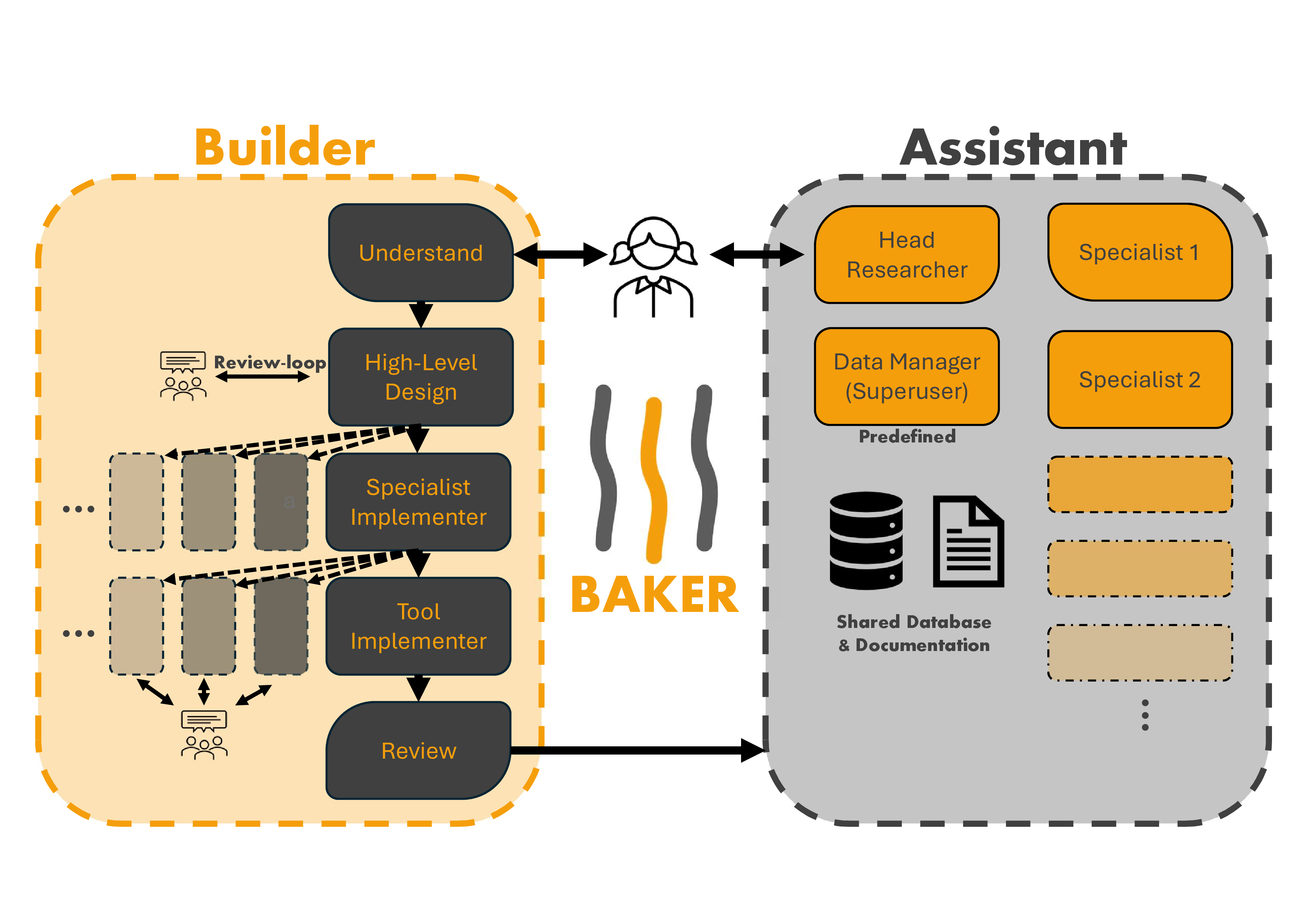}
    \caption{\textbf{Overview of the BAKER framework.}
    The system comprises a \textit{Builder} module that automatically designs, implements, and reviews specialized research assistants, and an \textit{Assistant} module that interacts with the user and manages execution. Each assistant is initialized with a predefined Data Manager node that oversees shared databases and documentation, while the Builder autonomously spawns all additional specialized nodes and tools through iterative review loops.}
    \label{fig:baker-descr}
\end{figure}

BAKER is proposed as a fully automated framework for spawning specialized research assistants. The workflow begins with the user answering a set of questions about the experiment or project to ensure that BAKER accurately \textbf{understands} the task (Fig.~\ref{fig:baker-descr}, \textbf{Builder}). Based on this input, BAKER constructs a multi-agent assistant through four stages.

First, a strategist–critic loop produces a \textbf{high-level design} of the assistant, distributing responsibilities across nodes and assigning appropriate tools to each. Second, multiple \textbf{specialist} node-cards are implemented in parallel, with each node specifying data permissions, available tools, and guidelines for interoperability with other nodes. Third, all required \textbf{tools} are implemented within a coder–critic loop to enforce compliance with database schemas, naming conventions, and cross-tool compatibility. Finally, a \textbf{review} phase analyzes the completed assistant to identify bottlenecks or integration issues and ensures that these are corrected before deployment.

Once construction is complete, the user is prompted by the newly generated \textbf{assistant} (Fig.~\ref{fig:baker-descr}) to proceed with their work. The only pre-defined node is a \textit{Data Manager} with superuser privileges, responsible for purging and pruning datasets in the shared database. All agents operate over a common vector database accessible to every tool, maintaining a unified information environment. Each node is also initialized with a minimal default toolset for database interaction.

Using BAKER, the team generated a simple agentic workflow for LLM-assisted Bayesian optimization in under five minutes—a process that would typically require a skilled LLM developer several hours to days. This result demonstrates BAKER’s potential to accelerate discovery and to democratize the design of agentic research assistants.

\subsection*{Future Work}
At present, BAKER operates with Python execution and databases hosted locally. Future development will focus on converting the full pipeline to generate complete Model Context Protocol (MCP) servers and connecting these to a user-friendly interface where agents can be instantiated at the press of a button. In terms of workflow scope, immediate goals include (i) enabling the Builder to store highly successful tools in a database for reuse in subsequent generations, (ii) allowing automatic generation of literature-search specialist agents, and (iii) directly integrating recent models for automated tool generation and refinement.

\subsection*{Open-source Materials}
All code is available on GitHub: \github{https://github.com/mattiasutancykeln/Baker}.



\section{\textbf{PolyPredictor}: Multimodal Representation of Polymers Using Large Language Model Embeddings}\label{sec:polypredictor}

%
%

Polymers are inherently multi-order systems whose macroscopic properties are governed by multiple interacting factors, including chemical structure, crystalline order, chain length and stoichiometry, phase behavior, and orientation \cite{D3PY00395G}. Current state-of-the-art approaches typically rely on one-dimensional PSMILES representations, which often fail to capture chain-level and microstructural features that critically influence polymer performance.  

The PolyPredictor team explored the use of multimodal embeddings generated by large language models (LLMs) to address these limitations. Starting from a PSMILES input, the workflow generates three complementary embeddings that encode chemical, structural, and conformational information. These embeddings are fused into a continuous multimodal representation and used as input to a graph-based learning model. The approach is evaluated on prediction of the glass transition temperature ($T_g$), a property that is highly sensitive to polymer chemistry, crystallinity, and conformation. The model is trained on a synthetic dataset of 10,000 polymers and validated on a separate set of 7,000 polymers with experimentally verified $T_g$ values.

\subsection*{Results}
\begin{figure}[h!]
    \centering
    \includegraphics[width=\linewidth]{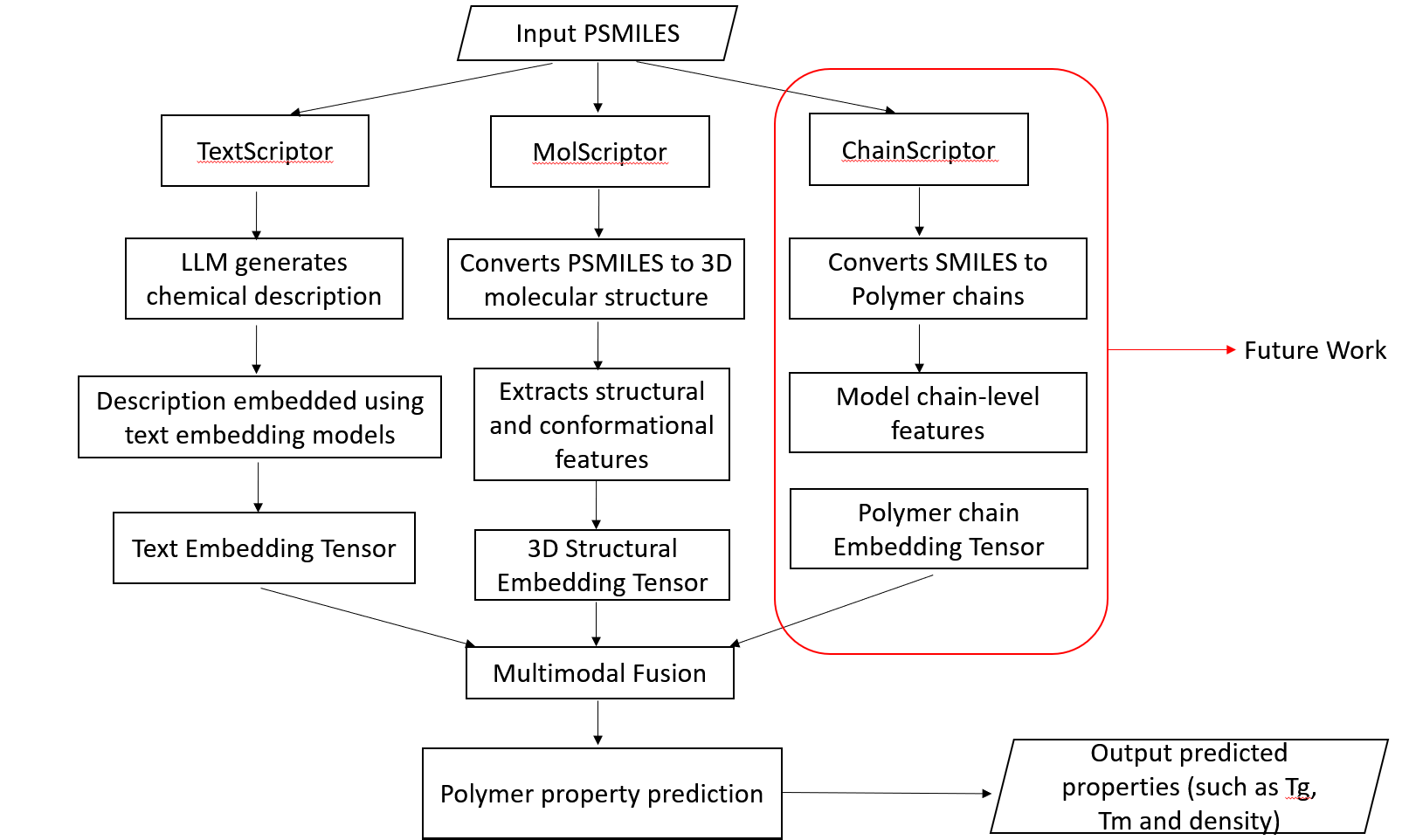}
    \caption{Overview of the PolyPredictor workflow.}
    \label{fig:ppred-descr}
\end{figure}

An overview of the PolyPredictor workflow is shown in Fig.~\ref{fig:ppred-descr}.

To construct a chemical description, a LangChain agent powered by a commercial Gemini 2.5 Pro LLM is employed. The agent is guided by detailed system prompts and few-shot examples to produce structured, natural-language descriptions of the polymer repeat unit. These descriptions are converted into vector embeddings using OpenAI’s \texttt{text-embedding-3-large} model, with multiple embedding dimensions (768, 1536, and 4096) explored.  

Molecular-level information is captured using the Uni-Mol representation, which produces a 1536-dimensional embedding encoding three-dimensional geometry and conformational features. All embeddings are projected into a shared latent space, where a gated fusion mechanism combines and updates the representations. The fused embeddings are then passed to a regression neural network and trained using the specified hyperparameters. To assess the contribution of each modality, additional models were trained using individual embeddings only. Performance comparisons are summarized in Table~\ref{tab:embedding-results}.

\begin{table}[h!]
\centering
\renewcommand{\arraystretch}{1.15}
\begin{tabular}{l l l l c c c c}
\toprule
\textbf{Embed inputs} &
\textbf{Embed size} &
\textbf{LLM} &
\textbf{Pred model} &
\textbf{Test MAE} &
\textbf{Test R\textsuperscript{2}} &
\textbf{Val MAE} &
\textbf{Val R\textsuperscript{2}} \\
\midrule
Text, Structural & {[}4096, 1536{]} & Llama3 & RNN        & 19.6 & 0.82 & 19.3 & 0.84 \\
Text, Structural & {[}768, 1536{]}  & Gemini & RNN        & 25.7 & 0.78 & 22.7 & 0.78 \\
Text             & {[}768{]}        & Gemini & RNN        & 26.3 & 0.71 & 24.3 & 0.70 \\
Text, Structural & {[}1536, 1536{]} & Gemini & RNN        & 24.2 & 0.76 & 25.1 & 0.68 \\
Text, Structural & {[}1536, 1538{]} & Gemini & RNN + GBR  & 18.8 & 0.84 & 18.3 & 0.87 \\
\bottomrule
\end{tabular}
\caption{Performance comparison of embedding configurations and prediction models for glass transition temperature ($T_g$) estimation.}
\label{tab:embedding-results}
\end{table}

To further improve predictive accuracy, an ensemble model combining a regression neural network with a gradient boosting regressor was evaluated, yielding the strongest overall performance, as reflected in the table.

\subsection*{Future Work}
Future work will focus on developing fine-tuned LLMs capable of generating more accurate chemical descriptions from truly multimodal inputs, including both PSMILES strings and three-dimensional structure files of repeat units. While the current embeddings capture repeat-unit-level chemical and structural information, there is substantial opportunity to extend this approach to chain-level representations. For example, LLMs could be used to generate natural-language descriptions of polymer chains derived from repeat units \cite{guo2021polygrammargrammardigitalpolymer}, or chain structures could be generated via molecular dynamics simulations and subsequently embedded.  

In addition, the semi-crystalline nature of many polymers plays a critical role in determining material properties. Large language models may be leveraged to predict polymer crystallinity and incorporate this information directly into the downstream property prediction models.

\subsection*{Open-source Materials}
\textbf{Code:} \github{https://github.com/554181320angela-lang/Poly-predictor}\,;
\textbf{Demo video:} \youtube{https://youtu.be/VM0r47ZCnyw}\,;

%


\section{Natural-Language-Driven Closed-Loop Optimization for Robotic Liquid-Handling Systems}
\label{sec:sdl-smart}




Automated experimentation has accelerated materials discovery, yet most self-driving laboratories (SDLs) still rely on platform-specific scripts and interfaces. This fragmentation limits interoperability and accessibility for researchers.  
The SDLsmart team demonstrates a natural-language-driven optimization workflow that connects a large language model (LLM) to a robotic liquid-handling system through the Model Context Protocol (MCP) and the IvoryOS orchestration layer \cite{mcp2024, zhang_ivoryos_2025}. The system enables users to specify experimental goals directly in natural language, which are automatically translated into executable workflows for laboratory hardware.


\subsection*{Results}

\begin{figure}[h!]
    \centering
    \includegraphics[width=\linewidth]{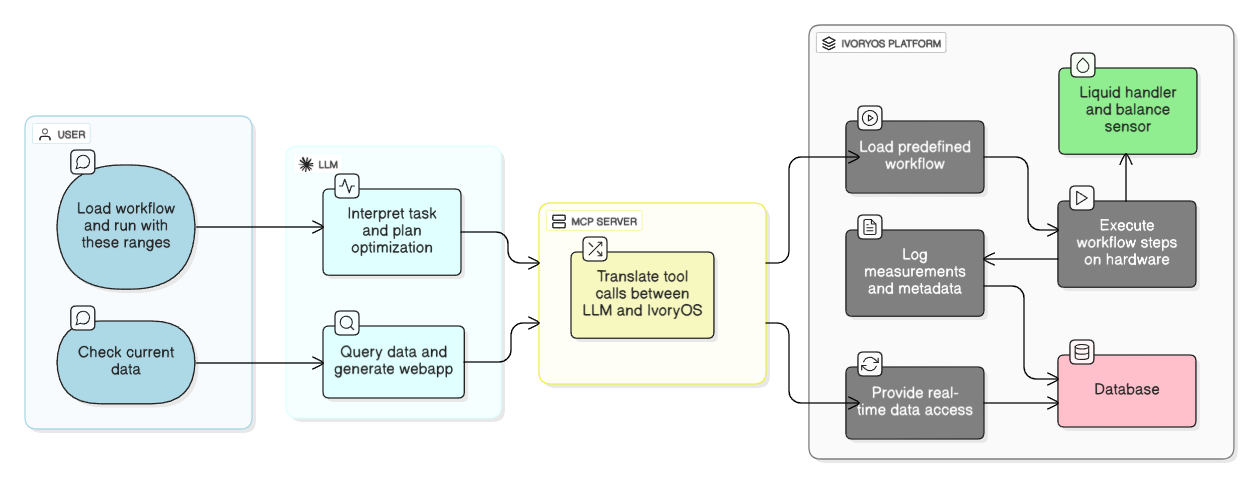}
    \caption{Closed-loop optimization enabled through MCP–IvoryOS integration. An LLM issues natural language commands via the MCP server to IvoryOS, which loads and executes workflows on connected hardware while logging experimental results to a database accessible to the LLM in real time.}
    \label{fig:sdlsmart}
\end{figure}

The targeted use case focused on liquid-handling accuracy optimization, a common challenge in automated synthesis workflows. An LLM agent was prompted to “optimize the mobile liquid handler accuracy in 60 trials,” leveraging predefined workflows available on the robotic platform (Figure~\ref{fig:sdlsmart}). Using MCP tools such as \texttt{platform-info} and \texttt{run-workflow-campaign}, the agent autonomously constructed a closed-loop optimization process with six tunable parameters, including air gaps, aspiration and dispense speeds, and delay times.  

The SDLsmart team employed the Ax Bayesian optimization framework to guide experimental selection across trials. The system executed all experiments, logged results, and visualized intermediate data without any manual scripting. This demonstrates that natural language can serve as a direct interface for adaptive, closed-loop experimentation on physical laboratory hardware.


\subsection*{Future Work}

Future work will focus on enabling LLM-driven workflow generation from scratch and incorporating LLM-in-the-loop parameter suggestion and decision-making. These extensions aim to allow the same natural language interface to orchestrate complex, multi-step synthesis workflows with increased autonomy and adaptability.


\subsection*{Open-source Materials}

Code is available on GitLab:  
\gitlab{https://gitlab.com/heingroup/llm-hackathon-2025}




\section{Unified Natural Language Control of Lab Modules}
\label{sec:unlclm}




Integrating large language models (LLMs) with legacy scientific and industrial hardware remains challenging, as many such systems lack modern application programming interfaces. The MCP4SDL team presents an interoperability framework that enables an LLM to act as a natural language interface for controlling complex laboratory systems. This capability is realized through IvoryOS, which exposes available hardware modules to the LLM via a Model Context Protocol (MCP) server \cite{mcp2024, zhang_ivoryos_2025}.


\subsection*{Results}

\begin{figure}[h!]
    \centering
    \includegraphics[width=\linewidth]{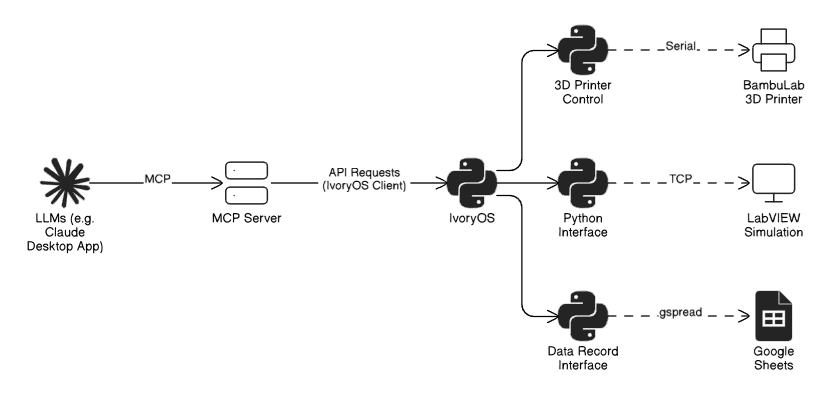}
    \caption{Interoperability architecture illustrating how an LLM communicates through the IvoryOS MCP server to Python-based modules controlling 3D printers, LabVIEW simulations, and automated data logging.}
    \label{fig:mcp4sdl}
\end{figure}

The MCP4SDL workflow enables researchers to perform complex experimental tasks using simple natural language instructions. As shown in Figure~\ref{fig:mcp4sdl}, the architecture allows a large language model (e.g., Claude) to interact with external laboratory tools through the Model Context Protocol server. The MCP server bridges the LLM to IvoryOS, which automatically discovers, summarizes, and exposes available methods from heterogeneous Python scripts.

These scripts interface with diverse systems, including a LabVIEW simulation environment, a 3D printer controlled via serial communication, and Google Sheets used for automated data logging. This modular execution chain—LLM → MCP Server → IvoryOS → Hardware—eliminates the need to re-engineer existing legacy infrastructure, enabling immediate compatibility with AI-driven laboratory automation.


\subsection*{Future Work}

Future work will extend this framework toward closed-loop, interoperable orchestration with integrated safety monitoring and autonomous decision-making. Incorporating real-time feedback and safety constraints will enable LLM agents to operate laboratory hardware with greater robustness and reliability.


\subsection*{Open-source Materials}

Code is available on GitHub:  
\github{https://github.com/ivoryzh/MCP4SDL}





\section{F.A.D.E: A Fully Agentic Drug Engine}\label{sec:FADE}




Drug discovery remains prohibitively expensive and time-consuming, often requiring more than a decade and billions of dollars to bring a single therapeutic to market. While computational approaches have been introduced to accelerate this process, they face critical limitations, including proprietary barriers that restrict community access, fragmented pipelines that demand specialised expertise across domains, and AI-driven methods that frequently lack rigorous physics-based validation. These challenges create high barriers to entry and slow innovation in life-saving drug development.  

The F.A.D.E team addresses these limitations through a fully automated, open-source, multi-agent workflow that democratizes drug discovery. Users can submit natural language queries—such as “Identify drug candidates targeting receptor \textbf{X}”—and the system autonomously orchestrates the full pipeline, spanning target identification, structure retrieval, de novo ligand generation, property screening, and binding affinity prediction.


\subsection*{Results}

The F.A.D.E platform implements a three-stage pipeline for identifying small-molecule drug candidates directly from natural language queries. The workflow begins with a hierarchical database search that prioritizes available structural information for the specified target. When a ligand-bound structure is available, the binding site is directly extracted and passed to the candidate generation stage. If only an apo structure exists, binding sites are identified using physics-based approaches \cite{leguilluox2009}.  

In cases where no experimental structure is available, the target structure is predicted using Boltz-based models \cite{passaro2025boltz2, wohlwend2024boltz1}. Following binding site identification, drug-like molecules are generated using a diffusion-based generative model \cite{arne2024} and subsequently ranked using QSAR predictions and binding affinity calculations \cite{passaro2025boltz2, wohlwend2024boltz1}.  

This workflow was validated on two representative targets—EGFR (PDB ID: 7SI1) and CRBP1 (PDB ID: 5H9A)—demonstrating the system’s ability to produce viable candidate molecules across distinct protein classes.


\begin{figure}[h]
    \centering
    \includegraphics[width=\linewidth]{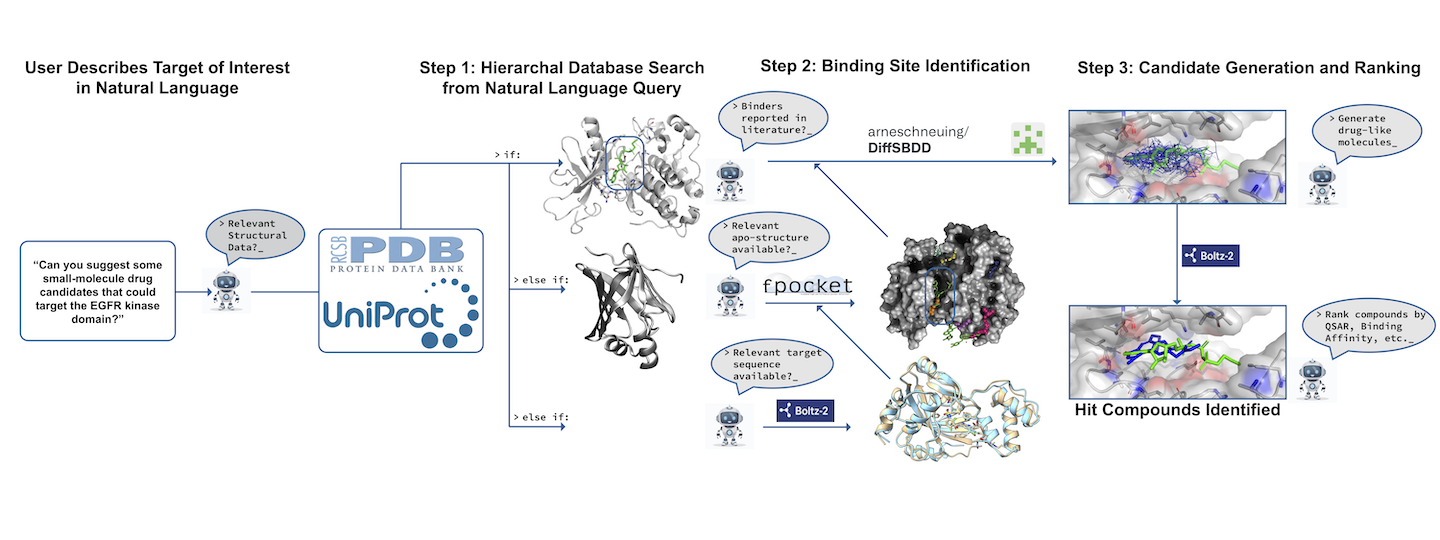}
    \caption{\textbf{FADE workflow for natural language-driven drug candidate discovery.} User queries describing targets of interest are processed through three sequential stages: (1) hierarchical database search for structural data or sequences; (2) binding site identification; and (3) computational generation and ranking of drug-like molecules using QSAR and binding affinity metrics to identify hit compounds.}
    \label{fig:FADE-pipeline}
\end{figure}

\subsection*{Future Work}

Future work will focus on expanding the validation and applicability of the F.A.D.E platform. Planned efforts include evaluating performance across a broader and more diverse set of benchmark targets, systematic comparison with established computational drug discovery methods, and validation against available experimental binding data to further assess predictive reliability.


\subsection*{Open-source Materials}

Code is available on GitHub:  
\github{https://github.com/Naveen-R-M/F.A.D.E}





\section{MatFOMGen: A Collaborative Tool for Identifying Domain-Specific Chemical and Materials Figures-of-Merit for Autonomous Labs}\label{sec:MatFOMGen}




MatFOMGen is a tool for autonomous and self-driving laboratories designed to rapidly identify appropriate computational targets, or figures-of-merit (FOMs), across diverse sub-domains of materials science and chemistry. These FOMs serve as objective functions toward which autonomous materials modeling, discovery, and synthesis workflows can be directed.  

Self-driving laboratories have attracted significant academic \cite{ceder_alab, alabos} and private-sector interest in recent years due to their potential to accelerate the typically years-long research and development cycles preceding the deployment of novel materials in industrial applications. A key differentiator in the competitiveness of an autonomous laboratory is its ability to pivot rapidly toward new computational targets driven by the heterogeneous requirements of industrial stakeholders. For example, a single self-driving lab may be expected to support both petrochemical applications and the development of high-entropy alloys using the same underlying software and hardware infrastructure.  

The RealWorldChem team designed MatFOMGen to address this materials-computational challenge by enabling rapid, domain-aware identification and formalization of relevant figures-of-merit.


\subsection*{Results}

MatFOMGen begins with a user specifying a domain of interest using natural language. An LLM-based agent then conducts a targeted literature search across the chemistry and materials science corpus and returns a set of relevant computational figures-of-merit along with a comprehensive list of consulted references. Each identified FOM is accompanied by a concise description, and each referenced publication is assigned a relevance score on a 0–1 scale.  

Users may then select a subset of FOMs for deeper analysis. Following selection, MatFOMGen generates, in sequential stages: (1) detailed technical descriptions of the selected FOMs, including the appropriate mathematical formulation, and (2) executable Atomic Simulation Environment (ASE) functions \cite{larsen2017atomic} to compute the requested FOMs. The generated ASE functions explicitly expose a placeholder for the \texttt{Calculator}, corresponding to the level of theory used to evaluate the FOM. This design choice reflects the reality that many autonomous laboratories deploy proprietary or custom machine-learning interatomic potentials, which must be evaluated within application-specific industrial contexts.  

\begin{figure}[h]
    \centering
    \includegraphics[width=0.65\linewidth]{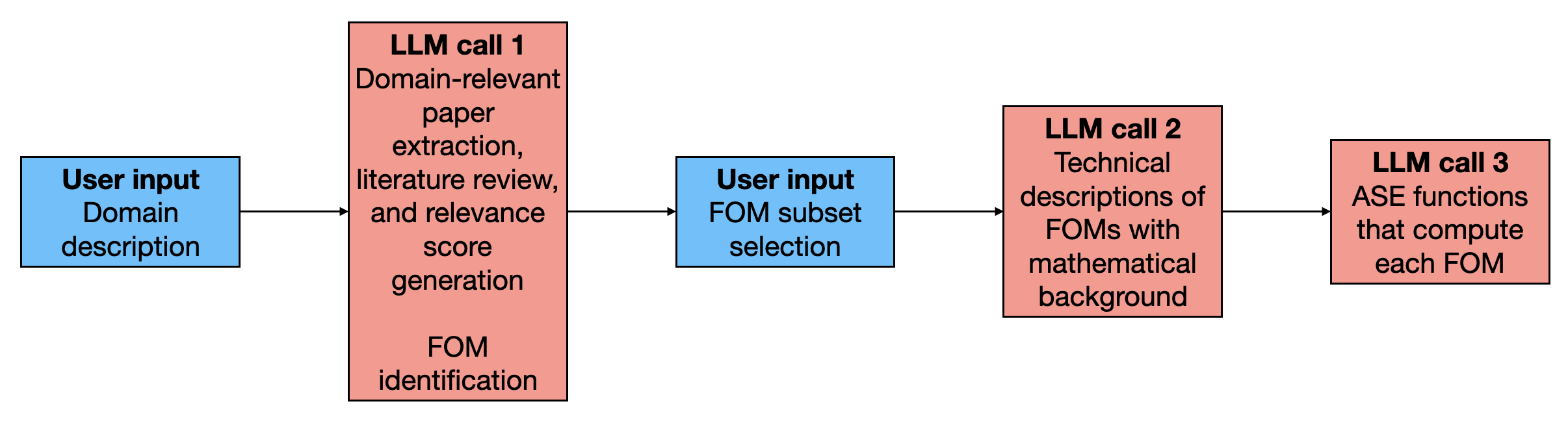}
    \caption{MatFOMGen workflow.}
    \label{fig:matfomgen_workflow}
\end{figure}

The full MatFOMGen workflow is summarized in Figure~\ref{fig:matfomgen_workflow}.

MatFOMGen employs LLMs in a modular, agentic, human-in-the-loop architecture. The literature review stage, mathematical formulation stage, and ASE function generation stage each invoke distinct LLM calls. This modularity allows users to control model selection and token allocation per task. In practice, smaller models were found to be sufficient for rapid literature review and FOM enumeration, while larger models yielded higher fidelity in mathematical descriptions and ASE code generation.

MatFOMGen was implemented using the Anthropic API and Streamlit.

\subsection*{Future Work}

A primary limitation of the current MatFOMGen pipeline is the lack of formal validation for LLM-generated ASE functions. Future work will explore additional LLM-based reflection and refinement stages to improve code reliability. Another promising extension is the use of fine-tuned LLMs to achieve higher-accuracy function generation in domain-specific settings.


\subsection*{Open-source Materials}

The source code, documentation, and examples for MatFOMGen are available on GitHub:  
\github{https://github.com/gopal-iyer/MatFOMGen}





\section{DFTPilot: Automation and Previewing of DFT Calculation Setups}\label{sec:DFTPilot}




Computational and experimental materials research have both progressed rapidly, yet the two communities remain only loosely connected. Density functional theory (DFT) \cite{hohenberg1964inhomogeneous, kohn1965self} exemplifies this gap: running DFT calculations reliably often requires substantial expertise in convergence behaviour, exchange–correlation functional selection, and numerical parameter tuning \cite{squires2025guidelines}. These requirements limit the accessibility of DFT beyond expert user groups and slow the transfer of computational insight into experimental workflows.

The DFTPilot team introduces \textbf{DFTPilot}, an intelligent assistant designed to bridge this gap by leveraging recent large language models (LLMs) \cite{brown2020language, openai2024gpt4technicalreport} to interpret DFT input and output files, connect new calculations to prior runs, and preview target properties. DFTPilot combines text-based reasoning with structured scientific data through retrieval-augmented generation (RAG) \cite{lewis2020retrieval, izacard2020distilling} and graph-based regression models \cite{xie2018crystal, chen2019graph, himanen2020dscribe}. The system allows users to query previous VASP calculations, identify similar materials, and estimate properties such as band gaps or total energies while accounting for the chosen functional \cite{perdew1996generalized, becke1988density}.

DFTPilot retrieves relevant OUTCAR files, predicts whether a calculation is likely to converge under current settings, and suggests updated INCAR parameters based on prior experience. The overarching goal is to help non-experts work more effectively with DFT tools such as VASP \cite{kresse1993ab, kresse1996efficient}, making first-principles modelling more accessible for routine materials research.

\subsection*{Dataset}

DFTPilot is currently trained and indexed on a dataset of 426 DFT calculations collected from earlier ablation and convergence studies. These calculations include standard structural relaxations performed with multiple exchange–correlation functionals (GGA, GGA+U, and hybrid functionals) and span systems ranging from 4 to 120 atoms per unit cell. The dataset primarily comprises metallic systems and wide-bandgap oxides.

Each entry includes the initial structure, INCAR and POSCAR inputs, and the resulting OUTCAR file. To support both retrieval and prediction tasks, three complementary embeddings were constructed: (i) SentenceTransformers embeddings \cite{reimers2019sentencebert} for INCAR and OUTCAR text, (ii) SOAP descriptors \cite{himanen2020dscribe} for structural similarity, and (iii) feature vectors for a crystal graph convolutional neural network (CGCNN) \cite{xie2018crystal} regressor.

The dataset was split by chemical and structural class, functional choice (GGA, GGA+U, HSE, PBE0), and calculation complexity (single-point, full relaxation, and spin–orbit coupling enabled). While this setup allows evaluation of generalisability, the dataset remains limited for narrow-bandgap semiconductors and materials with alternative topologies, such as layered or amorphous systems, which currently constrains predictive coverage.


\subsection*{Results}

DFTPilot integrates LLMs with VASP \cite{kresse1993ab, kresse1996efficient} and associated pre- and post-processing tools to interpret and automate DFT workflows. The RAG system links new user queries to prior DFT calculations, enabling identification of structurally or electronically similar materials. It also provides rapid previews of OUTCAR files and estimates material properties directly from the input structure and INCAR parameters.

A crystal graph neural network trained on the available dataset achieves a mean absolute error of approximately 0.36 eV for bandgap prediction when restricted to materials with RAG similarity scores exceeding 75\%. This performance is consistent with prior CGCNN studies \cite{xie2018crystal} and is notable given the limited dataset size and the partial incorporation of INCAR information in the current prototype.

Across multiple dataset splits, the framework generalises well to structurally distinct test cases and calculation types. Combined with multimodal embeddings, these results demonstrate that LLM-based systems can deliver interpretable predictions and actionable guidance on convergence behaviour directly from natural language queries.


\begin{figure}[h]
\centering
\includegraphics[width=0.8\linewidth]{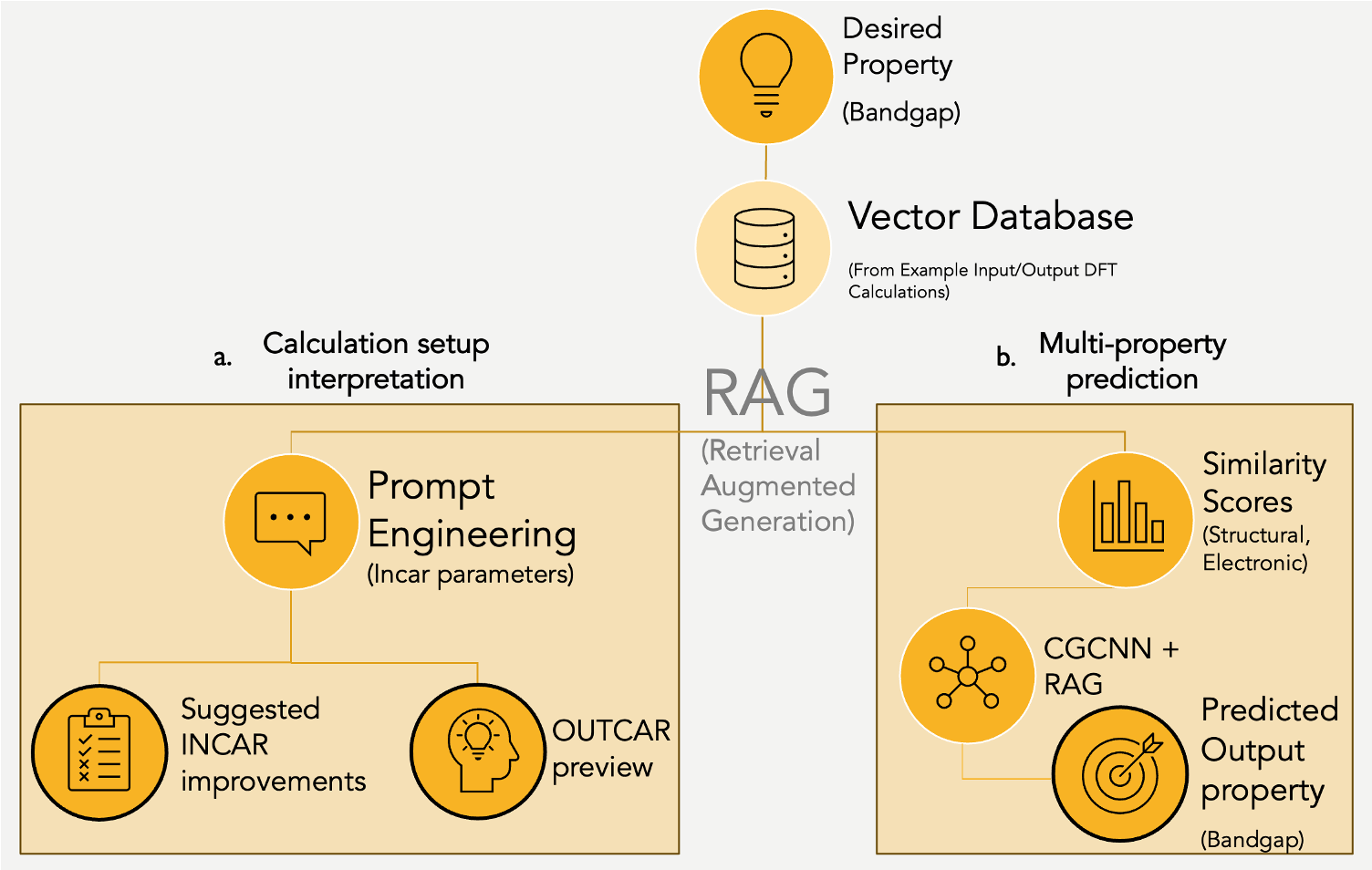}
\caption{Schematic workflow for DFTPilot. The user specifies the target property and material system. The model returns \textbf{(a)} suggested INCAR updates and an OUTCAR preview, and \textbf{(b)} CGCNN-based property predictions using multimodal similarity metrics from the vector index.}
\label{fig:DFTPilot-pipeline}
\end{figure}

\subsection*{Future Work}

Future extensions will expand the dataset to include a broader range of materials classes, calculation types, and levels of theory. Increased structural and chemical diversity is expected to improve both the similarity index and the accuracy of property predictions. To further extend chemical coverage, the DFTPilot team plans to incorporate data from public repositories such as NOMAD \cite{draxl2019nomad} and the Materials Project \cite{jain2013commentary}.

Additional planned developments include extending property prediction beyond band gaps to lattice parameters, dielectric tensors, and mechanical or electronic observables. These enhancements will enable more informative previews and support a more comprehensive workflow from calculation setup to property estimation.


\subsection*{Open-source Materials}

Code, pre-trained weights, and the RAG index are available on GitHub:  
\github{https://github.com/chiku-parida/DFTPilot}\,; Demo video: \youtube{https://www.linkedin.com/feed/update/urn:li:activity:7372418332953763840/}\,.





\section{Parse Patrol: Dual-Mode Scientific Parsing Infrastructure via MCP Servers}\label{sec:Parse_Patrol}




Parsing scientific files into structured data is strongly dependent on evolving specifications, which may exist at the format, schema, or ontological level. This dependency renders parser infrastructures brittle and labor-intensive to maintain, as changes to either source or target specifications often require manual updates to existing parsers. Such challenges are common in materials science and chemistry databases and in large-scale data consortia.

Recent advances in large language models (LLMs) have opened new opportunities for automating parser development, as modern models can reason over heterogeneous scientific file formats and adapt dynamically to changing specifications. Effective deployment of LLMs, however, requires a robust interface for connecting models to external tools. The Model Context Protocol (MCP), introduced in late 2024 \cite{mcp2024, mcp_def}, provides a standardized mechanism for agents to discover and invoke external tools, access resources, and reuse predefined prompts, and has rapidly become a widely supported industry standard.

The Parse Patrol team presents an AI-assisted workflow for rapidly discovering, testing, and deploying parsers that conform to user-defined specifications. Parse Patrol leverages MCP to integrate community-developed parsers into a unified interface, enabling agents to iteratively evaluate and refine parser choices against target schemas. The same infrastructure supports two complementary usage modes: (i) \textit{Discovery Mode}, in which an agent interactively tests parsers for schema conversion, and (ii) \textit{Direct Import Mode}, in which the same parsers are exposed as Python modules for use in production workflows.


\subsection*{Results}

Automated parser generation is hindered by sparse documentation of scientific specifications, hallucinations outside model training distributions, long test cycles, and fragile software architectures. Parse Patrol mitigates these challenges by building on existing community parsers rather than attempting full parser synthesis. The framework focuses on expanding source-specification coverage and parser design options, while MCP provides a model-agnostic interface through which individual parsers can be exposed as tools and orchestrated by agents during selection, testing, and schema conversion.

While MCP standardizes tool invocation, it does not prescribe how tools should be organized. To address this limitation, Parse Patrol introduces a \textit{hierarchical protocol} for structuring MCP servers. Each parser is implemented as an independent MCP server, enabling focused development and targeted testing. These parser servers are automatically registered with a central, user-facing server that aggregates available tools and resources and provides prompts for testing, evaluation, and production deployment. User specifications can be supplied directly through chat interfaces or as external files, while test cases may be retrieved via database MCP servers.

Although MCP tools enable interactive testing and ad hoc parsing tasks, production workflows often require importable software components. Parse Patrol therefore exposes each parser simultaneously as an MCP server and as a Python module. To ensure that agents correctly leverage these modules when generating code, the framework combines prompt-level guidance with explicit exposure of module paths and import syntax as MCP resources.

\begin{figure}[h]
    \centering
    \includegraphics[width=0.8\linewidth]{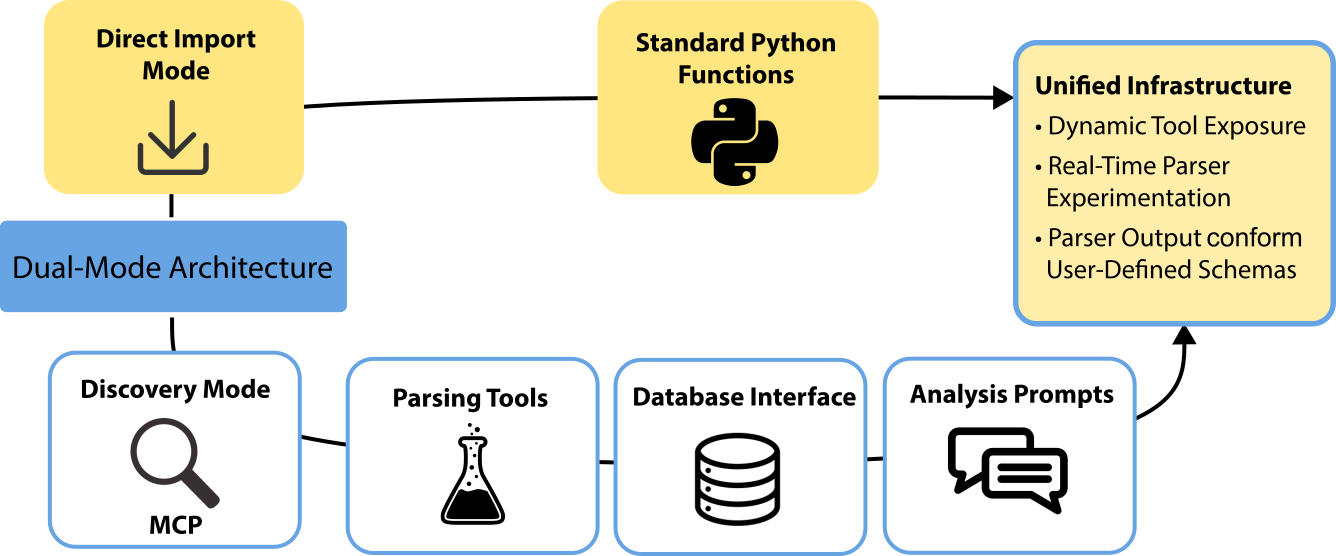}
    \captionsetup{width=\linewidth}
    \caption{
    Schematic depiction of the \textit{dual-mode} design in Parse Patrol.
    \textbf{Lower branch:} Discovery Mode provides a single MCP interface to multiple parser and database servers, enabling agents to iteratively design parsers that conform to user-defined specifications.
    \textbf{Upper branch:} Direct Import Mode exposes the same tools as Python modules for frictionless integration into production code.
    \textbf{Both branches} are unified under a single architectural framework.
    }
    \label{fig:parse-patrol}
\end{figure}

This \textit{dual-mode design} (Figure~\ref{fig:parse-patrol}) presents nontrivial engineering challenges, as MCP servers and Python packages follow different distribution paradigms. MCP servers are typically registered via the MCP Registry \cite{mcp2024registry}, whereas Python modules are distributed through the Python Package Index (PyPI) \cite{pypi}. Parse Patrol bridges this gap by maintaining a unified codebase that satisfies both frameworks, using a shared tool schema to ensure consistent interfaces across interactive and production contexts. To the authors’ knowledge, no existing framework provides this form of dual-mode parser deployment.


\subsection*{Future Work}

At the time of submission, Parse Patrol is under active development. While the core objective is to provide a flexible and extensible computational parser toolkit, future extensions are anticipated. In particular, incorporation of Parse Patrol into the NOMAD parser ecosystem is currently under consideration \cite{nomad_lab, draxl2019nomad}.


\subsection*{Open Source Materials}

The open-source code is available on GitHub:  
\github{https://github.com/ndaelman-hu/parse-patrol} (development version). Stable releases are available via version tags; the hackathon submission corresponds to \texttt{v0.0.2-beta}\,; Demo video:  
\youtube{https://www.youtube.com/watch?v=fSAyi5ubkR0}\,.






\section{Catalyst Assistant: Evidence-traced LLM Agent for $\mathrm{NH_3}$ Decomposition Catalysts}
\label{sec:catalyst-assistant}




Catalyst Assistant is an evidence-bound AI co-researcher designed to accelerate the discovery of catalysts for ammonia ($\mathrm{NH_3}$) decomposition. The system proposes and ranks catalyst candidates, plans synthesis and testing workflows, and automatically generates characterization visuals such as XRD and BET plots by fusing curated literature, structured datasets, and external APIs (e.g., the Materials Project).  

The Catalyst Assistant team designed the platform using context-engineered modules that explicitly minimize hallucinations through evidence tracing and verification. Large language models are customized with guardrails intended to support academic researchers, industrial practitioners, and students alike. The resulting workflow enables faster and safer catalyst screening, producing structured, reproducible outputs suitable for both research and educational use.


\subsection*{Results}

The Catalyst Assistant architecture demonstrates how a large language model operates within a structured agentic framework to address user queries through a systematic workflow. A frontend interface enables user interaction, allowing researchers to submit queries and datasets while receiving structured outputs such as tables, reports, visualizations, and explanatory summaries.

The backend integrates three coordinated modules. The \textit{Knowledge module} aggregates heterogeneous information sources, including datasets, scientific PDFs, and external APIs. The \textit{Context Engineering module} applies structured prompting strategies, routing rules, verification guardrails, and data-integration pathways. The \textit{Reasoning module} evaluates contextual consistency, validates evidence, and checks data integrity. Through agentic routing, the language model connects these components while leveraging chat history to maintain contextual continuity across interactions.

This architecture combines retrieval, coordination, and validation mechanisms to generate accurate, task-aligned responses grounded in traceable evidence, supporting reliable catalyst discovery workflows.


\begin{figure}[H]
    \centering
    \includegraphics[width=0.7\linewidth, height=0.4\textheight, keepaspectratio]{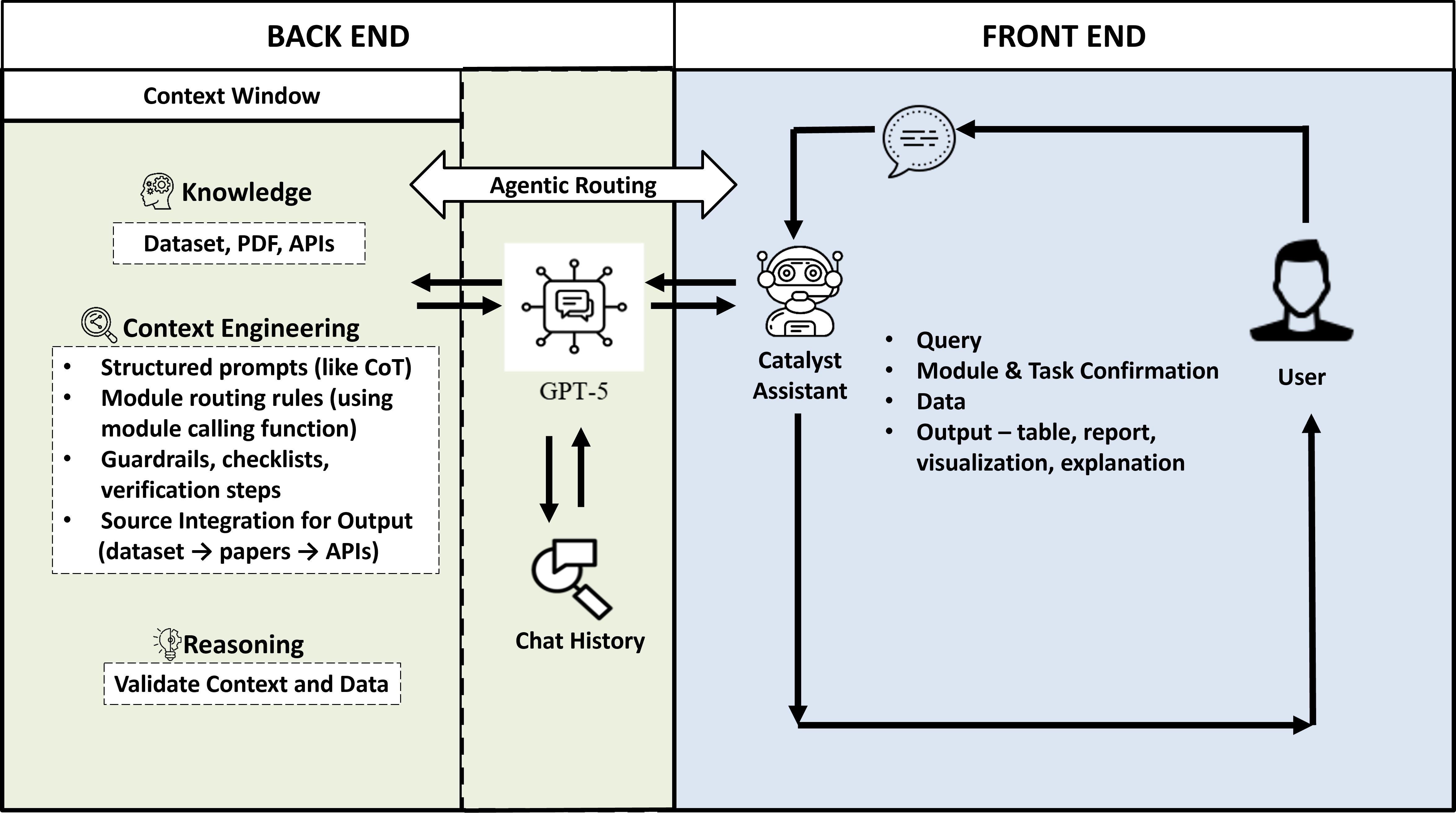}
    \caption{System architecture of Catalyst Assistant.}
    \label{fig:catalyst-assistant-architecture}
\end{figure}

\subsection*{Future Work}

Future development will focus on hardening the pipeline with a dedicated data-extraction and plotting service, alongside a hosted API to improve figure accuracy and reproducibility. Additional extensions include uncertainty-aware candidate ranking and automated synthesis and testing planners. These enhancements will enable broader exploration of non-precious catalyst materials by integrating additional materials databases and delivering experiment-ready outputs.


\subsection*{Open-source Materials}

The Catalyst Assistant LLM agent is available on the \href{https://chatgpt.com/g/g-68c09896f11c81918e86b7ddcbad47e3-catalyst-assistant}{GPT Store}\,; Demo video: \youtube{https://youtu.be/4uqrp6kbK4s}\,.





\section{Data-Driven Prediction of Thin Film Properties in Physical and Chemical Deposition Methods}\label{sec:ThinFilmAI}




ThinFilm.ai is a multimodal artificial intelligence framework for predicting thin film quality by fusing structured deposition data with unstructured textual information from the scientific literature. The ThinFilm.ai team curated a corpus of 100 research papers, with particular emphasis on studies reporting characterization data such as X-ray photoelectron spectroscopy (XPS), scanning electron microscopy (SEM), and X-ray diffraction (XRD).

A supervised learning pipeline employs three classification models trained on data extracted using the GPT-4 API to predict key thin-film quality metrics, including crystallinity, impurity content, and surface roughness. In parallel, an unsupervised pipeline generates domain-specific MatSciBERT embeddings from the \emph{Results} sections of the literature. These text embeddings are concatenated with structured deposition features and clustered to uncover latent relationships between processing conditions and film quality. KeyBERT is subsequently applied to extract representative keyphrases, enabling interpretable thematic labeling of each cluster with respect to fabrication parameters and material outcomes.

The final platform, hosted using Streamlit, integrates quantitative predictions with qualitative insights to provide a unified decision-support tool for materials researchers.


\subsection*{Results}

\begin{figure}[H]
    \centering
    \includegraphics[width=\linewidth, height=0.4\textheight, keepaspectratio]{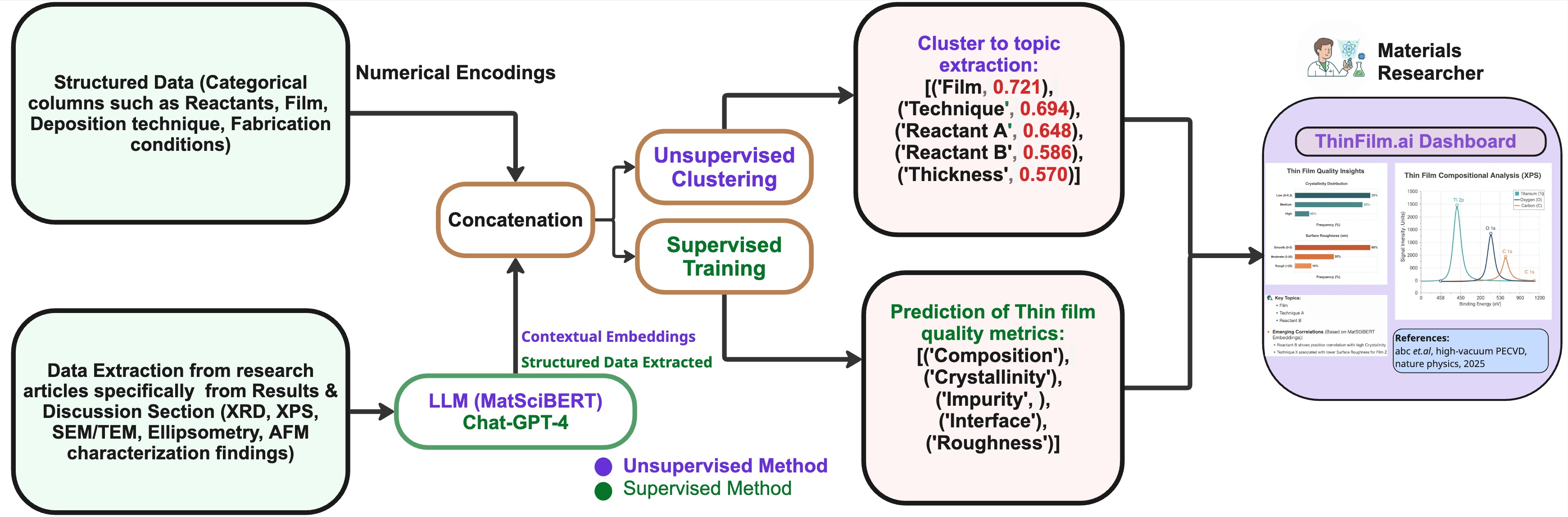}
    \caption{Data extraction methods and model process flow in ThinFilm.ai.}
    \label{fig:thinfilm}
\end{figure}

Figure~\ref{fig:thinfilm} illustrates the ThinFilm.ai workflow. The framework addresses a key challenge in experimental materials science: predicting thin film quality from deposition recipes prior to extensive laboratory experimentation. The Streamlit-based application delivers real-time predictions of crystallinity, surface roughness, impurity composition, and interface quality, enabling rapid screening of deposition conditions.


\subsection*{Future Work}

Future work will focus on expanding the dataset by incorporating a broader range of experimental studies and refining preprocessing pipelines. The ThinFilm.ai team plans to extend the feature space to include additional deposition parameters such as pressure, temperature, and deposition duration to improve predictive accuracy. Further architectural enhancements, including the integration of knowledge graphs, will be explored to better represent complex deposition processes. Model performance will also be improved through systematic fine-tuning and training of more robust predictive architectures.


\subsection*{Open-source Materials}

Code and pretrained weights are available on GitHub: \github{https://github.com/ram123-debug/mat-chem-llm-hackathon.git}\,; Demo video: \youtube{https://youtu.be/dazb9VQNvOY}\,.








\section{Scaffold Conscious Agent for Learning \& Exploration}\label{sec:SCALE}

The SCALE team focuses on the design and optimization of functional small molecules, which underpin advances in drug discovery, materials development, and chemosensory science. Despite major progress in deep generative modeling, most existing approaches still struggle to maintain chemical validity, preserve pharmacophoric scaffolds, or produce synthetically feasible compounds \cite{sanchez2018inverse}. At the same time, large language models (LLMs) have demonstrated powerful reasoning and symbolic manipulation capabilities, enabling new opportunities for chemistry-aware design frameworks \cite{elton2019deep}. However, their direct application to molecular generation remains limited by hallucinations, lack of physical grounding, and the absence of adaptive feedback loops.

To address these challenges, the SCALE team developed SCALE (Scaffold-Conscious Agent for Learning and Exploration), an LLM-driven molecular optimization framework that couples reasoning-based molecular edits with cheminformatics guardrails and lightweight physics-aware scoring. SCALE enables scaffold-preserving exploration of chemical space for diverse applications, including drug discovery, fragrance design, and repellent or agrochemical screening, all within an interpretable and iterative optimization loop.

%

\subsection*{Results}
The SCALE framework operates as an adaptive, closed-loop optimization system that unites the reasoning capacity of large language models (LLMs) with the quantitative rigor of cheminformatics and physics-based scoring. The process begins with one or more seed molecules that define the structural scaffolds and property targets of interest. These seeds constrain the search space, ensuring that subsequent modifications retain core pharmacophores or functional groups relevant to biological or sensory function. SCALE can therefore operate flexibly across application domains such as drug discovery, fragrance and flavor chemistry, or repellent and agrochemical design.

An LLM controller acts as the generative engine, proposing scaffold-preserving edits such as R-group substitutions, heteroatom replacements, or ring decorations. Unlike statistical generative models, the LLM uses explicit textual reasoning, incorporating feedback from previous rounds to refine its chemical intuition. Each generated molecule undergoes a multi-layered guardrail evaluation that verifies SMILES validity, eliminates PAINS and reactive motifs, and estimates synthetic accessibility. Molecules that meet these cheminformatics criteria are then analyzed with lightweight physical descriptors such as MMFF94 strain energy, QED, SA score, and logP to assess structural stability and drug-likeness. These filters collectively prevent the propagation of chemically implausible or synthetically inaccessible candidates.

To enable rapid and uncertainty-aware property estimation, SCALE employs a Random Forest (RF) surrogate model trained on molecular descriptors and precomputed physicochemical features. The surrogate predicts mean property values ($\mu$) and associated uncertainties ($\sigma$), which are combined using an Upper Confidence Bound (UCB = $\mu$ + $\kappa\sigma$) criterion to balance exploitation of promising molecules and exploration of uncertain regions of chemical space \cite{he2023structured}. Candidates exceeding a dynamic UCB threshold are retained, while the remainder are discarded or used to update the model’s exploration bias. Feedback from this layer guides the LLM’s next sequence of scaffold edits, forming a self-correcting optimization loop that continues until convergence criteria—such as plateaued property improvement, diversity saturation, or iteration limits—are reached. Through this integration of LLM reasoning, chemical guardrails, and physics-informed uncertainty modeling, the SCALE framework establishes an interpretable and efficient pathway for hit identification and lead optimization in both pharmaceutical and chemosensory molecular design.

\begin{figure}[H]
    \centering
    \includegraphics[width=\linewidth, height=0.4\textheight, keepaspectratio]{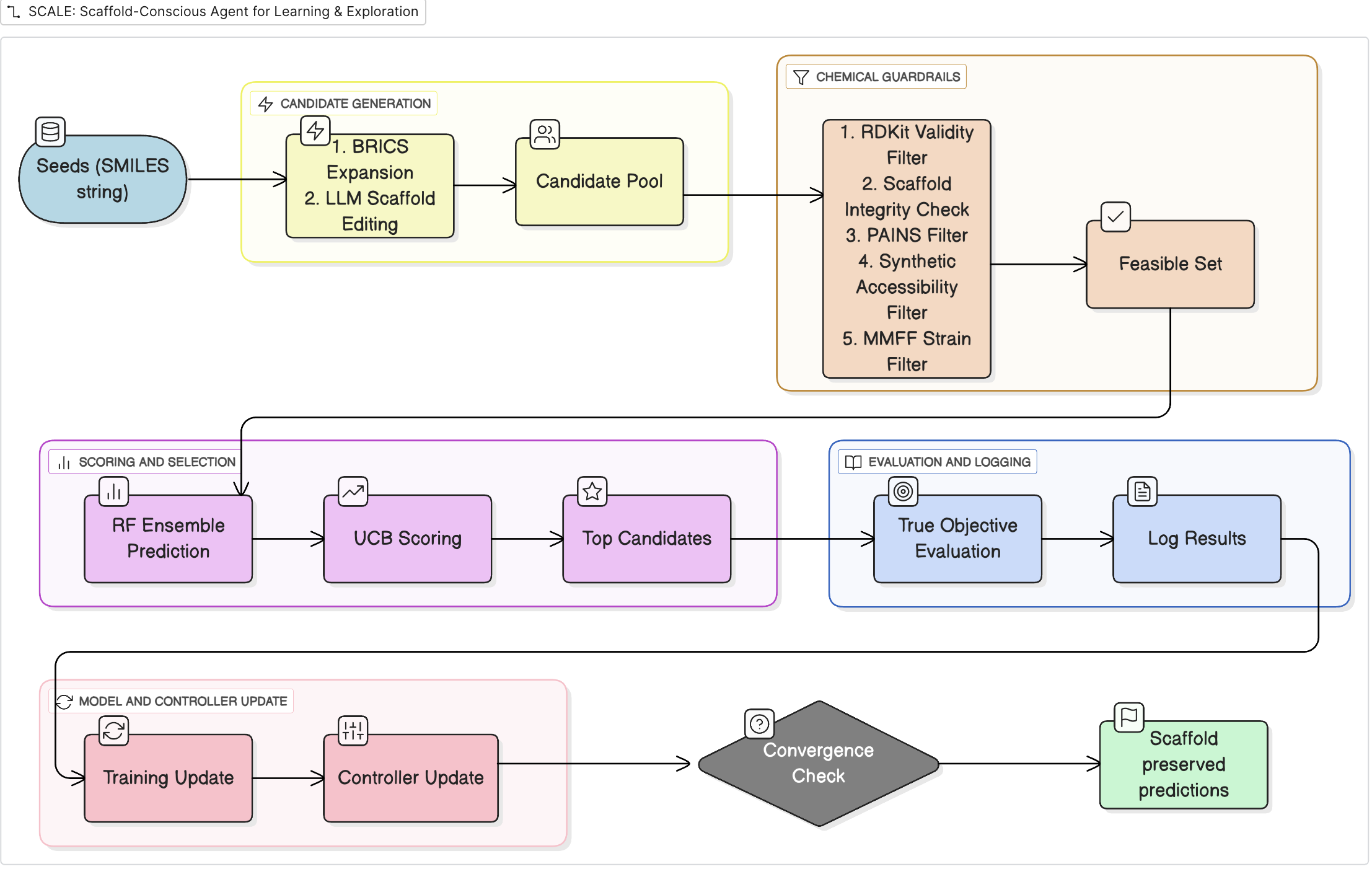}
    \caption{SCALE workflow diagram.}
    \label{fig:scale}
\end{figure}


\subsection*{Future Work}
Future extensions of SCALE will focus on enhancing both chemical accuracy and autonomy by integrating higher-fidelity physics and adaptive learning. Incorporating semiempirical or DFT-level calculations (e.g., GFN2-xTB, $\omega$B97X-D) into the surrogate model would improve the reliability of property predictions beyond empirical descriptors, while active learning loops could dynamically retrain the Random Forest as the explored chemical space expands. Replacing the RF surrogate with a graph neural network or a multi-task ensemble would enable simultaneous optimization of multiple properties, including target affinity, toxicity, and volatility. Fine-tuning the LLM controller on chemically curated corpora or reaction-based transformations could further reduce hallucinations and improve context-aware molecular reasoning. Additionally, coupling SCALE with automated synthesis planning and experimental validation pipelines could transform it into a fully closed-loop discovery platform. Benchmarking SCALE against established generative frameworks such as REINVENT, MolDQN, or GFlowNet would quantitatively assess its efficiency and generalizability across chemical domains.

\subsection*{Open Source Materials}
\textbf{Code:} \github{https://github.com/schandy2211/scale}\,; \textbf{Demo video:} \youtube{https://youtu.be/k0Y7rfM3rLY?si=TOZxu3DYb_vBUStO}\,.







\section{LLM as Aerogel Research Assistant}\label{sec:aerogel}

The L.A.R.A team addresses the challenge that aerogels—highly open-porous nanostructured materials—remain slow to develop despite their exceptional properties. Aerogels are among the most versatile porous materials, typically exhibiting extremely low densities, high porosity, and low thermal conductivities \cite{aegerter2011aerogels}. However, progress in aerogel research is hindered by complex multi-scale synthesis pathways and fragmented knowledge scattered across the literature. Recent advances in large language models (LLMs) have opened new possibilities for automated scientific reasoning. The L.A.R.A framework leverages this opportunity by combining a fine-tuned LLaMat model \cite{mishra2025foundationallargelanguagemodels} with a materials knowledge graph and a modular tool ecosystem, enabling structured hypothesis generation, synthesis planning, and characterization workflows. While tailored specifically to aerogel science, the framework is designed to be extensible to related porous material classes such as hydrogels and xerogels.


\subsection*{Results}

L.A.R.A consists of three core components: (1) a fine-tuned language model specialized for aerogel synthesis and characterization, (2) a knowledge-graph interface built on MatKG \cite{venugopal2024matkg} for structured reasoning, and (3) a tool ecosystem supporting literature search, molecular simulations, and image-based characterization.

The model is fine-tuned on more than one thousand expert-curated aerogel synthesis examples using LoRA parameter-efficient training, with a maximum sequence length of 1024 tokens and a learning rate of \(10^{-5}\). Knowledge-graph integration maps scientific questions to materials entities, properties, and synthesis relationships, enabling grounded hypothesis generation. The tool ecosystem includes microscopy image segmentation, retrieval-augmented generation (RAG)-based literature search, and molecular dynamics (MD) simulation modules orchestrated through a multi-agent interface.

\begin{figure}[h!]
    \centering
    \includegraphics[width=0.85\textwidth]{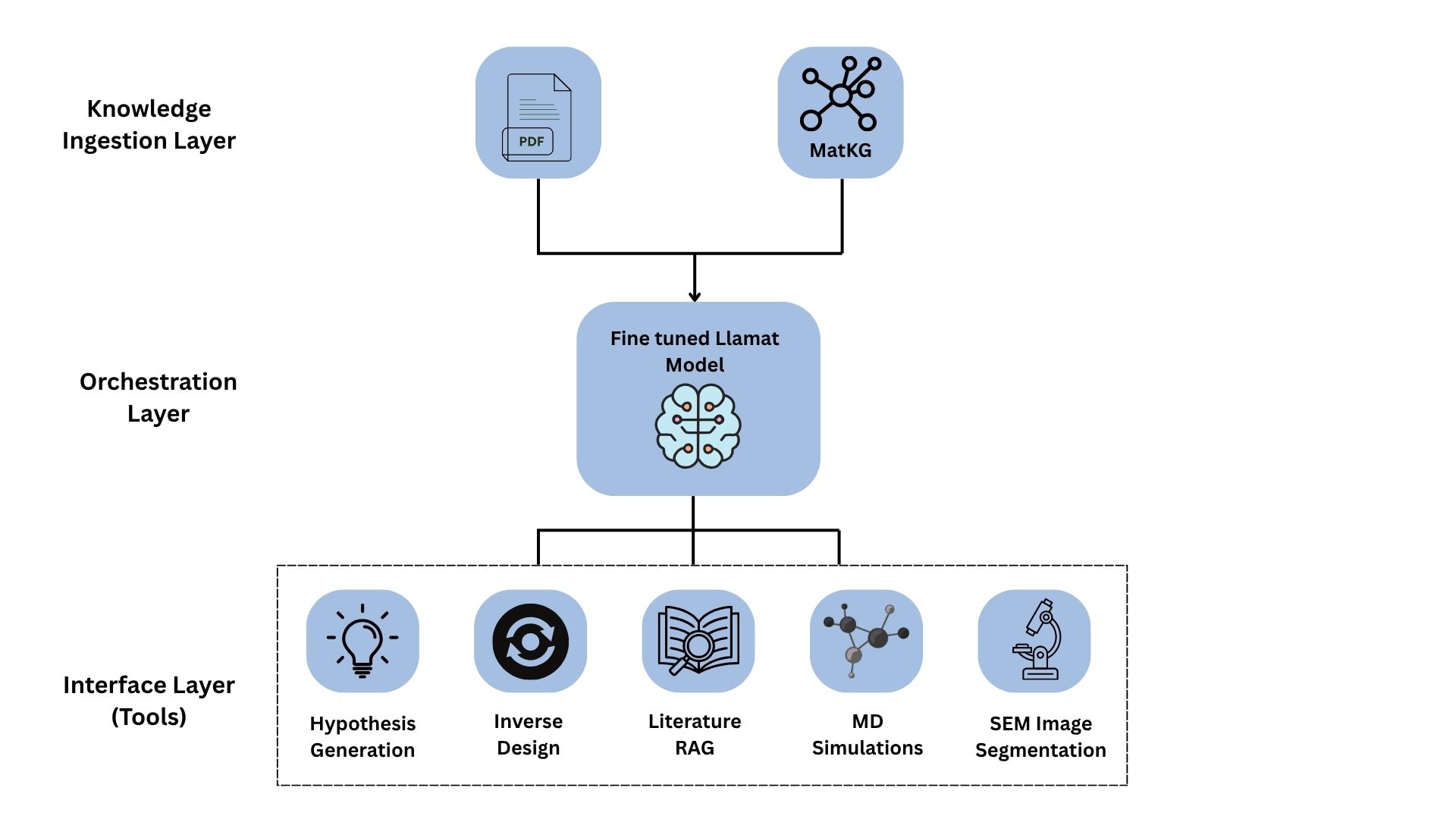}
    \caption{Architecture and workflow of L.A.R.A. The fine-tuned model determines whether to respond using the knowledge-ingestion layer—leveraging either direct fine-tuning or the knowledge graph—or to invoke one of the tools available in the interface layer.}
    \label{fig:lara}
\end{figure}

The fine-tuned model demonstrates strong performance in scientific reasoning and synthesis-route generation for aerogels. Benchmarked against expert-annotated queries, the L.A.R.A framework produces relevant hypotheses and plausible synthesis routes consistent with established sol–gel chemistry. It also supports inverse-design queries in which target properties are specified directly; for example, requesting porosity above 0.95 combined with a minimum electrical conductivity returns multiple synthesis pathways accompanied by mechanistic scientific rationales. A complete research workflow is illustrated in Figure~\ref{fig:lara}.

In addition to domain-specific reasoning, L.A.R.A incorporates an agentic decision layer that determines, for each user query, whether internal reasoning is sufficient or whether external tool invocation is required. This decision is guided by both the semantic content of the user prompt and structured descriptions of available tools. Conceptual questions—for instance, those concerning sol–gel reaction mechanisms—are handled directly by the LLM. In contrast, prompts requiring empirical grounding or numerical evaluation (e.g., literature searches on RF aerogels, SEM image analysis, or molecular dynamics simulations for silica clustering) are routed to the appropriate external tools.

This routing emerges from a representation-alignment process in which user requests and tool descriptions are projected into a shared latent space, allowing the system to assess their compatibility. If the required evidence cannot be generated through internal reasoning alone, the query is delegated to the tool whose operational description best matches the inferred objective. Table~\ref{tab:model_comparison} summarizes key differences in inference and reasoning capabilities between L.A.R.A and ChatGPT (GPT-5.1) for hypothesis generation and inverse design tasks in aerogel research.

\begin{table}[h!]
\centering
\caption{Comparison of ChatGPT (GPT-5.1) and L.A.R.A responses.}
\label{tab:model_comparison}
\begin{tabular}{|p{3.5cm}|p{5cm}|p{5cm}|}
\hline
\textbf{Evaluation Criteria} & \textbf{ChatGPT (GPT-5.1)} & \textbf{L.A.R.A} \\
\hline
Response Philosophy & Educational and conceptual, focused on understanding principles & Hypothesis-driven scientific analysis with research-oriented insights \\
\hline
Output Structure & Sequential stage-based frameworks (3–4 stages) & Multiple ranked hypotheses with confidence intervals \\
\hline
Technical Specificity & Conceptual identification of materials and methods & Specific parameters, including temperature ranges, time frames, and chemical formulas \\
\hline
Ethics \& Safety & Refuses actionable lab procedures and numerical conditions & Discusses safety concerns and processing challenges within technical context \\
\hline
Quantitative Information & Avoids specific numerical parameters & Includes numerical values (e.g., 400–1200\,$^\circ$C, several hundred bars, confidence intervals from 0 to 1) \\
\hline
Scientific Depth & Conceptual explanation of mechanisms and property relationships & Detailed discussion of molecular mechanisms, interfacial interactions, and structure \\
\hline
Practical Applicability & Non-actionable, educational content & Partially actionable, suitable for researchers familiar with laboratory protocols \\
\hline
\end{tabular}
\end{table}


\subsection*{Future Work}

Future extensions of the L.A.R.A framework will focus on expanding the fine-tuning dataset to additional aerogel families. While the current system employs a fixed tool setup, a natural next step is to mediate interactions through a Model Context Protocol (MCP) layer. Moreover, the MatKG currently lacks comprehensive entity coverage and relationship definitions for aerogel-specific use cases, motivating targeted graph expansion and curation. From an agentic perspective, the present implementation supports only single-step tool calls; future versions will incorporate multi-tool, multi-turn workflows that enable iterative reasoning, tool chaining, and autonomous planning toward defined research objectives.

\subsection*{Open-Source Materials}

The open-source implementation of L.A.R.A Module~2, including model fine-tuning scripts, tools, and demonstrations, is available at: \github{https://github.com/sugannathan/L.A.R.A-LLMs-as-Aerogel-Research-Assistants/tree/main}







\section{Agentic AI for Autonomous Research and Simulation of Differential Equation Models}\label{sec:odeforge}

The ODE FORGE team focuses on modeling dynamic systems with Ordinary Differential Equations (ODEs), a cornerstone of research in materials science and chemistry. Despite their importance, translating theoretical models from the literature into rigorous, runnable computational simulations remains a major bottleneck. This process often requires substantial programming expertise, creating friction between domain experts and rapid experimentation. ODE Forge addresses this challenge by automating the research-to-analysis workflow. By leveraging a multi-agent Large Language Model (LLM) framework, the system bridges natural-language scientific queries and immediately runnable, interactive simulations, enabling researchers to prioritize scientific inquiry over implementation details.


\subsection*{Methodology}

ODE Forge employs a two-phase agentic workflow (Figure~\ref{fig:odeforge_architecture}) composed of conversational research and autonomous coding and execution. Each phase is implemented as a LangGraph \cite{langchain2023} graph, with nodes representing specialized agents.

\subsection*{Phase 1 – Conversational Research}

The workflow begins with an interactive dialogue in which the user describes the target system, for example requesting a simulation of the Lotka–Volterra equations with specific rate constants. A large language model (LLM) agent queries both a user-provided embedded paper library and external web sources to construct a comprehensive model specification. This specification includes the governing differential equations, relevant parameters and initial conditions, solver preferences such as Runge–Kutta or stiff solvers, and required analyses or visualizations. When relevant literature is available, a retrieval-augmented generation (RAG) mechanism \cite{lewis2020retrieval} prioritizes the trusted local corpus to preserve domain-specific fidelity. The resulting structured specification is serialized as a JSON file and passed directly to the autonomous coding phase.

\subsection*{Phase 2 – Autonomous Coding and Execution}

\begin{figure*}[t]
    \centering
    \begin{minipage}{0.49\textwidth}
        \centering
        \includegraphics[width=\linewidth]{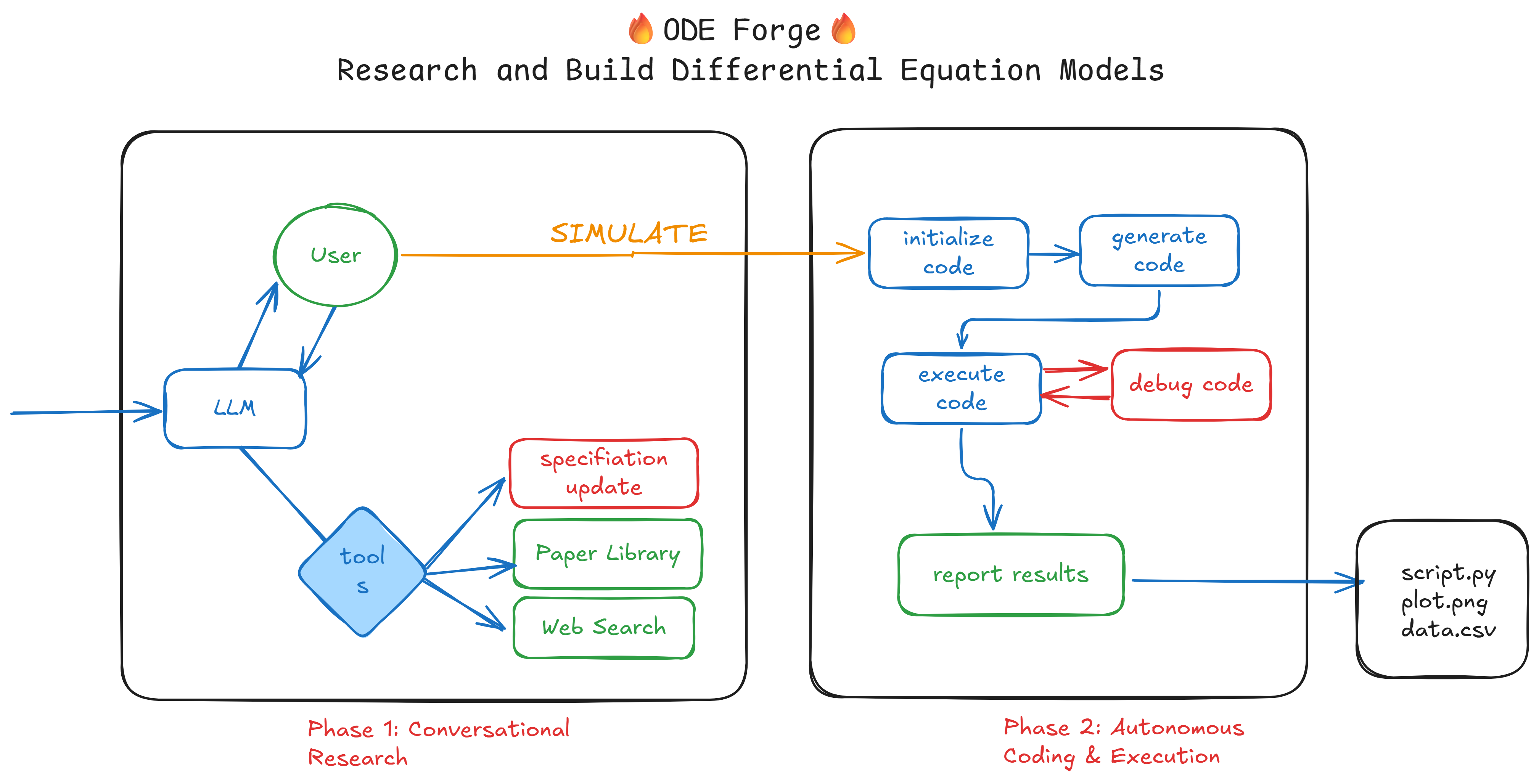}
    \end{minipage}
    \hfill
    \begin{minipage}{0.49\textwidth}
        \centering
        \includegraphics[width=\linewidth]{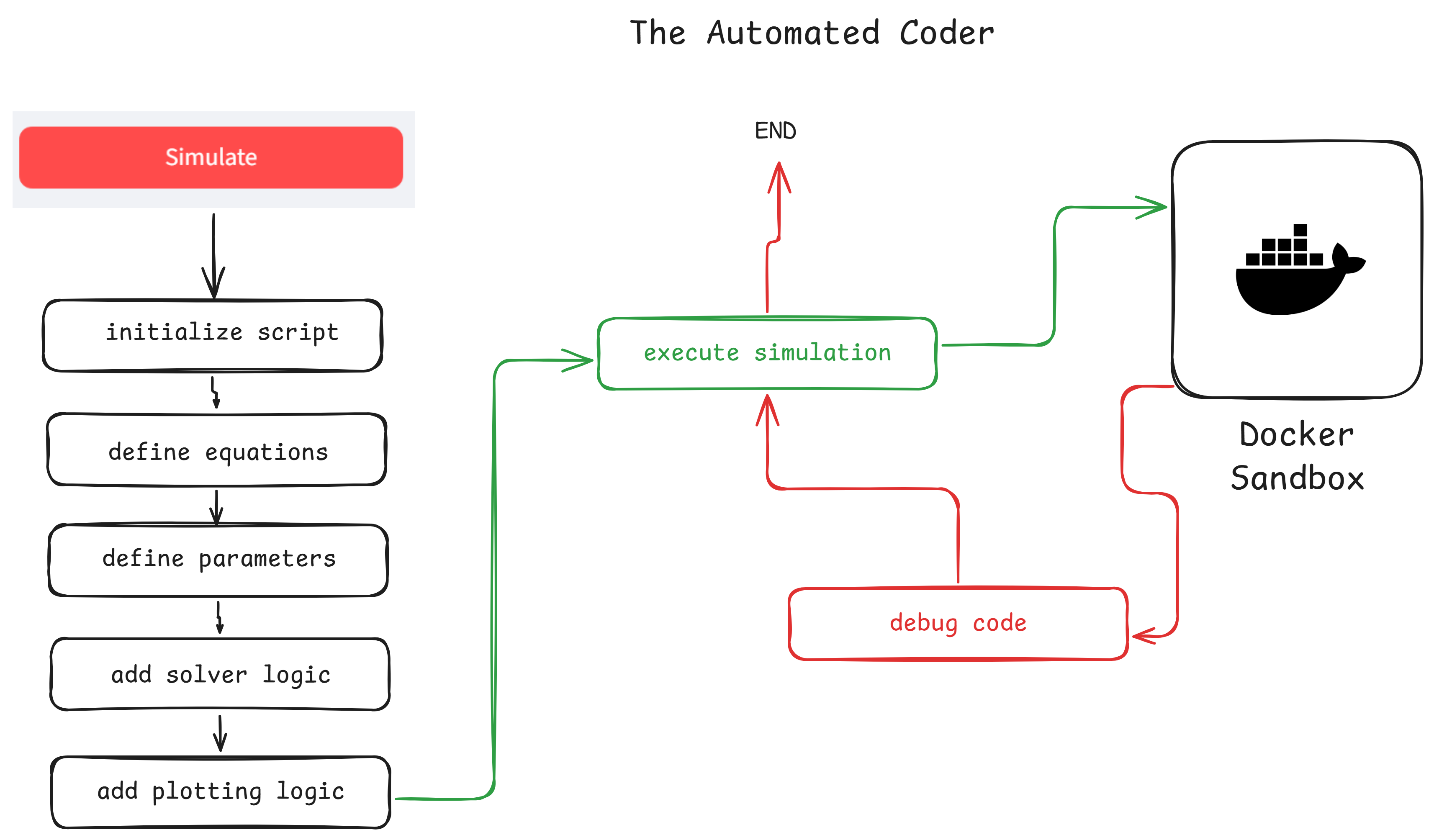}
    \end{minipage}
\caption{Workflow of ODE Forge showing (left) the two-phase agentic pipeline for research and model construction, and (right) the internal automated coding process for script generation, debugging, and execution within a Docker sandbox.}
\label{fig:odeforge_architecture}
\end{figure*}

In the second phase (Figure~\ref{fig:odeforge_architecture}), an automated coding agent transforms the structured specification into a complete, executable Python script. The agent initializes the project environment, defines all equations and parameters, integrates solver and plotting routines, and executes the script within a Docker sandbox to ensure safety and reproducibility. During execution, a self-healing debug loop continuously monitors the sandbox output. When errors occur, the agent interprets the traceback, edits the code, and re-runs the script autonomously until successful completion. The final artifacts—\texttt{script.py}, \texttt{plot.png}, and \texttt{data.csv}—are generated automatically and exported to a results dashboard for inspection and post-processing.


\subsection*{Technical Implementation}

ODE Forge is implemented in Python using LangGraph to orchestrate a multi-agent workflow over a structured state. This state tracks the full conversation, the evolving problem specification—including model descriptions, symbolic equations, parameters, initial conditions, and simulation settings—as well as the generated solver code, execution status, and debugging information. Maintaining a unified state enables smooth transitions from interactive research to fully automated code generation and simulation.

\captionsetup{type=figure,width=\textwidth}

During the conversational phase, an LLM agent incrementally builds the specification by asking targeted questions and invoking specialized tools as needed. These tools include a web researcher (via Tavily search), a local RAG expert that queries a Chroma vector store built from user-uploaded PDFs \cite{lewis2020retrieval}, and a specification tool that commits validated information back into the structured specification. The local library is persisted on disk, allowing reuse across sessions without rebuilding the index.

Once the specification is confirmed, control passes to the coding phase. The system constructs a Python script in stages: starting from a template, translating symbolic equations into SymPy expressions, converting textual parameters and initial conditions into executable assignments, and generating SciPy solver calls alongside plotting and CSV export logic \cite{DBLP:journals/corr/abs-1907-10121}. The completed script is executed inside an isolated Docker image. If execution fails, the captured traceback is returned to the model, which proposes revisions; this closed-loop repair cycle repeats up to a fixed retry limit.

\subsection*{Discussion}

The ODE FORGE team demonstrates how agentic AI can function as an active scientific collaborator \cite{Decardi_Nelson_2024,Boiko2023} rather than a passive assistant. Beyond interpreting natural-language prompts, the framework performs structured reasoning that integrates symbolic mathematics, literature retrieval, and iterative verification. This architecture represents a reproducible paradigm for future AI laboratories in which conversational specification replaces manual scripting, autonomous debugging ensures robustness, and sandboxed execution preserves scientific rigor.

While current implementations focus on ordinary differential equation systems, the same principles can be extended to partial differential equations, stochastic simulations, and optimization problems by incorporating domain-specific solver modules and tool adapters. Current limitations stem primarily from reliance on LLM reasoning for equation parsing and the absence of multi-physics coupling support.

\subsection*{Open-source Materials}

The code for ODE Forge is available at: \github{https://github.com/souvikta/ODEForge}




\section{MP-LLM: Materials Project Query Interface}
\label{sec:mp-llm}



Accessing large scientific databases, such as the Materials Project (MP) \cite{jain2013commentary}, typically requires a detailed understanding of the MP API client syntax as well as knowledge of the available API endpoints, creating a barrier to entry. The MP-LLM team addresses this challenge by introducing MP-LLM, an interface that translates natural language queries into structured API calls. This tool democratizes access to materials data by enabling researchers to query the Materials Project database using conversational language.

\begin{figure}[H]
    \centering
    \includegraphics[width=\linewidth, height=0.3\textheight, keepaspectratio]{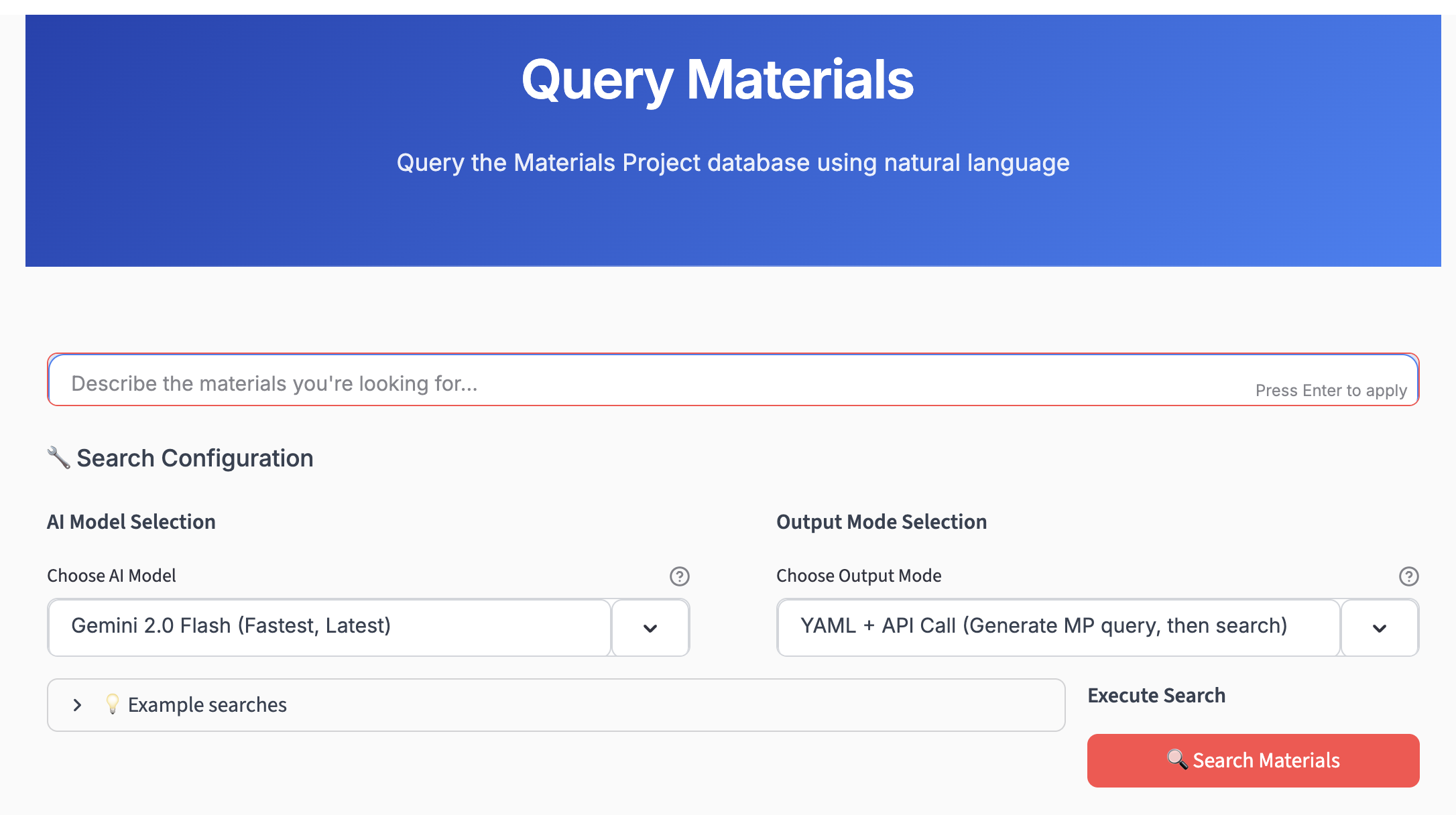}
    \caption{User interface to perform Materials Project queries.}
    \label{fig:mp-llm-ui}
\end{figure}

\subsection*{Results}
The system provides a simple web interface (Figure~\ref{fig:mp-llm-ui}) where users input natural language prompts, such as ``find all stable oxides with a band gap over 2 eV.'' This query is processed by a user-selected model, which generates a structured YAML query for the official Materials Project API. Two distinct methods were evaluated: \textit{direct} generation, where the model attempts to output material IDs, and \textit{tool-augmented} generation, where the model produces a formal API query.

\begin{figure}[H]
    \centering
    \includegraphics[width=\linewidth, height=0.3\textheight, keepaspectratio]{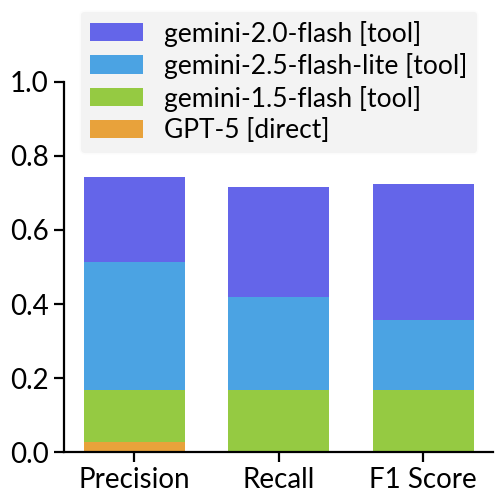}
    \caption{Benchmark results for various model choices. Tool-augmented methods are generally more performant than direct generation methods across tested queries.}
    \label{fig:mp-llm-benchmark}
\end{figure}

Benchmark results (Figure~\ref{fig:mp-llm-benchmark}) show that tool augmentation is critical for this task. The \textbf{Gemini 2.0 Flash [tool]}~\cite{Gemini} method achieved the highest performance, with a balanced F1 score of 72.3\% (74.3\% precision, 71.4\% recall). In contrast, \textit{direct} generation approaches such as \textbf{GPT-5 [direct]}~\cite{openai2024gpt4technicalreport} performed poorly, achieving 0\% recall and only 2.5\% precision, demonstrating that direct ID generation is unsuitable for reliable querying. Overall, the results show that tool-augmented approaches provide a more robust and effective strategy for natural language interaction with materials databases.

\subsection*{Future Work}
Future work will focus on natural language–driven visualization, converting prompts into chart specifications directly linked to Materials Project API results. Planned features include suggesting appropriate defaults, enabling interactive exploration, and supporting export of figures and reproducible code. Additional extensions include built-in post-processing triggered by short text instructions, such as filtering by stability thresholds, normalizing chemical formulas, aggregating by crystal system, computing simple derived properties, and flagging outliers, all with input validation and provenance tracking.

\subsection*{Open-source Materials}
The MP-LLM codebase is available on GitHub: \github{https://github.com/killiansheriff/MP-LLM}.





\section{PALS: Property Analogies with LLMs}\label{sec:PALS}



\indent \indent The ATOMS Lab team investigates whether large language models (LLMs) can reason about molecular and materials properties through analogical reasoning. LLMs can be trained to predict molecular properties via supervised learning \cite{jablonka2024leveraging} as well as in-context learning \cite{li2024empowering}. However, analogical reasoning for molecular property prediction remains underexplored \cite{segal2025known}. Here, the team presents preliminary experiments evaluating LLMs’ ability to construct and reason about analogies in molecules and materials.

\vspace{-0.3\baselineskip}
\subsection*{Results}

\subsection*{Crystal Structure}
\indent \indent Datasets of structural analogues were curated from the Materials Project \cite{jain2013commentary} using the StructureMatcher functionality of the Python library Pymatgen \cite{ong2013python}. Structural analogues maintain the same bond angles when one element is replaced with another. These analogues do not always share the same generic formula (e.g., ABC$_3$ for AcTiO$_3$ and HfZnO$_3$), although the generic formula was kept constant in this study.

\begin{figure}[ht]
    \centering
    \includegraphics[width=0.9\textwidth]{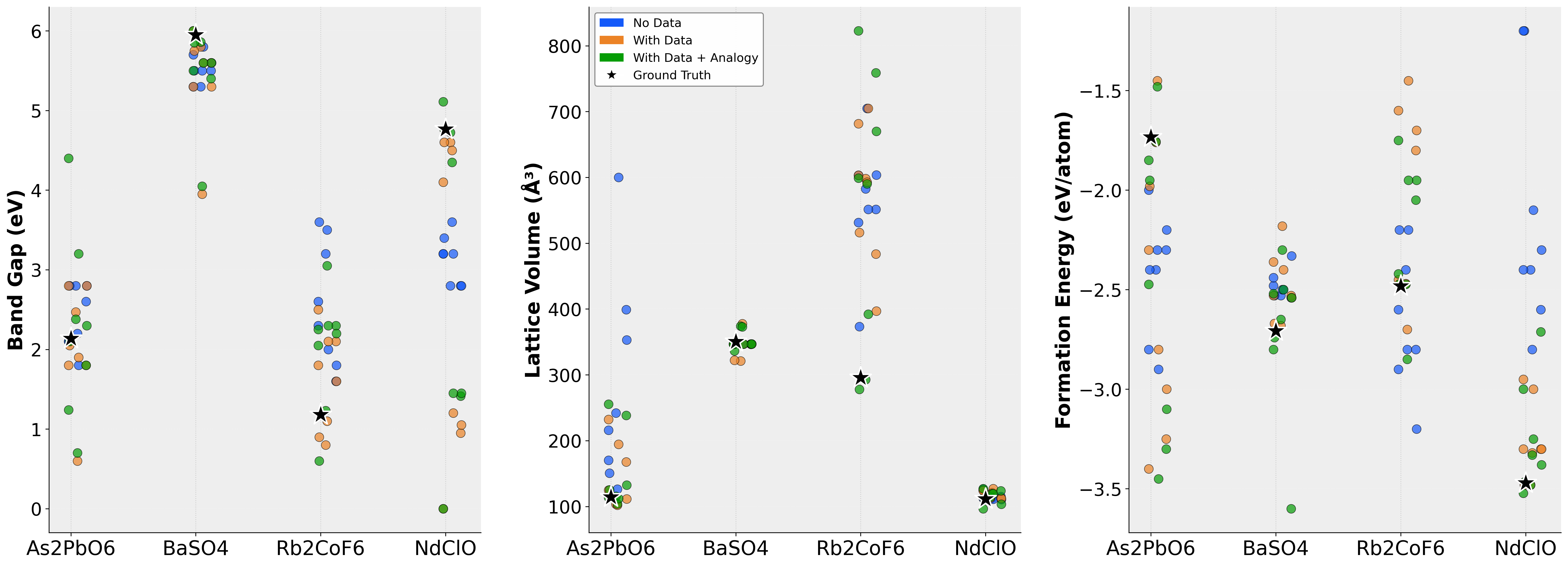}
    \caption{Predictions for four materials, comparing different prompting methods. A table of summary statistics is available on GitHub: \github{https://github.com/ahaibel/mp-property-analogies/blob/main/hackathon_submission_resources/mae_across_methods_properties.csv}}
    \label{fig:umbc_cluster_plots_all_properties}
    \vspace{-1\baselineskip}
\end{figure}

Three categories of prompts were tested using a commercial LLM (GPT-5-mini). The ``no data'' prompt asked directly for material properties. The ``baseline'' and ``analogical'' prompts provided datasets of structural analogues as context. For the baseline prompt, any prediction strategy was allowed, whereas the analogical prompt instructed the LLM to form an A : B :: C : D analogy. For example, one response constructed the analogy Sb$_2$SnO$_6$ : Sb$_2$PbO$_6$ :: As$_2$SnO$_6$ : As$_2$PbO$_6$ by exchanging the B-site cation while keeping the A-site cation analogous. In practice, the LLM frequently failed to follow instructions, instead relying on interpolation across exemplars. Prediction accuracy varied substantially across materials and properties (Figure~\ref{fig:umbc_cluster_plots_all_properties}). Providing data clearly improved performance, though the specific benefit of analogical prompting was less consistent.

\begin{figure}[ht]
    \centering
    \includegraphics[width=0.85\textwidth]{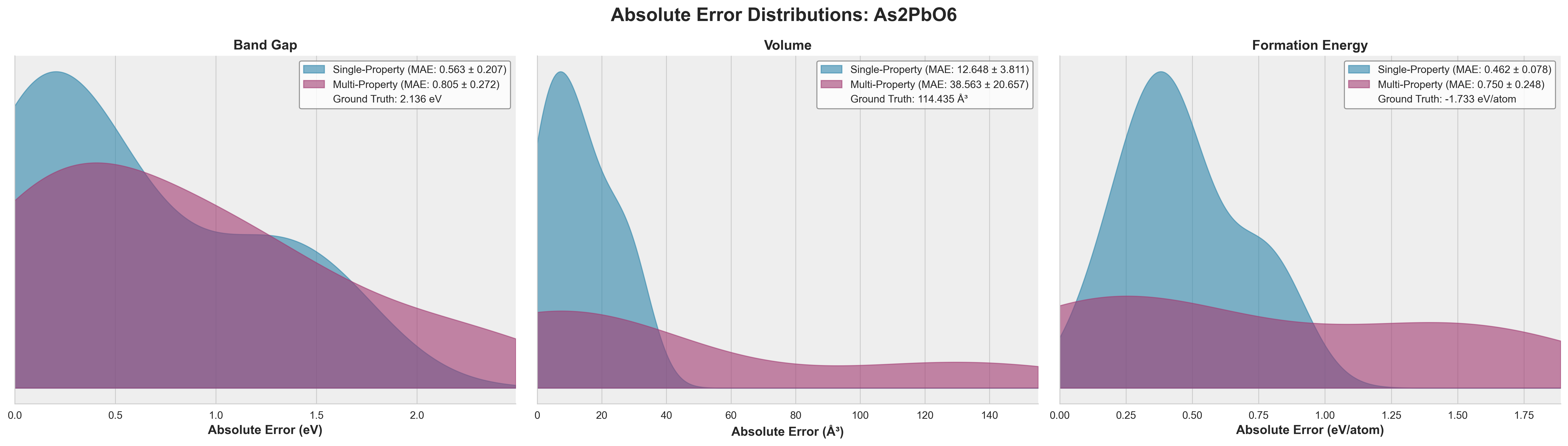}
    \caption{Model error distributions for single-property and multi-property prediction.}
    \label{fig:umbc_As2PbO6_distributions}
    \vspace{-1.2\baselineskip}
\end{figure}

\vspace{0.3\baselineskip}
Initial experiments asked the LLM to predict multiple properties simultaneously. Subsequent tests allowed the model to construct separate analogies for individual properties. Results are consistent with improved performance for single-property prediction, as reflected by more favorable error distributions (Figure~\ref{fig:umbc_As2PbO6_distributions}).

\subsection*{Chemosensory Prediction}
\begin{wrapfigure}{r}{0.4\textwidth}
\vspace{-\baselineskip}
\centering
\begin{tabular}{lccc}
\toprule
\textbf{Property} & \textbf{Baseline} & \textbf{Analogical} \\
\midrule
Ammonia  & 13.2 & 10.5 \\
Cold     &  9.1 &  7.9 \\
Decayed  & 12.8 & 11.1 \\
Familiar &  7.6 &  7.0 \\
Fish     & 12.2 & 11.1 \\
Pleasant & 11.9 & 10.2 \\
Strength & 11.9 & 12.6 \\
\midrule
\textbf{Overall} & \textbf{11.2} & \textbf{10.1} \\
\bottomrule
\end{tabular}
\captionof{table}{MAE for LLM scent prediction.}
\label{tab:umbc_scent_mae}
\end{wrapfigure}

\indent \indent LLMs were also evaluated for predicting psychophysical scent ratings using data from \cite{keller2016olfactory}, where human participants rated molecular scents across multiple descriptors. A total of 100 molecules were tested for seven descriptors (Table~\ref{tab:umbc_scent_mae}). For each test molecule, the remaining dataset of 480 molecules was provided as context to support analogical reasoning.

\vspace{0.3\baselineskip}
Both baseline and analogical prompts were evaluated. GPT-5-mini performed poorly overall, with performance worse than a simple mean-based predictor. However, analogical prompting achieved a modest reduction in MAE compared to the baseline approach (10.1 vs.\ 11.2 across descriptors).

\vspace{-0.3\baselineskip}
\subsection*{Future Work}
\indent \indent While LLMs appear capable of selecting exemplars and reasoning about their properties, their ability to perform true analogical reasoning remains unclear. Molecular and materials property prediction offers a promising domain for studying this capability. Key future directions include improving instruction adherence, developing methods to evaluate analogy quality, and designing experiments that distinguish between knowledge derived from training data and information provided in context.

\vspace{-0.3\baselineskip}
\subsection*{Open-source Materials}
Code is available on GitHub (run \texttt{main.py} with command-line arguments): \github{https://github.com/ahaibel/mp-property-analogies}\,;
Demo Video: \youtube{https://www.youtube.com/watch?v=Fboa8sOo3w0}





\section{CaMEL-RAG: A Retrieval-Augmented Generation Framework for Catalytic Screening}
\label{sec:camel-rag}



Screening heterogeneous catalysts for sustainable energy storage applications, such as green hydrogen production or CO$_2$ reduction, traditionally requires expensive density functional theory (DFT) calculations and expert knowledge of computational chemistry workflows \cite{run-ping2019hydrogenation, cho2025expanding}. The Code4Catalysis-KFUPM team introduces \textbf{Ca}talysis \textbf{M}odel \textbf{E}nhanced by \textbf{L}LM-RAG (\textbf{CaMEL-RAG}), a retrieval-augmented generation framework that translates natural-language queries into actionable catalytic insights.

\begin{figure}[h]
    \centering
    \includegraphics[width=0.8\linewidth]{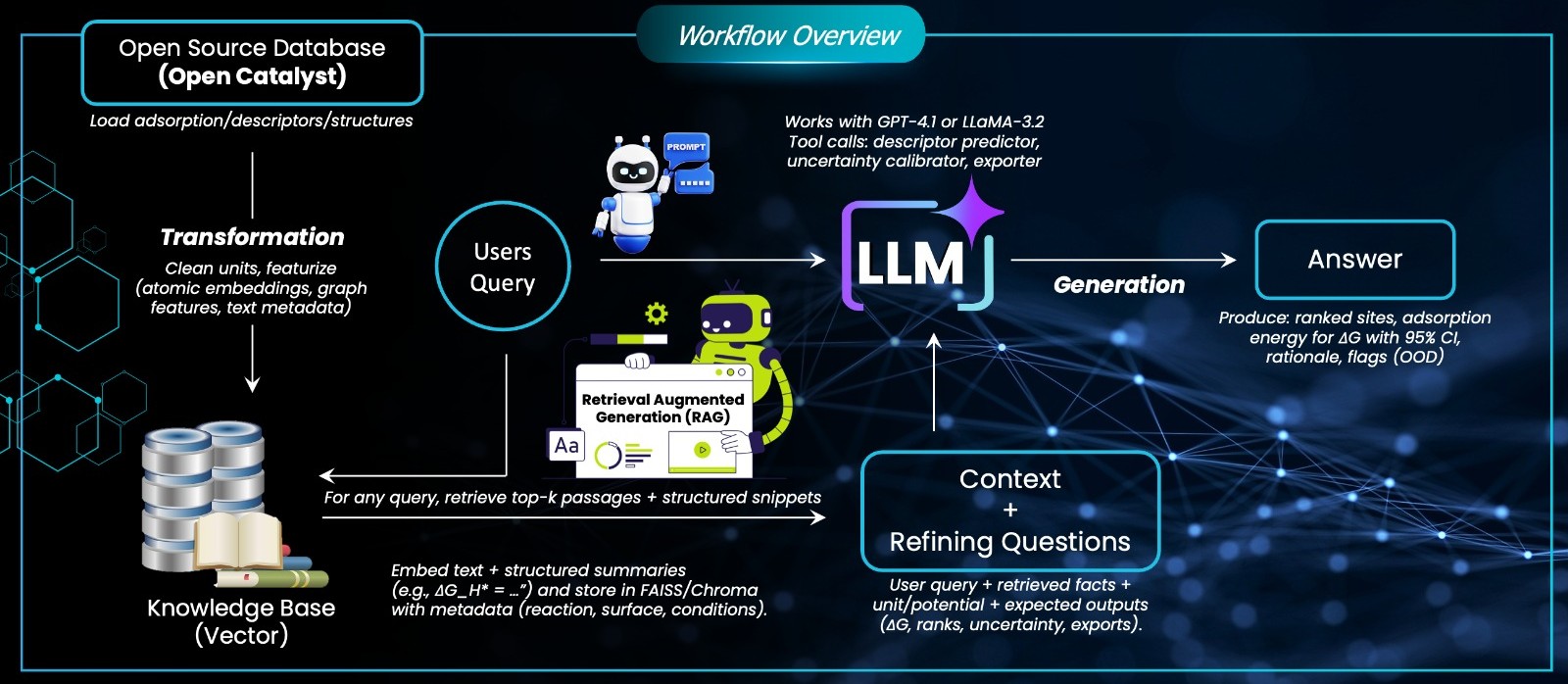}
    \caption{\textbf{CaMEL-RAG} framework for catalysis prediction.}
    \label{fig:camel_rag_Workflow}
\end{figure}

The CaMEL-RAG architecture is inspired by the hierarchical retrieval and orchestration concepts of CHORUS \cite{ahmed2025chorus}, adapted specifically for heterogeneous catalysis. The framework leverages the Open Catalyst Dataset \cite{zitnick2020introduction}, which contains structured records describing the slab, surface site, adsorbate, and corresponding adsorption energy. Since the dataset lacks intrinsic hierarchy, a flat vector representation was employed instead of CHORUS’s multi-level memory.

Each structured record was converted into a natural-language description retaining complete system information and treated as an independent document. Approximately 100{,}000 such descriptions were embedded using a Sentence-Transformer model to construct the CaMEL-RAG knowledge vector. During inference, a natural-language query triggers retrieval of semantically similar documents, which are re-ranked for relevance before being passed as contextual input to a large language model. The overall workflow is illustrated in Figure~\ref{fig:camel_rag_Workflow}.

\begin{figure}[h]
    \centering
    \includegraphics[width=0.9\linewidth]{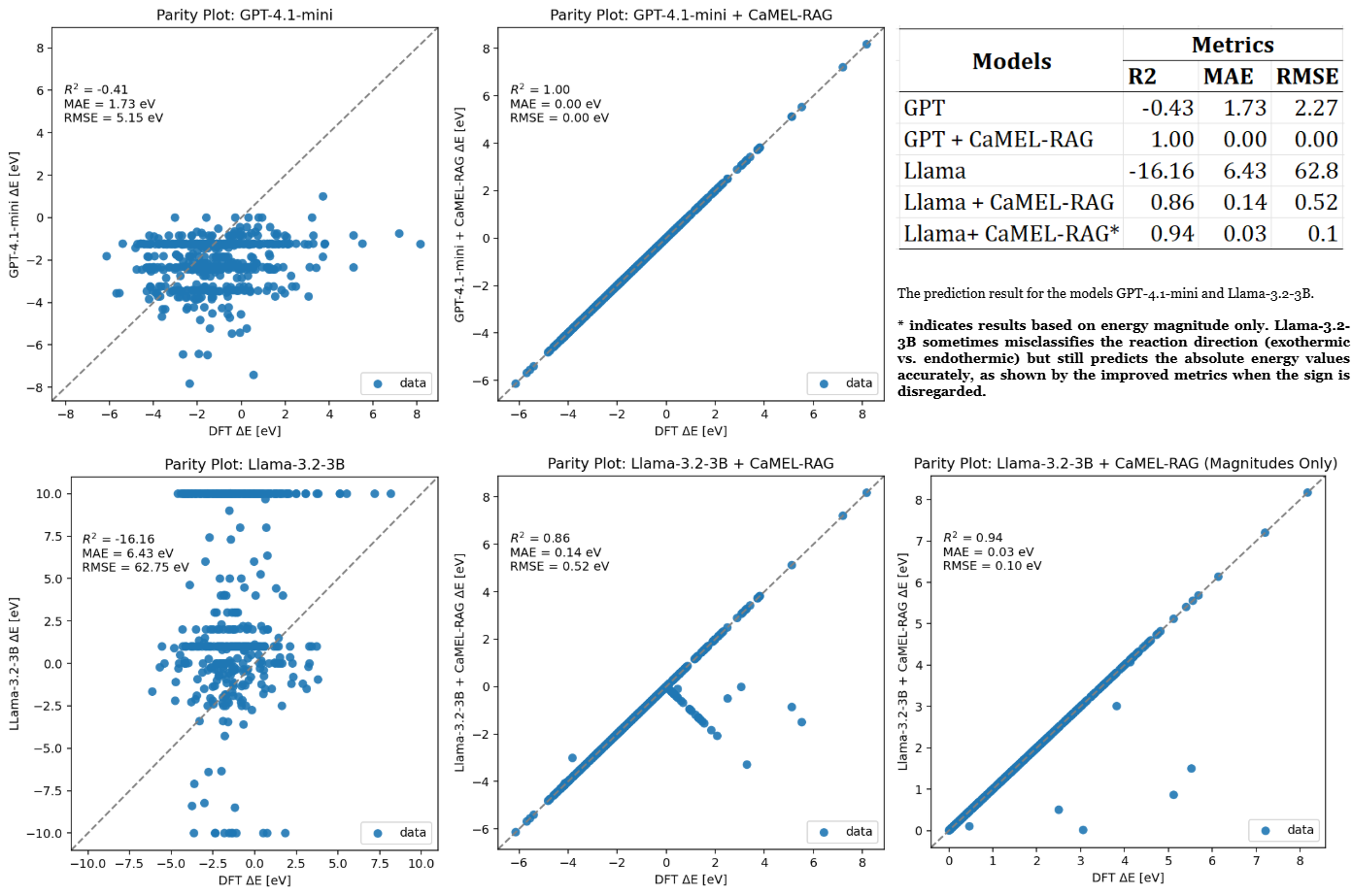}
    \caption{Performance comparison of baseline LLMs and CaMEL-RAG-enhanced models for adsorption-energy prediction.}
    \label{fig:camel_rag_result}
\end{figure}

\subsection*{Results}
CaMEL-RAG was evaluated for adsorption-energy prediction using two large language models: \texttt{gpt-4o-mini} and \texttt{LLaMA~3.2--3B}. Model predictions were compared against DFT-calculated adsorption energies using the mean absolute error (MAE) and coefficient of determination ($R^2$).

As shown in Figure~\ref{fig:camel_rag_result}, baseline LLMs exhibited substantial performance degradation under zero-shot inference, particularly for numerically intensive, domain-specific queries. In contrast, CaMEL-RAG significantly improved predictive accuracy for both evaluated models by providing relevant contextual information. The \texttt{gpt-4.1-mini} model achieved strong agreement with reference DFT values, correctly capturing both magnitude and sign of adsorption energies. While \texttt{LLaMA~3.2-3B} reproduced reasonable magnitudes, it occasionally failed to predict the correct sign, highlighting challenges in reasoning about reaction energetics.

\subsection*{Future Work}
Future extensions of CaMEL-RAG will explore hierarchical retrieval mechanisms to better capture relationships among catalyst families, surface facets, and adsorbate types. Additional directions include incorporating multimodal data (e.g., atomic structures or simulation outputs), fine-tuning open-weight language models on catalysis-specific corpora, and extending the framework toward generative catalyst design and automated reaction-pathway prediction.

\subsection*{Open-source Materials}
All data and code associated with this project are available on GitHub: \github{https://github.com/ashikiut/CaMEL-RAG/}





\section{SuperconLLM: Multi-Agent Framework for Automating Superconductivity Literature Knowledge Extraction}\label{sec:SuperconLLM}



Extracting structured data from scientific literature is critical for accelerating materials discovery, yet manual curation remains labor-intensive and costly. Recent work has demonstrated the potential of large language models (LLMs) for materials science information extraction tasks \cite{Foppiano2024}, but most existing approaches evaluate named entity recognition (NER) and relation extraction (RE) in isolation, without demonstrating a fully automated paper-to-database pipeline. The Team\_Tc team developed SuperconLLM to enable end-to-end automation of superconductivity database construction through a four-stage multi-agent LLM framework.

\begin{figure}[h]
    \centering
    \includegraphics[width=1.0\linewidth]{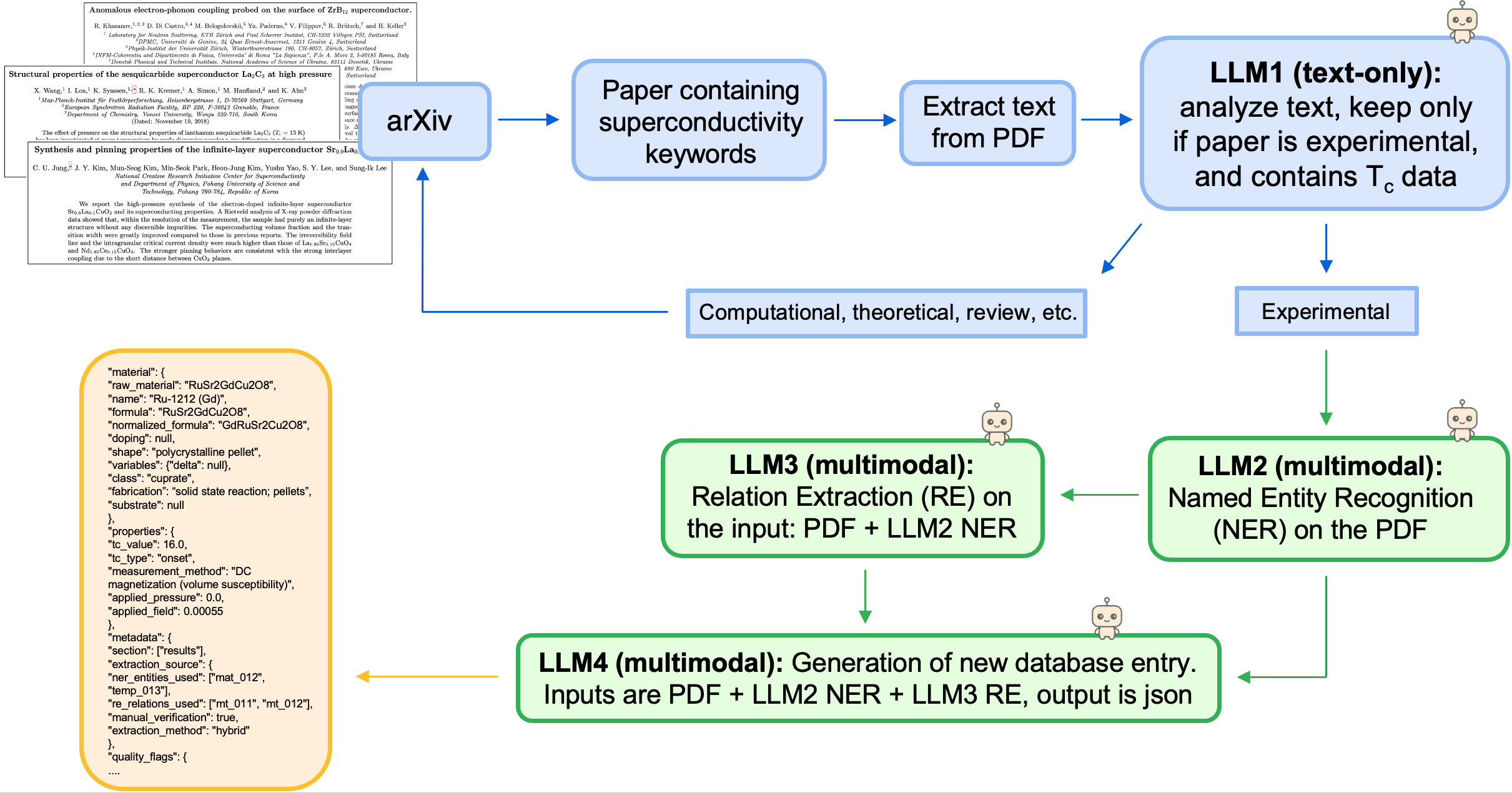}
    \caption{The SuperconLLM fully automated workflow, from arXiv papers to JSON records.}
    \label{fig:SuperconLLM_figure}
\end{figure}

\subsection*{Results}
SuperconLLM implements a pipeline (Figure~\ref{fig:SuperconLLM_figure}) that scans the arXiv database for relevant papers and transforms raw PDFs directly into structured JSON database entries. Superconductivity papers are sourced via the arXiv API and filtered using keyword matching (e.g., ``superconductor'', ``superconductivity'', ``superconducting'' in titles or abstracts). Each paper is then processed independently through a sequence of specialized LLM agents:

\begin{itemize}
    \item \textbf{LLM1 – Paper Screening (Qwen-3-235b-a22b-instruct-2507):} Text is extracted from PDFs using PyMuPDF and passed to a fast, text-only LLM via the Cerebras API to identify experimental papers reporting critical temperature ($T_{\text{c}}$) measurements. Purely theoretical, computational, and review papers are filtered out, and only qualifying papers proceed to subsequent stages.
    
    \item \textbf{LLM2 – Named Entity Recognition (Claude Sonnet 4):} Performs multimodal NER directly on PDFs (claude-sonnet-4-20250514 via the Anthropic API), identifying materials (chemical formulas and sample identifiers), temperatures ($T_{\text{c}}$ values and transition temperatures), and experimental conditions (pressure, magnetic field, synthesis parameters). Outputs structured JSON with entity spans, confidence scores, and page-level provenance.
    
    \item \textbf{LLM3 – Relation Extraction (Claude Sonnet 4):} Consumes the PDF alongside the NER output, establishing relationships between entities (e.g., linking specific materials to measured $T_{\text{c}}$ values under defined conditions). Produces JSON-formatted relation triples with supporting evidence and confidence scores.
    
    \item \textbf{LLM4 – Database Synthesis (Claude Sonnet 4):} Integrates information from the PDF, NER, and RE stages to generate final database-ready JSON entries. This stage performs document-level aggregation, chemical formula normalization, duplicate detection, and validation. Each record includes material composition, measured superconducting properties, experimental conditions, measurement methods, metadata, quality flags, and full provenance tracking.
\end{itemize}

Preliminary validation on recent arXiv papers demonstrated successful end-to-end extraction, producing database entries judged correct by human evaluation. However, quantitative evaluation of precision, recall, and F1 score against established baselines such as Grobid-superconductors \cite{Foppiano2023} remains future work.

\subsection*{Future Work}
Future efforts will focus on establishing quantitative benchmarks by evaluating current prompt configurations against Grobid-superconductors, with iterative prompt refinement until performance saturation is reached. Locally hosted open-weight models (e.g., Qwen~2.5~VL) will also be explored to reduce API costs and enable large-scale processing. If satisfactory performance is achieved, the framework will be scaled to the full arXiv superconductivity corpus to produce a comprehensive, continuously updatable open database.

A complementary ``one-shot'' approach, in which a single multimodal LLM performs NER, RE, and database synthesis in one pass, will also be investigated. While potentially more cost-efficient, this approach increases task complexity and may trade off extraction accuracy. Finally, the SuperconLLM framework could be adapted to automate dataset creation in other materials science domains.

\subsection*{Open-source Materials}
The codebase, including prompts for each LLM agent, is available on GitHub: \github{https://github.com/fpriante/SuperconLLM-Multi-Agent-Framework-for-Automating-Superconductivity-Knowledge-Extraction}





\section{Catalyze: AI-Powered Multi-Agent Chemistry Assistant for Lab Automation and Safety Analysis}\label{sec:Catalyze}



Laboratory automation in chemistry and materials science traditionally requires extensive programming expertise and platform-specific knowledge, creating significant barriers for researchers seeking to leverage liquid-handling robots and automated workflows. The Catalyze team addresses this challenge by introducing an intelligent multi-agent AI system that democratizes access to lab automation through natural-language interfaces. The platform combines specialized AI agents with comprehensive chemical databases to provide context-aware assistance across research workflows, protocol generation, dual-platform automation (OpenTrons Python and Dynamic Devices Lynx C\#), and integrated safety analysis. By bridging conversational AI and laboratory hardware control, Catalyze enables researchers to generate validated automation scripts and safety-compliant protocols without deep programming expertise.

\begin{figure}[h]
    \centering
    \includegraphics[width=0.95\linewidth]{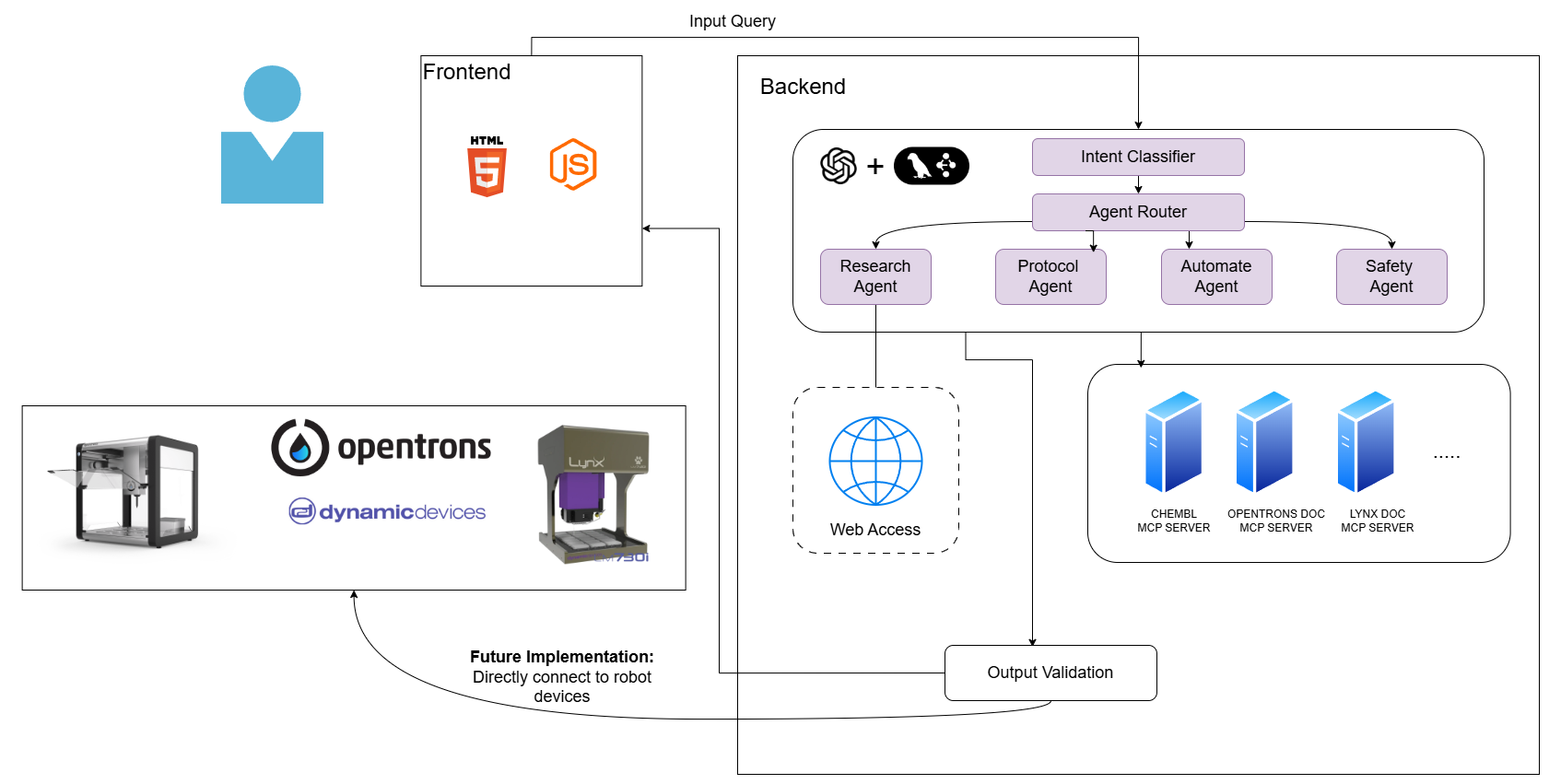}
    \caption{End-to-end architecture of Catalyze, illustrating agent orchestration from user query to validated automation script generation and platform integration.}
    \label{fig:catalyze_workflow}
\end{figure}

\subsection*{Results}
Catalyze employs a modular, agent-based architecture comprising five specialized AI agents coordinated by a router and pipeline manager (Figure~\ref{fig:catalyze_workflow}). The Research Agent integrates ChEMBL’s suite of tools and the PubChem database for compound and target analysis. The Protocol Agent generates detailed laboratory procedures, including safety checks, material lists, and professionally formatted protocols. The Automation Agent enables interactive code generation for OpenTrons OT-2 (Python) and Dynamic Devices Lynx (C\#), producing validated scripts with error handling and appropriate liquid-class parameters. The Safety Agent performs hazard assessments with MSDS integration, while the PDF Analysis Module leverages GPT-4o to extract and consolidate information from scientific papers via drag-and-drop uploads.

The system is implemented using Python~3.12+, Flask~3.1+, LangChain, and LangGraph, with a lightweight HTML/CSS/JavaScript frontend. It connects to the ChEMBL MCP Server for chemical data access and includes platform-specific code generators with built-in validation. For example, when prompted to ``generate code for serial dilution,'' the Automation Agent requests the target platform and returns a complete, executable script with appropriate imports, metadata, and validation logic.

The user interface supports responsive dark and light themes, real-time contextual processing, and seamless handling of both text and PDF inputs. Figure~\ref{fig:catalyze_workflow} illustrates the full workflow, from query routing through agent execution, MCP integration, validation, and response delivery. This scalable architecture facilitates rapid extension to additional agents and laboratory automation platforms.

\subsection*{Future Work}
Future development will focus on expanding multi-PDF support for concurrent document analysis, integrating advanced molecular structure visualization with interactive 3D viewers, and enabling real-time collaborative protocol editing. Planned extensions include direct hardware integration for closed-loop experimental feedback and expanded automation platform support beyond OpenTrons and Lynx to systems such as Hamilton STAR and SPT Labtech. Additional priorities include persistent user data storage with Redis-based caching, development of a public API for third-party integrations, and containerization into microservices with automated CI/CD pipelines to improve deployment and scalability.

\subsection*{Open-source Materials}
Code and documentation are available on GitHub: \github{https://github.com/srustisain/mit-catalyze}





\section{CAMEL: Causal Analysis of Materials Extracted from Literature}
\label{sec:camel}


Understanding how processing, structure, and properties are causally connected remains a central goal in materials science. While large language models (LLMs) have enabled automated text mining, extracting explicit causal relationships from the literature remains challenging. Such causal understanding is critical for building interpretable process–structure–property maps that support inverse design and hypothesis generation.

\begin{figure}[h]
    \centering
    \includegraphics[width=0.8\linewidth]{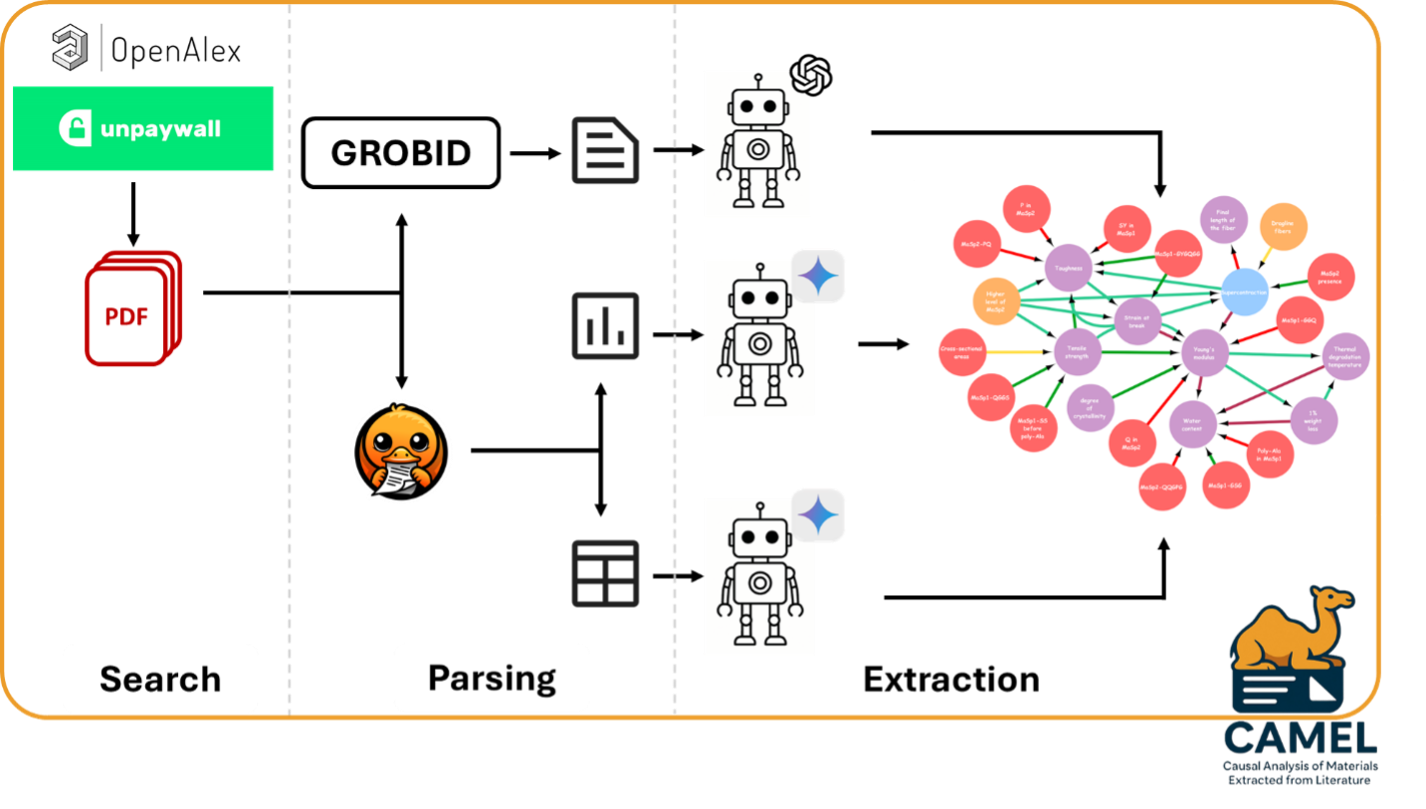}
    \caption{CAMEL workflow. Open-access papers are collected via OpenAlex and Unpaywall, then parsed into text, figures, and tables using GROBID and Dockling. Independent multimodal LLM agents process language, visual, and tabular data to extract causal relationships, which are merged into a unified knowledge graph.}
    \label{fig:camel_workflow}
\end{figure}

\subsection*{Results}
CAMEL addresses causal relationship extraction by using multimodal LLMs to identify and organize causal and correlational knowledge from scientific literature into interpretable graph structures. The CAMEL pipeline integrates literature retrieval, multimodal document parsing, and LLM-based entity–relation extraction into a fully automated workflow. Independent LLM agents—leveraging models such as GPT-5 or Gemini~2.5—process textual, visual, and tabular content separately, and the extracted information is merged into a unified causal knowledge graph linking material processes, structures, and properties.

The framework was demonstrated on spider silk literature, a representative hierarchical biomaterial exhibiting complex structure–property coupling. CAMEL successfully combined information from text, figures, and tables to construct causal graphs capturing relationships such as the influence of spinning conditions and $\beta$-sheet content on tensile strength. Compared to a text-only baseline, CAMEL achieved over $90\%$ correctness and substantially higher coverage of relevant relationships, highlighting the advantages of multimodal reasoning for causal extraction.

\subsection*{Future Work}
Future extensions will focus on ontology alignment, uncertainty quantification, and integration with simulation and experimental databases. These developments aim to enable scalable, literature-informed materials discovery supported by interpretable causal models.

\subsection*{Open-source Materials}
Code and documentation are available on GitHub: \github{https://github.com/gourav-k/LLM4Causal.git}





\section{ZeroMAT: Zero-training MATerial Autonomous Analysis with Large Language Model and Retrieval-Augmented Generation}\label{sec:ZeroMAT}



Materials property prediction faces a critical bottleneck in which machine learning models demand substantial computational resources and extensive training data, while most materials datasets remain small and incomplete \cite{schmidt2019recent, butler2018machine}. Large language model (LLM)-based approaches lack direct access to physical features required for bandgap prediction, struggle with missing values, and often require costly retraining for each new property \cite{borlido2020exchange}. The ZeroMAT team presents ZeroMAT, which integrates TabPFN’s zero-shot learning capability \cite{hollmann2025accurate} with retrieval-augmented generation (RAG) \cite{lewis2020retrieval} to create a zero-training framework for materials property prediction. ZeroMAT achieves superior predictive performance (R$^2$ up to 0.8261, a 43\% improvement over baseline TabPFN), while delivering 30$\times$ faster processing and 2--4$\times$ reduced GPU memory usage compared to fine-tuned LLM approaches.

\subsection*{Results}
\begin{figure}[h]
    \centering
    \includegraphics[width=0.90\linewidth]{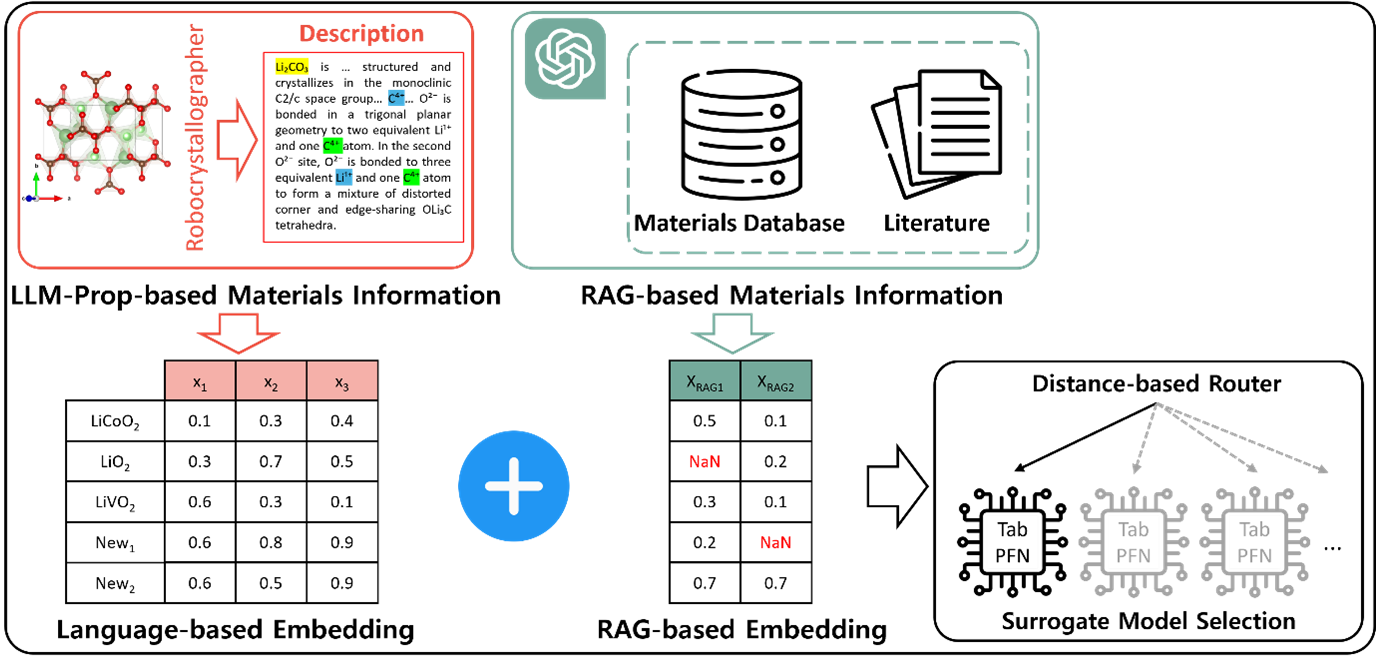}
    \caption{ZeroMAT framework architecture.}
    \label{fig:ZeroMAT}
\end{figure}

ZeroMAT integrates three complementary components to enable zero-training materials property prediction (Figure~\ref{fig:ZeroMAT}). First, domain-aware textual representations are generated using Robocrystallographer to create crystallographic descriptions for each material. These descriptions are encoded into 512-dimensional embeddings via LLMProp and subsequently reduced to 100 dimensions using principal component analysis (PCA) for computational efficiency. Second, these representations are augmented through retrieval-augmented generation by querying GPT-4 with property-specific prompts (e.g., ``What are the most important features for predicting bandgap in perovskites?''). Relevant material properties are then retrieved from the Materials Project database, providing additional chemical and structural context.

To enable scalable deployment across datasets of varying size, ZeroMAT employs chemistry-aware clustering using k-means to partition the dataset and train separate TabPFN surrogate models for each cluster. This strategy exploits TabPFN’s strong performance on focused subsets (up to 500 dimensions and 10{,}000 data points) while extending applicability to larger databases. During inference, predictions are routed based on the target material’s distance to cluster centroids, selecting the most chemically relevant surrogate model. This distance-based routing maintains computational efficiency while ensuring relevant training context for each prediction.

Experimental evaluation using bandgap data from the Materials Project \cite{jain2013commentary} demonstrates that ZeroMAT delivers substantial improvements in both accuracy and efficiency (Table~\ref{tab:ZeroMAT}). Traditional fine-tuning approaches exhibit poor performance and high computational cost, requiring approximately 38 minutes of training time and $\sim$8~GB of GPU memory. In contrast, ZeroMAT with RAG integration achieves a 43\% improvement in R$^2$ while processing in just 78 seconds with less than 1~GB of memory. When scaled to a large Materials Project bandgap dataset containing 40{,}000 materials, ZeroMAT maintains high accuracy and memory efficiency through chemistry-aware clustering, a regime where conventional approaches fail. These results position ZeroMAT as a paradigm shift in materials property prediction, demonstrating that zero-training frameworks can outperform resource-intensive fine-tuning by up to 30$\times$ in speed while delivering superior predictions.

\begin{table}[h]
\centering
\small
\caption{Performance comparison of different approaches for bandgap prediction on Materials Project datasets.}
\label{tab:ZeroMAT}
\begin{tabular}{lcccccc}
\toprule
\textbf{Approach} & \textbf{Dataset} & \textbf{Training} & \textbf{GPU} & \textbf{R$^2$} & \textbf{MAE} \\
 & \textbf{Size} & \textbf{Time} & \textbf{Memory} & \textbf{Score} & \textbf{(eV)} \\
\midrule
LLM-Prop + Fine-tuning & $<$10k & 38 min & $\sim$8 GB & 0.388 & 0.800 \\
LLM-Prop + TabPFN & $<$10k & 62.34 s & $<$1 GB & 0.579 & 0.656 \\
LLM-Prop + TabPFN + RAG & $<$10k & 78.32 s & $<$1 GB & 0.826 & 0.365 \\
LLM-Prop + TabPFN + RAG & 40k & 157.40 s & $<$1 GB & \textbf{0.988} & \textbf{0.006} \\
\bottomrule
\end{tabular}
\end{table}

\subsection*{Future Work}
Future directions include integrating uncertainty quantification to support closed-loop autonomous materials discovery, extending the RAG framework to incorporate materials knowledge graphs for hierarchical and multi-scale property prediction, and developing an inverse design module that leverages ZeroMAT’s efficiency to explore chemical spaces and identify optimal compositions. These advances aim to transform forward property prediction into scalable, generative materials design.

\subsection*{Open-source Materials}
The ZeroMAT codebase is available on GitHub: \github{https://github.com/Ahri111/ZEROMAT}





\section{ULNA: Unstructured Lab Notebook Assistant}
\label{sec:ulna}


For polymeric materials, small changes in synthesis and processing procedures can considerably reshape topology and final properties, even for the same chemistry \cite{rubinstein2003polymer}. In research labs, these procedural shifts often surface as qualitative observations such as “the resin turned milky halfway through” or “I left the sample cooling on the bench overnight.” The ULNA team builds on previous Hackathon efforts on using LLMs to improve lab notebooks \cite{jablonka202314, zimmermann202534} and uses GPT-5 not only to convert such narrative, unstructured experiment logs into structured, searchable data but also to expose how fuzzy, human-level details influence polymer outcomes.


\subsection*{Results}
The ULNA team provided 20 audio recordings, transcribed using Whisper \cite{radford2023robust}, of a user narrating their procedure while setting up Differential Scanning Calorimetry (DSC) experiments on reactive polymer resins, together with the corresponding LIMS output files containing reaction enthalpy, onset temperature, and the raw heat-flow versus temperature data. A dedicated prompt (available on the GitHub repository) guided GPT-5 to structure the data on demand and then analyze it.

\textit{Structure from recordings}: The model successfully extracted 315 distinct entries with greater than 99\% accuracy, missing only two due to incomplete transcription. This result is consistent with previous demonstrations of speech-to-structured-data conversion \cite{zimmermann202534}. The system also captured experimental corrections (“I aimed to measure 9.5 g, but it’s actually 9.48 g”), procedural differences such as the order of adding comonomers, and even computed molar ratios directly from the chemistry data sheet containing molecular weights, without explicit instruction.

\textit{Analysis from recordings}: When prompted to identify which variables affected reaction enthalpy, GPT-5 recognized both expected structured factors (such as monomer type) and subtle unstructured procedural trends. For example, it linked notes about undissolved catalyst particles or “insoluble residue left after sonication” to reduced enthalpy. As another example, it highlighted that an experimentalist who mentioned “I left for lunch before running the DSC” produced a run with lower heat release, interpreted as sample aging due to background reaction during the intervening time. These findings show that fuzzy, narrative data can reveal cause–effect patterns in DSC results that structured tables alone cannot.

The application of ULNA is particularly relevant for polymers, where small differences in handling or processing can influence network formation, cross-link density, or chain topology. The results suggest that capturing and analysing such subtle, human-level process details—often absent from structured datasets—may help explain why polymers prepared under nominally identical conditions sometimes show measurable differences in resin reactivity or final material structure. ULNA thus points toward a path for integrating unstructured procedural context into the quantitative understanding of polymer behaviour.

\subsection*{Future Work}
Next steps will test ULNA in live polymer experiments to link narrated procedures with other measured physical properties. Beyond manual audio recordings, incorporating vision-based detection approaches will take this a step further by observing procedural steps and resulting phenomena that the human researcher did not notice. This could speed up the research process by identifying trends more quickly and eliminating mistakes. Embedding this approach in routine lab workflows could reveal how subtle process variations influence polymer reactivity and structure, improving reproducibility and helping inform future research directions.


\subsection*{Open-source Materials}
Code available on GitHub: \github{https://github.com/iarretche/Unstructured-Lab-Notebook-Assistant-ULNA-}






\section{Multilingual Multimodal Materials Information Extraction (MuMMIE)}
\label{sec:mummie}



Information extraction (IE) is an important part of the materials discovery pipeline. Global research efforts toward materials discovery are documented across text, tables, and figures in patents, handbooks, and research papers written in multiple languages \cite{puccetti2021simple}. Due to the absence of standardized conventions for reporting materials composition and properties in scientific documents, extracting and compiling such information into structured databases remains challenging \cite{hira2025matskraft, circi2024well, kosonocky2024mining}. The \textsc{MuMMIE} team addresses this problem by introducing a curated benchmark of patent PDFs in six languages (EN, RU, CN, JP, KR, FR), paired with ground-truth annotations for materials composition and properties. The team also proposes composition-level, property-level, and combined metrics for evaluating the performance of IE tools on this challenging multilingual and multimodal dataset.


\begin{figure}[h]
    \centering
    \includegraphics[width=0.8\textwidth]{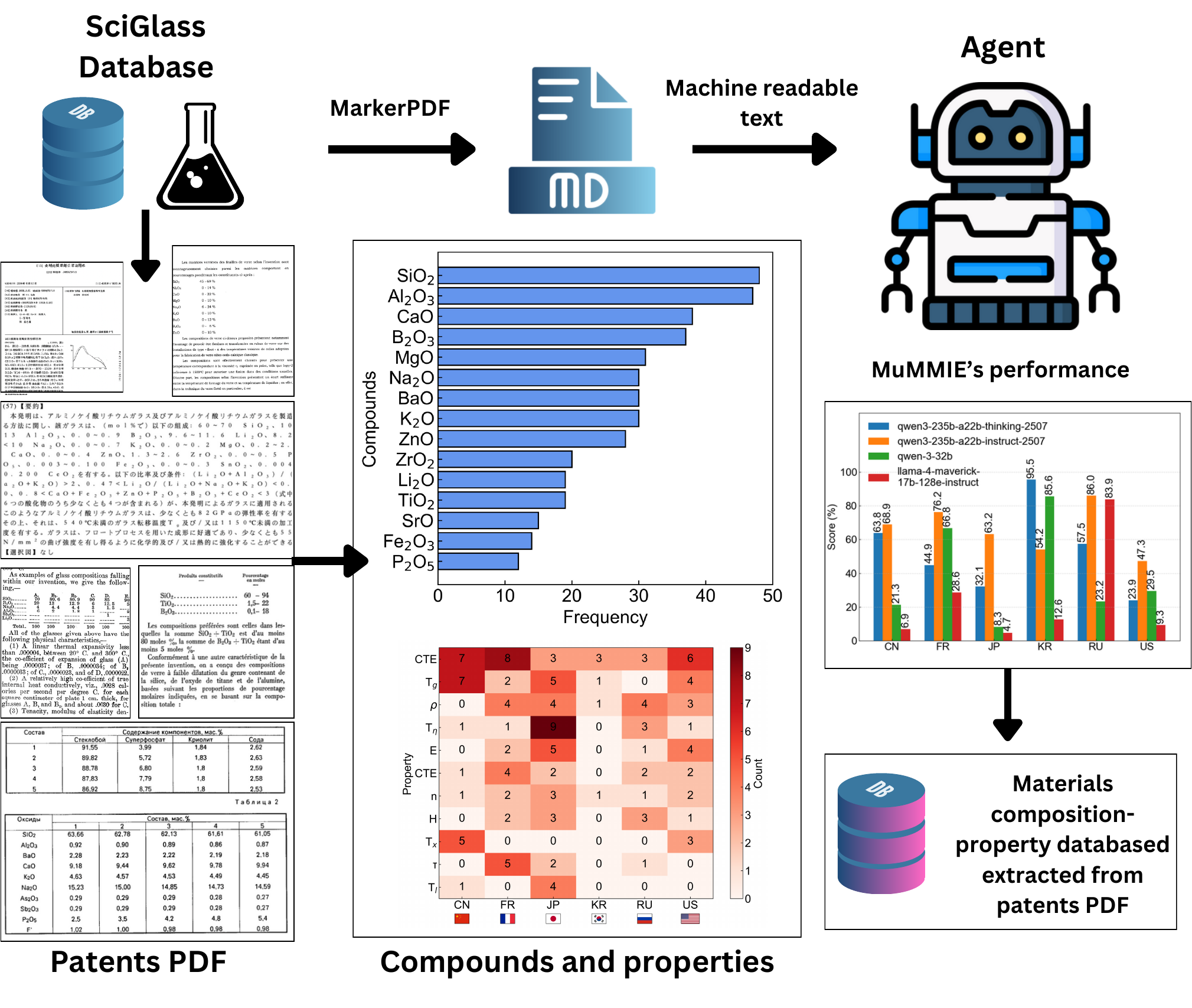}
    \caption{Workflow of MuMMIE model pipeline.}
    \label{fig:mummie}
\end{figure}

\subsection*{Results}
In the multilingual patent corpus spanning Chinese, Russian, French, Japanese, Korean, and English, the team observed that while chemical compound names often remain consistent across languages, the associated property labels vary widely. This inconsistency makes it difficult to build unified, machine-readable datasets. The primary objective of MuMMIE is to leverage large language models (LLMs) to accurately extract materials compositions and properties from documents written in different languages.

The current workflow focuses on patent data processing. Patent PDFs are first converted into Markdown format using the \texttt{marker} library to prepare text for downstream analysis. For information extraction, four LLMs were evaluated to identify and link composition elements or compounds with their associated material properties. The extracted data were then analyzed and visualized to highlight the top 20 compositions and their corresponding property distributions within the corpus (Fig.~\ref{fig:mummie}).


\subsection*{Evaluation}
\subsubsection*{Metrics}
Composition extraction is evaluated on a per-patent basis by converting both reference annotations and model predictions into sets of tuples $(c,v,u)$, where $c$ denotes the component name, $v$ the numeric value rounded to two decimal places, and $u$ the unit. Two families of metrics are computed. Let $S_g$ denote the number of gold-labelled patents and $S_p$ the number of patents for which the system produced at least one prediction. For patent $i$, with gold set $G_i$ and predicted set $P_i$:
\begin{align}
\mathrm{Prec}_{\text{exact}} &= \frac{\#\{i:\,P_i = G_i\}}{S_p},\quad
\mathrm{Rec}_{\text{exact}}  = \frac{\#\{i:\,P_i = G_i\}}{S_g},\quad
\mathrm{F1}_{\text{exact}}   = \frac{2\,\mathrm{Prec}_{\text{exact}}\,\mathrm{Rec}_{\text{exact}}}{\mathrm{Prec}_{\text{exact}}+\mathrm{Rec}_{\text{exact}}}.
\end{align}

\noindent\textbf{Macro (Exact)} reports the mean of per-patent exact scores.  
\noindent\textbf{Pairwise (tuple-level):} Let $G=\bigcup_i G_i$ and $P=\bigcup_i P_i$. Define $\mathrm{TP}=|P\cap G|$, $\mathrm{FP}=|P\setminus G|$, and $\mathrm{FN}=|G\setminus P|$:
\begin{align}
\mathrm{Prec}_{\text{pairs}} &= \frac{\mathrm{TP}}{\mathrm{TP}+\mathrm{FP}},\quad
\mathrm{Rec}_{\text{pairs}}  = \frac{\mathrm{TP}}{\mathrm{TP}+\mathrm{FN}},\quad
\mathrm{F1}_{\text{pairs}}   = \frac{2\,\mathrm{Prec}_{\text{pairs}}\,\mathrm{Rec}_{\text{pairs}}}{\mathrm{Prec}_{\text{pairs}}+\mathrm{Rec}_{\text{pairs}}}.
\end{align}

\noindent\textbf{Macro (Pairs)} averages pairwise scores over patents, while \textbf{Micro (Pairs)} computes scores over corpus-level true positives, false positives, and false negatives.


\begin{table}[t]
\centering

\begin{minipage}{0.48\textwidth}
\centering
\caption{Overall evaluation across all languages.}
\label{tab:overall}
\footnotesize
\setlength{\tabcolsep}{3pt}
\begin{tabular}{lccccc}
\toprule
\textbf{Model} & \textbf{F1 (Exact)} & \textbf{F1 (Pairs)} & \textbf{P} & \textbf{R} & \textbf{F1} \\
& (\%) & (\%) & (\%) & (\%) & (\%) \\
\midrule
Llama-4 17B & 19.68 & 32.29 & 74.74 & 17.56 & 28.43 \\
Qwen-3 32B  &  8.16 & 27.43 & 67.99 & 20.26 & 31.22 \\
Qwen-3 235B (I) & \textbf{26.63} & \textbf{62.56} & \textbf{71.38} & \textbf{62.56} & \textbf{66.68} \\
Qwen-3 235B (T) & 22.35 & 41.95 & 88.48 & 30.56 & 45.43 \\
\bottomrule
\end{tabular}
\end{minipage}
\hfill
\begin{minipage}{0.48\textwidth}
\centering
\caption{Per-language Micro (Pairs) F1 (\%).}
\label{tab:lang}
\footnotesize
\setlength{\tabcolsep}{4pt}
\begin{tabular}{lcccc}
\toprule
& \textbf{Llama-4} & \textbf{Qwen-3} & \multicolumn{2}{c}{\textbf{Qwen-3 235B}} \\
\cmidrule(lr){4-5}
& \textbf{17B} & \textbf{32B} & \textbf{Instr.} & \textbf{Think.} \\
\midrule
CN & 6.90 & 21.28 & \textbf{68.86} & 63.81 \\
FR & 28.57 & 66.76 & \textbf{76.24} & 44.93 \\
JP & 8.26 & 4.69 & \textbf{63.17} & 32.07 \\
KR & 12.61 & 85.58 & 54.17 & \textbf{95.52}$^\dagger$ \\
RU & 83.94 & 23.24 & \textbf{85.97} & 57.47 \\
US & 9.29 & 29.50 & \textbf{47.33} & 23.86 \\
\bottomrule
\end{tabular}
\end{minipage}

\end{table}

\subsection*{Overall Summary}
Table~\ref{tab:overall} summarizes corpus-level outcomes across all 53 patents. Among the evaluated models, Qwen-3 235B (Instruct) achieves the strongest overall performance, attaining the highest Macro F1 scores (Exact and Pairs) as well as the best Micro (Pairs) F1.

\subsection*{Per-Language Results}
Table~\ref{tab:lang} reports Micro (Pairs) F1 scores per language. Qwen-3 235B (Instruct) leads performance in most languages, including Chinese, French, Japanese, Russian, and English. Korean results peak with Qwen-3 235B (Thinking), though this result reflects reduced coverage due to missing JSON outputs.

\subsection*{Model Comparison}
\textbf{Best overall:} Qwen-3 235B (Instruct) achieves the highest Macro F1 (Exact and Pairs) and the best Micro (Pairs) F1 (66.68\%).  
\textbf{Language strengths:} Qwen-3 235B (Instruct) leads in CN, FR, JP, RU, and US, while KR performance is strongest for Qwen-3 235B (Thinking).  
\textbf{Underperformers:} Qwen-3 32B exhibits low Macro F1 (Exact) despite moderate Micro (Pairs) F1, and Llama-4 underperforms in CN, JP, and US but remains competitive in RU.

\subsection*{Future Work}
Future directions include expanding supported data modalities by incorporating figures and schematic images in addition to text, enabling richer capture of scientific information. The team also plans to broaden model evaluation by testing additional multilingual and domain-adapted LLMs to improve robustness and cross-lingual accuracy.

\subsection*{Open-source Materials}
Code available on GitHub: \github{https://github.com/zakidotai/MuMMIE}\,;
Demo Video: \youtube{https://x.com/DCirci/status/1966658394363699590}\,;

\section{Towards an Automated System for Smarter Electrolyte Design via Machine Learning Methods}\label{sec:electrolyte_ml}


The JH\_sqr team proposes an automated machine-learning-driven framework to select solid electrolytes that maximize battery capacity given specific anode/cathode materials and operating conditions. Battery capacity (charge storage) is treated as the target reward, while factors such as battery chemistry, temperature and pressure, C-rate, state-of-health (SoH), depth-of-discharge (DoD), manufacturing quality (including humidity percentage), electrode thickness, cell area, contact resistance, particle size, and porosity percentage are modeled as the system state.

Using the language of reinforcement learning (RL), the state space $\mathbb{S}$ is defined as the set of all possible tuples of the form (anode/cathode, energy density, temperature, pressure, C-rate, SoH, DoD, numeric values describing manufacturing quality, size). The action space $\mathbb{A}$ consists of all possible solid electrolyte candidates obtained via a high-throughput screening process. This setup can be formulated as a finite-horizon Markov decision process (MDP), where the reward $r_h(s,a)$ at each cycle $h$ corresponds to the battery capacity achieved for state $s \in \mathbb{S}$ using electrolyte $a \in \mathbb{A}$. The goal is to compute a policy $\hat{\boldsymbol{\pi}} = \left( \hat{\pi}_h : h \in [H] \right) \in \left( \mathbb{S} \to \Delta(\mathbb{A}) \right)^H$ that maps states to electrolyte formulations so as to nearly maximize cumulative capacity. Since data collection is limited to pre-existing electrochemical experiments, the team adopts an offline RL approach, learning the policy solely from a batch dataset of cycling trajectories.


\subsection*{Results}

\paragraph{Phase \RNum{1}: Assemble experimental data.}
Existing electrochemical datasets from experiments and literature are aggregated with the aid of large language models (e.g., ChatGPT and Perplexity). In parallel, high-throughput imaging data—including microscopy, X-ray computed tomography, and spectroscopy maps of electrodes—are collected. Computer vision models are trained to extract quantitative features from these images. For example, contrastive learning methods such as CLIP or DINO are used to learn meaningful representations linking image features to semantic descriptors. Alternatively, vision transformers (ViTs) or convolutional neural networks (CNNs) can segment structures (particles, dendrites, pores) and compute metrics such as crack size or tortuosity. These image-derived features are appended to the electrochemical dataset.

\paragraph{Phase \RNum{2}: High-throughput screening.}
This phase defines the action space $\mathbb{A}$ by identifying electrolyte candidates with superior properties. Robotic high-throughput experimentation (HTE) platforms are deployed to formulate hundreds of non-aqueous mixtures and measure key properties such as ionic conductivity, oxidative stability, and Coulombic efficiency. Machine-learning models (e.g., neural networks or random forests) are trained in parallel, and the best-performing model is used to predict new high-conductivity formulations, closing the experiment–prediction loop. Active learning strategies combined with Bayesian optimization may also be employed to iteratively propose solvent mixtures that maximize target properties. The resulting high-quality electrolyte candidates form the action space for RL.

\paragraph{Phase \RNum{3}: Generative modeling.}
To augment limited experimental data, the team synthesizes imitative cycling trajectories using generative models. The collected HTE trajectories constitute a batch dataset $\mathcal{D}$ generated under an unknown behavioral policy of the finite-horizon MDP. A generative model, such as a score-based diffusion model, is trained to learn the joint distribution of $(\textnormal{state}, \textnormal{action}, \textnormal{reward})$ sequences. Once trained, the model samples a large synthetic dataset $\widetilde{\mathcal{D}}$ that statistically mimics $\mathcal{D}$. The combined dataset $\mathcal{D} \cup \widetilde{\mathcal{D}}$ is then used to strengthen offline policy learning.

\paragraph{Phase \RNum{4}: Policy learning via offline RL.}
Offline RL algorithms are applied to learn a deterministic near-optimal policy $\hat{\boldsymbol{\pi}} = \left( \hat{\pi}_h : h \in [H] \right) \in \left( \mathbb{S} \to \mathbb{A} \right)^H$. Candidate methods include off-policy policy-gradient approaches, value iteration, and Q-learning augmented with pessimism under uncertainty. The learned policy outputs an electrolyte $\hat{\pi}_h(s) \in \mathbb{A}$ that maximizes expected capacity at each cycle step $h$ for a given state $s \in \mathbb{S}$. In practice, feeding battery state data into the policy yields a recommendation for the optimal electrolyte.

\paragraph{Phase \RNum{5}: Experimental verification.}
The learned policy is validated through real experiments. For selected battery states $s \in \mathbb{S}$, cells are assembled using the solid electrolyte $\hat{\pi}_h(s)$ recommended by the policy, and their capacities are measured. These results are compared against cells using alternative electrolytes at the same cycle step. Successful validation is demonstrated if the policy-selected electrolyte yields near-maximum capacity.

\begin{figure}[h!]
    \centering
    \includegraphics[width=\linewidth]{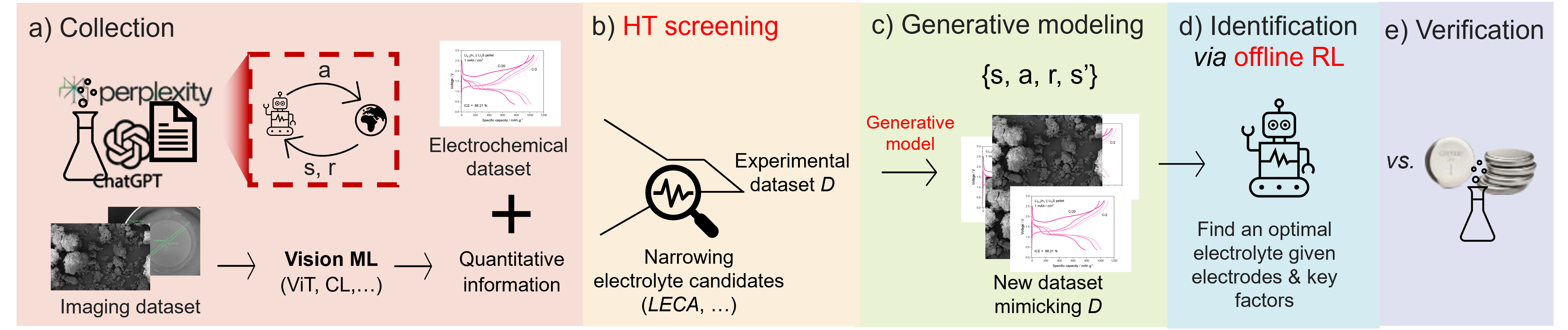}
    \caption{Overview of the automated electrolyte discovery system via offline reinforcement learning.}
    \label{fig:automated_system_offline_RL}
\end{figure}

\subsection*{Future Work}
Future extensions include multi-objective optimization that simultaneously considers capacity, stability, cost, and other practical constraints relevant to battery design. Improving the generative modeling stage by exploring more expressive or conditional generators is another open direction, as this could yield higher-fidelity synthetic data. Finally, continued expansion of the dataset is critical: incorporating more diverse chemistries, wider operating conditions, and advanced molecular representations (such as graph neural networks or transformer-based materials models) may further enhance performance.

\subsection*{Open-source Materials}
Code is available on GitHub: \github{https://github.com/jaydenlee97/2025-LLM-Hackathon-for-Applications-in-Materials-and-Chemistry/tree/main/code}





\section{Once Upon a Time in Chromatography: Adaptive Denoising and Peak Discovery}
\label{sec:benzene-boyz}


The \textsc{BENZENE BOYZ} team presents a statistically principled pipeline for comprehensive two-dimensional gas chromatography coupled with time-of-flight mass spectrometry (GC$\times$GC--TOFMS), which produces dense two-dimensional chromatograms where chemical compounds appear as local intensity peaks \cite{Mondello2025, Marriott2012}. Accurate and well-calibrated peak detection is a prerequisite for downstream compound identification and quantification \cite{Huang2022, Nieer2024}.  

\begin{figure}[t]
    \centering
    \begin{minipage}{0.48\linewidth}
        \centering
        \includegraphics[width=\linewidth]{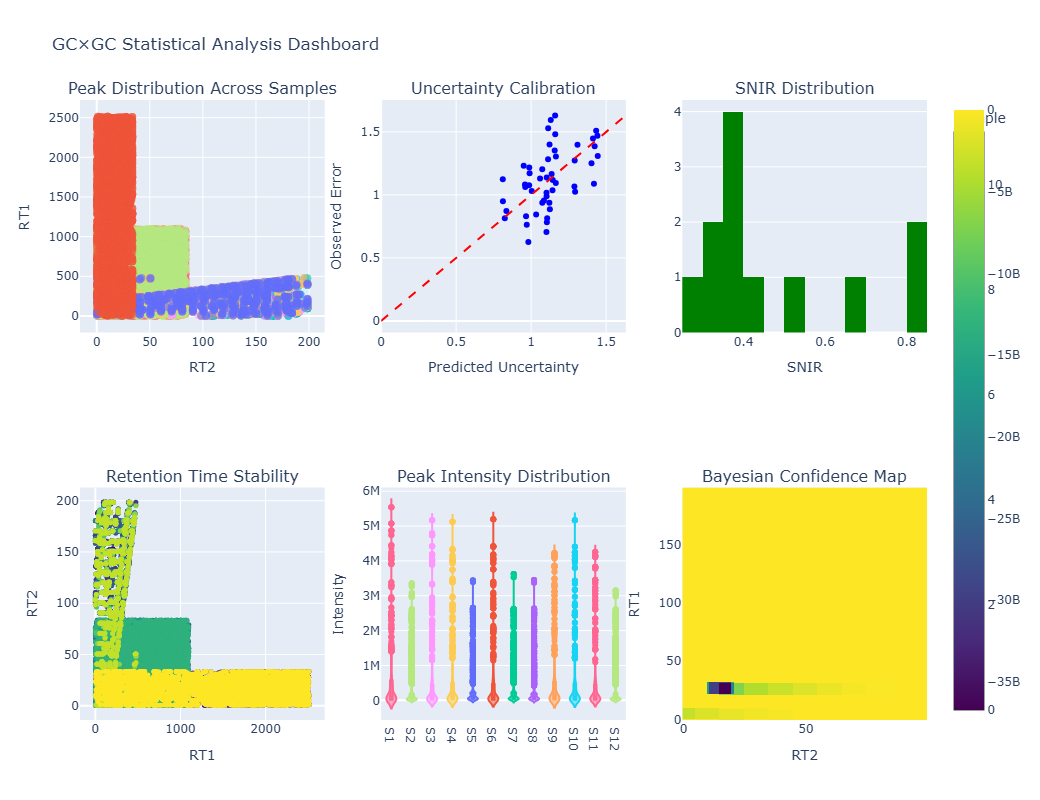}
        \caption{Bayesian probability heatmap}
        \label{fig:bayesian_heatmap}
    \end{minipage}
    \hfill
    \begin{minipage}{0.48\linewidth}
        \centering
        \includegraphics[width=\linewidth]{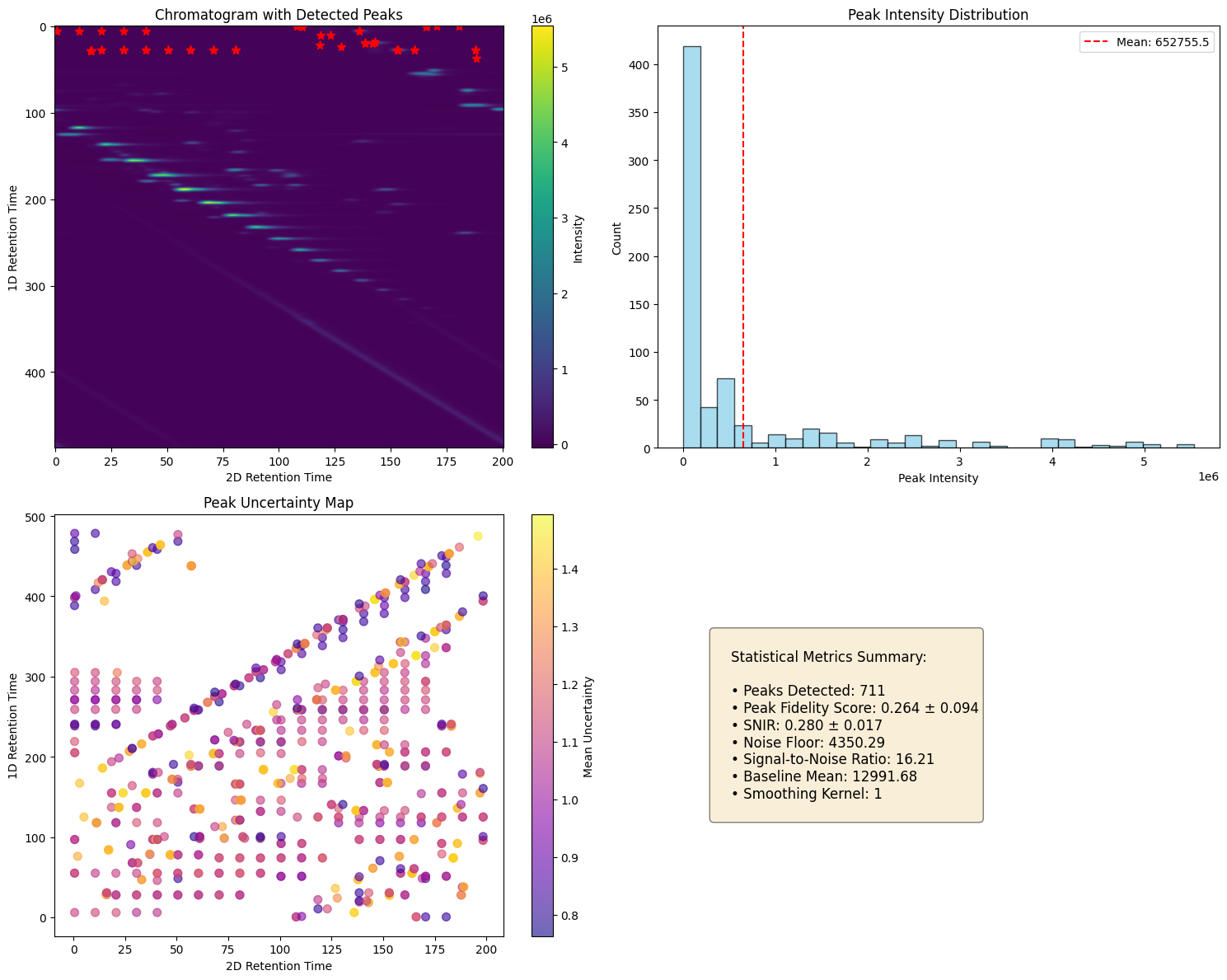}
        \caption{Statistical dashboard}
        \label{fig:stat_dashboard}
    \end{minipage}
\end{figure}

The team proposes a unified framework that (i) robustly preprocesses GC$\times$GC data, (ii) produces peak estimates with quantified uncertainty, (iii) evaluates detector performance using novel metrics, and (iv) provides human-interpretable summaries via vision--language models (VLMs). Key contributions include a practical local Bayesian peak estimator based on prior sampling and likelihood scoring, robust preprocessing using ALS baseline subtraction and MAD-based noise scaling, new performance metrics (Peak Fidelity Score and Signal-to-Noise-and-Interference Ratio) with bootstrap confidence intervals, and a simple PAC--Bayes–motivated bound linking reconstruction risk and model complexity.


\subsection*{Methodology}

\begin{algorithm}
\caption{Local posterior approximation (per window)}
\begin{algorithmic}[1]
\For{each window \(W\) in \(\widetilde{Y}_{\mathrm{sm}}\)}
  \If{\(\max(W) > \tau\)}
    \State \(\mu_0 \gets \arg\max(W)\)
    \State Draw \(\mu^{(s)} \sim \mathcal{N}(\mu_0,\sigma_p^2 I)\) for \(s=1,\dots,N\)
    \State Compute \(\mathcal{L}(\mu^{(s)})\)
    \State Compute \(\widehat{\mu}, \widehat{\Sigma}, \overline{\mathcal{L}}\)
    \If{heuristic on \(\overline{\mathcal{L}}\) holds}
      \State Emit peak descriptor \(p_k\)
    \EndIf
  \EndIf
\EndFor
\end{algorithmic}
\end{algorithm}


\subsection*{Results}
The pipeline is evaluated on synthetic GC$\times$GC chromatograms generated by summing two-dimensional Gaussian peaks with random locations, amplitudes, and widths under additive Gaussian noise. This controlled setting enables direct measurement of localization accuracy, uncertainty calibration, and signal-to-noise ratio improvements. For reproducibility, default hyperparameters include a prior sample count \(N=500\), bootstrap iterations \(B=100\), ALS smoothing parameter \(\lambda=10^6\), and confidence level \(0.95\). Adaptive smoothing is applied throughout the analysis.

\begin{table}[ht]
\centering
\begin{minipage}{0.32\linewidth}
    \centering
    \textbf{Final Summary Report}\\[4pt]
    \begin{tabular}{@{}ll@{}}
    \toprule
    Files processed & 12 \\
    Total peaks detected & 20549 \\
    Average Peak Fidelity Score & 0.217 \\
    Average SNIR & 0.472 \\
    Average Noise Floor & 5309.79 \\
    Average SNR & 14.61 \\
    PAC--Bayes bound & 1.0000 \\
    \bottomrule
    \end{tabular}
\end{minipage}
\hfill
\begin{minipage}{0.32\linewidth}
    \centering
    \textbf{Detailed Performance Metrics}\\[4pt]
    \begin{tabular}{@{}ll@{}}
    \toprule
    Avg peaks/sample & 1712.4 \\
    Mean uncertainty & 1.130 \\
    Uncertainty std.\ dev. & 0.182 \\
    Min uncertainty & 0.741 \\
    Max uncertainty & 1.502 \\
    \bottomrule
    \end{tabular}
\end{minipage}
\hfill
\begin{minipage}{0.32\linewidth}
    \centering
    \textbf{Results Summary}\\[4pt]
    \begin{tabular}{@{}ll@{}}
    \toprule
    PAC--Bayes bound & 1.0000 \\
    Statistical confidence & 95\% \\
    Bootstrap iterations & 1000 \\
    MCMC samples & 5000 \\
    \bottomrule
    \end{tabular}
\end{minipage}
\caption{Combined comparison of summary statistics and performance metrics.}
\label{tab:summary_combined}
\end{table}


\subsection*{Discussion \& Limitations}
The detector estimates posterior uncertainty via prior sampling with likelihood scoring rather than full Markov chain Monte Carlo (MCMC), favoring computational efficiency but potentially underrepresenting tail behavior or multimodality. The isotropic Gaussian peak template may mischaracterize asymmetric or skewed chromatographic peaks; more expressive models such as skewed Gaussians or Voigt profiles could improve fidelity. In addition, the PAC--Bayes bound relies on a heuristic Kullback--Leibler proxy; a rigorous formulation would require explicit priors and posteriors over parameterized peak models.

\subsection*{Conclusion}
The \textsc{BENZENE BOYZ} team introduces a compact GC$\times$GC--TOFMS analysis pipeline that integrates robust preprocessing, a lightweight Bayesian peak estimator with calibrated uncertainty, and quantitative performance metrics for objective evaluation. The approach balances statistical rigor with computational efficiency, making it well suited for rapid prototyping. Future directions include full MCMC posterior inference, richer asymmetric peak models, integration with mass-spectral identification, and a formal PAC--Bayes analysis with explicit priors and posteriors.

\subsection*{Open-source Materials}
Code available on GitHub: \github{https://github.com/SubramanyamSahoo/LLM-Hackathon-for-Applications-in-Materials-and-Chemistry-2025}





\section{Sol-Agent: An LLM-Driven Framework for Predicting Sol-Gel Synthesizability of OER Catalysts}\label{sec:sol-agent}


The Sol-Agent team focuses on sol--gel synthesis, a widely used solution-based route for preparing inorganic materials such as metal oxides, metal complexes, and hybrid composites. Its ability to precisely control composition, homogeneity, and nanostructure makes sol--gel synthesis particularly valuable for catalysis, including oxygen evolution reaction (OER) catalysts. However, the breadth and fragmentation of sol--gel literature pose significant challenges for systematically extracting, organizing, and reusing synthesis knowledge, especially when linking synthesis routes to material structure, composition, and properties.

To address this challenge, the team developed Sol-Agent, an LLM-driven synthesis agent designed to extract and summarize sol--gel synthesis recipes from the literature and construct a structured, queryable knowledge base. As illustrated in Fig.~\ref{fig:two_images_simple_SOLGEL}, detailed experimental procedures are extracted from published works, cleaned, normalized, and segmented into semantically meaningful text chunks. These chunks are embedded using the LLaMA~3.1 (8B) model, fine-tuned through the Hugging Face and Unsloth ecosystem for efficient domain adaptation. The resulting embeddings are indexed and stored in a FAISS database, enabling fast similarity-based retrieval and contextual linking across diverse synthesis reports. This pipeline provides an intelligent interface for identifying synthesis patterns, parameter trends, and knowledge gaps in sol--gel research, thereby accelerating data-driven materials discovery and synthesis optimization.


\subsection*{Results}

\begin{figure*}[ht]
    \centering
    \includegraphics[width=0.48\textwidth]{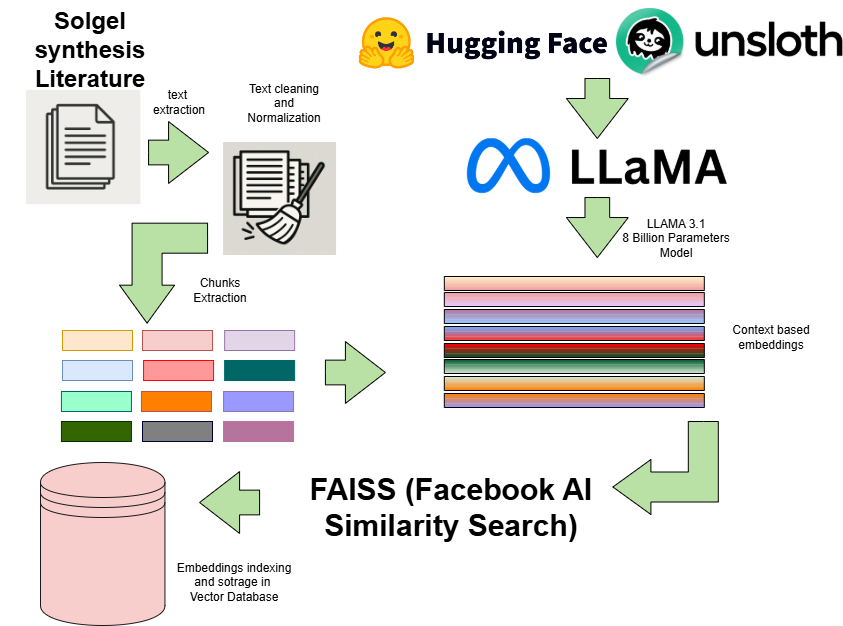}
    \hfill
    \includegraphics[width=0.5\textwidth]{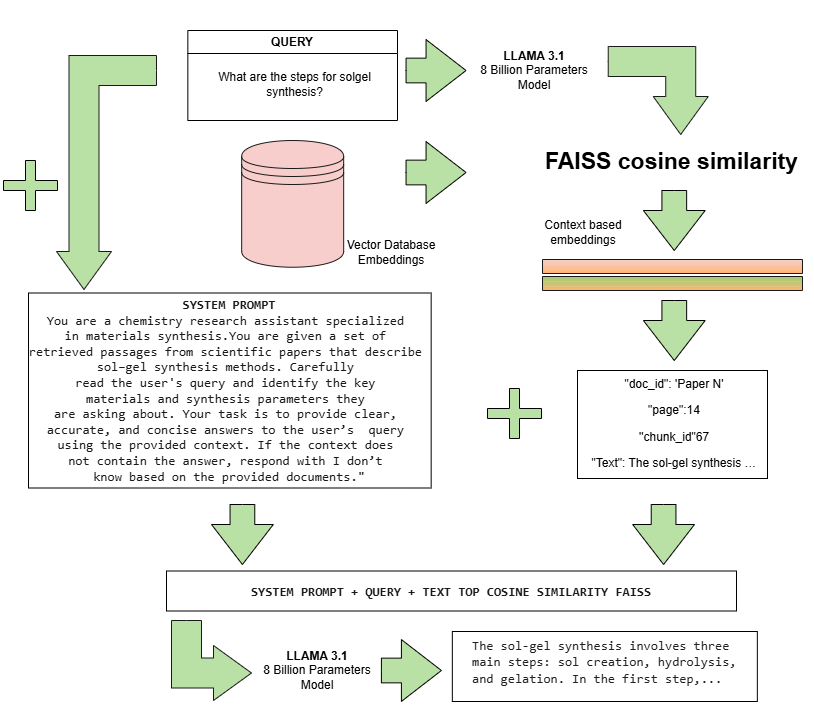}
    \caption{Vector database construction and retrieval-augmented generation (RAG) workflow for Sol-Agent.}
    \label{fig:two_images_simple_SOLGEL}
\end{figure*}

To accelerate knowledge extraction for materials synthesis, the Sol-Agent team developed a streamlined LLM-based pipeline that processes sol--gel literature, including Refs.~\cite{SolGelPaper1,SolGelPaper2,SolGelPaper3,SolGelPaper4,SolGelPaper5,SolGelPaper6,SolGelPaper7,SolGelPaper8}, selected to cover a broad range of material systems. These documents are transformed into a searchable vector database that supports efficient retrieval and automated reasoning. As shown in Fig.~\ref{fig:two_images_simple_SOLGEL}, text is extracted, cleaned, and segmented into chunks, embedded using LLaMA~3.1 (8B), and indexed with FAISS to enable rapid semantic search. This approach supports downstream tasks such as inverse synthesis design and autonomous literature triage, demonstrating a scalable strategy for mining domain-specific knowledge from unstructured text.

To enable context-aware question answering, a retrieval-augmented generation workflow is implemented on top of the vector database. User queries are embedded using the same model as the literature corpus, and cosine similarity search via FAISS retrieves the most relevant text chunks. These passages are combined with a system prompt that defines the assistant as a domain-specific synthesis expert, guiding generation toward concise, literature-grounded responses. The final answers are produced by the LLaMA~3.1 model, minimizing hallucination while preserving scientific accuracy. Table~\ref{tab:table_SolAgent} presents representative example queries and responses generated by Sol-Agent, illustrating coherent and scientifically consistent outputs. Additional evaluations and queries can be performed using the publicly available open-source implementation.

\begin{table}[h!]
\centering
\begin{tabular}{|p{4cm}|p{10cm}|}
\hline
\textbf{Question} & \textbf{Sol-Agent Response} \\
\hline
What are the steps for sol--gel synthesis?
&
The sol--gel synthesis involves three main steps: sol creation, hydrolysis, and gelation. In the first step, precursor molecules are dissolved in a solvent to create a sol, determining the composition of the final material. During hydrolysis, the precursor interacts with water molecules to form metal hydroxides. Finally, gelation leads to the formation of a metal hydroxide network, resulting in a gel. \\
\hline
What is the overpotential of Na\textsubscript{3}[Ru\textsubscript{2}(\(\mu\)CO\textsubscript{3})\textsubscript{4}]?
&
The overpotential for Na\textsubscript{3}[Ru\textsubscript{2}(\(\mu\)CO\textsubscript{3})\textsubscript{4}] is 235~mV at 20~mA~cm\textsuperscript{-2}. \\
\hline
\end{tabular}
\caption{Example queries and Sol-Agent responses.}
\label{tab:table_SolAgent}
\end{table}

\subsection*{Future Work}
Future directions include expanding the literature corpus to further reduce the risk of hallucination, incorporating quantitative metrics to rigorously assess the accuracy and scientific validity of generated responses, and evaluating Sol-Agent across a wider range of foundation models. Systematic comparisons of retrieval-augmented generation behavior under different model architectures, embedding strategies, retrieval schemes, and chunking parameters will provide deeper insight into the robustness and generalizability of the framework.

\subsection*{Open-source Materials}
Code and pretrained weights are available on GitHub: \github{https://github.com/rafaelespinosacastaneda/Hackaton-for-Applications-in-Materials-Science-and-Chemistry}





\section{RedoxFlow: An Agentic Workflow for Preparing Simulations in High-throughput Redox Potential Screening}\label{sec:RedoxFlow}



The RedoxFlow team addresses high-throughput screening of organic electrolytes by focusing on redox potential, a fast thermodynamic descriptor that quantifies a molecule’s tendency for oxidation or reduction \cite{min4020345}. Redox potentials are widely used via the Nernst relation for rapid ranking of organic electrolytes \cite{pelzer2017effects}, estimating cell voltages \cite{lin2016redox}, and evaluating electrochemical stability limits prior to detailed kinetic analysis \cite{wedege2016organic,hasan2023quinones}. The team presents an agentic workflow that autonomously prepares density functional theory (DFT) inputs for NWChem \cite{valiev2010nwchem} to compute Nernstian redox potentials under implicit solvation using the Computational Hydrogen Electrode (CHE) framework \cite{masood2023comparative,singh2025sulfonated}. The agent extracts thermodynamic quantities, applies fixed algebraic corrections, and computes redox potentials in a deterministic manner.

Candidate electrolytes are generated \emph{de novo} as SMILES strings using GP-MoLFormer-Uniq \cite{ross2025gp}, filtered for realistic and tractable structures, and paired with predicted reduction products. Reactant–product conformers are optimized, and corresponding NWChem DFT input scripts are produced automatically. While simulations themselves are launched externally, the workflow fully automates molecular generation, input preparation, and post-processing, serving as a proof-of-concept for scalable, high-throughput redox screening.


\subsection*{Results}

\textbf{(i) Reactant generation.}
Candidate reactants are generated \emph{de novo} as SMILES strings using moderate-temperature sampling. Outputs are RDKit-canonicalized \cite{landrum2013rdkit}, deduplicated, and sanity-checked for valence, charge balance, and disconnected fragments. Rule-based pruning filters syntactically valid and chemically plausible structures, for example excluding highly strained ring motifs. For DFT tractability, chemistry is restricted to C/N/O/F. To preserve generative yield, elements in the raw outputs are downcast within periodic groups while preserving connectivity (e.g., S$\rightarrow$O, Se$\rightarrow$O, P$\rightarrow$N, B/Si$\rightarrow$C, Cl/Br/I$\rightarrow$F); molecules that cannot be mapped are rejected. Reactant acceptance criteria include RDKit sanitization, the presence of at least one heteroatom, bounds on heavy-atom count, exclusion of blacklisted motifs, and a specified size range. Canonical SMILES and InChIKey hashing with persistent memory prevents duplicate reactants across runs.

\textbf{(ii) Predicted reduced products.}
Reduced products are generated by applying a curated set of RDKit SMARTS transformations. For each reactant $A$, reaction rules are applied recursively up to depth $D$, after which products are sanitized, canonicalized, and deduplicated. Each rule encodes a proton-coupled electron transfer (PCET) template,
$\mathrm{A} + x\,e^- + y\,\mathrm{H}^+ \rightarrow \mathrm{B}$,
where $x$ is the electron count and $y$ the implicit proton balance specified in rule metadata. For this proof-of-concept, only cases with $x=y$ are considered. These interpretable reaction rules motivate future extensions toward more context-aware product prediction.

\textbf{(iii) Conformer search.}
For each reactant and its predicted products, up to $N$ three-dimensional conformers are generated using Open Babel (\texttt{--gen3d --best --conformer --nconf $N$}). Each conformer is minimized with the MMFF94 force field using \texttt{obenergy}, and the lowest-energy conformer is retained. The selected geometries are written to XYZ format, providing consistent, energy-minimized starting points for subsequent DFT calculations. RDKit and Open Babel sanitization and deduplication ensure coherence with the persistent molecular memory.

\textbf{(iv) Input script preparation.}
NWChem input files are generated deterministically via a Python class whose attributes map one-to-one to NWChem directives (e.g., \texttt{charge}, \texttt{geometry}, \texttt{task}). For the present demonstration, tractable defaults such as the PBE exchange–correlation functional are used. This programmatic construction guarantees reproducible, scheduler-ready inputs aligned with the CHE redox thermocycle, without relying on heuristic or free-form LLM-generated input decks.

\textbf{(v) Redox calculation (CHE).}
Redox potentials are computed using a fixed Born–Haber CHE cycle \cite{singh2025sulfonated}. For each state, the solution-phase free energy $G^{\mathrm{sol}}$ is assembled from the optimized gas-phase electronic energy, thermal corrections (zero-point energy, enthalpy, entropy), and a COSMO solvation correction \cite{klamt2011cosmo}. For an aqueous $x$-electron PCET step,
$A + x\,e^- + x\,\mathrm{H}^+ \rightarrow B$,
the reaction free energy is
$\Delta G_{\mathrm{rxn}} = G^{\mathrm{sol}}(B) - G^{\mathrm{sol}}(A) - \tfrac{x}{2} G^{\mathrm{sol}}(\mathrm{H_2})$,
and the corresponding redox potential is
$E = -\Delta G_{\mathrm{rxn}}/(xF)$,
where $F$ is Faraday’s constant. The thermocycle is encoded as a fixed algebraic template, and required quantities are extracted deterministically from known locations in simulation outputs via substring matching.


\subsection*{Future Work}
Planned extensions include integration with workload schedulers (e.g., SLURM), migration to a relational database backend for improved scalability, exposure of finer DFT controls such as dielectric constant and temperature, and development of a constrained natural-language front end. Additional directions involve charge-based reduction-site inference, synthesizability score filtering, user-configurable motif blacklists, and cautious expansion beyond C/N/O/F chemistries (e.g., inclusion of Cl, P, or S). Confining LLM usage primarily to molecular proposal is motivated by reduced token overhead and improved robustness for large-scale deployments.

\subsection*{Open-source Materials}
Code is available on GitHub: \github{https://github.com/CGruich/RedoxFlow} \cite{RedoxFlow_Zenodo_2025}





\section{AutoFeatSci: Automated Feature Engineering for Materials Science}\label{sec:AutoFeatSci}

%

The The Featurizers team addresses a central bottleneck in materials-science modeling workflows: feature engineering. Traditional workflows rely heavily on manual extraction of domain knowledge from the literature (e.g., metallurgy, phase transformations, defect mechanisms, thermodynamics) combined with experimental observations or simulations. Scientists typically read papers, formulate hypotheses, handcraft features, train models, and iteratively refine them by comparing predictions with experiments. While effective, this process is highly manual, time-consuming, and fundamentally constrained by human capacity to design informative descriptors.

To overcome these limitations, the team proposes AutoFeaSci, an agentic system that automates feature engineering through a multi-agent large language model (LLM) pipeline. AutoFeaSci integrates literature-grounded reasoning with data-aware interpretation to accelerate hypothesis testing in feature space. The system is designed to rapidly generate, evaluate, and iterate on physics-informed descriptors, bridging the gap between raw experimental or simulation data and predictive machine-learning pipelines.

The key components of AutoFeaSci include: (i) a literature-grounded reasoning module that synthesizes insights from materials-science publications to extract governing mechanisms, relevant descriptors, and domain hypotheses; (ii) a data-aware interpretation module that contextualizes dataset variables, maps raw features to their physical meaning, and links them to extracted domain knowledge; (iii) a feature-engineering framework that cross-references scientific insights with dataset semantics to construct physically meaningful descriptors beyond naive transformations; (iv) systematic feature evaluation using predictive modeling and statistical metrics such as RMSE and $R^2$, combined with feature-importance analysis; and (v) a closed-loop optimization process in which evaluation feedback is fed back into the system to iteratively refine feature hypotheses and converge toward an optimized, task-aligned feature space.


\begin{figure}
    \centering
    \includegraphics[width=0.9\linewidth]{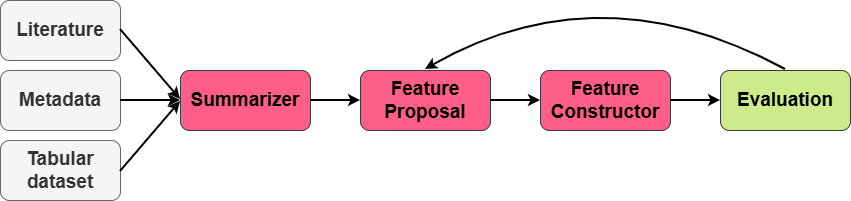}
    \caption{Workflow of the AutoFeaSci multi-agent featurization system. Literature, metadata, and tabular datasets are processed by a Summarizer agent to extract domain mechanisms. A Feature-Proposal agent generates candidate descriptors, which are instantiated by the Feature-Constructor agent and evaluated in downstream machine-learning models.}
    \label{fig:featurizer}
\end{figure}

\subsection*{Results}
Figure~\ref{fig:featurizer} illustrates the iterative AutoFeaSci workflow. The system takes three primary inputs—scientific literature, research metadata, and tabular datasets—which are first processed by a Summarizer agent specialized in extracting mechanistic insights from multimodal scientific sources. These distilled insights are passed to a Feature-Proposal agent, which formulates candidate feature hypotheses and outlines transformation strategies such as combining, normalizing, or deriving interactions among existing variables. Proposed features are then handed to a Feature-Constructor agent, which acts as a Python code generator and executor, translating proposals into concrete feature-engineering operations and augmenting the dataset.

The enriched dataset is evaluated using a tree-based gradient-boosting model implemented with the H2O package \cite{H2OAutoML20}. AutoFeaSci collects performance metrics including mean squared error (MSE), root mean squared error (RMSE), and $R^2$, and reports them back to the Feature-Proposal agent as feedback to guide subsequent iterations. This loop continues until the user determines that the train–test performance satisfies the desired criteria.

At the current stage, the team has successfully implemented and executed the complete AutoFeaSci pipeline, demonstrating smooth end-to-end operation across several initial iterations. However, robustness has not yet been fully validated. After approximately three iterations, both training and test accuracy degraded. Inspection of the Feature-Proposal agent’s reasoning revealed instances of severe hallucination, such as proposing a “porosity fraction” defined as the ratio between experimentally measured alloy density and simulated alloy density—an ill-defined descriptor for predicting alloy yield strength in the given context. These observations highlight the need for an additional review or validation mechanism to filter or correct proposed features before downstream use. Although such safeguards could not be implemented within the hackathon timeframe, they represent a critical direction for future work. Despite these limitations, AutoFeaSci establishes a functional proof-of-concept that can extract domain knowledge, guide feature engineering, and iteratively refine descriptors with the goal of improving predictive performance.


\subsection*{Open-source Materials}
Code and pipeline are available on GitHub: \github{https://github.com/guannant/LLM_auto_featurization}





\section{MAGE: Materials Agent for Generative and Evaluative Design}\label{sec:MAGE}

%

The MAGE team addresses a long-standing challenge in materials science: the discovery and design of new materials with targeted functional properties. Generative models such as variational autoencoders, generative adversarial networks, and diffusion models have shown strong promise for accelerating materials discovery \cite{park2024has}. However, these approaches typically require significant technical expertise and model-specific inputs, limiting their accessibility through natural language interaction. To overcome this barrier, the team introduces MAGE (Materials Agent for Generative and Evaluative Design), an agentic framework that enables intuitive natural language interaction with materials property prediction and structure generation. MAGE interprets user prompts, autonomously selects which functions to invoke, executes the appropriate generative or predictive models, and returns results in natural language.


\subsection*{Results}

\begin{figure}[!t]
    \centering
    \includegraphics[width=0.75\textwidth]{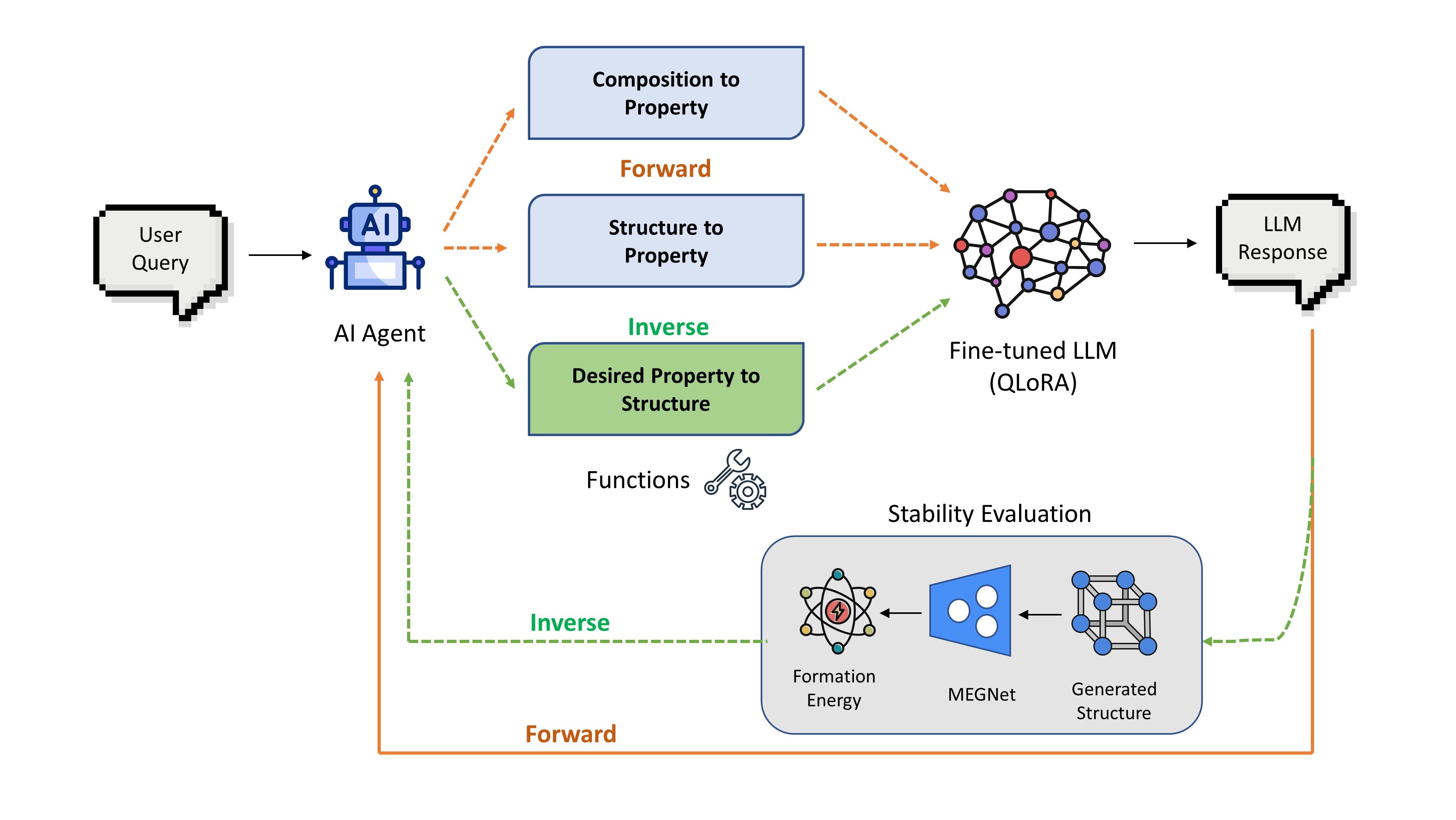}
    \caption{MAGE workflow. The agent interacts with the user and invokes the appropriate function based on the user prompt. A unified fine-tuned LLM supports both property prediction and inverse design. For inverse design, the system generates crystal structures and evaluates their thermodynamic stability via formation-energy prediction.}
    \label{fig:mage_1}
\end{figure}

Figure~\ref{fig:mage_1} illustrates the MAGE workflow, which integrates an intelligent agent with a fine-tuned large language model (LLM) to support both forward and inverse materials design \cite{choudhary2024atomgpt}. MAGE is implemented using the Google ADK framework for agent orchestration and Streamlit for the user interface. Users submit queries through an interactive interface, where an AI agent (Gemini-2.0-Flash) dynamically determines whether each query corresponds to a forward-design task (composition-to-property or structure-to-property) or an inverse-design task (desired property-to-structure), and routes the query to the appropriate function. For inverse design, the system generates candidate crystal structures corresponding to the target property and subsequently evaluates their thermodynamic stability by predicting formation energies using a materials graph neural network model \cite{chen2022universal}.

For both predictive and generative tasks, the team fine-tuned the Mistral-7B-v0.3 model using Quantized Low-Rank Adaptation (QLoRA) \cite{mohanty2025crystext}. Training was performed on the Matbench bulk modulus dataset \cite{dunn2020benchmarking} using an 80:20 train–test split. From the training data, diverse instruction-tuning datasets were constructed to cover: (i) prediction of bulk modulus from composition, (ii) prediction of bulk modulus from crystal structure (CIF), and (iii) generation of crystal structures targeting a specified bulk modulus value. The fine-tuned model was evaluated on both predictive accuracy and generative quality.

For the prediction task, MAGE achieved $R^2 \approx 0.85$ with a mean absolute error of approximately 15~GPa for both composition-based and structure-based inputs on the held-out test set. For the inverse design task, the team uniformly sampled target bulk modulus values in the range 10–400~GPa and generated 1000 candidate structures. Structural validity was assessed by enforcing minimum interatomic distances of 0.5~\AA{} and unit cell volumes greater than 0.1~\AA$^3$. Compositional validity was evaluated using SMACT rules for charge neutrality and chemical plausibility \cite{davies2019smact}. The generated structures exhibited 95\% structural validity and 85\% compositional validity.

\subsection*{Future Work}
Future developments will enhance MAGE’s inverse design capabilities by incorporating reinforcement learning with multi-objective reward functions \cite{mohanty2025crystext}. These rewards will jointly evaluate structural and compositional validity, thermodynamic stability, novelty, and proximity to target property values. The team also plans to extend MAGE to handle multiple properties simultaneously, including shear modulus, band gap, and other mechanical and electronic properties, while exploring alternative foundation models for fine-tuning.

\subsection*{Open-source Materials}
Code and a live demo are available on GitHub: \github{https://github.com/truptimohanty/MAGE}





\section{BASIS: Bulk and Surface Interface Simulations with Intelligent Systems}\label{sec:BASIS}

%

The BASIS team presents \textbf{BASIS} (Bulk and Surface Interface Simulations with Intelligent Systems), an agentic AI framework for automating bulk and surface density functional theory (DFT) simulations. The system integrates intelligent agents with literature mining to autonomously extract relevant information from materials databases and scientific publications, recommend suitable computational parameters, prepare simulation inputs, and execute calculations for crystalline solids and surface systems. By combining agentic reasoning with first-principles simulation workflows, BASIS aims to reduce the manual overhead traditionally required for setting up and validating DFT studies.

\begin{figure}[H]
    \centering
    \includegraphics[width=0.9\linewidth]{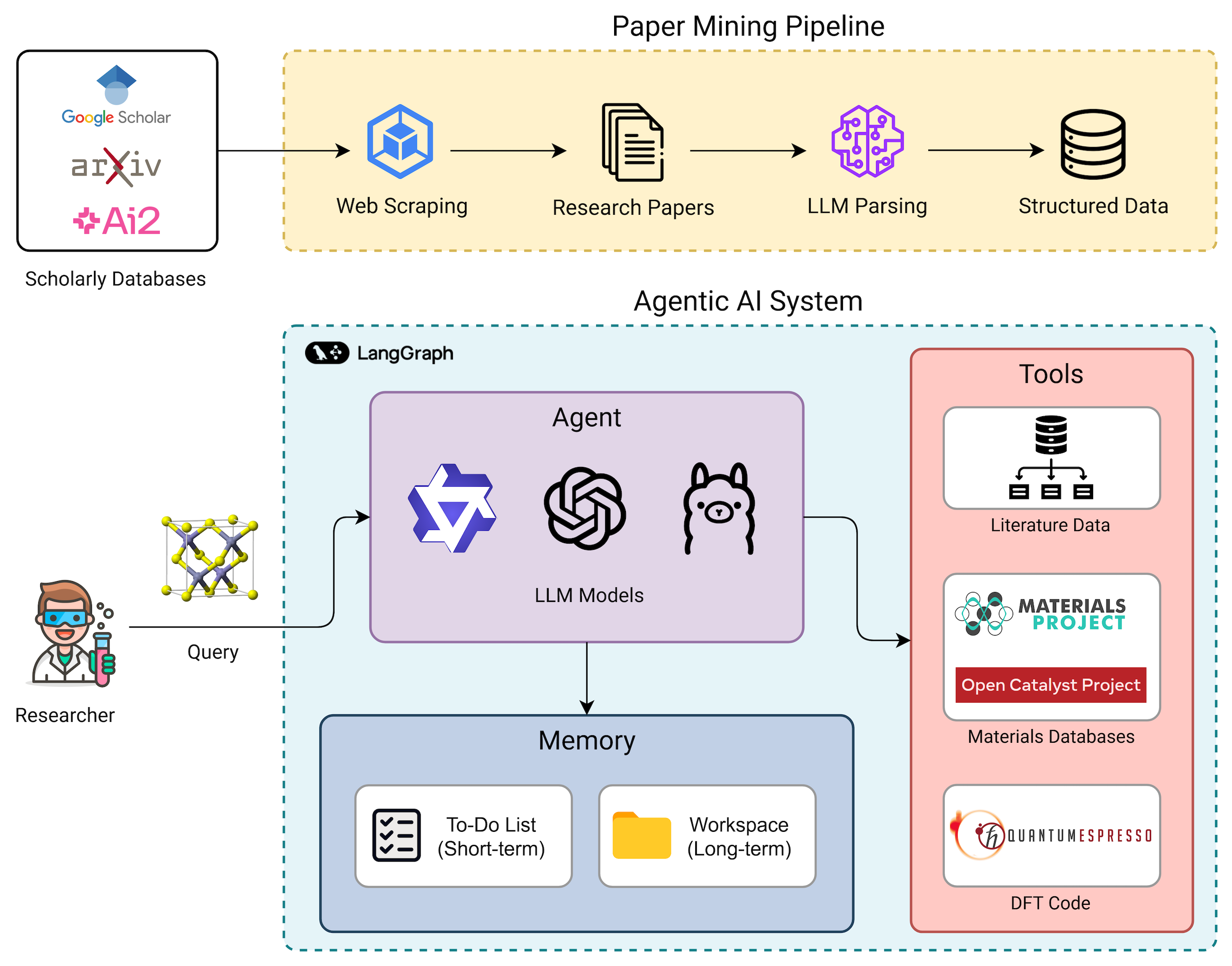}
    \caption{Architecture diagram of \texttt{BASIS}.}
    \label{fig:LLM-framework}
\end{figure}


\subsection*{Data Collection and Analysis}
Computational datasets are collected from publicly available computational and scholarly sources. Bulk and surface structure data are obtained from the Materials Project via its free API and from the Open Catalyst Database (OC20), providing a diverse set of crystalline materials and surface configurations. Scholarly datasets are gathered using web-scraping techniques applied to the \texttt{Asta Scientific Corpus}, enabling extraction of computational parameters and methodological details directly from literature identified by Digital Object Identifiers (DOIs). This combined strategy streamlines access to large-scale datasets and associated properties while grounding simulation setups in established literature practices.

\begin{figure}
    \centering
    \includegraphics[width=0.8\linewidth]{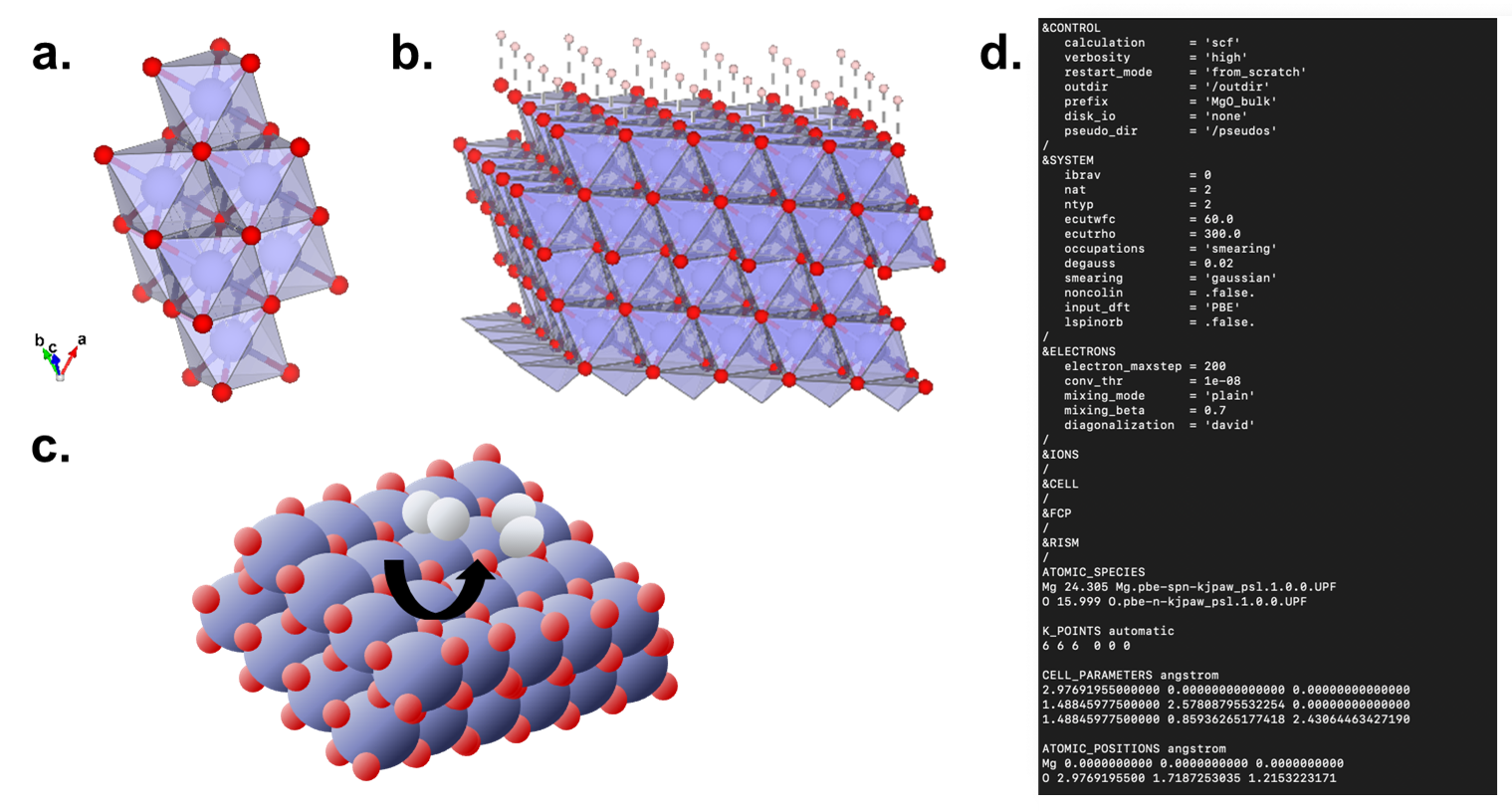}
    \caption{Overview of the DFT workflow performed by \texttt{BASIS}. (a) Generation of a bulk MgO crystal, (b) construction of the MgO(111) surface, and (c) representative surface configurations for evaluating hydrogen adsorption and reaction pathways. (d) The agent automatically produces Quantum ESPRESSO input files, enabling scalable batch simulations validated against literature benchmarks.}
    \label{fig:structure-generation}
\end{figure}


\subsection*{Future Work}
Future development of BASIS will focus on expanding system autonomy, interoperability, and simulation capabilities. Planned enhancements include a multi-agent framework capable of deeper autonomous research and self-validation of results, integration of additional first-principles methods and molecular dynamics simulations, and the adoption of iterative, agentic reinforcement-learning-based training to further improve performance and reliability across diverse materials simulation tasks.

\subsection*{Open-source Materials}
Code and data are available on GitHub: \github{https://github.com/abir0/dft-agent}





\section{Titanarium: A Digital Terrarium for Scientist-Persona Debate}\label{sec:Titanarium}

%

The Titanarium team introduces Titanarium, a virtual arena in which materials-scientist-inspired personas and humans engage in autonomous, topic-driven debates. Rather than treating large language models solely as tools for optimization or problem solving, Titanarium frames AI-to-AI interaction itself as an interactive experience. Multiple LLM nodes, each instantiated with a distinct character profile—such as personas inspired by historical figures like Mendeleev, Curie, and Pauling—participate alongside optional human chat nodes. Users may either actively join the discussion or observe it unfold, analogous to watching an ecosystem evolve within an aquarium or terrarium.

From an architectural perspective, Titanarium comprises multiple MCP servers, each hosting a persona or a human node, a centralized LLM backend served via \texttt{llama.cpp} and Docker, and a Streamlit-based user interface that orchestrates and visualizes debate rounds. Once the persona servers and backend are running, users can trigger automated discussion rounds or participate interactively through a ``Human Chat'' panel. Although the current implementation is limited to a local demonstration environment and is not intended for production or research-grade deployment, it demonstrates that multi-agent, character-driven conversations can be made concrete, repeatable, and interactively explorable with modest computational infrastructure.


\subsection*{Results}
\begin{figure}[!t]
    \centering
    \includegraphics[width=0.75\textwidth]{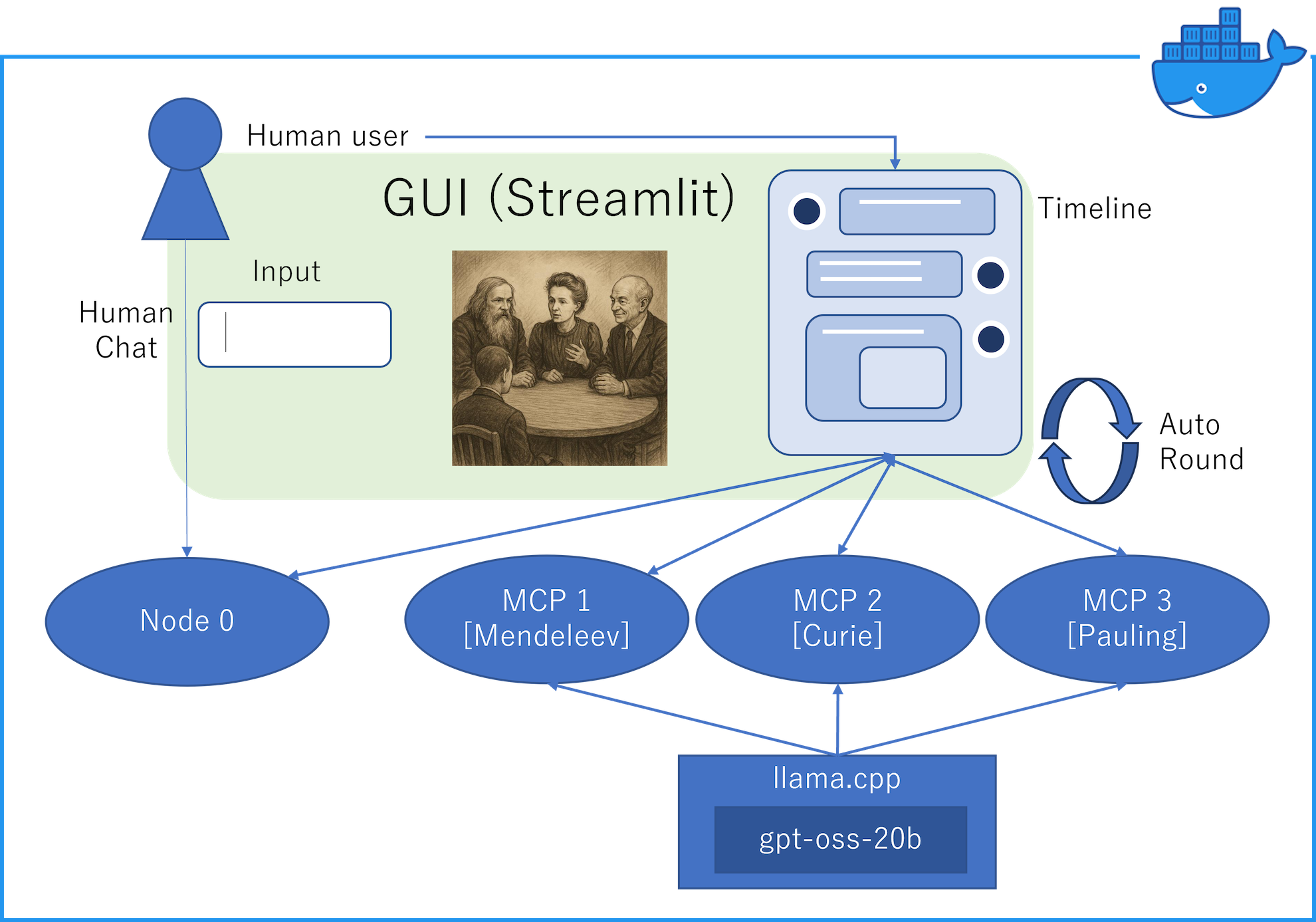}
    \caption{Titanarium working prototype showing multi-agent scientist-persona debate.}
    \label{fig:titanarium}
\end{figure}

A working prototype was implemented, as shown in Fig.~\ref{fig:titanarium}, capable of running multi-round autonomous debates among three scientist personas and an optional human node. After launching the \texttt{llama.cpp}-backed LLM server and the persona MCP servers, the Streamlit interface allows users to initiate an ``Auto Round'' and observe real-time natural-language exchanges among the personas. The system uses GPT-oss 20B in GGUF format \cite{gpt_oss_20b_gguf} as the base model. Each agent operates with its own configuration and seeded persona data, resulting in clearly distinguishable styles, preferences, and argumentative tendencies.

In addition to autonomous debate, Titanarium supports direct human interaction through a ``Human Chat'' mode, enabling a user to message an individual persona and receive a context-aware response routed through the same LLM backend. The prototype was developed under hackathon constraints and assumes execution in a local environment, but nonetheless demonstrates the feasibility of orchestrating multi-agent debates with consistent persona behavior and real-time visualization.

\subsection*{Future Work}
Several extensions are required before Titanarium can move beyond a proof-of-concept. On the interaction side, a key direction is integration with existing social networking and communication platforms, enabling AI debates to appear directly within familiar user environments such as channels, threads, or group chats. This would necessitate robust authentication, rate limiting, message routing, and fine-grained control over when and how personas contribute.

Richer persona modeling is another priority. Instead of relying on simple seed prompts and patches, future versions should define personas using structured belief graphs, long-term memory, and configurable reasoning styles that evolve across sessions rather than resetting at each run. From a technical and safety standpoint, the current dependence on a single local LLM endpoint limits robustness. Supporting multiple models, fallback strategies, or remote inference services would improve reliability. Additional policy layers, safety filters, and topic-specific tools—such as retrieval over scientific corpora—would further constrain debates and reduce risk if the system were deployed beyond a controlled demo.

Finally, Titanarium currently lacks quantitative evaluation. Future work should incorporate instrumentation for logging and analytics, metrics for debate quality and viewpoint diversity, and user studies to assess engagement and usefulness. Such evaluations will clarify whether the ``digital terrarium'' paradigm offers a distinct and valuable experience compared to conventional single-model chat interfaces.

\subsection*{Open-source Materials}
Code is available on GitHub: \github{https://github.com/JSR-ISM-Smart-Chemistry-Lab/Titanarium}





\section{Can Large Language Models Predict Concrete Materials Properties?}\label{sec:LLM4ConProp}

%

The LLM4ConProp team investigates whether large language models (LLMs) can be leveraged for quantitative materials property prediction, with a focus on concrete—the most used manufactured material worldwide and a major contributor to resource consumption and CO$_2$ emissions. Predicting materials properties is central to accelerating materials design and discovery. While machine learning (ML) approaches have demonstrated strong performance, they typically depend on structured datasets and are often constrained by data scarcity. In contrast, LLMs are pre-trained on massive text corpora, encode broad human knowledge, and can be used without coding expertise, making them a fundamentally different and potentially complementary paradigm for materials research.

This work explores three central questions: (i) whether LLMs can achieve prediction accuracy comparable to state-of-the-art ML models under zero-shot and few-shot settings, (ii) whether expert domain knowledge can be effectively infused into LLM prompts to improve prediction performance, and (iii) whether LLMs can generate high-quality synthetic data to augment limited datasets and enhance the performance of ML models.


\subsection*{Results}

\begin{figure}[!htb]
\centering
\includegraphics[width=\linewidth]{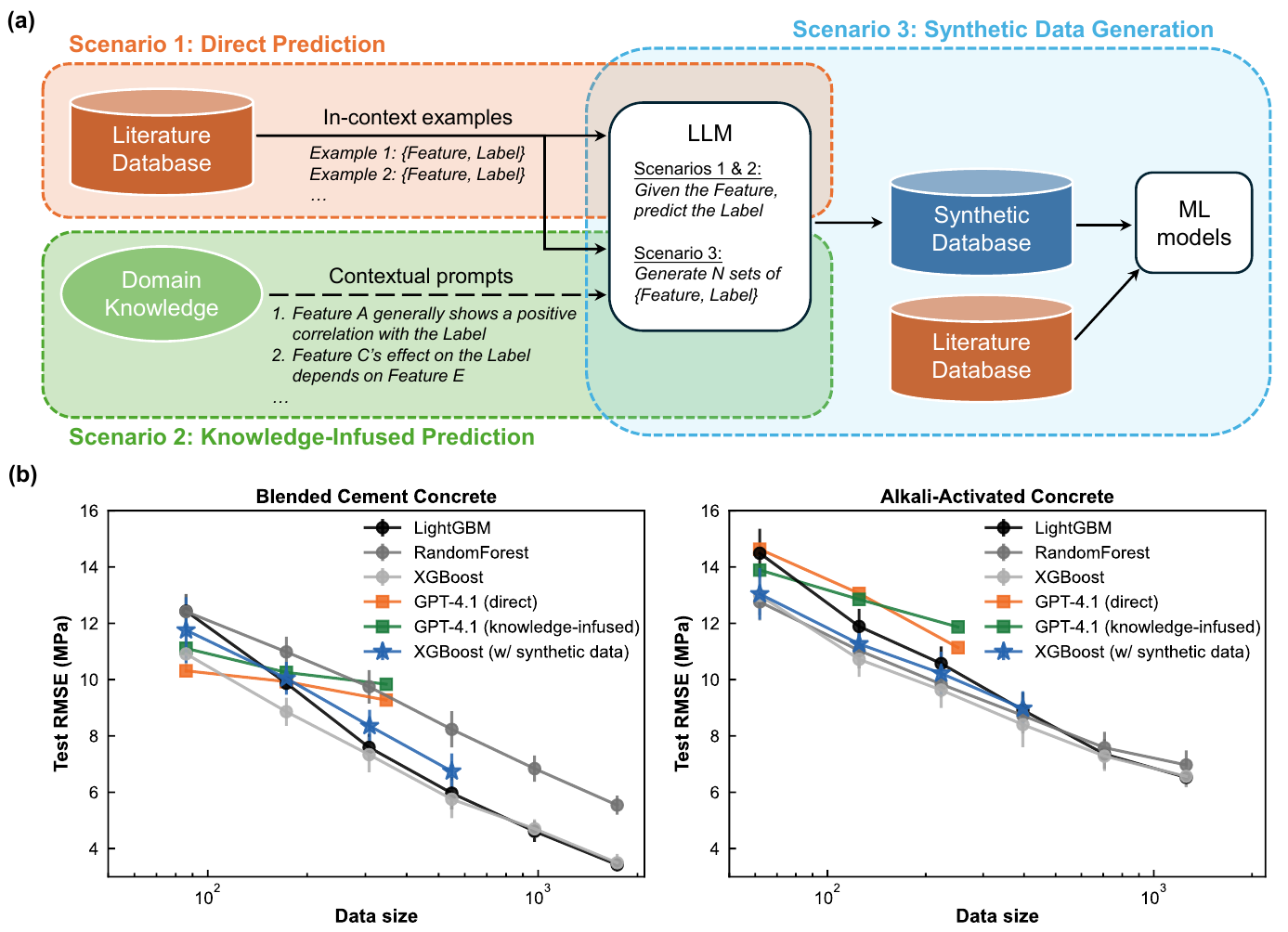}
\caption{Evaluation of large language models (LLMs) for concrete property prediction.
(a) Three evaluation scenarios: direct prediction, knowledge-infused prediction, and synthetic data generation. Prompt examples are shown in italics.
(b) Prediction accuracy as a function of data size for the two datasets, where data size refers to the amount of training data for machine learning (ML) models or the number of in-context examples for LLMs. For the synthetic data generation scenario, LLM-generated data were used to augment ML training datasets; data size reflects the amount of real data used, to which an equal amount of synthetic data was added for training. LLM results are shown for a single run, whereas ML model results are averaged over 10 runs with different random seeds; error bars indicate standard deviations. RMSE denotes root mean square error.}
\label{fig:llm4conprop}
\end{figure}

To address these questions, the team designed three evaluation scenarios for LLMs, as illustrated in Fig.~\ref{fig:llm4conprop}a: (1) direct property prediction under zero-shot and varying few-shot settings, (2) knowledge-infused prediction by incorporating expert domain knowledge into prompts, and (3) synthetic data generation to augment training datasets for ML models. Performance was benchmarked on two curated datasets of concrete compressive strength: a blended cement concrete dataset containing 2,171 records \cite{Imran2023} and an alkali-activated concrete dataset with 1,572 records \cite{Torres2023}, representing two major classes of sustainable concrete materials. For LLM-based evaluations, GPT-4.1 was selected as the representative model. ML baselines included tree-based methods—random forests \cite{Breiman2001}, XGBoost \cite{Chen2016a}, and LightGBM \cite{Ke2017}—which represent state-of-the-art approaches for tabular materials data.

Overall, both LLM and ML models exhibited higher performance with increasing data size (Fig.~\ref{fig:llm4conprop}b). At similar data sizes, the performance of the LLM model was comparable to that of ML models, while zero-shot LLM predictions exhibited lower accuracy (root mean square errors of 12.7 MPa for the blended cement concrete dataset and 23.7 MPa for the alkali-activated concrete dataset). Incorporating textual domain knowledge through prompting did not appear to further improve LLM performance, suggesting that the in-context examples may already capture the most relevant information, and more advanced representations of domain knowledge may be required to meaningfully enhance quantitative property prediction. Similarly, augmenting ML training datasets with LLM-generated synthetic data did not improve the predictive performance of ML models, which may reflect limitations in the quality or diversity of the generated data and highlights opportunities for further improvement.


\subsection*{Future Work}

Our preliminary findings suggest several promising research directions. Future work will explore more advanced prompting strategies, retrieval-augmented generation to incorporate external materials knowledge, and domain-specific fine-tuning to better ground LLMs in materials science. In addition, hybrid modeling frameworks that integrate LLM reasoning with data-driven ML approaches may more effectively leverage the complementary strengths of both paradigms.





\section{Nanoparticles Image Analysis with ms2nano}
\label{sec:ms2nano}
%

The ms2nano team focuses on automating the analysis of transmission electron microscopy (TEM) images for nanocrystal characterization. The synthesis of nanocrystals is notoriously difficult to control, requiring materials-science intuition combined with the precision of organic chemistry. Although TEM remains the gold standard for morphological characterization, manual image analysis is slow, subjective, and poorly suited to large datasets. To address this limitation, the team applies computer vision and machine-learning techniques to enable reliable, high-throughput extraction of quantitative morphological information from nanoparticle images.


\begin{figure}[t]
    \centering
    \includegraphics[width=0.75\textwidth]{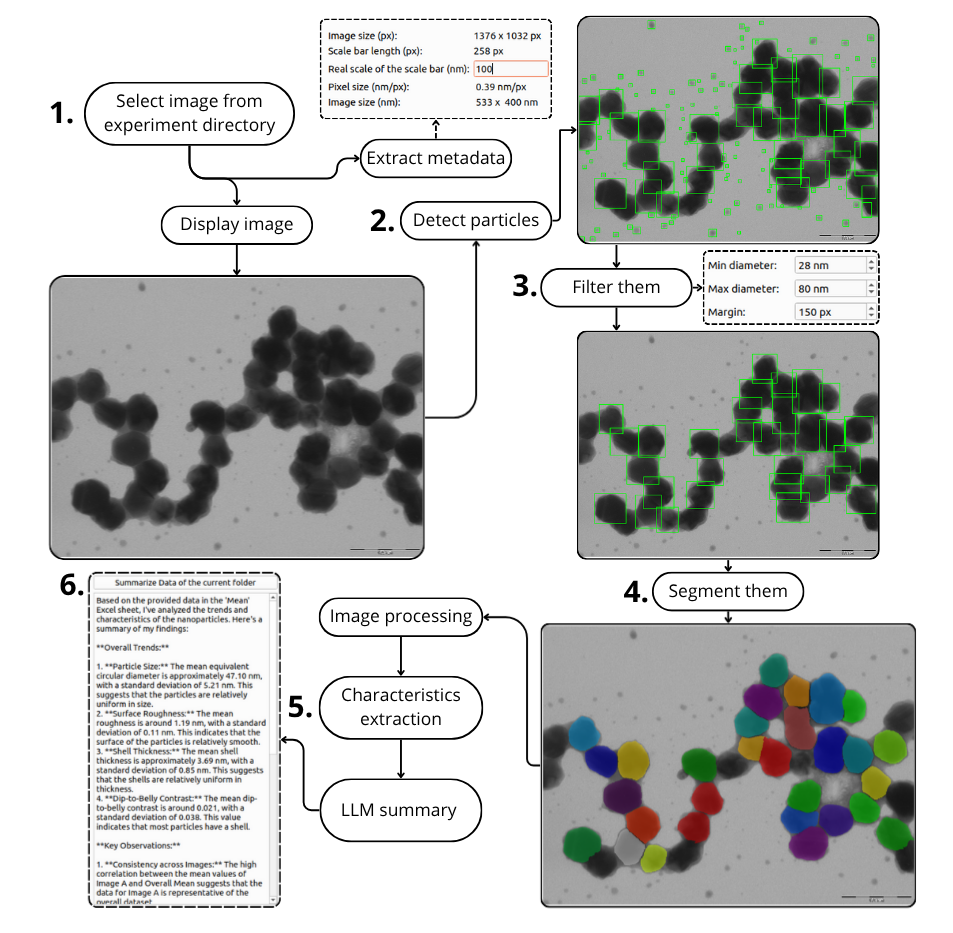}
    \caption{Complete nanoparticle analysis and LLM-driven insight generation workflow. The pipeline includes image selection and metadata extraction, particle detection and filtering, segmentation, extraction of morphological descriptors (roughness, shell thickness, porosity), dataset summarization, and final interpretation through an integrated LLM layer.}
    \label{fig:ms2nano}
\end{figure}

\subsection*{Results}

Building on prior systematic synthesis studies \cite{bastus2020}, the ms2nano team analyzes nanoparticles using three performance-critical morphological parameters: \textbf{roughness}, \textbf{shell thickness}, and \textbf{porosity}. The resulting tool, \textbf{ms2nano}, provides an automated workflow that leverages computer vision and machine learning to rapidly quantify these characteristics, supporting data-informed predictions of material behavior.

The technical approach extends a previous framework for scanning transmission electron microscopy (STEM) image analysis \cite{GENC2025114116} through several key enhancements:
\begin{enumerate}
    \item \textbf{Refined segmentation:} Post-segmentation utilities allow users to manually add or remove nanoparticles, enabling expert correction when automated detection is imperfect.
    \item \textbf{Data structuring:} Improved metadata handling and a clear directory structure for raw images, processed images, and statistical outputs streamline the overall analysis pipeline.
    \item \textbf{Size filtering:} User-defined nanoparticle size filters adapt the analysis to system-specific knowledge and experimental constraints.
    \item \textbf{LLM integration:} An integrated large language model layer (Llama~3 via Ollama) generates direct, interpretive insights from the extracted morphological data, helping users contextualize trends and results.
\end{enumerate}

\subsection*{Future Work}

Future development of ms2nano will focus on integrating a retrieval-augmented generation (RAG) system to enable flexible querying of analysis results, incorporating active learning to continuously improve segmentation and measurement accuracy through user feedback, and expanding the set of supported morphological parameters to capture a broader range of nanoparticle features.

\subsection*{Open-source Materials}

Code and demo are available on GitHub: \github{https://github.com/icn2-ai/m2snano-llm}





\section{DynaAgent: A Modular Multi-Agent Framework for Autonomous Protein--Ligand Molecular Dynamics Simulations}\label{sec:DynaAgent}

%

The LIAC-DynaAgent team addresses a key bottleneck in biomolecular modeling: the technical complexity of setting up and executing molecular dynamics (MD) simulations for proteins and protein–ligand systems. MD simulations are indispensable for probing biomolecular structure, dynamics, and function \cite{hollingsworth2018molecular, karplus2002molecular, schlick2021biomolecular, schlick2011biomolecular}, yet their practical adoption is limited by the difficulty of parameterization, input preparation, and software configuration \cite{lemkul2024introductory}. Although agentic large language models (LLMs) have recently shown promise in autonomously executing multi-step scientific workflows \cite{zou2025agente, campbell2025mdcrow, chandrasekhar2025automating}, they have not previously been demonstrated for fully autonomous protein–ligand MD simulations.

\begin{figure}[ht]
    \centering
    \includegraphics[width=0.8\linewidth]{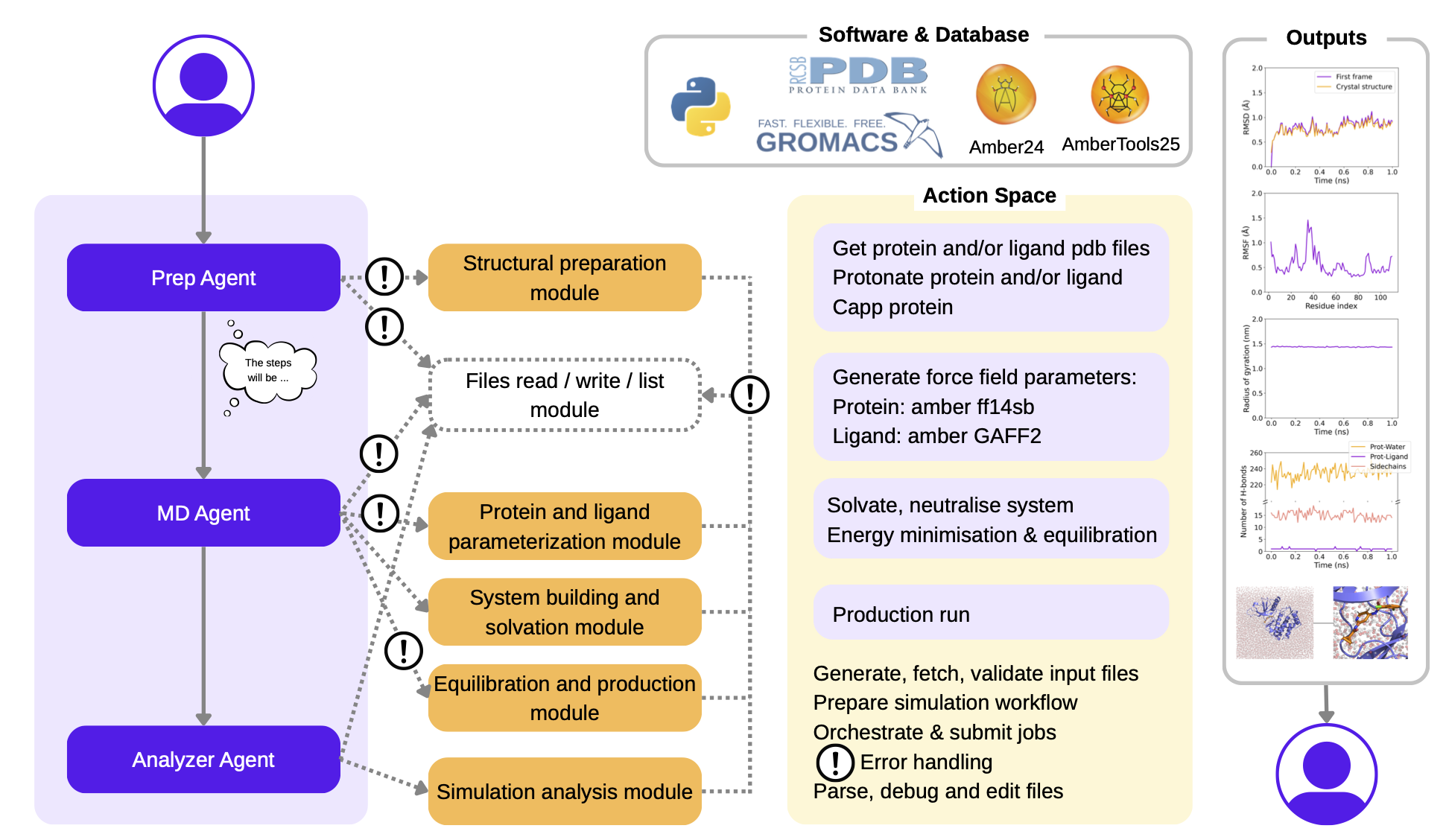}
    \caption{Overview of the \textbf{DynaAgent} architecture. The \texttt{PrepAgent} constructs a context-aware simulation plan, the \texttt{MDAgent} executes the plan with error-corrective reasoning, and the Analyser interprets the resulting trajectories. Available tools are shown in the action space.}
    \label{fig:dynaagent_agentic_system}
\end{figure}

To address this gap, the team introduces \textbf{DynaAgent}, a modular, multi-agent LLM framework that autonomously designs, executes, and analyzes complete MD workflows for protein-only and protein–ligand systems. As illustrated in Fig.~\ref{fig:dynaagent_agentic_system}, the pipeline spans parameter selection, simulation execution, automatic error correction, and post-simulation analysis. The agent can write and modify files, invoke predefined Python and bash tools, and operate entirely within a sandboxed directory. The workflow is organized into three coordinated agents: a Planner Agent (\texttt{PrepAgent}), a Molecular Dynamics Agent (\texttt{MDAgent}), and an Analyser Agent. Together, these agents interface with AmberTools \cite{case2023ambertools}, GROMACS \cite{abraham2015gromacs}, and custom utilities to autonomously complete MD tasks.


\subsection*{Results}

The performance and generalizability of \textbf{DynaAgent} were evaluated on eight distinct systems: five protein–ligand complexes and three protein-only systems. These test cases span diverse protein structures and ligand configurations, challenging the agent to adapt its planning and execution to system-specific requirements. All inputs were provided directly as Protein Data Bank (PDB) identifiers \cite{burley2025updated}, as specified in the user prompts. A built-in literature search tool enabled the agent to select appropriate temperatures and simulation parameters consistent with prior studies.

\textbf{DynaAgent} successfully completed 1.0~ns production MD simulations for all five protein–ligand systems. Analysis of the resulting root-mean-square deviation (RMSD), root-mean-square fluctuation (RMSF), and radius of gyration profiles indicated equilibrated trajectories. Inspection of generated input and output files confirmed correct system parameterization and valid trajectory generation. Example results for the PDB:5UEZ system are shown in Fig.~\ref{fig:dynaagent_agentic_system}.

Agent performance was quantified along two dimensions: efficiency and accuracy. Efficiency was defined as the ratio of successful tool calls to total tool calls,
\[
\text{efficiency} = \frac{\text{number of successful tool calls}}{\text{number of tool calls}},
\]
reflecting how effectively the agent minimized unnecessary iterations. Accuracy was defined as the ratio of successfully completed tasks to the minimum number of tasks required for a correct setup,
\[
\text{accuracy} = \frac{\text{number of successful tool calls}}{\text{minimum required tool calls}}.
\]

\begin{figure}[htbp]
    \centering
    \begin{subfigure}[b]{0.48\textwidth}
        \centering
        \includegraphics[width=\linewidth]{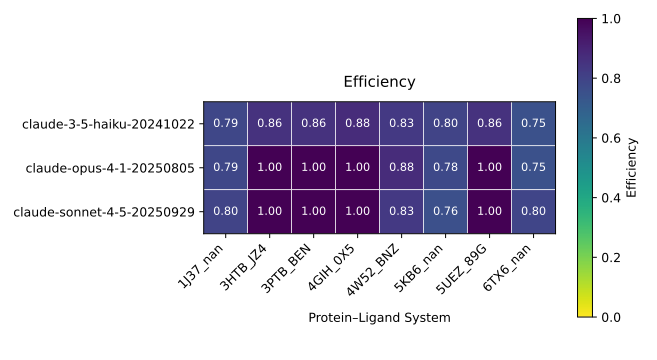}
        \caption{Efficiency of different LLMs, defined as the proportion of successful tool calls among all tool calls.}
        \label{fig:md_results_eff}
    \end{subfigure}
    \hfill
    \begin{subfigure}[b]{0.48\textwidth}
        \centering
        \includegraphics[width=\linewidth]{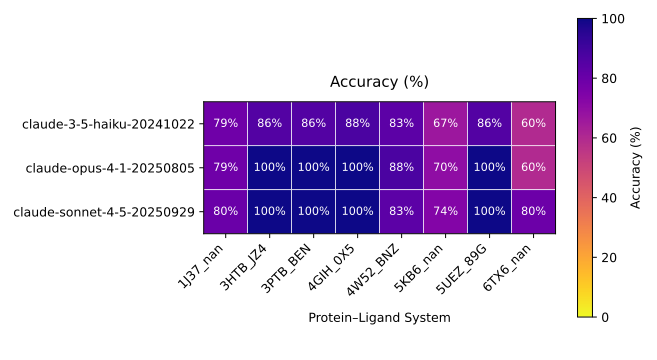}
        \caption{Accuracy of different LLMs, defined as the fraction of required tasks successfully completed.}
        \label{fig:md_results_acc}
    \end{subfigure}
    \caption{Comparison of efficiency and accuracy across different LLM backends.}
    \label{fig:md_results_combined}
\end{figure}

The framework was evaluated using three Claude models—Haiku~3.5, Opus~4.1, and Sonnet~4.5—across all eight test cases (Fig.~\ref{fig:md_results_combined}). Haiku consistently underperformed relative to Opus and Sonnet. Both Opus and Sonnet achieved 100\% efficiency in four of the protein–ligand systems. Efficiency losses primarily arose from incorrect tool ordering or malformed inputs, which increased the number of iterations required. In terms of accuracy, Opus and Sonnet again outperformed Haiku, achieving 100\% accuracy in four protein–ligand systems. For the fifth system (PDB:4W52 \cite{merski2015homologous}), accuracy reached 83\%, while overall task completion still reached 100\%, demonstrating effective error recovery.

\subsection*{Future Work}

Future extensions of \textbf{DynaAgent} will incorporate binding-affinity estimation tools to compute and compare $\Delta\Delta G$ values from protein–ligand MD trajectories, with benchmarking against the Open Free Energy platform. Errors encountered in protein-only simulations will be further analyzed and addressed through improved tooling. Additional system classes, including DNA, membrane proteins, and multimeric complexes with multiple ligands, will be introduced. Finally, systematic analysis of error-handling behavior across different agent backends will be conducted to better understand robustness and generalizability.

\subsection*{Open-source Materials}
Code and tools are available on GitHub: \github{https://github.com/schwallergroup/MDAgent}






\section{CrysTalk: An LLM-Agent System for Natural Language--Driven Crystal Structure Editing}\label{sec:CrysTalk}

%

The CrysTalk team addresses a persistent bottleneck in computational materials research: the preparation and modification of crystal structures for first-principles and machine-learning-based simulations. Although modern electronic-structure and molecular-dynamics workflows are increasingly accessible to experimental researchers, editing initial crystal structures still typically requires coding expertise and nontrivial software setup. At the same time, recent large language models (LLMs) demonstrate strong capabilities in contextual understanding, reasoning, and executable code generation. Motivated by these developments, the team aims to democratize computational simulation by introducing an LLM-powered agent that enables crystal structure editing through natural language.


\subsection*{Results}

\begin{figure}[h]
    \centering
    \includegraphics[width=0.80\textwidth]{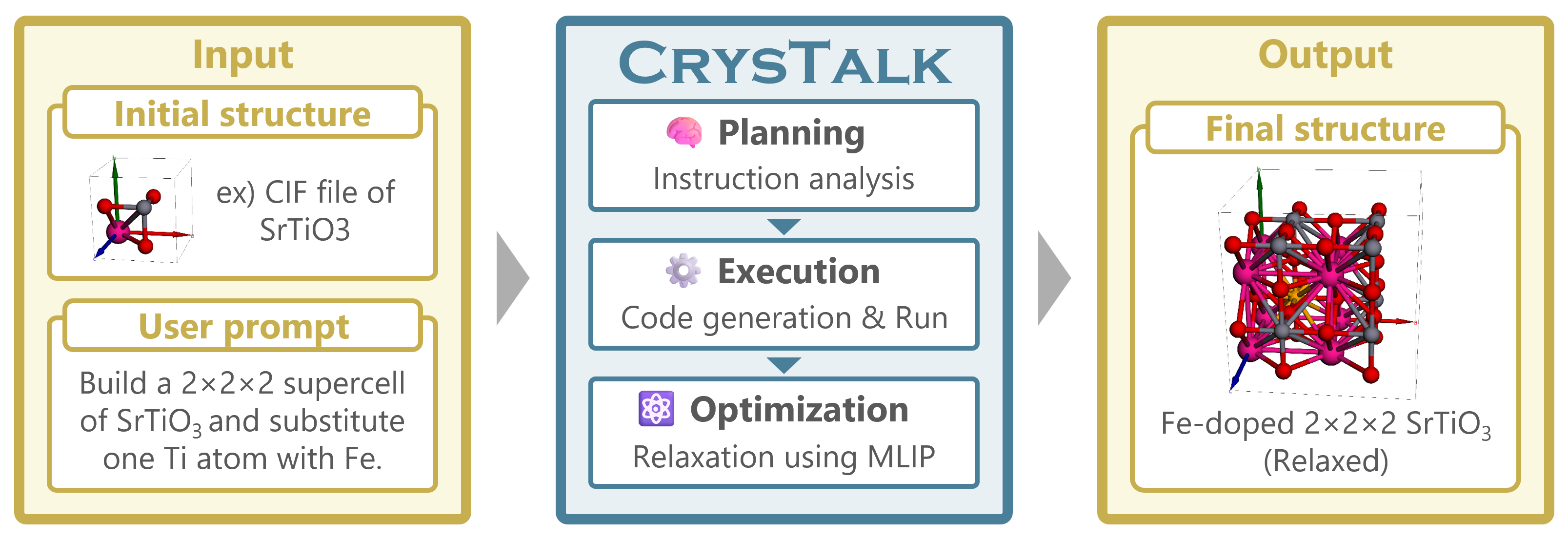}
    \caption{Workflow of CrysTalk. Given an input structure file and a user prompt, the agent performs planning, execution, and optional optimization to generate the final edited crystal structure.}
    \label{fig:crystalk}
\end{figure}

The CrysTalk team developed \textbf{CrysTalk}, an agent system that automates crystal structure editing from natural-language instructions, as illustrated in Fig.~\ref{fig:crystalk}. In this workflow, a user provides an initial crystal structure file (CIF or POSCAR) together with a natural-language prompt. The LLM agent then autonomously executes a sequence of steps including information retrieval, task planning, code generation and execution, reflection or retry, and optional structural optimization. Upon completion, the system outputs the edited structure file.

The workflow begins with user interaction through either a command-line interface or a graphical user interface implemented with Gradio. The input structure is parsed using \texttt{pymatgen} and visualized in real time. The agent interprets the user instruction and, when ambiguities arise, retrieves relevant terminology and background information using the Tavily search tool. Based on this context, the agent decomposes the instruction into subtasks and formulates an execution plan.

Each subtask is executed by a dedicated Coding Agent operating in a sandboxed environment, which generates and runs Python scripts to perform the requested structural manipulations. After completing all edits, the system can optionally relax the resulting structure using a machine-learning interatomic potential such as MACE~\cite{batatia2023foundation} to obtain an energetically stable configuration. The final structure is then exported in standard formats, including POSCAR and CIF, and a three-dimensional visualization is displayed in the GUI.

CrysTalk is designed as a modular agent system, with clear separation between planning, execution, reflection, and optimization components. Safety is ensured by sandboxing all generated code, and transparency is maintained by logging all inputs, outputs, and intermediate steps. Using this framework, the team confirmed that CrysTalk can successfully perform a wide range of crystal structure editing tasks, including supercell construction, slab generation, and elemental substitution.


\subsection*{Future Work}

Future work will focus on quantitative evaluation of CrysTalk to rigorously assess the reliability and correctness of its structure manipulation capabilities, which have so far been evaluated primarily qualitatively. Enhancing the agent’s understanding of crystallographic concepts is expected to enable more complex and sophisticated operations. In addition, integrating literature APIs and the Materials Project API, as well as providing interfaces that allow other LLM agents to invoke CrysTalk as a tool, will further advance autonomous and language-driven materials design.

\subsection*{Open-source Materials}
Code and a demo are available on GitHub: \github{https://github.com/MatAgentHub/crystalk}





\section{SpectroBot: One-Click FTIR/UV--Vis Analysis with Literature-Supported Interpretation}
\label{sec:spectrobot}

%

The SpectroBot team tackles a persistent translational bottleneck in chemistry and materials science: the reliable interpretation of laboratory spectra. While FTIR and UV--Vis measurements are routinely collected, existing software often stops at visualization, leaving peak identification, functional assignment, and reporting to manual interpretation that varies widely across users and laboratories. SpectroBot is designed to standardize this process by combining automated spectral analysis with literature-supported interpretation, reducing the full workflow—upload, analyze, and interpret—to a matter of minutes while preserving scientific rigor.


\subsection*{Results}

SpectroBot ingests CSV files containing either FTIR data (wavenumber in cm$^{-1}$ versus transmittance) or UV--Vis data (wavelength in nm versus absorbance). For FTIR spectra, the analysis pipeline detects troughs in transmittance, estimates peak width and prominence, and evaluates candidate functional groups against a curated ruleset derived from standard FTIR reference tables. For UV--Vis spectra, the pipeline identifies the absorption maximum ($\lambda_{\max}$), computes the full width at half maximum (FWHM), generates a publication-ready plot, and summarizes key spectral features.

Each analysis produces two machine-verified outputs: (i) a formatted PDF containing clean plots, tables, and annotations, and (ii) a structured JSON bundle that records detected peaks, thresholds, and scoring evidence. An interpretation stage then reads the generated PDF/JSON artifacts (along with a CSV preview) to produce a concise, branded ``SpectroBot---AI Interpretation'' report. This narrative includes in-text citations and a final bibliography drawn strictly from curated references bundled with the software, ensuring that all interpretations are literature-supported rather than speculative.

\begin{figure}[!ht]
    \centering
    \includegraphics[width=\linewidth]{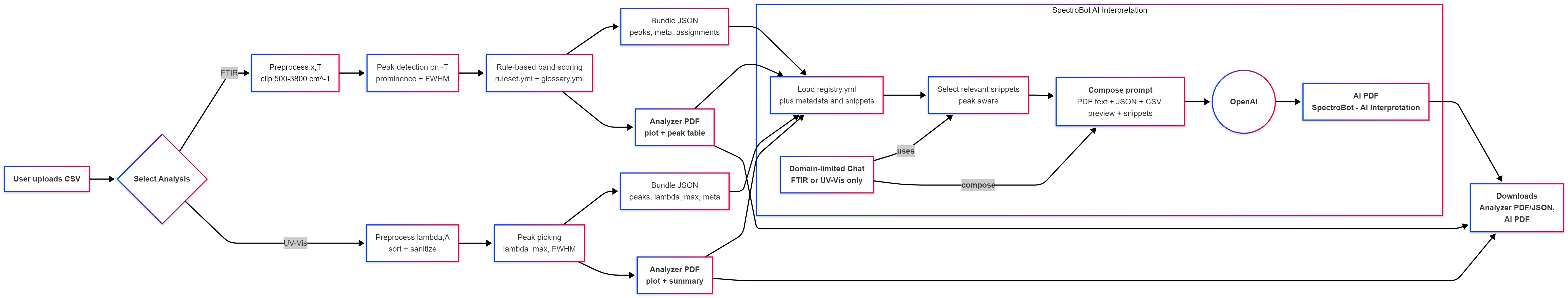}
    \caption{SpectroBot workflow. A user uploads a CSV file; the FTIR or UV--Vis analyzer generates verified PDF and JSON artifacts; and the interpretation component synthesizes a literature-supported report alongside a focused, domain-restricted chat interface.}
    \label{fig:spectrobot}
\end{figure}

The end-to-end workflow is illustrated in Fig.~\ref{fig:spectrobot}. After uploading a CSV file, the appropriate analyzer generates the PDF and JSON outputs, which are then consumed by the interpretation module to synthesize a concise, citation-backed report. The user interface exposes both artifacts and provides a restricted chat environment limited to FTIR and UV--Vis questions. In validation tests using cellulose-rich fibers and polymer membranes, the FTIR module reliably recovered hallmark vibrational bands, including O--H stretching at $3340$--$3100\ \mathrm{cm}^{-1}$, C--H stretching near $\sim2900\ \mathrm{cm}^{-1}$, carbonyl features around $\sim1735\ \mathrm{cm}^{-1}$ when present, and glycosidic C--O--C modes in the $1160$--$1050\ \mathrm{cm}^{-1}$ range, consistent with reported literature. The UV--Vis module similarly identified primary absorption bands and their bandwidths, enabling comparisons across formulations and processing conditions.


\subsection*{Future Work}

Future development will focus on expanding the FTIR rule base to cover a broader range of polymers and biomaterials, adding optional Raman spectroscopy support using the same JSON contract, and integrating lightweight retrieval mechanisms to access larger reference libraries while maintaining local, citation-controlled interpretation behavior.

\subsection*{Open-source Materials}
A code prototype is available on GitHub: \github{https://github.com/MUzair20/spectrobot}.  
A working model is available at: \url{https://spectrobot.streamlit.app/}





\section{Collaboration Learning AI-Agents to Accelerate Multidisciplinary Projects}\label{sec:MindMesh}

%

The MindMesh team addresses a common bottleneck in multidisciplinary scientific projects: ineffective communication arising from differences in background knowledge, terminology, and learning styles. In such settings, researchers often spend significant effort translating concepts across disciplines rather than advancing the research itself. Motivated by these challenges, the team developed MindMesh, a system that generates personalized AI learning agents tailored to individual users’ knowledge bases and learning preferences. By adapting explanations to the way each researcher best understands information, MindMesh aims to streamline collaboration and accelerate progress in multidisciplinary environments.


\subsection*{Results}

The first MindMesh prototype was developed to facilitate multidisciplinary collaboration through personalized AI learning agents. The workflow begins with a short questionnaire designed to capture two key dimensions of each user: a \emph{knowledge profile}—including academic background, domain expertise, and technical familiarity—and a \emph{learning profile}, such as preferences for explanations using analogies, visual representations, equations, or direct textual descriptions. Based on these inputs, the system instantiates a personalized AI agent that communicates scientific concepts in a manner aligned with the user’s individual learning style.

Users interact with their personalized agent through a front-facing chat interface, receiving explanations, clarifications, and guidance that are dynamically adapted to their profile. Although the current prototype focuses on individual agents, the design establishes the foundation for a centralized “mother agent” capable of synthesizing project goals, translating information between agents, and coordinating tasks across a team. This architecture demonstrates the feasibility of leveraging personalized AI agents to reduce communication friction, improve mutual understanding, and accelerate collaborative problem-solving in multidisciplinary scientific projects.

\begin{figure}[h]
    \centering
    \includegraphics[width=1.0\linewidth]{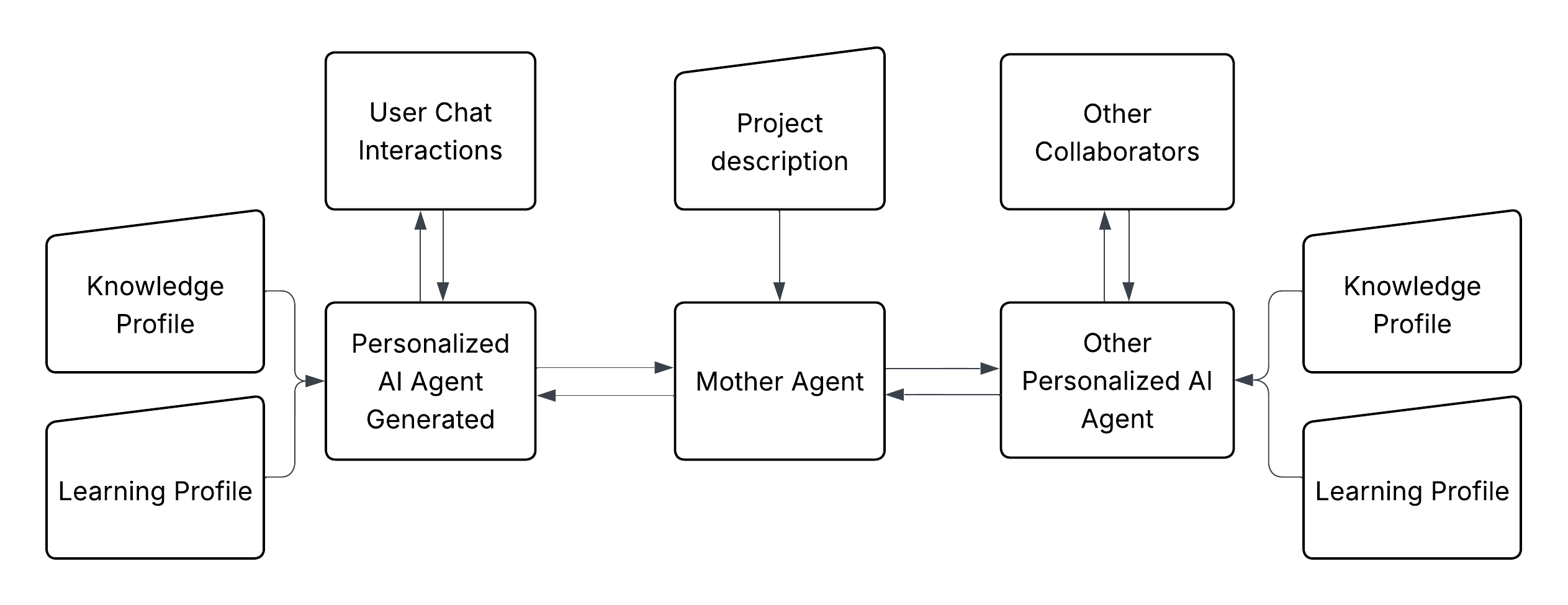}
    \caption{Workflow of the personalized agents in MindMesh, illustrating the generation of user-specific learning agents from knowledge and learning profiles.}
    \label{fig:MindMesh-pipeline}
\end{figure}


\subsection*{Future Work}

Future developments will focus on integrating multiple personalized agents into a centralized coordination framework, or “mother agent,” to manage team workflows and mediate communication across disciplines. Additional directions include extending MindMesh to classroom education and science communication settings, and conducting systematic evaluations to quantify the impact of personalized AI agents on collaborative efficiency, learning outcomes, and project velocity in real-world multidisciplinary teams.

\subsection*{Open-source Materials}
A code prototype is available on GitHub: \github{https://github.com/estefaniavazquez/MindMesh}





\section{SyntheSeek: Agentic Workflow for Synthesis Method Generation}\label{sec:SyntheSeek}

%

The SyntheSeek team addresses a persistent challenge in materials chemistry: synthesis knowledge is largely embedded in unstructured research papers, making it difficult for researchers to efficiently identify, compare, and adapt existing synthesis methods. As a result, chemists often rely heavily on personal experience and intuition when designing new synthesis routes. SyntheSeek introduces an agentic large language model (LLM) workflow that retrieves, structures, and adapts synthesis procedures from the scientific literature, enabling on-demand access to prior knowledge while accounting for user-specific constraints such as available equipment and target material properties.


\subsection*{Results}

\begin{figure}[h]
    \centering
    \includegraphics[width=0.75\linewidth,trim={0 32cm 0 40cm},clip]{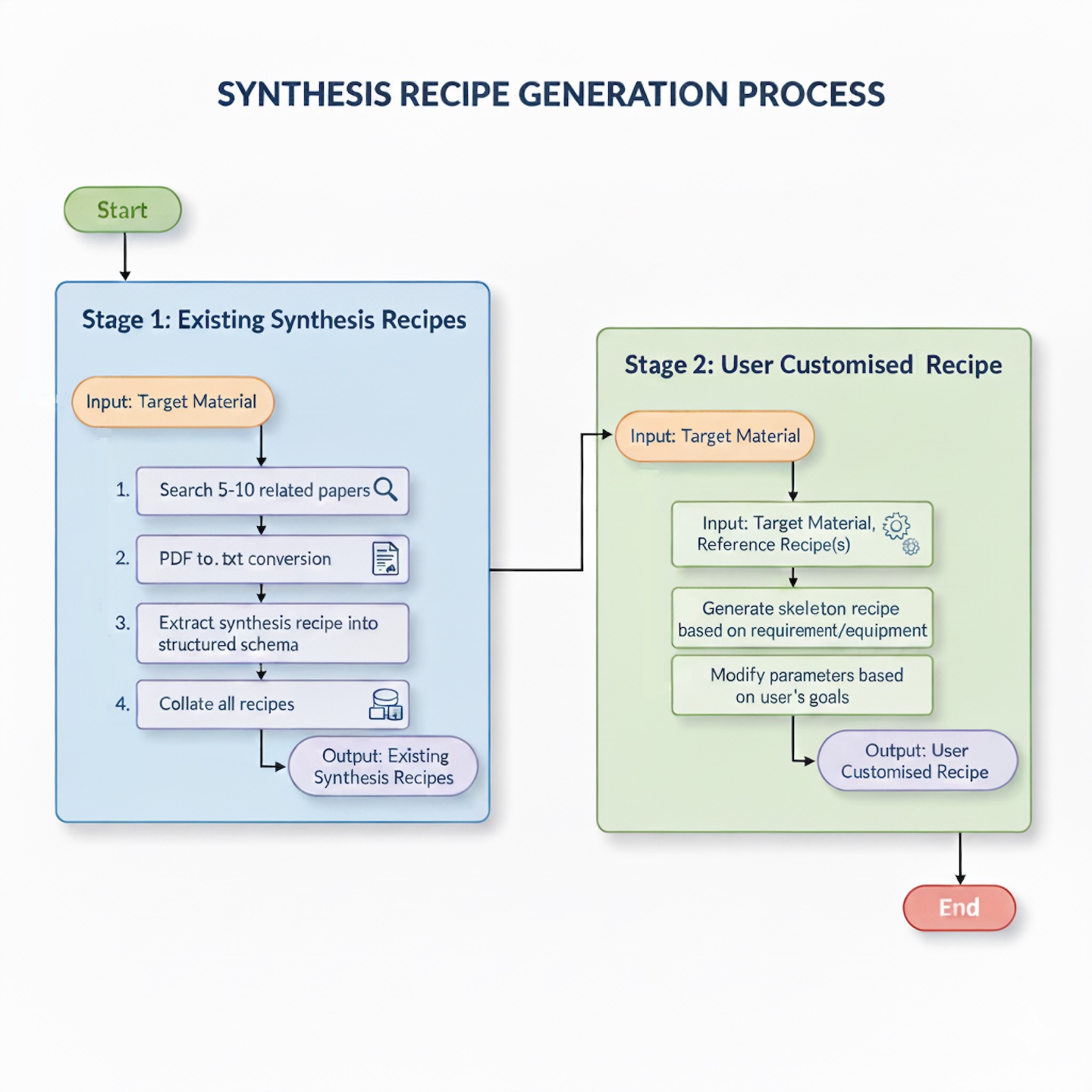}
    \caption{SyntheSeek two-stage synthesis recipe generation workflow.}
    \label{fig:syntheseek_workflow}
\end{figure}

Figure~\ref{fig:syntheseek_workflow} illustrates the two-stage workflow implemented in SyntheSeek for generating synthesis procedures tailored to a user-specified target material. 

In \textbf{Stage~1}, the agent performs literature-driven synthesis retrieval. It identifies approximately 5--10 relevant publications, converts their PDF files into text, and extracts synthesis protocols into a unified, structured schema. These extracted procedures are normalized across publications and consolidated into a comparable format, which is then presented to the user as a curated set of reference synthesis methods.

In \textbf{Stage~2}, the system generates a customized synthesis plan by integrating the curated reference procedures with user-defined constraints and objectives. First, a template ``skeleton'' procedure is constructed based on the user’s available equipment and the general methodological patterns identified in Stage~1. Subsequently, parameters within each synthesis step—such as reagent ratios, temperatures, processing times, and optional variations—are adaptively adjusted to align with the user’s target properties and experimental goals. The final output is a coherent, actionable synthesis outline tailored to the specific laboratory context.


\subsection*{Future Work}

Future work will focus on leveraging the structured synthesis outputs from Stage~1 to build a comprehensive synthesis-method database. Such a resource would be valuable in its own right, addressing the current lack of openly available, structured synthesis knowledge. Beyond direct retrieval, this database could support training of embedding models for unsupervised clustering and improved literature retrieval. These embeddings could also serve as downstream inputs to decoder-based or diffusion-based generative models, enabling more advanced synthesis method generation and exploration.

\subsection*{Open-source Materials}

Code and the novel synthesis schema developed in this work are available on GitHub: \github{https://github.com/alexchen5/syntheseek}





\section{Validated Rapid Adsorption Probe Interaction Discovery System}\label{sec:V-RAPIDS}

%

The V-RAPIDS team introduces the \textbf{Validated Rapid Adsorption Probe Interaction Discovery System (V-RAPIDS)}, a lightweight agentic workflow designed to democratize qualitative dry-lab simulations of probe–target–substrate interactions. Traditional atomic simulation workflows often require extensive expertise and hours to days of computation, creating a barrier for researchers who need rapid, qualitative insights. V-RAPIDS addresses this challenge by delivering referenceable interaction energies and optimized geometries from minimal user input, enabling fast exploratory analysis without the overhead of full first-principles simulations.

\begin{figure}[t]
    \centering
    \includegraphics[width=0.70\linewidth]{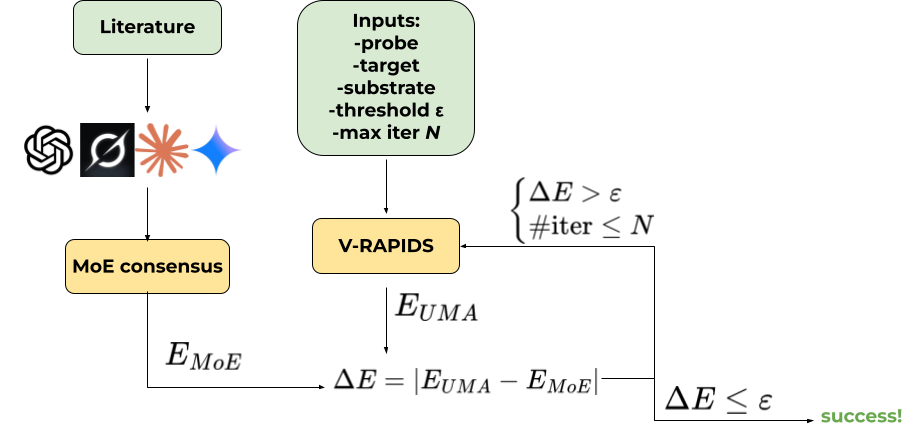}
    \caption{Overview of the V-RAPIDS workflow, illustrating UMA-based structure optimization followed by literature-backed validation using a mixture-of-experts approach.}
    \label{fig:v-rapids-1}
\end{figure}

V-RAPIDS is built on Meta FAIRChem’s Universal Models for Atoms (UMA) \cite{wood2025umafamilyuniversalmodels}, which provide rapid surrogate predictions of atomic interactions. From a user-specified probe, target, and substrate, the system generates optimized structures (exported as \texttt{.vasp} files) and computes adsorption energy, vacuum interaction energy, and total three-component binding energy. The current implementation supports a growing set of substrates, including two-dimensional materials (e.g., graphene, MoS$_2$, black phosphorus, Si, ZnO) and metal–organic frameworks (e.g., Co-HHTP, Cu-HHTP, Ni-HHTP). For three-component systems, V-RAPIDS employs a sequential optimization strategy: first optimizing the probe on the substrate, then introducing the target molecule and re-optimizing the combined system to capture realistic binding configurations.

While UMA enables orders-of-magnitude speedups relative to density functional theory (DFT), its quantitative accuracy gap relative to high-fidelity methods is well documented \cite{wood2025umafamilyuniversalmodels}. To address this limitation, the team incorporates a validation layer that benchmarks UMA-derived energies against literature-reported DFT results using a mixture-of-experts (MoE) framework powered by large language models. This MoE aggregates consensus estimates from multiple LLM providers, including OpenAI, xAI, Anthropic, and Google Gemini, each tasked with retrieving and interpreting relevant literature data for comparable systems.

\begin{figure}[h]
    \centering
    \includegraphics[width=0.70\linewidth]{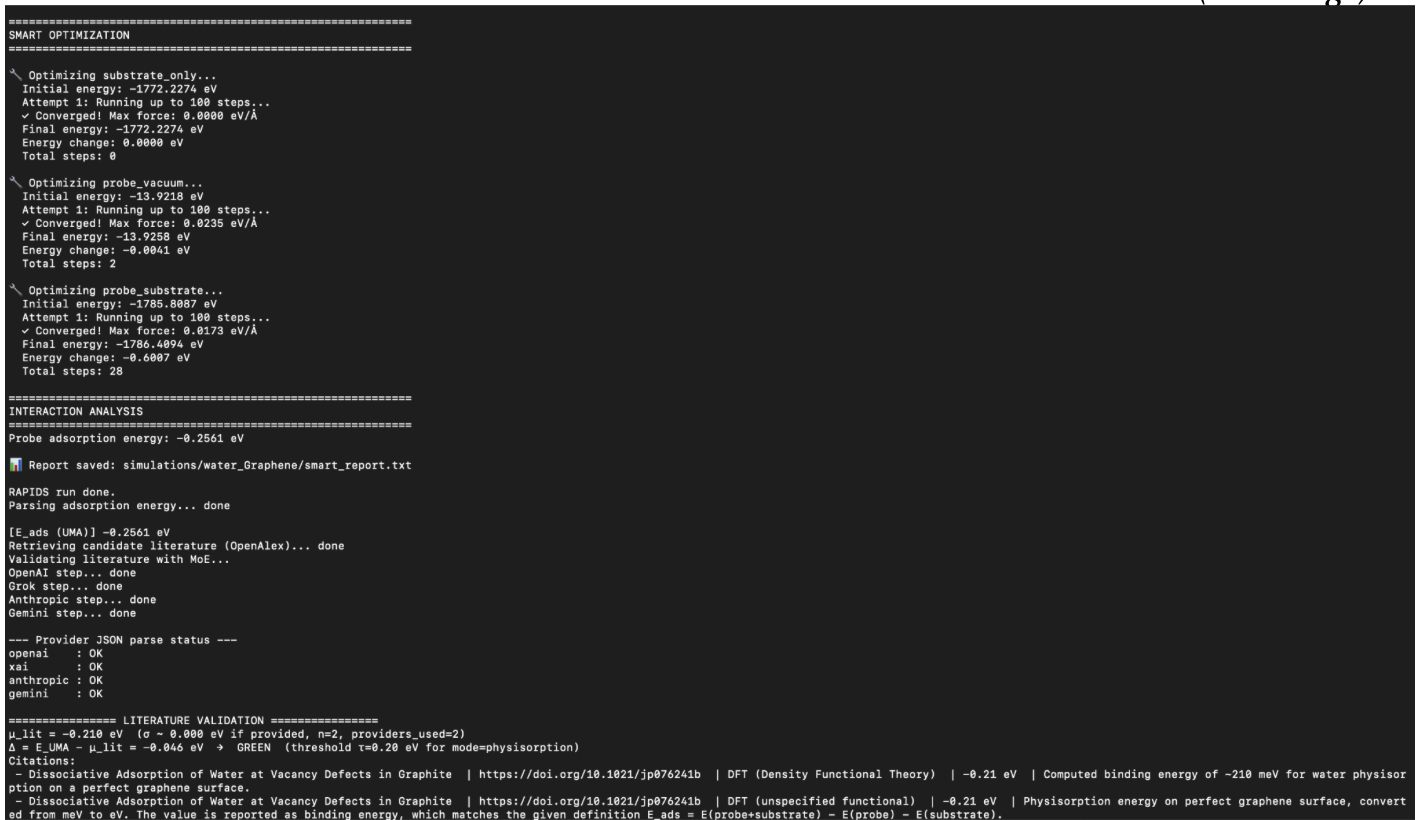}
    \caption{Representative V-RAPIDS output for the water–graphene system, including optimized geometries and computed interaction energies.}
    \label{fig:v-rapids-2}
\end{figure}

If the consensus literature-based energy agrees with the UMA prediction within a user-defined tolerance $\varepsilon$, the result is accepted. Otherwise, V-RAPIDS iteratively refines the simulation setup—adjusting parameters such as force convergence criteria, simulation cell dimensions, or molecular placement—and re-runs the calculation until agreement is achieved or a maximum number of iterations $N$ is reached. If convergence is not obtained, the system alerts the user for manual review. This hybrid approach combines the speed of UMA (minutes) with literature-validated reliability, offering a practical compromise between qualitative insight and scientific credibility.

\subsection*{Results}

Figure~\ref{fig:v-rapids-1} summarizes the end-to-end V-RAPIDS workflow. Given a probe–target–substrate specification, the system performs rapid UMA-based optimization, computes interaction energies, and validates the results against literature-derived references through the MoE consensus mechanism. Figure~\ref{fig:v-rapids-2} shows a representative output for a water–graphene system, demonstrating the generation of optimized structures and associated energy metrics suitable for qualitative interpretation and comparison.

\subsection*{Future Work}

Future development will focus on expanding the supported materials space by incorporating additional two-dimensional materials (e.g., h-BN and further transition metal dichalcogenides) and a broader range of metal–organic frameworks. Planned enhancements include automated hyperparameter tuning within the validation loop, tighter integration with external scientific databases such as the Materials Project to improve literature grounding, and extension of the framework to surface catalysis simulations. The team also aims to systematically benchmark the MoE-based validation strategy across diverse chemical systems to establish robust, system-specific tolerance thresholds.

\subsection*{Open-source Materials}
Code and usage instructions are available on GitHub: \github{https://github.com/ruiding-uchicago/auto_CPT_uma_simul}






\section{NOMAD RAGBOT: An AI Assistant for Navigating Distributed Community Knowledge}
\label{sec:nomad-ragbot}

Community knowledge is often fragmented across websites, forums, and chat platforms, making it difficult for users to locate reliable, up-to-date, and verifiable information. Although large language models (LLMs) can summarize such content, their responses may be hard for non-experts to validate, can incorporate outdated material, and often lack clear links to authoritative sources. To address these challenges, we developed \textbf{NOMAD RAGBOT}, a retrieval-augmented assistant grounded in verified documentation. As a representative use case, the system is applied to the NOMAD ecosystem—a comprehensive research data management platform for materials science \cite{scheidgen2023nomad}. By combining semantic retrieval with controlled, citation-aware generation, NOMAD RAGBOT delivers accurate, traceable answers that help users confidently explore relevant resources.

\subsection*{Results}

We implemented a retrieval-augmented generation (RAG) workflow \cite{lewis2020retrieval} over a corpus comprising more than 50 documentation sites of varying size and structure. Source documents in markdown or HTML format are ingested and transformed into vector representations using a custom embedding infrastructure based on the \texttt{Nomic-Embed-Text} model \cite{nomic2024embed}, deployed on a remote server. The resulting embeddings are stored in ChromaDB \cite{chroma2023db} to enable efficient semantic retrieval. A modular architecture cleanly separates document processing, indexing, retrieval, and conversational management, facilitating straightforward extension to additional data sources.

User queries are embedded and matched against the vector store via a retriever–reranker pipeline. Retrieved passages are semantically scored, and the most relevant chunks are passed to an LLM through an adaptive prompt template that emphasizes factual, context-grounded responses. Compared to direct LLM querying, this design substantially reduces hallucinations and improves clarity and reliability.

Multi-turn dialogue is supported through a conversational memory module that constructs context-aware prompts by combining prior chat history with newly retrieved evidence. The backend operates as a REST API with a dedicated \texttt{/ask} endpoint, enabling external integrations such as a Discord interface.

To further improve retrieval accuracy, we implemented a context-aware dynamic chunking strategy that respects markdown heading hierarchies. Each chunk preserves section and subsection titles and includes overlapping boundaries to maintain continuity. This structure-aware segmentation enhances retrieval precision and ensures that generated answers remain well grounded in their original context.

To support systematic improvement and rapid diagnosis of retrieval or generation failures, we added an evaluation dashboard that runs a gold-standard question set, reports pass rates under adjustable thresholds, allows filtering by source, and provides searchable test cases.

The prototype was demonstrated using a Gradio-based web interface \cite{gradio}, featuring a simple prompt box, example queries, and separate fields for responses and citations. In preliminary evaluations, the system provided coherent and relevant answers, including for advanced queries related to complex software development topics within the NOMAD ecosystem. Observed limitations include (i) insufficient prioritization of more mature or authoritative documentation sources, and (ii) occasional inaccurate combinations of information drawn from distinct sources.

\begin{figure}[h]
    \centering
    \includegraphics[width=\linewidth, keepaspectratio]{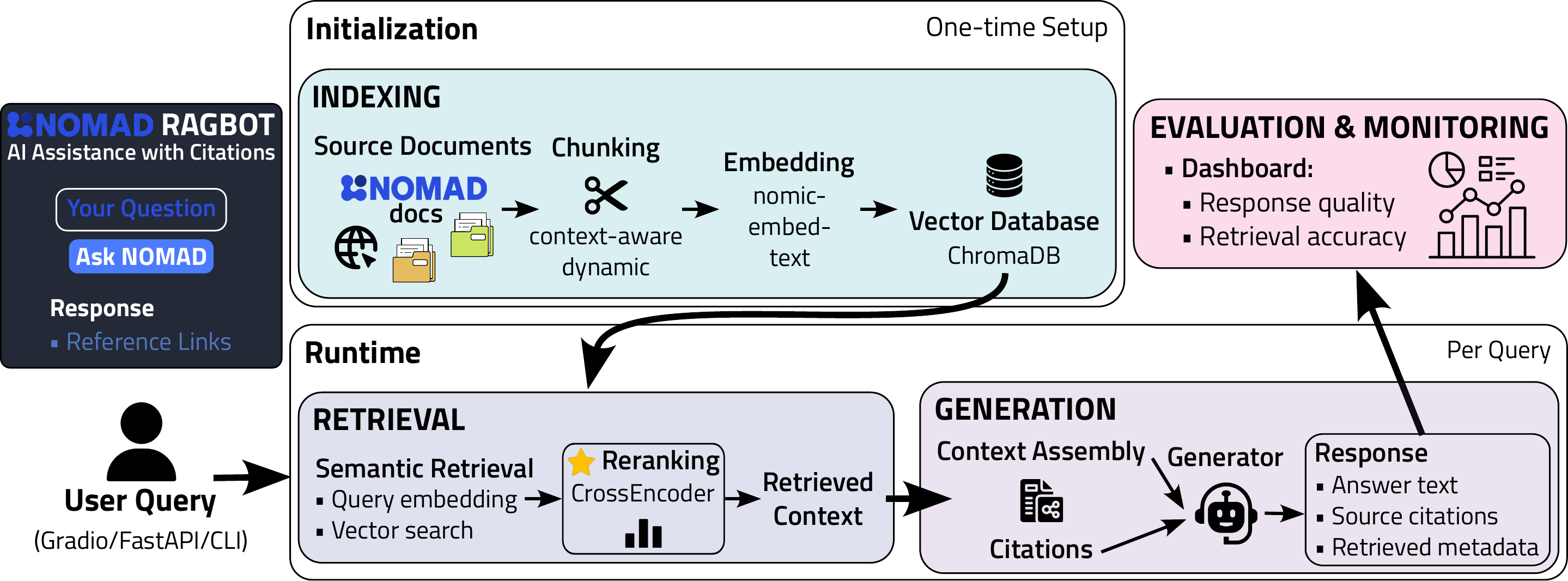}
    \caption{NOMAD RAGBOT workflow. The system performs (1) offline indexing with context-aware chunking, (2) semantic retrieval with CrossEncoder reranking, and (3) LLM-based answer generation with explicit citations. System performance is monitored via an evaluation dashboard.}
\end{figure}

\subsection*{Future Work}

Despite the immediate usability of the prototype, several impactful improvements remain. Planned enhancements include prioritizing primary documentation sources, enriching retrieval with structured metadata and code-aware chunking, and expanding the evaluation suite with task-oriented metrics. We also aim to extend coverage to additional NOMAD resources, including community discussions such as Discord chats, which introduce new challenges for context-preserving data segmentation. Ultimately, NOMAD RAGBOT will be integrated directly into NOMAD user interfaces and community platforms.

\subsection*{Open-source Materials}
The source code is available on GitHub: \github{https://github.com/FAIRmat-NFDI/nomad-bot-rag-docs-discord}. \\
An explanation and demo video are available on Zenodo: \href{https://doi.org/10.5281/zenodo.17604263}{https://doi.org/10.5281/zenodo.17604263}.





\section{MIDAS: Language Controlled Molecular Design and Analysis}\label{sec:MIDAS}




Structure-based drug design (SBDD) has emerged as a cornerstone of modern drug discovery, enabling the rational design of molecules that complement the three-dimensional structure of biological targets. Recent advances in deep learning, particularly diffusion models~\cite{ho2020denoisingdiffusionprobabilisticmodels,song2021scorebasedgenerativemodelingstochastic}, have demonstrated strong capabilities in generating novel molecular structures that satisfy geometric constraints imposed by protein binding pockets~\cite{liu2022generating3dmoleculestarget, xu2023geometriclatentdiffusionmodels, peng2025pocket2molefficientmolecularsampling}. However, purely geometric methods often operate as black boxes, making it difficult for molecule designers to guide generation toward specific chemical motifs or properties. Expert knowledge about desirable functional groups, pharmacophores, or physicochemical properties is also hard to specify explicitly.

\begin{figure}[h]
    \centering
    \includegraphics[width=0.60\linewidth]{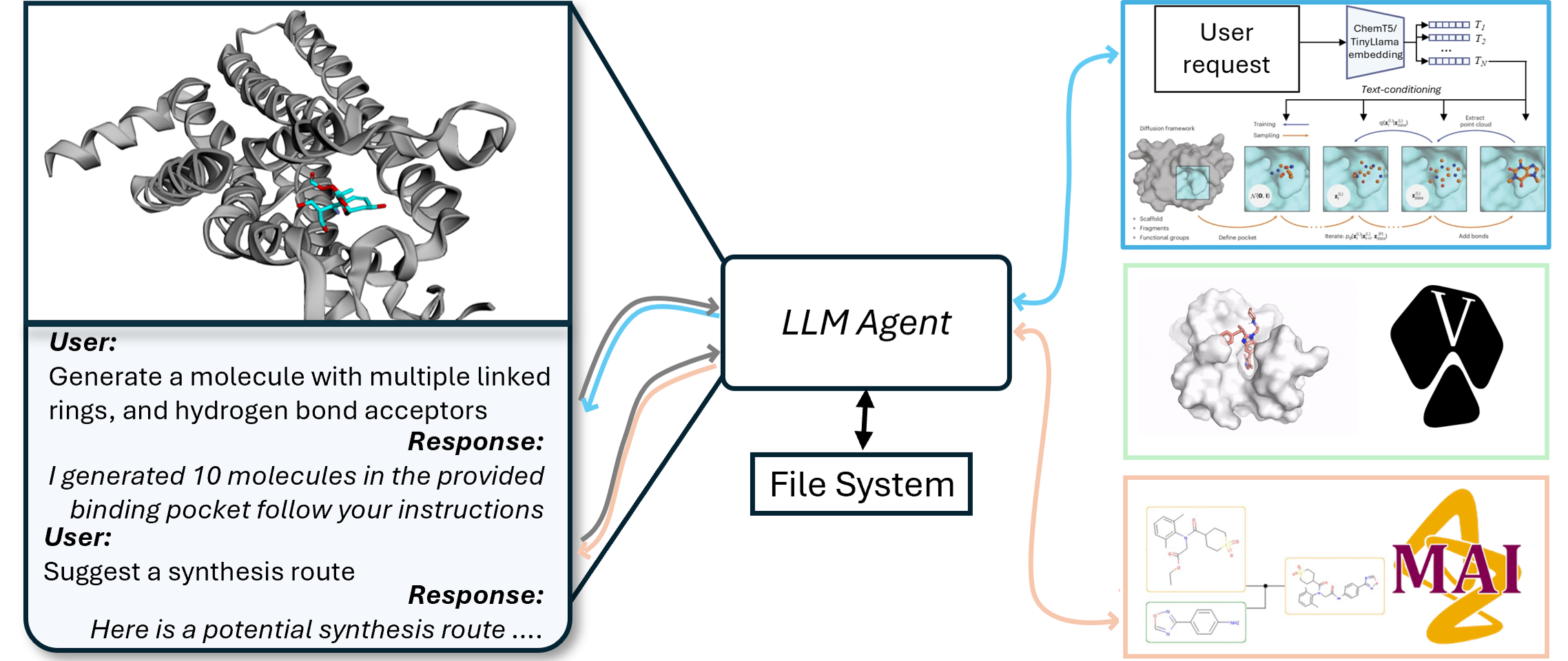}
    \caption{The conceptual diagram of Language Controlled Molecular Design and Analysis.}
    \label{fig:concept}
\end{figure}

The MIDAS team presents an approach that addresses these limitations by combining structure-based generation with natural language control. MIDAS (Molecular Intelligence \& Design Articulated by Semantics; Figure~\ref{fig:concept}) is an LLM-based agentic framework that uses tool calling together with a language-conditioned extension of equivariant diffusion models~\cite{schneuing2024structurebaseddrugdesignequivariant, hoogeboom2022equivariantdiffusionmoleculegeneration}. Unlike existing substructure-conditional generation methods that rely on domain-specific encodings and can be rigid or opaque to non-experts~\cite{fischer2025flowrflowingsparsedense, cremer2025flowrflowmatchingstructureaware}, conditioning on natural language enables direct use of chemical intuition, interpretable control over molecular properties, and iterative refinement through textual feedback. The team also integrates established cheminformatics tools for downstream analysis of generated molecules, including molecular docking and retrosynthesis.

\subsection*{Results}

\begin{figure}[h]
    \centering
    \includegraphics[width=0.60\linewidth]{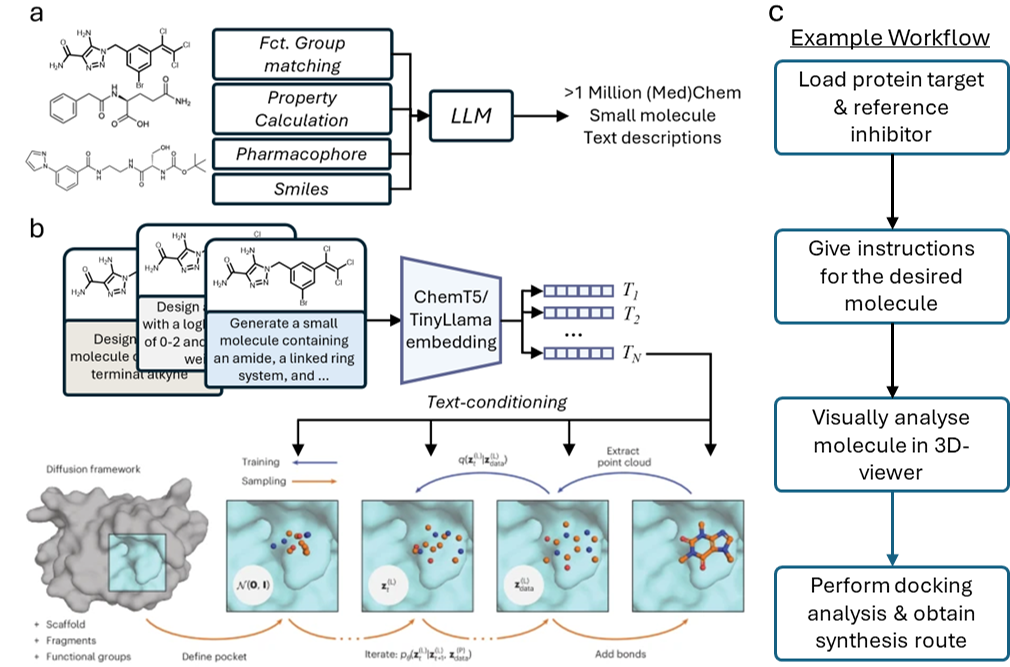}
    \caption{Conceptual overview of (a) the molecule–text description dataset, (b) text conditioning, and (c) the end-to-end design workflow.}
    \label{fig:training_data}
\end{figure}

At the core of MIDAS is an equivariant diffusion model conditioned on protein pocket geometries~\cite{hoogeboom2022equivariantdiffusionmoleculegeneration,satorras2022enequivariantgraphneural}, extended with natural-language-guided conditioning via FiLM~\cite{perez2017filmvisualreasoninggeneral}. Natural language instructions are encoded using either a pretrained T5-base model~\cite{christofidellis2023unifyingmoleculartextualrepresentations,JMLR:v21:20-074} or a TinyLLaMA encoder, as illustrated in Figure~\ref{fig:training_data}b. Pocket geometry information and textual instructions jointly guide three-dimensional molecular generation.

Language descriptions of small molecules were generated using GPT-3.5 for compounds in the CrossDocked dataset. For each molecule, descriptions covering functional groups, molecular properties, pharmacophores, and unguided “medicinal-chemist-like” narratives were produced. The resulting dataset comprises approximately one million molecule–text description pairs. The diffusion model was trained jointly on these textual descriptions and corresponding protein pocket geometries.

The trained diffusion model is exposed within MIDAS as a tool through a user interface that combines a chat-based interaction paradigm with a molecular viewer implemented using py3Dmol~\cite{rego20153dmoljs}. The system is implemented as a ChatAgent using LangChain and was evaluated using GPT-4o mini as the base language model. Additional tools support post-generation analysis, including pose and property analysis, structure similarity search, retrosynthesis, and iterative refinement via natural-language feedback, as shown in Figure~\ref{fig:training_data}c.

\subsection*{Future Work}

Future work will focus on improving the precision and reliability of language-conditioned molecular control. Planned directions include learning finer-grained mappings between textual attributes and local structural modifications, expanding the text–molecule training corpus, and enforcing stronger semantic–structural consistency during generation.

The team also aims to strengthen the connection between text prompts and protein context by grounding language in three-dimensional pocket features, enabling instructions such as “extend toward the hydrophobic pocket” or “avoid the catalytic residue.” Another direction is the integration of MIDAS into a closed-loop optimization workflow, in which generated molecules are evaluated using docking and property predictors and then refined through iterative language feedback.

Finally, the team plans to develop an interactive three-dimensional molecular design interface that allows users to highlight residues or regions in a protein structure and steer generation toward specific sites. Integration with automated synthesis and testing platforms is envisioned to enable rapid, closed-loop validation and to further close the gap between language-guided design and experimental feedback.

\subsection*{Open-Source Materials}

All code is available on GitHub:
\github{https://github.com/pagel-s/MIDAS.git}.
A short demo video is available at
\href{https://www.youtube.com/watch?v=7uQPwcDsr5U}{this link}.
Datasets and trained models can be accessed
\href{https://drive.google.com/drive/folders/1KYPGxqjo9HHVani3FrhEmbZvHSaf7dro}{here}.

\section{AdsKRK: An agentic atomistic simulation framework for surface science}\label{sec:AdsKRK}

Figuring out a stable configuration of adsorbates (using MLIP-driven relaxation \cite{batatia2023foundation}) on a hetero-catalytic surface is a trial-and-error manual process, placing a central importance on the quality of the starting configuration. Mimicking this expert-driven trial-and-error process, the LIAC-AdsKRK team developed AdsKRK as an iterative agentic loop that allows LLMs to learn from previous failed and successful attempts. The framework leverages tools encapsulated in the Autoadsorbate library \cite{fako2025simple} to generate a population of chemically plausible starting configurations.

\begin{figure}[!htbp]
    \centering
    \includegraphics[width=0.9\linewidth]{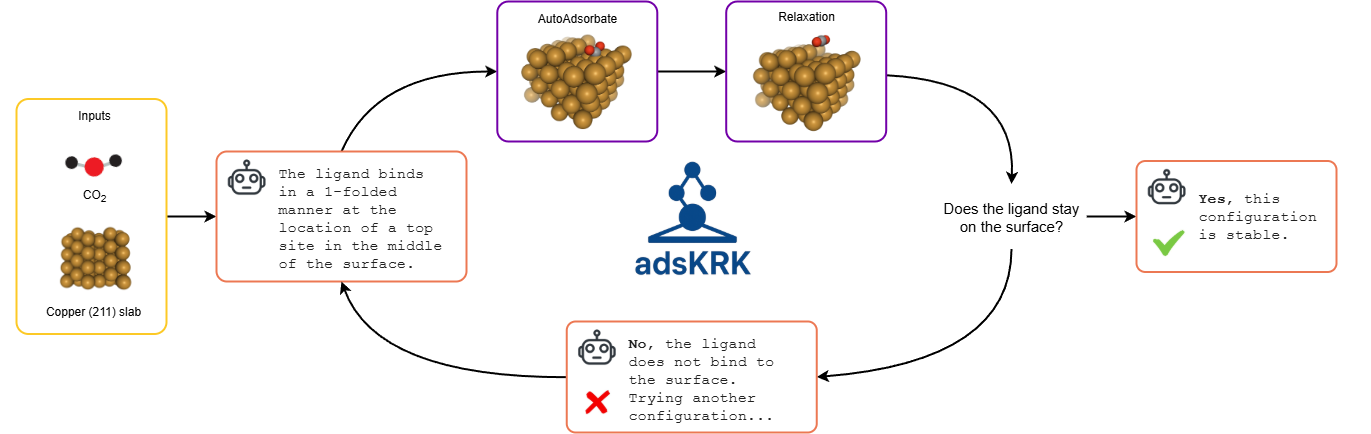}
    \caption{Conceptual overview of AdsKRK.}
    \label{fig:adskrk_concept}
\end{figure}

\subsection*{Results}
In its original implementation, the LIAC-AdsKRK team employed the CodeAct \cite{codeact2024} framework to enable a flexible trial-and-error workflow. Within CodeAct, the agent incrementally generates executable code that follows the instructions specified in the prompt, while the code-execution node returns the corresponding outputs. By iteratively repeating this generate--execute--observe loop, the team aimed to allow the agent to refine its actions and converge toward the desired objective. The team found that, in some cases, the agent was capable of solving tasks that were considerably complex. The major take-home message, however, is that full flexibility caused stochastic behavior with poor repeatability of tasks ($\sim$50\% success rate), a result important to share with the community. An additional observation is that the employed LLM model resorted to writing code to directly answer all geometric questions, without attempting approaches akin to a domain-expert chain-of-thought.

To address the issue of repeatability, the framework was significantly re-architected from the original hackathon prototype to enhance robustness and scientific accuracy. The team implemented a stateful, iterative StateGraph that enables the agent to systematically search for a globally optimal configuration by learning from a history of its previous simulation attempts, replacing the original linear agent. This new architecture introduces a critical Planner-Validator-Executor-Analyzer feedback workflow. The agent first plans a simulation configuration in a structured JSON format. This plan is then programmatically validated by a Python validator before execution, preventing chemically invalid simulations and forcing the LLM to correct its own plan. Furthermore, complex chemical logic, such as the generation of surrogate SMILES for different site types (ontop, bridge, side-on) and fixes for RDKit atom-ordering dependencies, was moved from brittle prompts into robust Python functions. Finally, the analysis module was upgraded to calculate quantitative metrics, including adsorption energy and site-slippage detection, providing a true scientific endpoint for the automated workflow.

\subsection*{Future Work}
Future work will focus on systematically performing benchmark tests on the agent against existing computational chemistry workflows, changing LLMs from ``closed-source'' models (e.g.\ Google's Gemini-2.5-Pro) to locally deployed ``open-source'' models (e.g.\ Alibaba Cloud's Qwen3 model family \cite{yang2025qwen3}, Moonshot AI's Kimi-K2-thinking), and introducing retrieval-augmented generation (RAG) on molecular databases into the agent workflow to optimize the decision-making process of the agent.

\subsection*{Open-source Materials}
The code used in this project can be found at: 
\github{https://github.com/schwallergroup/llm_adsorbate}





\section{AssemblAI: An LLM Agent for Designing Peptide Self-Assembly Protocols}\label{sec:AssemblAI}

\begin{figure}[htbp]
    \centering
    \includegraphics[width=0.95\linewidth]{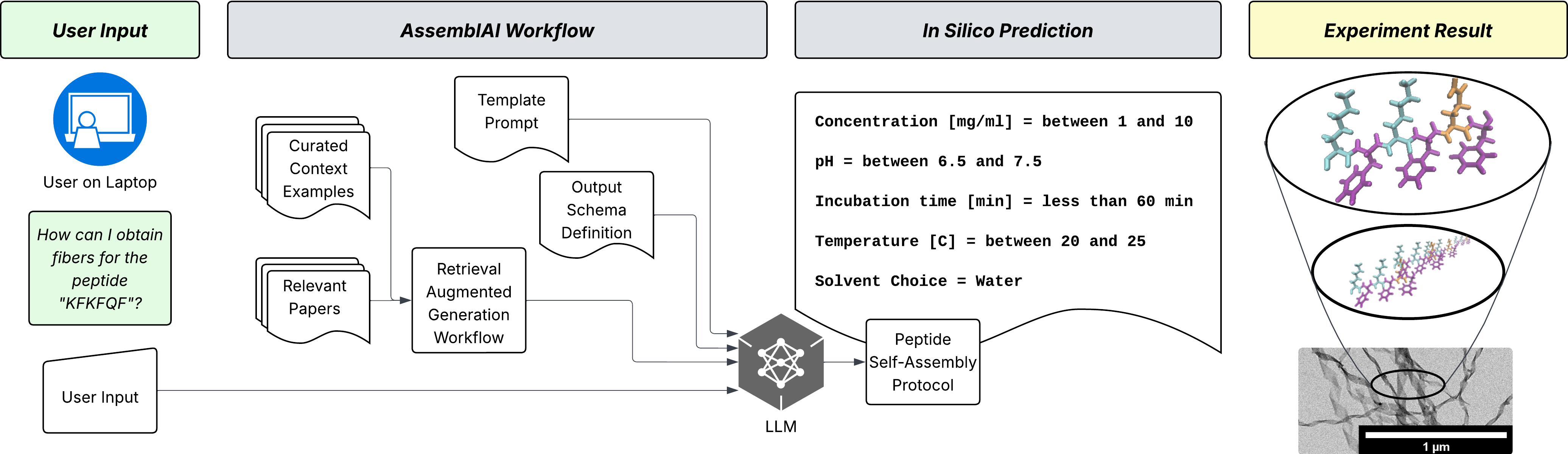}
    \caption{AssemblAI's workflow diagram. Users provide text input to generate a peptide self-assembly experimental protocol, experimental results are visually validated by transmission electron microscopy.}
    \label{fig:assemblai_workflow}
\end{figure}

Identifying optimal experimental protocols for peptide self-assembly is a significant bottleneck in materials discovery, often requiring extensive and time-consuming manual literature review. The AssemblAI team developed AssemblAI, an LLM-powered agent, to automate and accelerate this process. As described in Figure~\ref{fig:assemblai_workflow}, AssemblAI generates a complete experimental procedure based on a target peptide sequence and desired morphology (e.g., ``fiber'' or ``sphere''). The agent employs a Retrieval-Augmented Generation (RAG) workflow \cite{lewis2020retrieval}, retrieving context from a curated vector store of 989 experimental examples~\cite{MATHUR2021104391, doi:10.1126/sciadv.adv1971, doi:10.1021/acsnano.5c00670, Jankovic2023} to inform the LLM's protocol generation.

\subsection*{Results}
\begin{figure}[htbp]
     \centering
     \begin{subfigure}{0.45\textwidth}
         \centering
         \includegraphics[width=\linewidth]{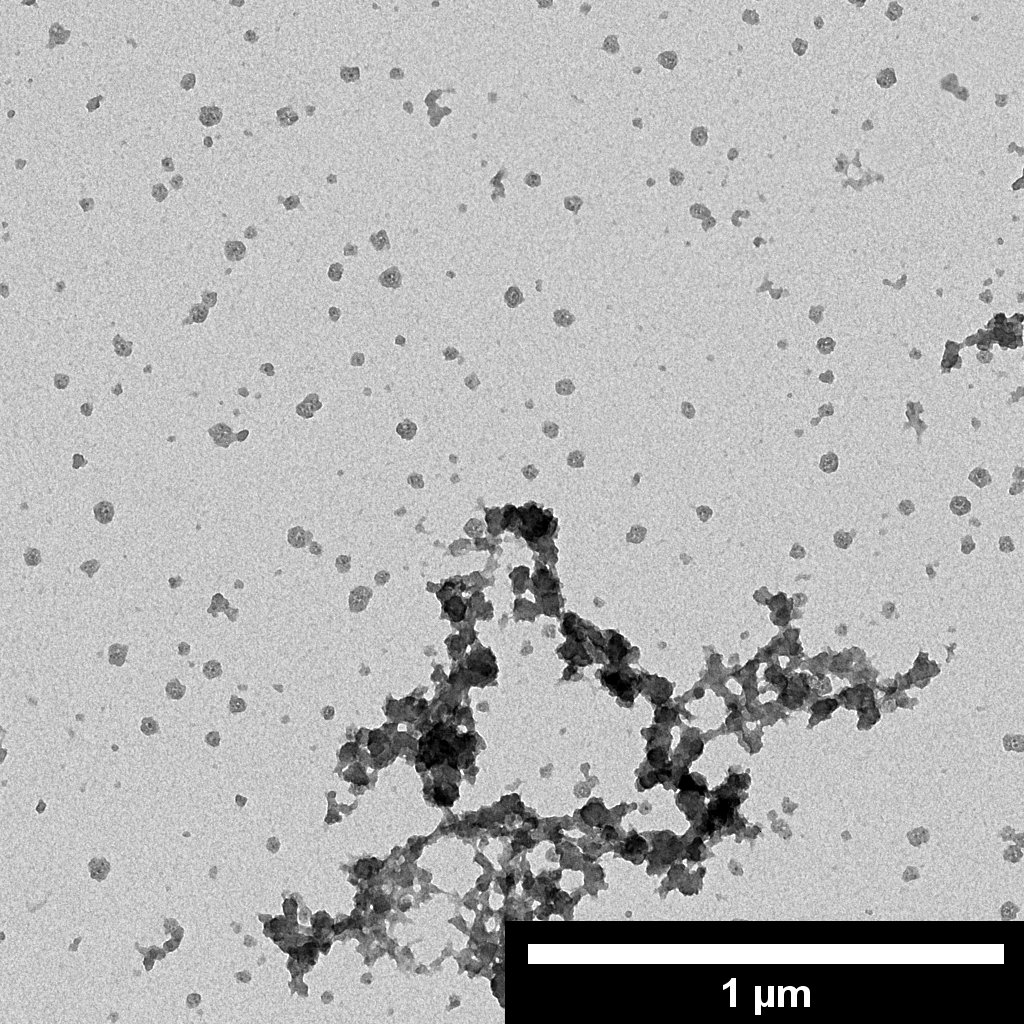}
         \caption{Spherical morphology of KFKFQF. Peptide concentration: 1 mg/mL, pH=7.4.}
         \label{fig:sphere_kfkfqf}
     \end{subfigure}
     \begin{subfigure}{0.45\textwidth}
         \centering
         \includegraphics[width=\linewidth]{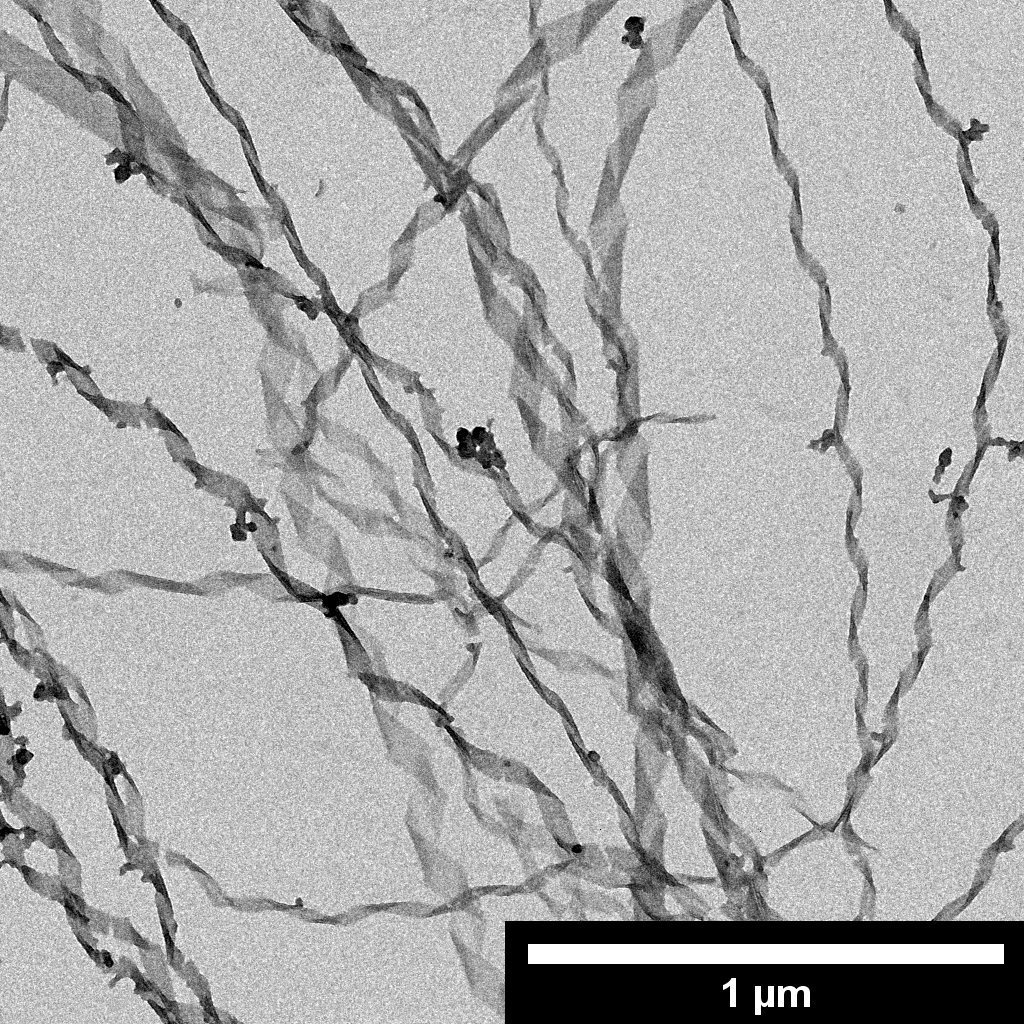}
         \caption{Fiberous morphology of KFKFQF. Peptide concentration: 10 mg/mL, pH=7.4}
         \label{fig:fiber_kfkfqf}
     \end{subfigure}
     \caption{Transmission electron microscopy images of the peptide KFKFQF after self-assembly experiments. Graciously provided by K. Kaygisiz.}
     \label{fig:kfkfqf_morphology}
\end{figure}

As a proof-of-concept, the team tested AssemblAI on the peptide `KFKFQF', which is known to form spheres at low concentrations and fibers at high concentrations (Figure~\ref{fig:kfkfqf_morphology}). Without prior knowledge of this specific system, the agent correctly identified concentration as the key experimental parameter controlling the morphological outcome, as seen in Figure~\ref{fig:assemblai_agentoutput}.

\begin{figure}[htbp]
\centering
\begin{minipage}{0.48\textwidth}
    \textbf{Agent outputs for Spherical KFKFQF:}

    Concentration [mg/ml] = between 0.1 and 1
    
    pH = between 7.0 and 8.0
    
    Incubation time [min] = less than 30 min
    
    Temperature [C] = between 20 and 37
    
    Solvent Choice = Water or PBS

\end{minipage}%
\begin{minipage}{0.48\textwidth}
    \textbf{Agent outputs for Fibrous KFKFQF:}
    
    Concentration [mg/ml] = between 1 and 10
    
    pH = between 6.5 and 7.5
    
    Incubation time [min] = less than 60 min
    
    Temperature [C] = between 20 and 25
    
    Solvent Choice = Water
\end{minipage}

\caption{Summarized agent outputs describing the self-assembly protocol of the peptide `KFKFQF` into a spherical and fibrous morphology.}
\label{fig:assemblai_agentoutput}
\end{figure}

\begin{figure}[htbp]
  \centering
  \includegraphics[width=0.85\linewidth]{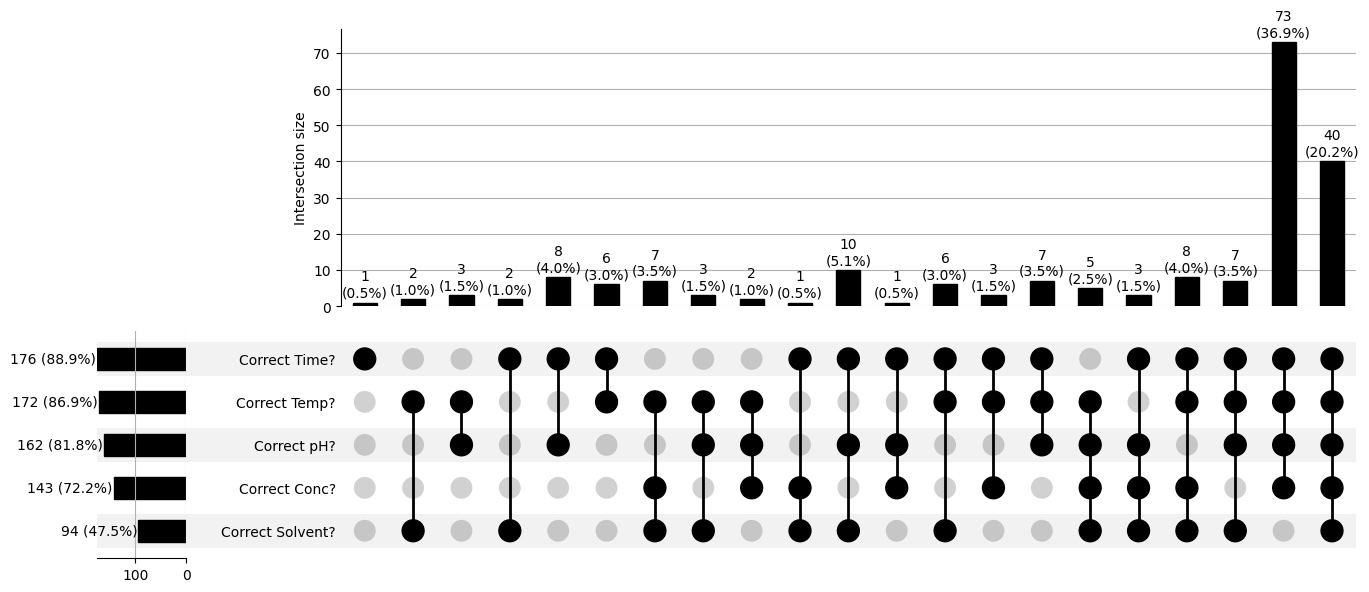}
  \caption{Benchmark performance of AssemblAI on a withheld test set (N=198). The plot shows the number of examples (Intersection Size) for which different combinations of experimental conditions were correctly predicted. 69\% of test data (e.g., the 40 + 73 + ... examples) were correctly predicted for at least 4 out of 5 conditions.}
  \label{fig:assemblai_upset}
\end{figure}

To quantify performance, the team benchmarked the agent on a withheld test set of 198 examples. The agent demonstrated strong predictive power, with 69\% of test cases having at least four of the five key experimental conditions (pH, concentration, temperature, time, solvent) correctly predicted (Figure~\ref{fig:assemblai_upset}).

\subsection*{Future Work}
The team plans to extend this work by benchmarking other LLMs, expanding the curated dataset, and deploying a more robust agent-based system to improve predictions, especially for solvent choice.

\subsection*{Open-source Materials}
All code and curated data are available on 
\github{https://github.com/ndharms/peptide-agent} and a video demo is available on 
\href{https://www.linkedin.com/posts/ndharms_hi-there-last-week-i-participated-in-activity-7374759205112655872-e6vl}{LinkedIn}.

\section{MaterialMind:A RAG+LLM Recommender for Materials Selection}\label{sec:MaterialMind}

\textbf The MaterialMind team presents a RAG-based materials selection system that retrieves evidence from technical literature, generates candidate materials, and produces ranked recommendations with explicit trade-offs and page-level citations. By combining constrained LLM reasoning with an independent-weight scoring model over performance, stability, cost, and availability, the system enables transparent and auditable engineering decisions.

Selecting materials for engineering applications requires balancing performance, stability, processing constraints, cost, and availability — a process traditionally reliant on extensive technical literature and expert judgment. As the volume of materials science publications grows, manual workflows become increasingly time-consuming, difficult to audit, and inconsistent. Recent advances in Large Language Models (LLMs) and Retrieval-Augmented Generation (RAG) enable evidence-grounded decision support by combining semantic retrieval with citation-aware reasoning; however, existing tools largely focus on summarization or property extraction rather than transparent, ranked recommendations. The MaterialMind team addresses this gap by converting unstructured materials literature into structured, explainable recommendations using RAG-constrained LLM reasoning and an independent-weight scoring model that preserves engineer intent without normalization, resulting in an auditable and defensible material-selection workflow.

\subsection*{System Overview}

MaterialMind is an AI-assisted materials selection system that combines Retrieval-Augmented Generation (RAG) with Large Language Models (LLMs) to produce evidence-grounded and auditable material recommendations. Given application constraints (e.g., environment, temperature, and mechanical limits), the system retrieves relevant technical literature using semantic vector search \cite{gao2023ragsurvey,yuan2024lamp}, extracts candidate materials grounded in scientific evidence \cite{cheng2025aidriven,zhou2025smalldata}, and generates ranked recommendations with transparent trade-offs and page-level citations.

The system follows a structured RAG pipeline in which documents are chunked, embedded, and indexed, and relevant passages are fused with user constraints to form a grounded prompt that constrains LLM reasoning \cite{gao2023ragsurvey,yuan2024lamp,xie2024darwin}. This design reduces hallucination and ensures scientifically valid outputs adapted for materials science applications \cite{cheng2025aidriven}.

Candidate materials are evaluated using four criteria — performance, stability, cost, and availability — with independently assigned weights in the range 0--100\%, without normalization, following modern multi-criteria decision-making formulations \cite{kes2023mcdmreview,liao2023datadrivenMCDM}. Utility values are derived from materials-design heuristics \cite{zhou2025smalldata}, and the composite ranking is defined as
\[
R(m) = \sum_i w_i \cdot u_i(m), \qquad R(m)\in[0,400].
\]

\subsection*{System Architecture}

\begin{figure}[h]
    \centering
    \includegraphics[width=0.70\linewidth]{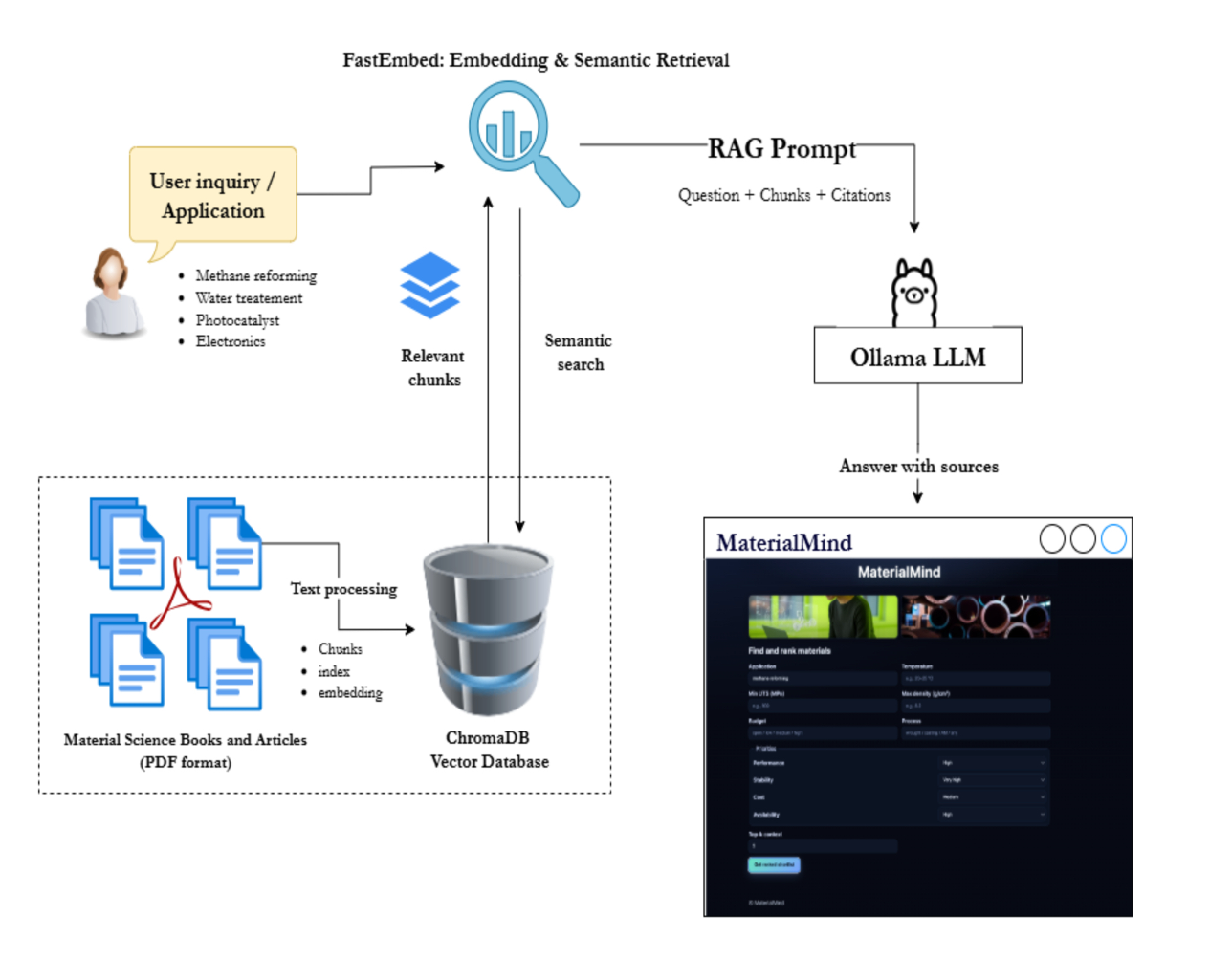}
    \caption{MaterialMind system architecture combining retrieval, reasoning, and scoring components.}
    \label{fig:materialmind_architecture}
\end{figure}

MaterialMind is implemented as an RAG-driven assistant combining semantic retrieval and structured LLM reasoning. Users specify environmental, thermal, mechanical, or chemical constraints, which guide evidence retrieval from embedded technical documents. Retrieved passages and page-level citations are injected into the RAG prompt, ensuring that generated outputs remain auditable. The system outputs shortlisted materials, supporting reasons, trade-offs, and numeric scores informed by current LLM capabilities in materials research \cite{xie2024darwin,cheng2025aidriven}. The overall architecture of MaterialMind is illustrated in Fig.~\ref{fig:materialmind_architecture}.

To support transparency, reproducibility, and community collaboration, all code, model configurations, and supplementary resources for \textit{MaterialMind} are publicly available:

\begin{itemize}
    \item \textbf{GitHub Repository (Source Code):}
    \github{https://github.com/AzizahAlq/MaterialMind}

    \item \textbf{Google Drive Folder (Datasets, PDFs, Documentation):}
    \href{https://drive.google.com/drive/folders/1IgP_zhMtMGHpANE5ZVRZzQUlVWQy-ufD}{Google Drive}
\end{itemize}

\subsection*{Conclusion}

MaterialMind transforms unstructured scientific literature into transparent, ranked material recommendations by combining RAG-based retrieval \cite{gao2023ragsurvey}, LLM reasoning tailored to materials science \cite{xie2024darwin,cheng2025aidriven}, and modern multi-criteria decision-making frameworks \cite{kes2023mcdmreview,liao2023datadrivenMCDM}. The system provides a reliable, auditable AI assistant for materials selection under real-world engineering constraints.




\section{ChemTutor-AI: Intelligent Chemistry Learning}\label{sec:chemtutor-ai}

Recent advancements in Large Language Models (LLMs) have significantly transformed how learners understand scientific concepts at both secondary and tertiary levels. Traditional chemistry education has long relied on repetitive experimental procedures with limited theoretical depth, an approach that historically catered to industrial needs rather than fostering conceptual mastery. To address this gap, the Triple-As team developed ChemTutor AI, a domain-specific LLM designed to enhance chemistry education and research. Unlike general-purpose models, ChemTutor AI is trained on highly curated datasets encompassing chemical reactions, molecular structures, thermodynamic principles, and advanced materials science concepts. This specialization enables the model to accurately interpret complex scientific queries and deliver context-aware, precise explanations, bridging theory and practice. Furthermore, ChemTutor AI integrates structured chemical ontologies and reaction networks, combines symbolic reasoning with neural architectures, and supports advanced tasks such as molecular property prediction and spectroscopic interpretation. These capabilities position ChemTutor AI as a novel and superior solution compared to existing models, offering research-grade performance while promoting deep conceptual understanding. By moving beyond rote memorization and procedural repetition, ChemTutor AI establishes a new paradigm that accelerates discovery, reduces trial-and-error costs, and strengthens the link between education and industrial innovation. ChemTutor AI is an innovative learning tool for chemistry and materials science, designed with a strong focus on accessibility, effective teaching methods, and user experience. Its clean and intuitive interface helps learners stay focused, navigate easily, and move smoothly through lessons. This platform makes studying both more engaging and more productive \cite{bretz2019evidence}.

\subsection*{Results}

ChemTutor AI is an AI-powered chemistry tutoring platform that generates personalized problems, provides interactive molecular visualizations, and adapts to each student's learning pace. It uses LLMs to create custom explanations, generate practice problems, and provide step-by-step solutions with visual aids. It has interactive 3D molecular models using 3.js for hands-on learning, real-time AI problem generation tailored to the difficulty level, an adaptive learning system that evolves with student performance, and visual explanation of solutions combining text and molecular graphics, all presented through a professional, award-winning UI design.

\subsection*{Pipeline Explanation}

\textbf{User Request}: The process begins when a user selects a topic (e.g., ``Quantum Chemistry'') and difficulty level (``Intermediate'') and clicks the ``Generate Problem'' button.

\textbf{Query Processing}: The system takes the user's input and refines it into a detailed query, incorporating the user's learning level from their profile to better tailor the search.

\textbf{Vector Database Retrieval}: The processed query is sent to a specialized vector database containing a vast collection of chemistry knowledge broken down into chunks and converted into numerical representations (\textbf{vectors}). The system retrieves the most relevant pieces of information based on the query, pulling from:

\textit{``\textbf{Chemistry Textbooks}''}: Core concepts, formulas, and explanations.

\textit{``\textbf{Scientific Literature}''}: Advanced topics, niche examples, and pedagogical papers.

\textit{``\textbf{Existing Problem Database}''}: Previously generated problems used as examples of structure, style, and difficulty.

\textit{``\textbf{Context Aggregation}''}: The top-ranked documents and data chunks retrieved from the database are gathered, forming a rich contextual foundation for the AI to work with.

\textbf{Prompt Engineering}: A sophisticated prompt is constructed combining the original user request with the retrieved context and specific instructions for the Large Language Model (LLM). For example:

\textit{``Given the following context on the Schrödinger equation and particle-in-a-box models, generate an intermediate-level quantum chemistry problem. The output must be a JSON object with fields for `title', `question', `solution', `concepts', and `hints'."}

\textbf{LLM Generation}: The engineered prompt is sent to the LLM. The model uses the provided context to generate a new, accurate, and relevant chemistry problem in the requested JSON format, avoiding hallucination by grounding its answer in the retrieved data.

\textbf{Output \& Storage}: The generated JSON is received by the application, and it is,

\textbf{Displayed in the App UI}: The problem is parsed and presented to the user in a clean, readable format.

\textbf{Saved to Problem Entity}: The new problem is saved to the database, enriching the knowledge base for future retrieval tasks.

\begin{figure}
    \centering
    \includegraphics[width=0.5\linewidth]{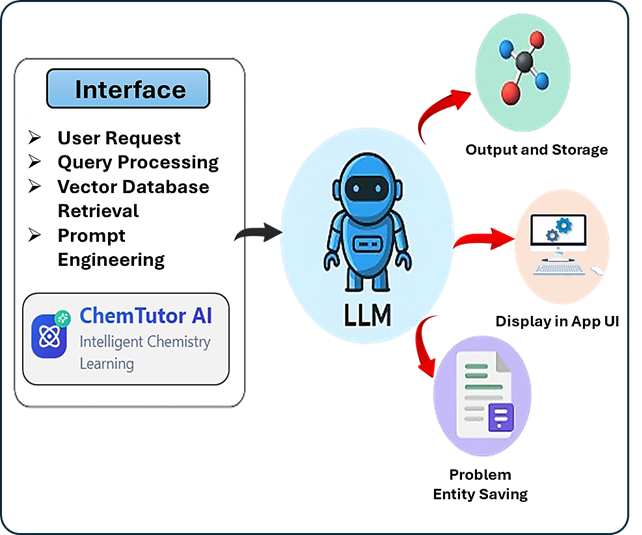}
    \caption{Workflow of ChemTutor AI and its future work perspectives.}
    \label{fig:ChemTutor_AI}
\end{figure}

\subsection*{Future Work}

Future enhancements include adding algorithmic problems for physical and analytical chemistry, incorporating LMS compatibility for grade synchronization, and adding multilingual support.

\subsection*{Open-source Materials}

Code and pretrained weights available on \href{https://huggingface.co/Abbasaabdul/AI_ChemTutor/tree/main}{HuggingFace}





\section{CrystaLenz: Agentic XRD Analysis Pipeline}
\label{sec:crystalenz}

X-ray diffraction (XRD) is a core tool for characterizing crystalline materials, linking structure, synthesis, and performance in fields such as electrochemical energy storage.\cite{acsaem5c00530} In practice, however, many research workflows still depend on fragmented scripts and manual steps for baseline correction, peak picking, profile fitting, and phase identification, making analyses difficult to reproduce and scale, especially when dealing with heterogeneous file formats and noisy data.

The CrystaLenz team addresses this gap by providing an end-to-end, modular XRD analysis pipeline. It ingests diverse instrument outputs, performs robust peak detection and Voigt-profile fitting, and extracts crystallite size and microstrain via Scherrer and Williamson--Hall methods. Leveraging pymatgen's \texttt{XRDCalculator} and the Materials Project API, CrystaLenz generates wavelength-consistent simulated patterns with HKL indexing and matches them to experimental peaks using adaptive tolerances. The system is designed to be agentic, extensible, and reproducible, with stateful hyperparameter control, provenance tracking, and structured JSON/plot outputs that can be extended to additional physicochemical characterization techniques in future work.

\subsection*{Results}

\begin{figure}[h]
    \centering
    \includegraphics[width=0.9\linewidth]{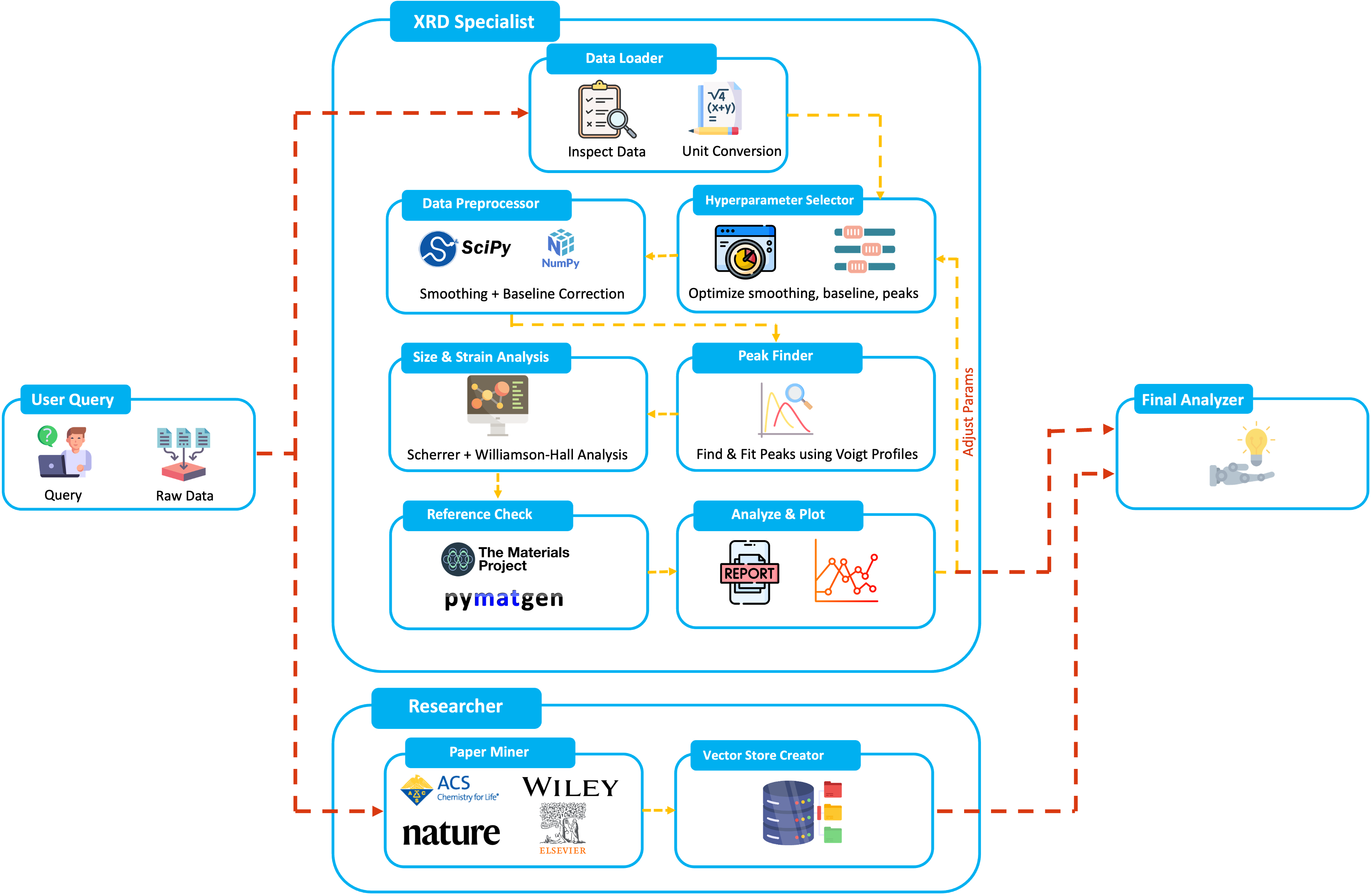}
    \caption{Overview of the CrystaLenz agentic XRD analysis workflow, including data loading, preprocessing, peak/size analysis, reference checking, and reporting.}
    \label{fig:CrystaLenz.png}
\end{figure}

CrystaLenz was implemented as a modular, agentic XRD workflow that connects raw data to structured outputs in a single loop (Figure~\ref{fig:CrystaLenz.png}). The system is organized into dedicated components for data loading, preprocessing, hyperparameter selection, peak finding, size/strain analysis, reference checking, and reporting. Each module exposes a clear interface so an XRD specialist can iteratively adjust parameters (e.g., smoothing level, peak thresholds, fitting ranges) while preserving a record of all choices.

On heterogeneous diffraction files from different instruments and noise levels, the pipeline consistently ingested raw patterns, performed robust peak detection and Voigt-profile fitting, and extracted crystallite size and microstrain using Scherrer and Williamson--Hall methods. By coupling pymatgen and the Materials Project API, CrystaLenz generated wavelength-consistent reference patterns with HKL indexing and matched them to experimental peaks using adaptive tolerances, producing stable and interpretable phase-identification results. The system outputs structured JSON summaries alongside publication-ready plots, enabling rapid inspection and downstream reuse.

In parallel, the team prototyped a ``paper miner'' and vector-store creator that aggregates XRD-relevant literature into an indexed knowledge base. This establishes a path for CrystaLenz to answer user queries using both newly analyzed diffraction data and curated knowledge from prior work, forming the basis for future retrieval-augmented XRD assistants.

\subsection*{Future Work}

Future work will focus on extending CrystaLenz beyond XRD to a broader suite of physicochemical characterization techniques. Because the system is built as a set of modular, well-defined components (data loaders, preprocessors, feature extractors, and analyzers), additional modules for methods such as Raman spectroscopy, SEM/TEM, and electrochemical measurements can be integrated without redesigning the core workflow.

By plugging these pipelines into the same agentic framework and provenance layer, CrystaLenz can evolve into a comprehensive, end-to-end characterization environment. In such a system, users would be able to ingest multi-modal data for a given material, run coordinated analyses across XRD and complementary techniques, and obtain a unified report that links structure, morphology, and performance, supporting faster, more reliable materials discovery and optimization.

\subsection*{Open-source Materials}

Code available on GitHub: \github{https://github.com/MJRaei/CrystaLenz}\;
Demo Video: \youtube{https://www.youtube.com/watch?v=uNsnd1BLsTs}





\section{Prototyping Autonomous Critical Materials Extraction}
\label{sec:acme}

\begin{figure}[!h]
    \centering
    \includegraphics[width=1\linewidth]{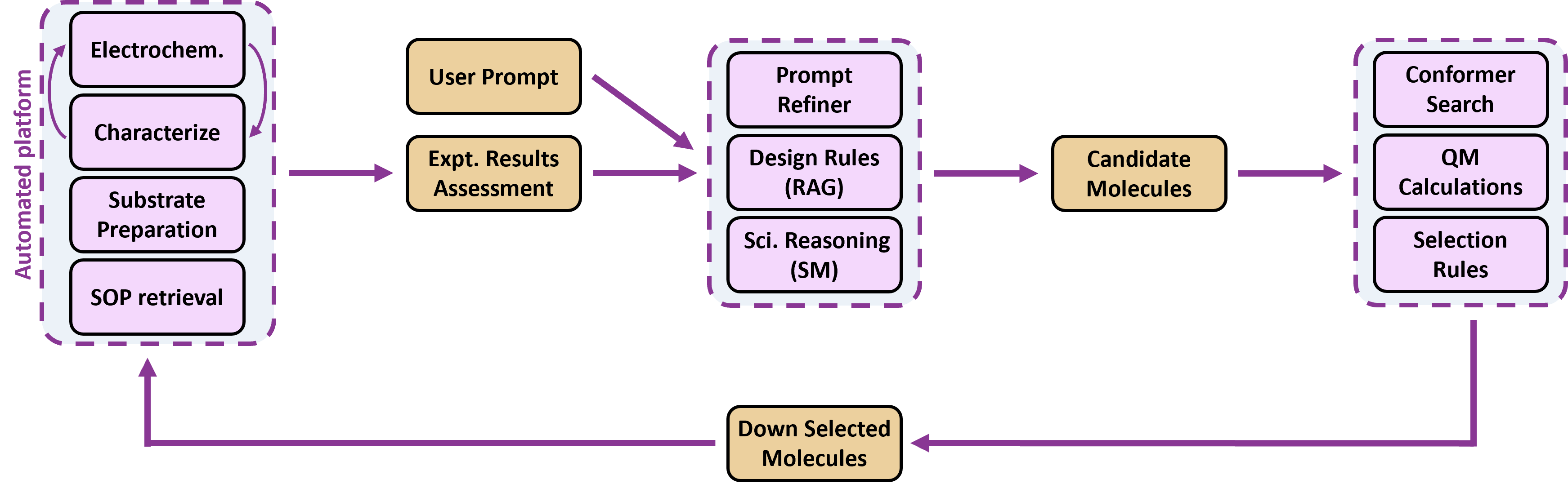}
    \caption{Overview of the closed-loop discovery platform (ACME).}
    \label{fig:wf_acme}
\end{figure}

The ACME team aims to connect molecular design, simulation, and experiment in a single closed loop. The ACME team built a small prototype of such a system focused on molecular discovery for critical materials extraction using electrochemical methods \cite{shukla2025alizarin}. The workflow, shown in Figure \ref{fig:wf_acme}, combines several AI and simulation components in one agentic framework. Reasoning LLMs interpret a user prompt, retrieve domain knowledge, and define chemically meaningful design rules. Generative models use these rules to propose candidate molecules, while predictive AI models and quantum chemical tools evaluate their properties. Experimental procedures enter the loop through modular plug-ins that translate the agent's plans into stepwise protocols and measurements. The full system operates as a feedback loop with multiple iterations. Computational and experimental results are analyzed by the reasoning agent, which updates the design rules, refines the search space, and proposes improved candidates. This structure illustrates how autonomous laboratories can couple generative, predictive, and experimental modules to accelerate molecular discovery.

\subsection*{Results}

Candidate generation begins with domain knowledge encoded through an in-house scientific reasoning model (Socratic AI reasoning model \cite{harb2025towards,harb2025hitchhiker}). A Retrieval-Augmented Generation (RAG) module accesses patterns, structural motifs, and property data from a curated knowledge base of protocol documents and peer-reviewed publications. The RAG module extracts known ligands, functional group motifs, and design rules for metal-ion binding, including constraints on charge balance, donor atoms, and coordination geometry. Given the retrieved context and user-defined target properties, the Socratic AI model constructs ligands or molecules that satisfy these design rules. Molecules that pass this reasoning stage are forwarded to computational downselection.

Computational screening evaluates the resulting molecules using metal-ion binding affinity scores from quantum mechanical calculations. The workflow reads a table of candidate molecules and target ions, builds 3D structures, and assembles each ligand--ion complex using conformational sampling at the semi-empirical extended tight-binding level GFN-xTB \cite{bannwarth2019gfn2}, as implemented in CREST \cite{pracht2020automated}. The metal ion binding affinity score is defined as the energy differential between the assembled complex and its separated constituents in implicit water solvent. These values are used to rank candidates. The QM stack is accessed through a Model Context Protocol (MCP) interface that enables the agent to invoke computational tools and downselect candidate pools.

Selected candidates are then evaluated through an agentic experimental workflow integrating synthesis, characterization, and electrochemical testing. Automated benchtop platforms characterize molecules following domain-expert designed prompts and SOPs. The LLM acts as an experiment pipeline designer, generating executable workflows from user inputs. Molecules are coated onto electrodes using automated spin coating, followed by X-ray fluorescence (XRF) and inductively coupled plasma mass spectrometry (ICP-MS) to quantify metal-ion concentrations. The agent ranks candidates based on metal concentration detected by XRF and ICP-MS. An automated electrochemistry platform modeled after ElectroLab from the Rodríguez-López laboratory at UIUC collects cyclic voltammograms for different electrode and electrolyte combinations to assess metal separation performance. The selectivity rankings are compared across candidates, and an optimal solution is identified based on user-defined criteria.

\subsection*{Future Work}

The prototype presented here is a first step toward an autonomous workflow for critical materials extraction and molecular discovery. Future work includes expanding the domain knowledge base beyond the current small set of references, with the RAG pipeline indexing a larger collection of peer-reviewed papers, protocol PDFs, and internal reports. Further development will upgrade the computational screening layer from semiempirical xTB to higher-fidelity quantum chemistry calculations such as density functional theory. Through the MCP interface, the agent will set up and run DFT, validate convergence, and use richer electronic-structure descriptors when ranking candidates. Finally, the experimental module will move from a conceptual plug-in to direct control of laboratory hardware, connecting the agent to robotic platforms for liquid handling, automated spin coating, electrochemical testing, and spectroscopy, using SOP-aware planning to generate executable workflows. Together, these extensions will turn the current prototype into a fully closed-loop system that couples literature, quantum chemistry, and real experiments for autonomous molecular discovery.

\subsection*{Open-source Materials}

Code available on GitHub: \github{https://github.com/HassanHarb92/ACME}






\section{HEAQuery: An Intelligent LLM Assistant for High-Entropy Alloy Research}\label{sec:HEAQuery}

High-entropy alloys (HEAs) are an exciting area of materials research, but the existing literature is highly fragmented across journals, experimental reports, and datasets. For researchers entering the field, finding reliable property trends, compositional relationships, and process--microstructure correlations can be overwhelming. To address this, the HEA Query team developed HEAQuery, an intelligent search engine designed to synthesize information from both scholarly papers and curated datasets, answering domain-specific questions about HEAs through a unified, data-driven approach.

\begin{figure}[h]
    \centering
    \includegraphics[width=0.85\linewidth]{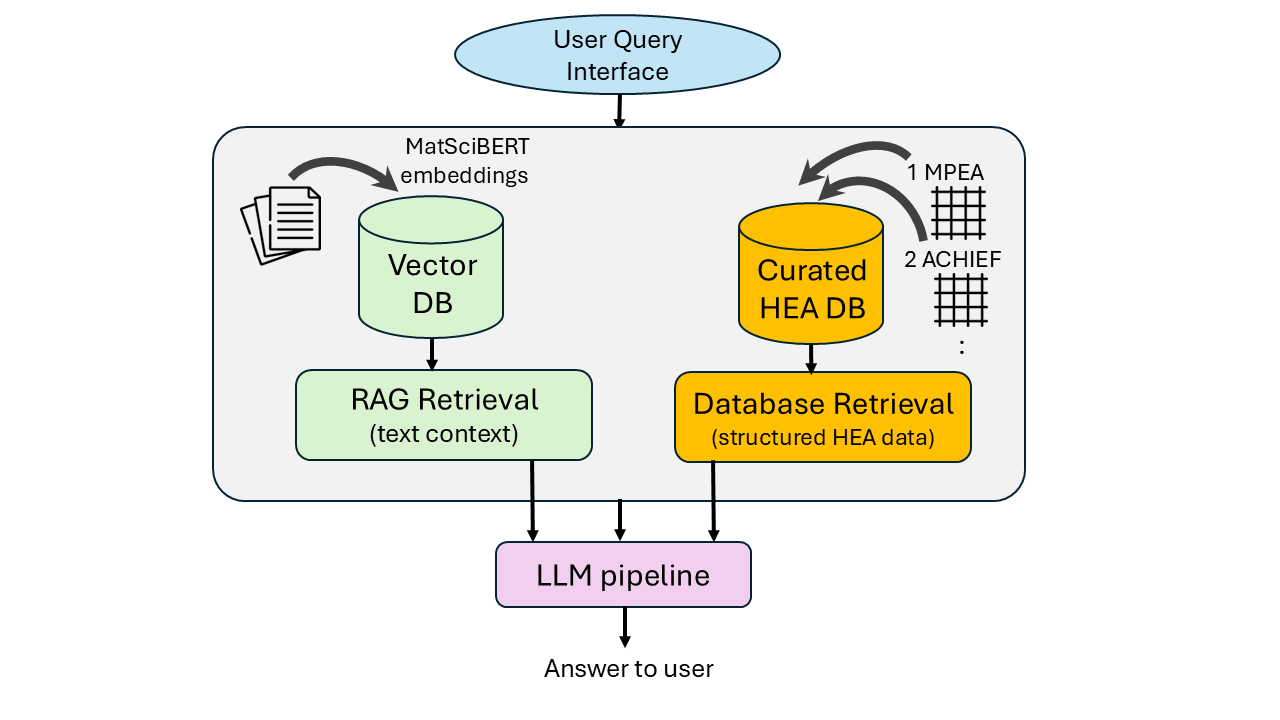}
    \caption{Workflow of HEAQuery.}
    \label{fig:heaquery}
\end{figure}

\subsection*{Results}

HEAQuery integrates diverse scientific literature and structured datasets into a cohesive LLM-powered assistant for HEA research. The system follows an end-to-end workflow that processes raw PDFs, cleans and harmonizes datasets, and enables natural-language querying using advanced semantic retrieval techniques.

The team began by processing over 3,500 HEA-related research papers from open-access repositories, including arXiv, Nature Communications, and Scientific Reports. Using PyMuPDF, text was extracted, noise (such as headers, footers, and DOIs) was removed, and content was structured into meaningful scientific sections. A GPU-accelerated BART model~\cite{DBLP:journals/corr/abs-1910-13461} then summarized each section into concise representations. These summaries were chunked with overlapping windows for context retention and encoded using the MatSciBERT model~\cite{GZKM2022}, generating vector embeddings. These embeddings were stored in a FAISS index, enabling rapid semantic search across the literature.

Simultaneously, the team curated and cleaned three public HEA datasets~\cite{BORG2020430,MACHAKA2021107346,precker20216403257}, covering mechanical properties, thermodynamic descriptors, and synthesis routes. The datasets were harmonized by standardizing column names, normalizing composition formulas using a chemistry-aware parser, and consolidating element-fraction columns. Inconsistent or incomplete entries were removed, and the data was unified across sources. This clean and unified dataset supports filtering by properties like phase structure, hardness, and yield strength.

The core of the HEAQuery system is its intelligent query engine. A natural-language parser extracts relevant constraints from user queries — such as composition, phase stability, or material properties — and applies these filters to the cleaned datasets. The FAISS index retrieves semantically relevant passages from the literature, and both structured and unstructured data are forwarded to a language model (GPT-2)~\cite{radford2019language} for answer synthesis. The LLM generates concise, evidence-backed responses that combine both dataset matches and text excerpts from the retrieved papers. The system is accessible through a user-friendly Gradio interface, which displays the results. The overall workflow of HEAQuery is illustrated in Figure~\ref{fig:heaquery}.

\subsection*{Future Work}

Future work will explore advanced LLM pipelines and alternative domain-specific embeddings to improve performance, alongside fine-tuning LLMs for HEA-specific terminology. The team will also investigate incorporating synthesis-condition predictions and structure-aware models to better support experimental design in HEA research.

\subsection*{Open-source Materials}

Code, datasets and documentation are available on GitHub: \github{https://github.com/staradutt/HEAquery/tree/main}





\section{PackSynth: Agent-Guided Workflow for Automated Molecular Simulation}\label{sec:PackSynth}

PackSynth is an agent-guided molecular simulation and analysis system designed to enhance the accessibility of atomistic modeling for researchers in materials science. Traditional simulation tools, such as LAMMPS~\cite{thompson2022lammps}, are powerful but demand extensive expertise in force fields, system construction, file formats, and analysis. These requirements create significant barriers, slowing research progress in areas involving metals, ceramics, polymers, and other diverse materials. The PackSynth team addresses these limitations by integrating natural language interaction with automated structure generation, simulation setup, and analysis. The system allows users to input a SMILES string, a polymer repeat unit, or a material name, then retrieves structural information, generates 3D geometries, prepares LAMMPS files, runs molecular dynamics simulations, and provides automated stability checks, RMSD analysis, and interactive 3D visualization. This unified framework reduces simulation setup time and offers an intuitive research tool for exploring a wide range of molecular and materials systems.

\subsection*{Results}

\begin{figure}[h]
    \centering
    \includegraphics[width=0.65\linewidth]{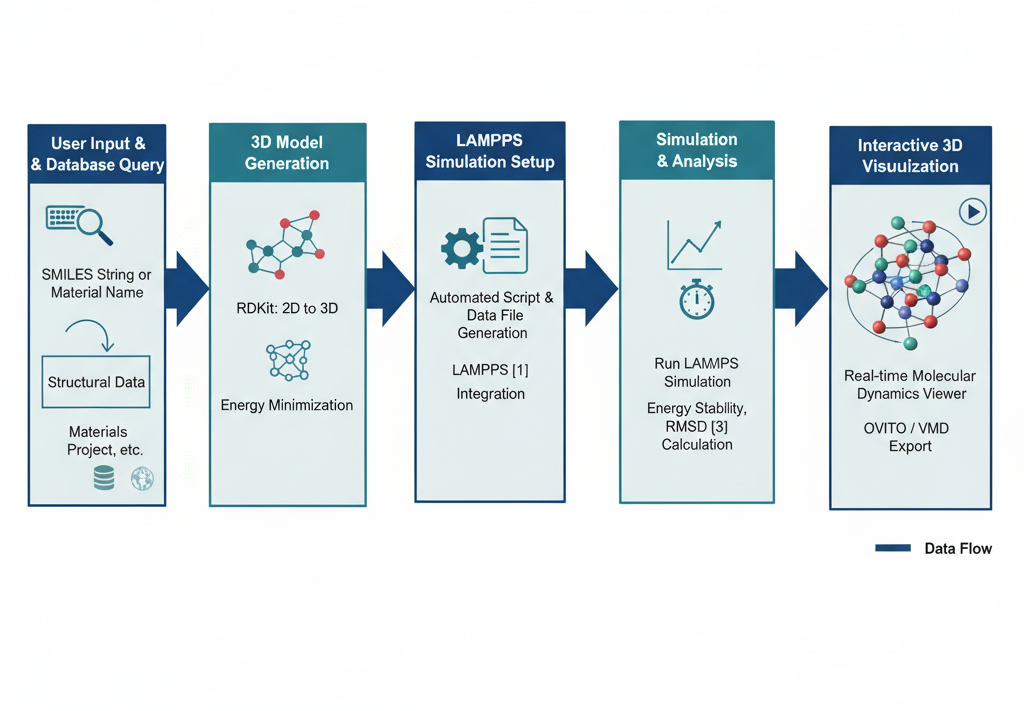}
    \caption{Automated workflow of PackSynth.}
    \label{fig:packsynth_workflow}
\end{figure}

PackSynth employs a modular workflow, depicted in Figure~\ref{fig:packsynth_workflow}, that integrates database retrieval, automated structure generation, programmatic LAMMPS input creation, simulation control, and post-processing.

The user provides an input (SMILES/Name), which the agent uses to fetch data from databases like the Materials Project. The system then uses RDKit~\cite{rdkit} to generate a 3D model, automatically prepares and runs the LAMMPS simulation, performs analysis (Energy, RMSD), and provides an interactive 3D visualization.

The workflow begins with Input Processing and Database Integration, where PackSynth accepts SMILES strings, polymer repeat units, or material names, querying databases like the Materials Project~\cite{jain2013commentary} for structural data. Next, 3D Structure Generation utilizes RDKit~\cite{rdkit} to create and energy-minimize initial 3D geometries. This is followed by automated LAMMPS File Creation, programmatically generating all necessary data files, topologies, and input scripts, thereby eliminating manual errors. For Simulation Execution, PackSynth runs LAMMPS~\cite{thompson2022lammps} molecular dynamics simulations through standard minimization, equilibration, and production runs, allowing user-specified parameters like temperature and pressure. Automated Analysis provides energy stability checks and RMSD calculations. Finally, Visualization and Output delivers real-time 3D visualization for interactive exploration, with options to export larger systems for tools like VMD~\cite{humphrey1996vmd} or OVITO~\cite{stukowski2010visualization}. This integrated, automated approach significantly reduces the time and expertise required for complex molecular simulations.

\subsection*{Future Work}

Several enhancements are planned to further improve PackSynth's accuracy, flexibility, and scalability. These include improved input validation to prevent unstable simulation conditions, multi-conformer generation to reduce uncertainty from 2D to 3D conversions, and an expanded potential library with automated selection (e.g., EAM, MEAM, Buckingham, ReaxFF) based on material composition. The team also plans advanced system builders for bulk, melt, or crystalline systems (e.g., Packmol integration), enhanced visualization capabilities for large-scale systems, and agentic troubleshooting and RAG integration to diagnose common simulation issues. Long-term goals include developing educational and training tools and extending capabilities for multi-scale modeling.

\subsection*{Open-source Materials}

Code, documentation, and examples are available on GitHub: \github{https://github.com/devanshu-777/PackSynth}





\section{Standardizing Material Property Data for ML-Ready Materials Datasets}\label{sec:ExpAlign}

LLMs enable large-scale extraction of materials data from the literature, but reported values often vary due to inconsistent measurement conditions and sample preparation. This inconsistency makes raw data hard to use in AI/ML workflows. The ExpAlign team uses LLMs to extract values and standardize key experimental metadata, with preliminary work focusing on extracting CO\textsubscript{2} gas permeability of polymers.

\subsection*{Results}

\begin{figure}[h]
    \centering
    \captionsetup{justification=justified, singlelinecheck=false}
    \includegraphics[width=1\linewidth]{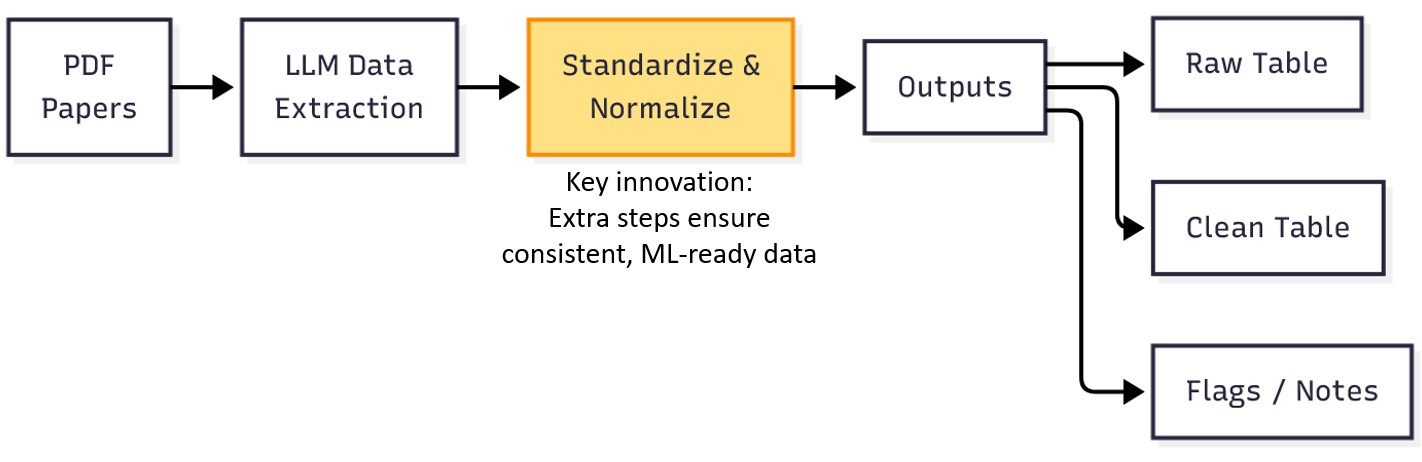}
    \caption{Workflow for data extraction and standardization.}
    \label{fig:ExpAlign_Workflow}
\end{figure}

The team developed an automated workflow to collect and standardize CO\textsubscript{2} permeability data from polymer membrane literature (Figure~\ref{fig:ExpAlign_Workflow}). Full-text PDFs were processed using an LLM to extract polymer names, CO\textsubscript{2} permeability values, and any clearly stated measurement conditions (such as temperature or pressure). The extracted data were then converted to consistent units and passed through an LLM-based name harmonization step to standardize polymer names across the entire dataset.

\begin{figure}[h]
    \centering
    \captionsetup{justification=justified, singlelinecheck=false}
    \includegraphics[width=0.7\linewidth]{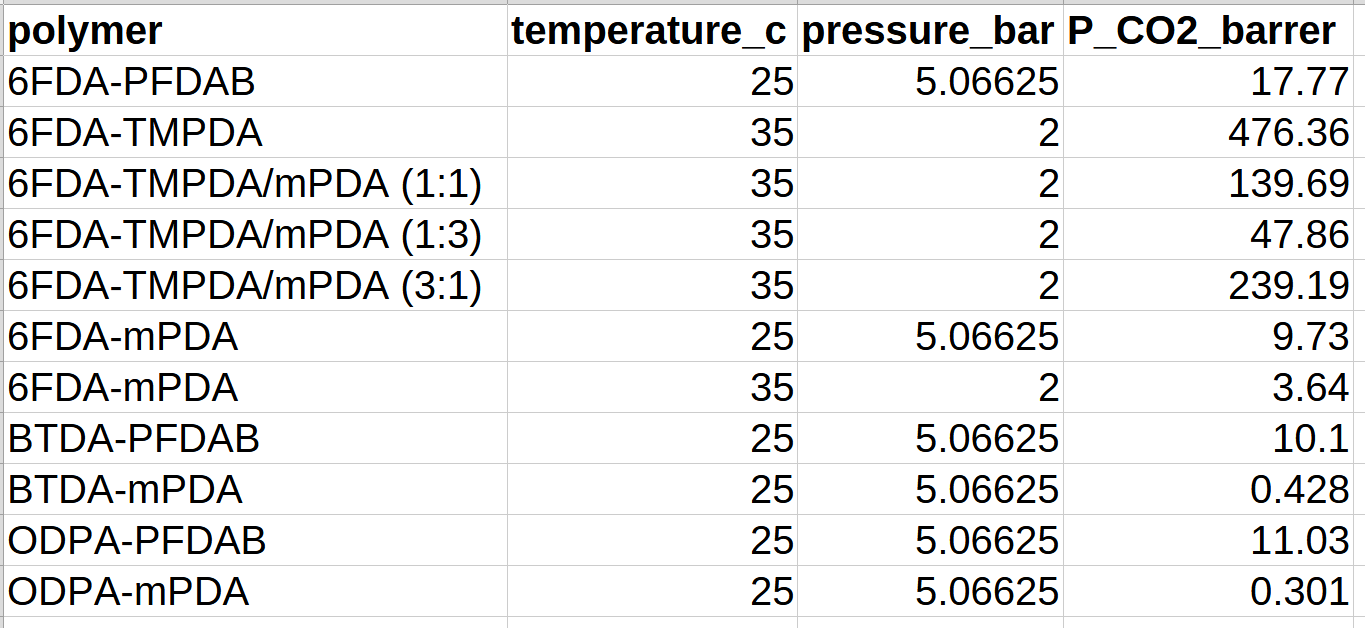}
    \caption{Example of final standardized dataset produced by the automated LLM workflow.}
    \label{fig:ExpAlign_Workflow_final_results}
\end{figure}

This procedure successfully resolved inconsistencies in how polymers were labeled. For example, 6FDA-mPDA appeared in different papers as ``6FDA-m-PDA''~\cite{kim2000incorporation} and ``6FDA-mPDA''~\cite{wang2008gas}; the workflow correctly merged these under a single canonical name. The resulting dataset includes (1) a condition-aware table summarizing median permeability values and highlighting cases with inconsistencies, and (2) a streamlined, ML-ready dataset containing polymer identities and their associated measurement conditions (Figure~\ref{fig:ExpAlign_Workflow_final_results}). These outputs provide a clean and consistent foundation for comparative analysis and data-driven modeling of polymer membrane performance.

\subsection*{Future Work}

Future work will expand condition and property standardization and incorporate journal or author reliability indicators when resolving highly inconsistent reports.

\subsection*{Open-source Materials}

Code is available on GitHub: \github{https://github.com/sptiwari/ExpAlign_paper}\,;
Demo Video: \youtube{https://www.youtube.com/watch?v=AUU9osunIuw}

\section{QSPHAgents: A Multi-Agent Framework for Qualitative Structure-to-Property Hypothesis}\label{sec:QSPHAgents}

The electronic density of states (DOS) is a crucial descriptor for materials, capturing how electronic states are distributed in energy. From the DOS, one can infer band gaps, distinguish metallic from insulating behaviour, and recognize features linked to superconductivity. Despite being a high-dimensional signal, extracting such qualitative insights requires connecting DOS features to underlying structure — a process that typically requires expert knowledge and iterative manual reasoning. This remains a bottleneck in many materials informatics workflows. To accelerate this step, the QSPHAgents team uses large language models (LLMs) \cite{singh2024,Lei2024} in a multi-agent framework capable of extracting features, analyzing trends, and generating hypotheses about a material's electronic behaviour from DOS. QSPHAgents achieves this by predicting qualitative DOS characteristics from structural descriptors using retrieval-augmented reasoning (RAG) and a generator--critic loop for iterative refinement.

\subsection*{QSPHAgents}

\begin{figure}[ht!]
    \centering
    \includegraphics[width=0.9\textwidth, height=0.37\textheight]{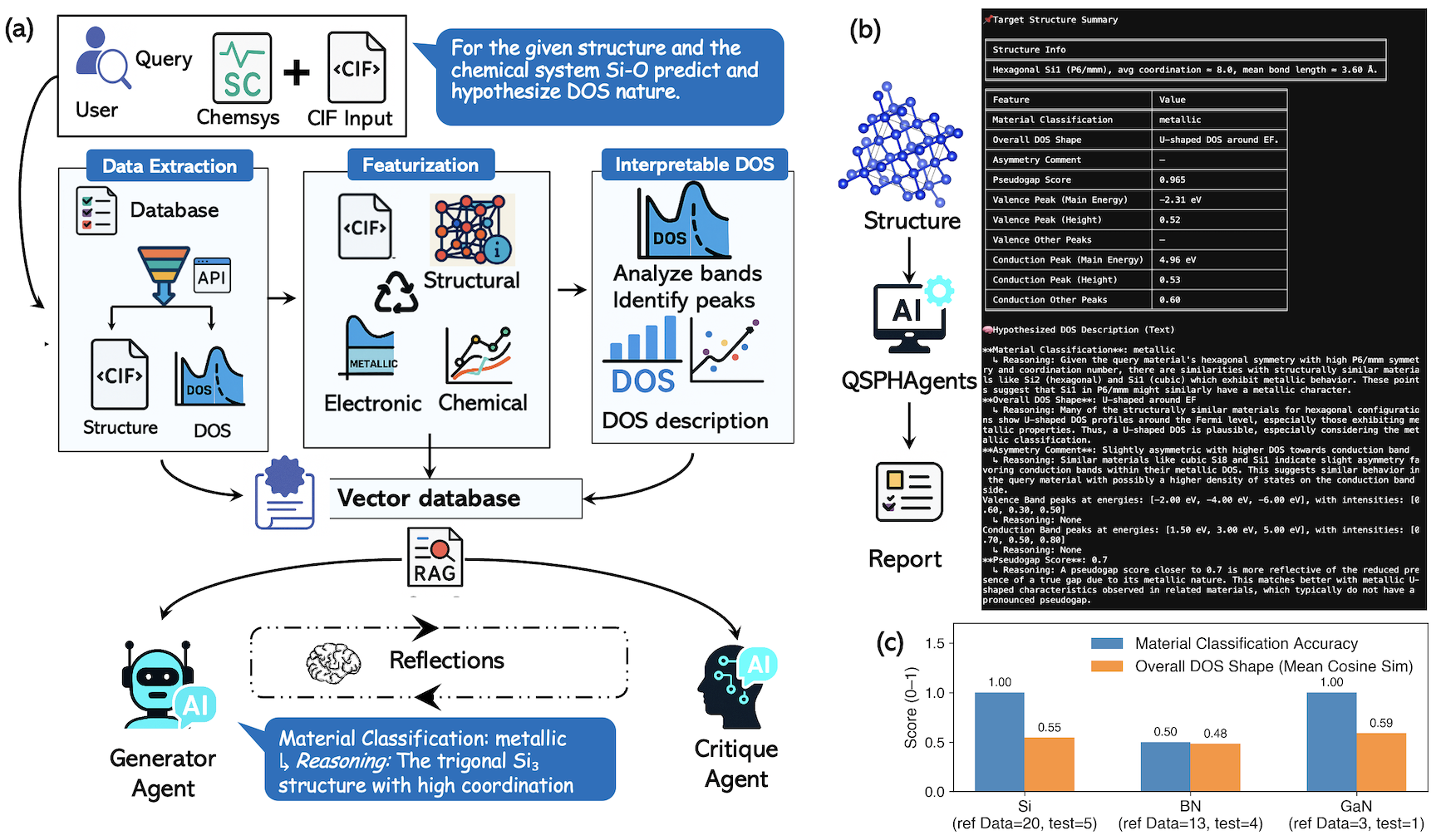}
    \captionsetup{width=\textwidth}
    \caption{(A) Overall workflow of the QSPHAgent framework for interpretable prediction of electronic density of states (DOS). (B) Representative DOS hypothesis generated for a given crystal structure by QSPHAgent. (C) Performance in material classification and overall DOS-shape prediction for the Si, BN, and GaN datasets.}
    \label{fig:qsph_single}
\end{figure}

The QSPHAgents framework consists of four main components (Fig.~\ref{fig:qsph_single}a). The first (``Data extraction'') retrieves data from materials databases via their APIs, collecting CIF files, DOS data, and chemical metadata for a given list of species. The second (``Featurization'') converts each structure into explainable feature vectors using descriptors that capture structural, local-environment, chemical, and electronic information relevant to the DOS. The third (``Interpretable DOS'') performs interpretable DOS extraction, identifying band gaps, pseudogaps, curvature and slope near the Fermi level, and qualitative signatures of metallic, semiconducting, insulating, or superconducting behaviour. This step outputs quantitative descriptors (e.g., $D(E_F)$, pseudogap score, asymmetry) and textual DOS summaries. The fourth (``Vector database'') integrates all information into a unified vector database, storing each material $i$ as $(\mathbf{f}_i, D_i(E), T_{\text{DOS},i})$. Retrieval (``RAG'') uses the L2 distance $d_i = \lVert \mathbf{f}^\ast - \mathbf{f}_i \rVert_2$ between a target feature vector $\mathbf{f}^\ast$ and stored entries to select the top-$k$ neighbours, which provide structural and behavioural context. Classification and regression models supply quantitative priors for validation to the critic agent. Finally, a generator--critic loop synthesizes the information: the generator proposes a DOS hypothesis informed by structure and neighbour context, while the critic refines it by removing inconsistencies to ensure a coherent and physically plausible prediction.

\subsection*{Results}

To evaluate performance, the team analyzed three representative systems — Si, BN, and GaN — using data extracted from the Materials Project \cite{jain2013commentary}. Dataset sizes ranged from 20 to 4 entries, and an 80:20 split was used to define the support set (7 nearest neighbors) and the test set. From the test-set reports, two metrics were measured: (i) material-classification accuracy (metallic vs.\ semiconducting) and (ii) semantic cosine similarity for DOS-shape prediction (Fig.~\ref{fig:qsph_single}c). Classification accuracy approached~1 with larger support sets, while DOS-shape similarity remained around~0.5, likely due to limited contextual information from the small support sets.

\subsection*{Future Work}

The team further aims to integrate causal reasoning and expand the materials database to provide richer contextual information for hypothesis generation. Plans also include incorporating graph neural networks that operate directly on atomic environments to improve quantitative inference. Causal analysis will help identify structural features that may link to superconductivity and refine the agent's predictive accuracy. In addition, a literature-summarization agent will be introduced to ground the critic's feedback in theoretical context. Inspired by recent advances in phonon DOS prediction using LLMs \cite{llm2024}, the team plans to extend QSPHAgents to include phonon DOS features by combining bonding analyses from tools such as LobsterPy \cite{Naik2024,Janine2022} with structural descriptors. This integration will enable hypotheses that jointly capture electronic and vibrational behavior. Finally, the team plans to deploy a lightweight web interface that allows users to upload structures, inspect retrieved neighbours, and interactively obtain DOS hypotheses.

\subsection*{Open-source Materials}

All code and demo notebooks are available on GitHub: \github{https://github.com/sbanik2/QSPHAgents}





\section{Generative Modeling of Stable Inorganic Crystals with GPT-OSS}\label{sec:CrysGen}

Discovering new inorganic materials is critical for advances in energy storage, catalysis, electronics, and structural applications. The space of possible crystal structures is astronomically large, yet only a small fraction of compositions and geometries are thermodynamically stable. Traditional discovery pipelines rely on density functional theory (DFT) calculations and expert-guided heuristics, which makes systematic exploration slow and computationally demanding. Recent work has shown that large language models (LLMs) can generate stable crystals when structures are encoded as text, using fine-tuned LLaMA models as the backbone \cite{gruver2024fine}.

Generative models that directly propose low-energy-above-hull candidates can substantially accelerate the discovery of metastable phases relevant to batteries, catalysts, and semiconductors. LLMs trained on crystal structures have been shown to generate plausible and often thermodynamically reasonable materials, offering a path to rapid exploration of composition--structure space without exhaustive DFT calculations \cite{gruver2024fine}. Text-based interfaces further enable conditional design, allowing models to be guided toward specific compositions, space groups, or property ranges, making LLMs flexible and controllable front-end tools for materials discovery \cite{Antunes2024}.

The CrysGen team fine-tunes gpt-oss-20b \cite{agarwal2025gpt} using LORA \cite{hu2022lora} for inorganic crystal generation. Following the crystal-as-text paradigm, lattice parameters, atomic identities, and fractional coordinates are encoded as newline-separated token sequences. Supervised fine-tuning is performed on large corpora of experimentally and computationally derived crystal structures, then preference-based optimization using energy-above-hull values is applied to encourage stability-aware generation. The resulting model supports unconditional sampling, text-conditional generation, and infilling of partial structures.

The experiments show that GPT-OSS learns key crystallographic regularities and generates a high fraction of structurally valid and thermodynamically reasonable candidates. Together with prior LLaMA-based results, this demonstrates that modern LLMs, when paired with simple text encodings, provide a powerful and scalable foundation for accelerated materials discovery.

\begin{figure}[ht!]
    \centering
    \includegraphics[width=1\textwidth, height=0.2\textheight]{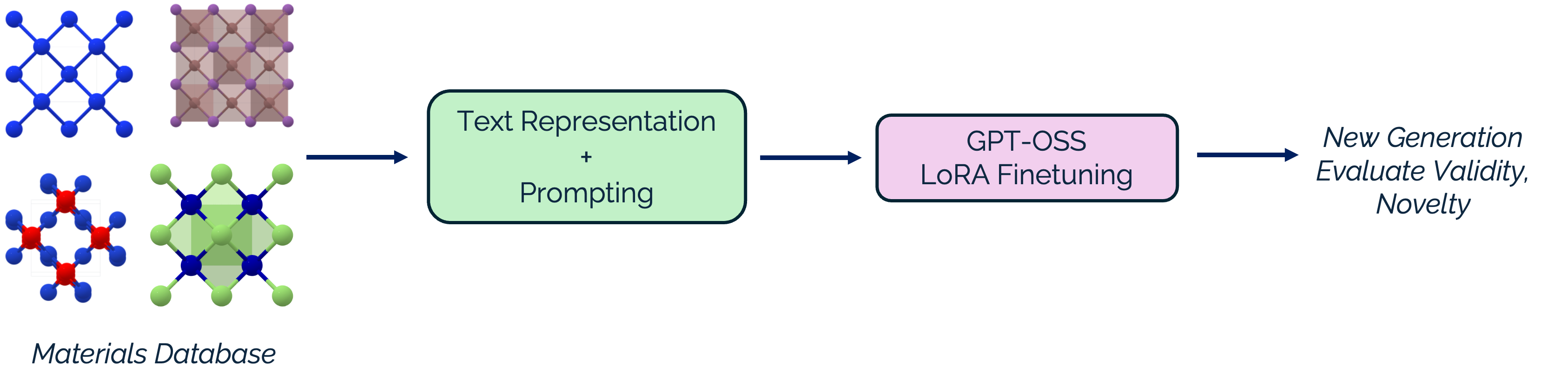}
    \captionsetup{width=\textwidth}
    \caption{Overview of the GPT-OSS--based materials generation framework. Starting from a database of inorganic crystal structures, each structure is converted into a CIF-like text sequence and task-specific prompting is applied. GPT-OSS is fine-tuned using LoRA adapters to learn crystallographic representations. The resulting model generates new inorganic materials, which are evaluated for validity and novelty.}
    \label{fig:crysgen}
\end{figure}

\subsection*{Results}

GPT-OSS was evaluated on unconditional crystal generation and compared against the LLaMA-2 baselines from Gruver et al.\ \cite{gruver2024fine}. Table~\ref{tab:gptoss_vs_llama} reports validity, coverage, and property-distribution metrics for all models. GPT-OSS achieves comparable structural and compositional validity, demonstrating that it reliably generates crystallographically consistent and chemically coherent structures. Its coverage scores indicate that the model reproduces key statistical patterns of real materials, while its property-distribution distances show strong alignment with empirical density and electronic-structure trends. Across several metrics, GPT-OSS performs competitively compared with the LLaMA-2 models, highlighting it as a strong and flexible generator for inorganic crystals.

\begin{table*}[ht!]
\centering
\begin{tabular}{l|cc|cc|cc}
\toprule
\textbf{Method} &
\multicolumn{2}{c}{\textbf{Validity Check}} &
\multicolumn{2}{c}{\textbf{Coverage}} &
\multicolumn{2}{c}{\textbf{Property Distribution}} \\
\midrule
\textbf{LLaMA-2} & Structural $\uparrow$ & Composition $\uparrow$ &
Recall $\uparrow$ & Precision $\uparrow$ &
wdist ($\rho$) $\downarrow$ & wdist ($N_{el}$) $\downarrow$
 \\
\midrule
7B ($\tau=1.0$)  & 0.918 & 0.879 & 0.969 & 0.960 & 3.85 & 0.96 \\
7B ($\tau=0.7$)  & 0.964 & 0.933 & 0.911 & 0.949 & 3.61 & 1.06 \\
13B ($\tau=1.0$) & 0.933 & 0.900 & 0.946 & 0.988 & 2.20 & 0.05 \\
13B ($\tau=0.7$) & 0.955 & 0.924 & 0.889 & 0.979 & 2.13 & 0.10 \\
70B ($\tau=1.0$) & 0.965 & 0.863 & 0.968 & 0.983 & 1.72 & 0.55 \\
70B ($\tau=0.7$) & 0.996 & {0.954} & 0.858 & 0.989 & 0.81 & 0.44 \\
\midrule
\textbf{GPT-OSS (ours)} \\
\midrule
20B ($\tau=0.7$) & 0.995 & 0.926 & 0.940 & 0.999 & 0.70 & 0.11 \\
20B ($\tau=1.0$) & 0.974 & 0.903 & 0.970 & 0.997 & 0.65 & 0.03 \\
\bottomrule
\end{tabular}
\caption{Evaluation of GPT-OSS compared to LLaMA-2 models on crystal generation tasks. Since the number of parameters for the GPT-OSS series and LLaMA-2 series is not the same, a direct apple-to-apple comparison cannot be performed.}
\label{tab:gptoss_vs_llama}
\end{table*}

\subsection*{Future Work}

Several extensions remain to strengthen the GPT-OSS materials generation pipeline. First, generated crystals were not evaluated using fast surrogates such as CHGNet \cite{deng2023chgnet}, nor were thermodynamic stability metrics (e.g., Ehull) computed. Building this stability-evaluation loop is a necessary next step, as these stability labels will be used to construct DPO preference pairs that distinguish lower-energy from higher-energy completions. Second, incorporating a stability-aligned DPO \cite{rafailov2023direct} stage will allow GPT-OSS to explicitly favor energetically plausible structures, enhancing the physical realism of generated crystals. Finally, future work includes expanding conditional generation capabilities and exploring property-guided prompting, making GPT-OSS a stronger and more reliable tool for accelerated inorganic materials discovery.

\subsection*{Open-source Materials}

Code and documentation available on GitHub: \github{https://github.com/RishikeshMagar/GPT-OSS-MAT} Demo video: \youtube{https://youtu.be/0XAuxwkscB8?si=9HXQQi-ZX6eDcEfD}





\section{YOLO vs. Multimodal LLMs for Automated Data Extraction from 2-D Plots: A Reproducible Benchmark on Synthetic Graphs}\label{sec:yolo-vs-mllm}

Quantitative results in scientific papers are frequently locked inside figures rather than prose or tables, making automated plot digitization a key enabler for machine-readable corpora and downstream meta-analysis. While traditional digitizers and handwritten heuristics can recover data, they are slow, brittle to stylistic variation, and hard to scale across thousands of PDFs. Recent multimodal large language models (MLLMs) promise an alternative: with carefully engineered instructions, they can identify axes, interpret tick marks, and localize plotted elements to approximate underlying data without task-specific training. PlotExtract exemplifies this approach, reporting> 90\% precision, \~90\% recall and $\lesssim$ 5\% coordinate error on two-axis plots using zero-shot prompting \cite{polak2025plotextract}. In parallel, one-stage visual detectors such as YOLO offer a complementary, high-throughput path: once trained, they run locally with low latency and zero per-image API cost, returning pixel-space point locations that can be mapped to data units via axis calibration \cite{redmon2016yolo,bochkovskiy2020yolov4}. The Alhaytham Eye team positions YOLO as a pragmatic baseline for 2-D plot digitization and compares it to a PlotExtract-style workflow on a controlled synthetic benchmark, evaluating precision, recall, coordinate error, runtime, cost, and success rate. The results indicate comparable accuracy on simple two-axis charts, with YOLO providing orders-of-magnitude throughput gains, suggesting a division of labor between lightweight detectors and reasoning-heavy MLLM pipelines.

\subsection*{Results}

The workflow begins by generating a synthetic dataset of 20 plots—line, scatter, and bar—with randomized axis ranges and point distributions, split 14/3/3 for train, validation, and test. Each image is 640×480 px with a predefined plotting area and axis ticks, and YOLO labels are derived by centering small bounding boxes on each plotted point. Using these labels, the team trains a YOLOv8n detector for 20 epochs (image size 640, batch size 4), selects the best checkpoint for evaluation, and measures per-image inference time in the same environment as the shared logs. For the multimodal baseline, the team approximates PlotExtract with a four-step chain-of-thought vision workflow applied to each plot, modeling latency and per-plot API cost on a Claude-class configuration and allowing accuracy statistics to vary around reported means; this baseline serves as a proxy rather than a full re-run of PlotExtract.

\begin{figure}[h]
    \centering
    \captionsetup{justification=justified, singlelinecheck=false}
    \includegraphics[width=0.7\linewidth]{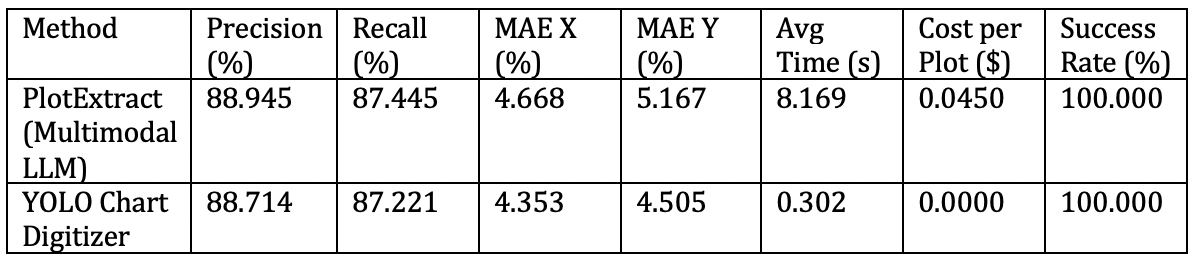}
    \caption{Table 1 — Method comparison (synthetic test set).}
    \label{fig:YOLO_vs_multimodal}
\end{figure}
YOLO and the LLM baseline show comparable accuracy on this synthetic benchmark, while YOLO is substantially faster (0.302 s vs. 8.169 s per plot, $\approx$27.0×) and incurs no API cost.

\textbf{Key takeaways.}
\begin{itemize}
    \item Accuracy: Both methods are \~89\% precision / \~87\% recall with \~4–5\% positional MAE.
    \item Throughput: YOLO is \~27× faster (0.302 s vs. 8.169 s per plot).
    \item Cost \& Privacy: YOLO runs locally with no API cost; LLM incurs \~\$0.045/plot.
\end{itemize}

\subsection*{Future Work}
First, the LLM baseline should be replaced with a fully executed PlotExtract run to eliminate modeling assumptions about cost and latency. Second, evaluations should include diverse plot types (e.g., multi‑series line charts, log scales, error bars, multiple y‑axes) and realistic publication artifacts (fonts, gridlines, noise, compression). Third, accuracy should be measured in data units after automatic axis calibration and scale inference—key steps for end‑to‑end digitization that were not benchmarked here. Finally, tests on public datasets would aid reproducibility and fair comparison.

\subsection*{Open-source Materials}
Code, datasets, and results are available on GitHub: \href{https://github.com/abdulazizashy5/Alhaytham_Eye}{\faGithub}





\section{VERA: AI-Powered Compliance Co-Pilot for Materials Testing Standards}\label{sec:VERA}

Materials testing compliance validation against standards like ISO 527~\cite{iso527} and ASTM D638~\cite{astmd638} is traditionally a manual, error-prone process requiring domain expertise to interpret lab results. VERA (Validation \& Explanation for Regulatory Assurance) is an AI-powered compliance co-pilot that automates this workflow by transforming messy lab data into instant, explainable PASS/FAIL decisions, combining deterministic rule checking with large language model intelligence for interpretability and remediation guidance.

\subsection*{Results}

\begin{figure}[h]
    \centering
    \includegraphics[width=0.9\linewidth]{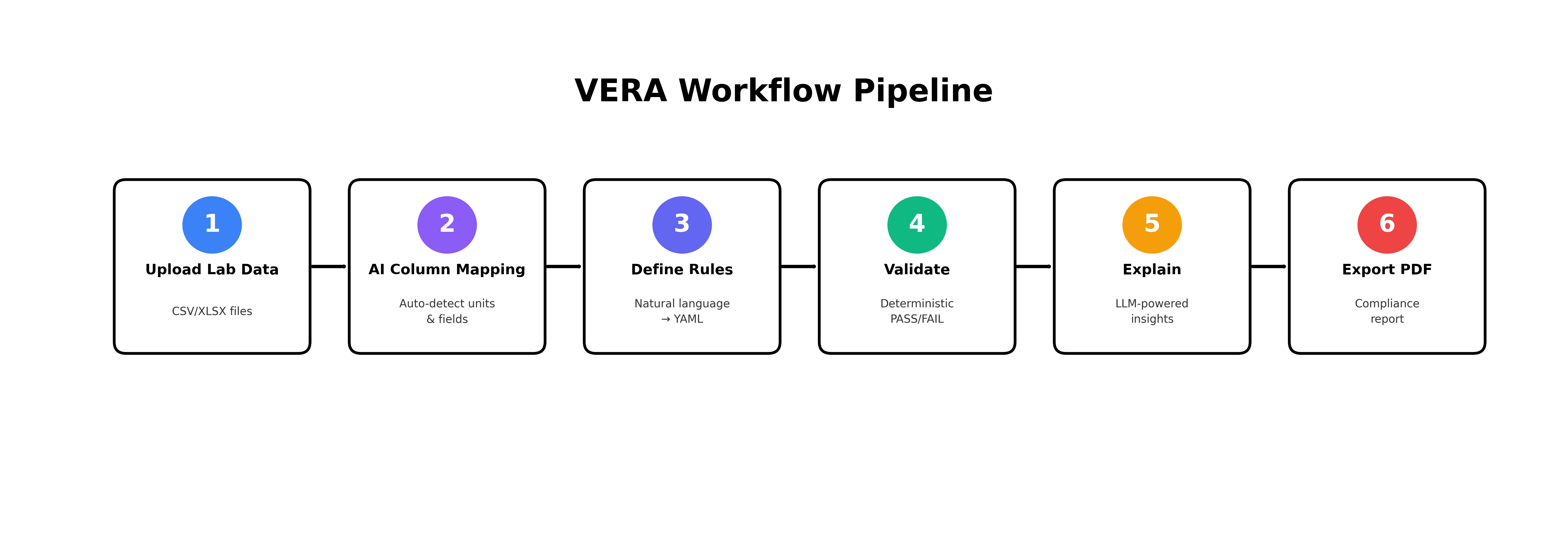}
    \caption{VERA workflow: from lab data upload to compliance validation and PDF report export.}
    \label{fig:vera-workflow}
\end{figure}

The workflow enables users to upload lab results in CSV or XLSX format, where VERA's AI automatically suggests column mappings and unit conversions. Users can define compliance rules using natural language prompts, which VERA converts into machine-readable YAML templates for reproducible validation. The system performs deterministic pass/fail validation with color-coded KPIs and generates LLM-powered explanations grounded in the actual test data, providing remediation tips for failed tests. Figure~\ref{fig:vera-workflow} illustrates the end-to-end pipeline from data upload to PDF report generation.

The system successfully validates tensile testing data against ISO 527 and ASTM D638 standards, automatically identifying unit mismatches and providing detailed explanations for non-compliance. VERA generates one-page PDF compliance reports suitable for regulatory submission, significantly reducing the time from raw data collection to compliant documentation.

\subsection*{Future Work}

The team plans to extend VERA to support additional materials testing standards, including ASTM E8 for metallic materials and ISO 178 for flexural properties. Integration with laboratory information management systems and real-time validation dashboards is also under development.

\subsection*{Open-source Materials}

Code and documentation available on GitHub: \href{https://github.com/puppalasaisrikar/VERA}{\faGithub} Demo video: \href{https://www.youtube.com/watch?v=69Uv0Yp_6XQ}{\faYoutube}





\section{MaterEase - Early Stage Materials Data Literature Survey using LLM}\label{sec:MaterEase}

Early-stage research in materials science and chemistry is sometimes hampered by the significant work necessary to conduct initial literature reviews, collect experimental data, and survey previous findings.  Researchers must negotiate scattered papers and fragmented databases, many of which are focused on restricted material families or specific property subsets, resulting in an inefficient and time-consuming approach \cite{jain2013commentary}.  This leaves little time for advanced scientific analysis, interpretation, and hypothesis building.  Conventional databases, such as the Materials Project, AFLOW, and OQMD, are useful, but they are frequently rigid, computationally heavy, or optimized primarily for ab initio data rather than experimentally focused, compositionally diverse systems such as high-entropy alloys, functional composites, or catalytic materials \cite{kaufmann2020searching}.  As a result, practitioners continue to rely significantly on manual extraction from a variety of sources, slowing both academic and industry development.

Recent breakthroughs in large language models (LLMs) and artificial intelligence for science present a transformative potential to alleviate this bottleneck.  Conversational interfaces driven by LLM can significantly speed up the retrieval, filtering, and summarizing of relevant studies by providing natural-language access to structured materials data \cite{lei2024materials,gromski2019explore}.  When paired with dynamic visualization tools like bar charts, scatter plots, and Ashby-style maps, these methods allow for faster trend identification and property comparison than standard static tools.  Open-source, community-driven data infrastructures increase access and involvement among students, academics, and industry experts.  Finally, the Alloyed Minds team aims to expedite discovery and learning by reducing the bottlenecks associated with early-phase literature exploration and allowing users to focus on producing insights rather than extracting data.

\subsection*{Results}

MaterEase integrates semantic search, AI reasoning, and data visualization into a unified framework for materials discovery. The system architecture consists of three core components: (1)\textbf{Semantic Retrieval Engine} using local sentence-transformers embeddings with FAISS vector database for efficient similarity search, (2) \textbf{Dual LLM Reasoning Pipeline} leveraging Google Gemini 2.0 Flash for comprehensive answer synthesis and Gemini 2.5 Flash for precise structured extraction of alloy compositions and processing methods, and (3)\textbf{Interactive Visualization Module} using Plotly for dynamic property comparisons and distribution analysis.

The semantic retrieval engine processes HEA datasets through intelligent chunking (10,000 character chunks with 1,000 character overlap) and generates vector embeddings locally, eliminating external API dependencies and ensuring data privacy. When a user queries the system (e.g., "Find HEAs with powder processing method and hardness above 600 HV"), the retrieval engine performs similarity search across the entire database, returning the top 500 most relevant entries. This context is then passed to the primary LLM (Gemini 2.0 Flash, temperature=0.2) which synthesizes a comprehensive, scientifically-grounded answer that includes alloy compositions, processing methods, and key properties.

The secondary LLM (Gemini 2.5 Flash, temperature=0.0) performs deterministic extraction, parsing the primary answer to extract structured information in JSON format: HEA compositions and processing methods. This extracted data is used to filter the original database, creating a focused subset for visualization. The visualization module automatically generates interactive charts including hardness distributions, processing method comparisons, microstructure analysis, and property relationships, enabling researchers to identify patterns and correlations at a glance.

\textbf{Performance and Validation:} MaterEase was evaluated on a comprehensive HEA database containing over 1,500 alloy entries with properties including hardness (HV), yield strength, density, microstructure \cite{miracle2017critical,zhang2014microstructures}, and processing methods. The system successfully processes natural language queries with sub-second retrieval times and generates accurate, contextually relevant responses. The dual LLM architecture ensures both comprehensive answers and precise data extraction, with the extraction module achieving high accuracy in identifying alloy compositions and processing methods from free-form text responses.

\begin{figure}[h]
    \centering
    \includegraphics[width=\linewidth]{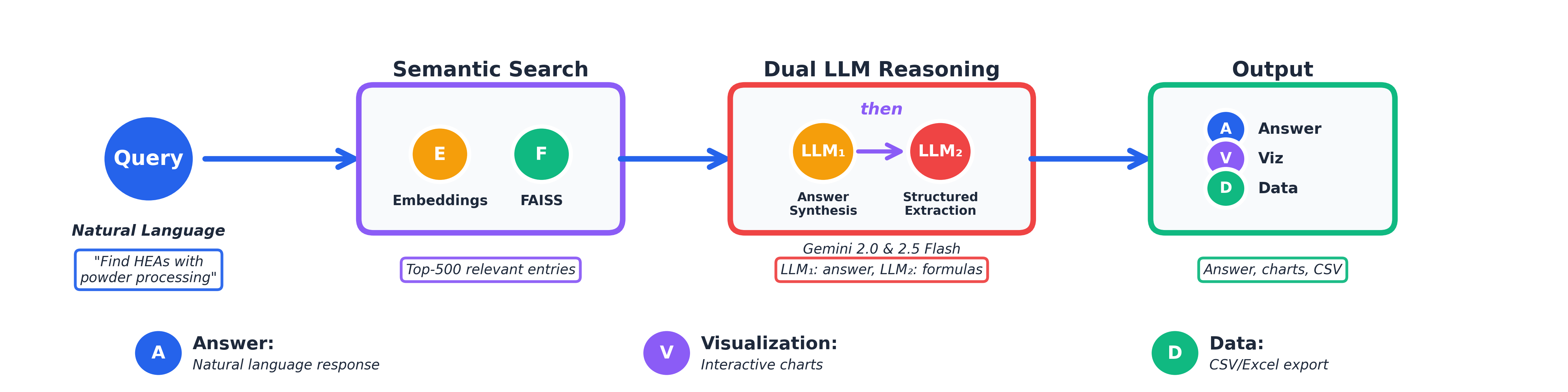}
    \caption{\textbf{The MaterEase Framework Architecture.} The complete workflow from natural language query to materials discovery and visualization.}
    \label{fig:materease_architecture_perfect}
\end{figure}


\subsection*{Future Work}

The team plans to expand MaterEase's capabilities across several dimensions to enhance its utility for materials science research. \textbf{Multi-modal Integration}: Extending the framework to incorporate experimental data, phase diagrams, and microscopy images, enabling queries that combine textual descriptions with visual material characteristics. \textbf{Experimental Validation Pipeline}: Developing automated workflows that connect MaterEase's predictions with laboratory synthesis and testing, creating a closed-loop system for hypothesis generation and validation. \textbf{Cross-domain Adaptation}: Extending the framework beyond high-entropy alloys to other materials classes including ceramics, polymers, and composites, with domain-specific knowledge bases and property schemas. \textbf{Real-time Knowledge Updates}: Implementing dynamic ontology learning to incorporate new research findings and maintain up-to-date knowledge bases continuously. \textbf{Enhanced Reasoning}: Integrating causal knowledge graphs to enable multi-step causal reasoning about material synthesis-property relationships, moving beyond retrieval to predictive hypothesis generation.

\subsection*{Open-source Materials}

Code is available on GitHub: \href{https://github.com/Areeb2735/MaterEase-Your-AI-Powered-Materials-Science-Assistant}{\faGithub}





\section{MatSciAgent: An Autonomous Coding Agent for Discovery in Materials Science and Chemistry}\label{sec:MatSciAgent}

\begin{figure}[h]
    \centering
    \includegraphics[width=\linewidth]{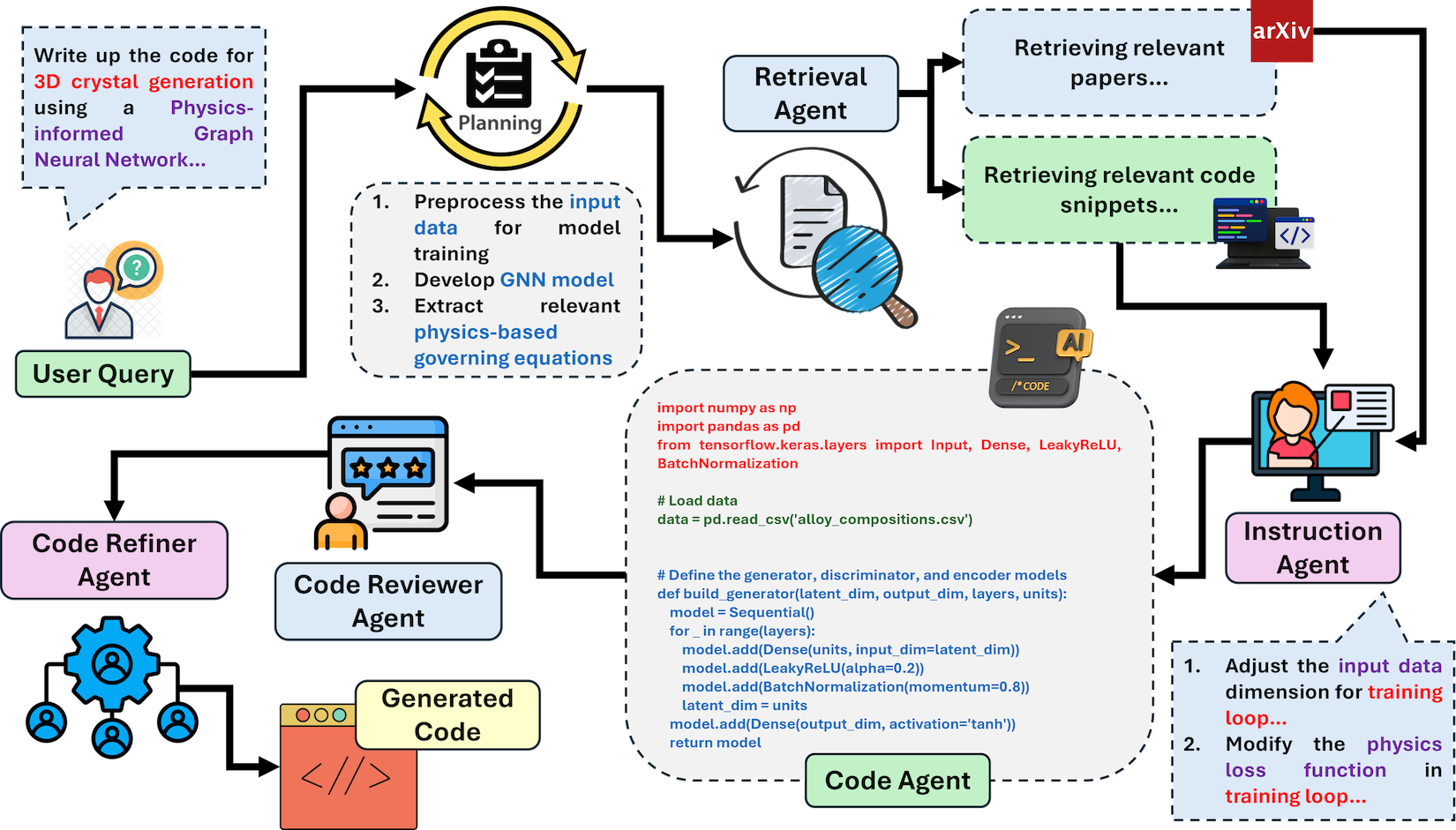}
    \caption{The end-to-end workflow of MatSciAgent for scientific code generation.}
    \label{fig:matsciagent-workflow}
\end{figure}

Code generation has accelerated rapidly in recent years, with frontier models posting strong results on standard benchmarks: for example, Claude-3.5-Sonnet reaches 92\% pass@1 on HumanEval, Claude-3-Opus achieves 86.4\% on MBPP, and GPT-4o-0513 reports 61.1\% on BigCodeBench \cite{jiang2024survey}. While these scores demonstrate impressive functional-level capabilities, they are often insufficient for end-to-end, system-level tasks where requirements are open-ended, multi-step, and user-specific. Recent work proposes multi-agent and large-scale generation frameworks to bridge this gap \cite{ishibashi2024self,roy2026autonomous,pratiush2025mic,bazgir2025multicrossmodal,bazgir2025matagent,bazgir2025proteinhypothesis}, but many systems still lack robust mechanisms for domain-aware reasoning and customization, leading to generic outputs that miss expert constraints \cite{huang2025deep,bazgir2025agentichypothesis,zimmermann202532,bazgir2025drug}. As shown in Figure~\ref{fig:matsciagent-workflow}, inspired by retrieval-augmented, research-oriented paradigms (e.g., "deep research" and "co-scientist" styles), the SciForge team introduces MatSciAgent, a domain-aware, agentic pipeline for materials science. MatSciAgent integrates targeted literature/code retrieval, knowledge synthesis, and iterative code planning/refinement to produce runnable, customized solutions that better align with real scientific workflows. The team built the system on LangGraph, which serves as the backbone orchestrating multi-step reasoning and inference toward code generation. The architecture is modular and model-agnostic: users can choose their preferred provider and model, currently OpenAI, Gemini, and local models via Ollama, with more backends planned. Core pipeline stages include query analysis, knowledge synthesis (literature/code retrieval + aggregation), code analysis, code generation, and iterative refinement, enabling customized, end-to-end, runnable solutions. Because domain-specific code-generation benchmarks for materials science are scarce, the team built its own. The team created question–answer tasks across multiple topics and plans to expand coverage and introduce graded difficulty levels for more comprehensive evaluation. To construct the benchmark, the team manually collected papers and their associated GitHub repositories, generated user queries from paper summaries, and curated ground-truth snippets by extracting the most essential code (e.g., model architectures, key scientific libraries). The dataset is JSON-formatted with metadata. For this hackathon, the team assembled 7 problems: Interpretable Alloy Design, Phonon Dynamics Reconstruction, Spectroscopy Data Compression, Inverse Pattern Design, Microstructure Fingerprinting Compression, Conditional Crystal Generation, and Inverse Crystal Design. The team expects to add more tasks in the near future. To avoid overestimating performance, the retrieval agents explicitly exclude the ground-truth paper and repository from the searchable corpus during evaluation (preventing data leakage). For scoring, an LLM-as-judge is used with a standardized rubric/prompt, which returns a 0–100 score for each task.

\subsection*{Results}

\begin{figure}[h]
    \centering
    \begin{subfigure}[b]{0.48\textwidth}
        \centering
        \includegraphics[width=\linewidth]{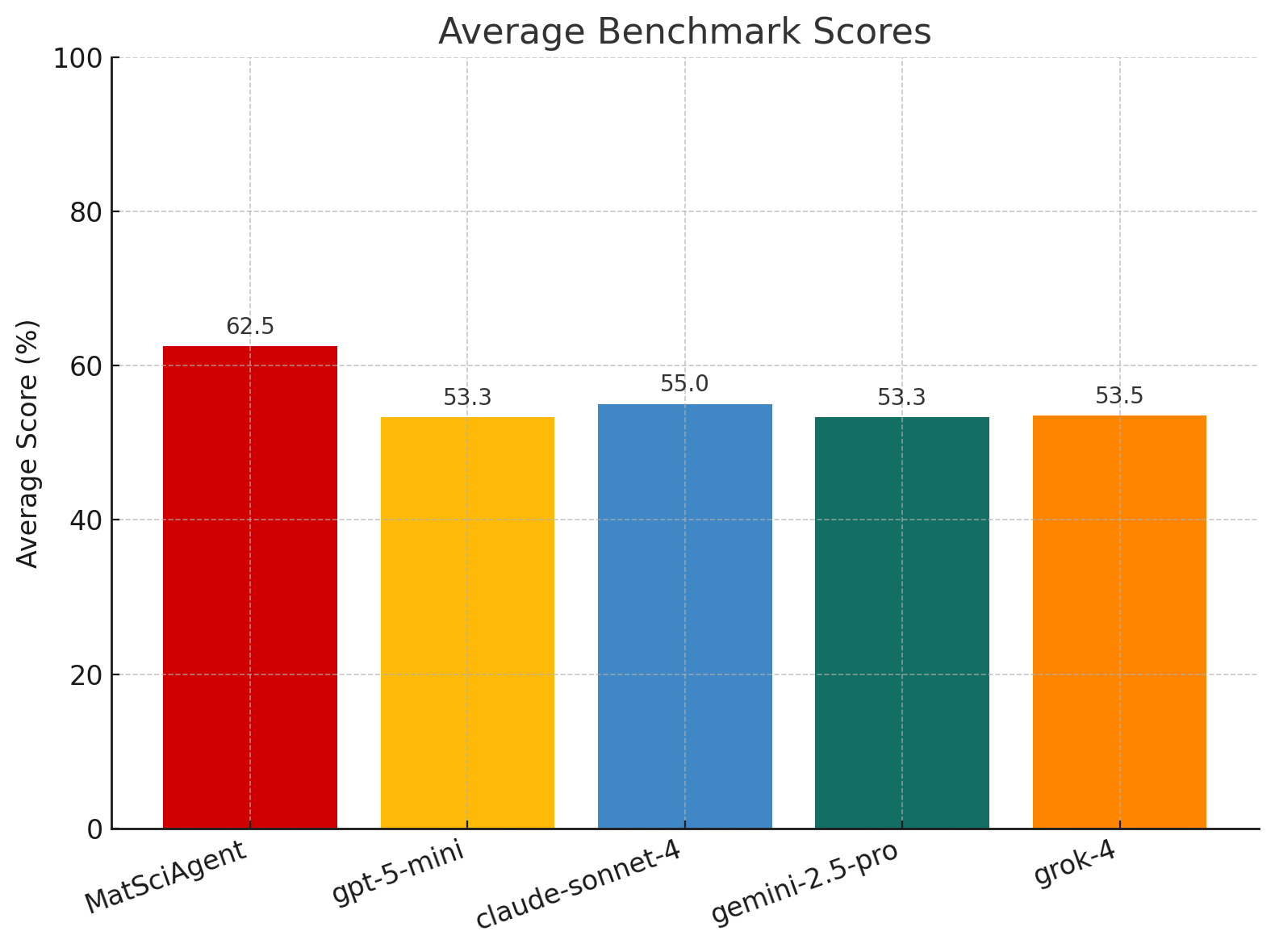}
        \caption{Comparison of Average benchmark scores between MatSciAgent and other models.}
        \label{fig:sciforge_benchmark}
    \end{subfigure}
    \hfill
    \begin{subfigure}[b]{0.48\textwidth}
        \centering
        \includegraphics[width=\linewidth]{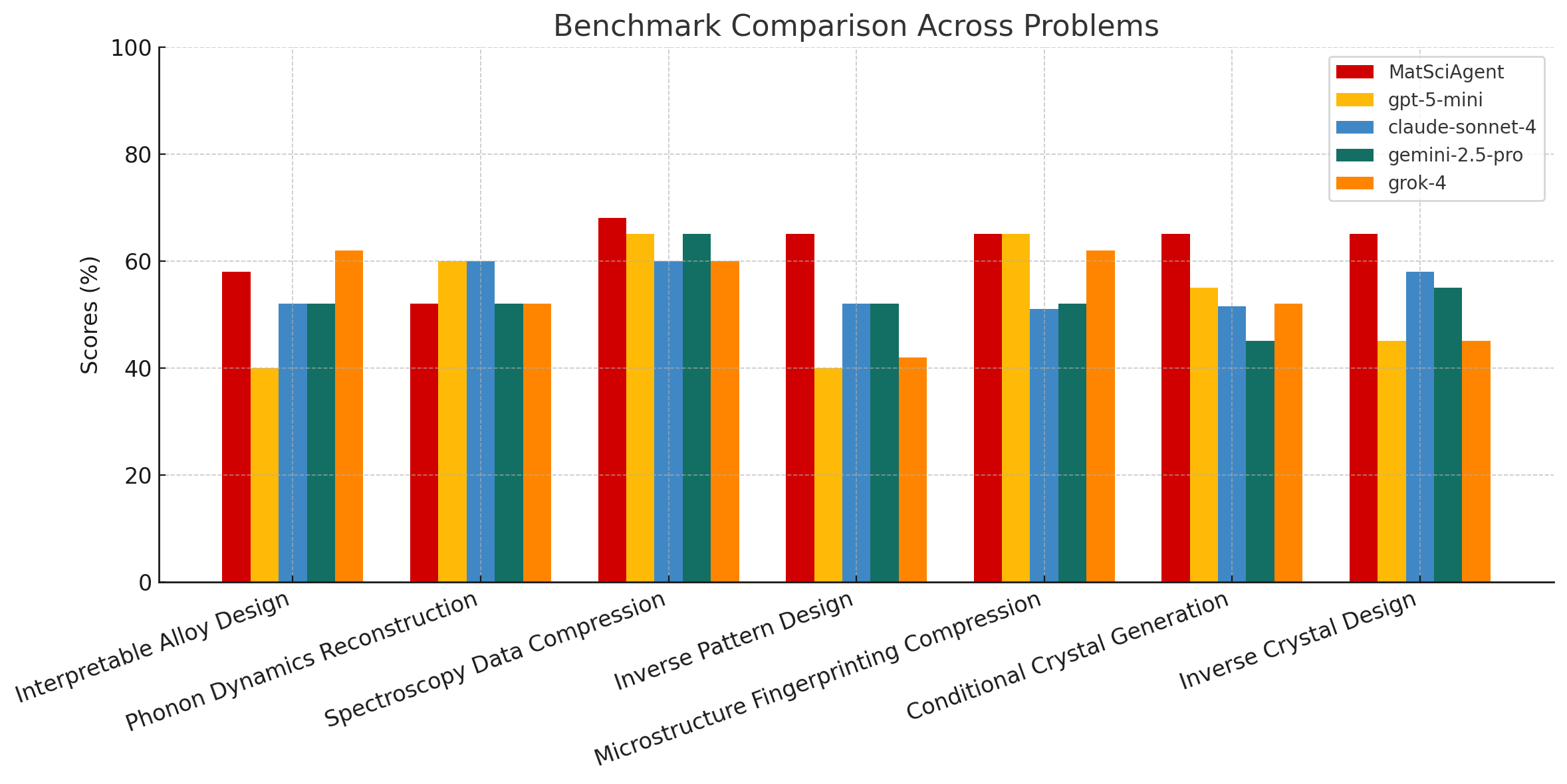}
        \caption{Accuracy of different LLMs and our agent per specific tasks covered through our benchmark.}
        \label{fig:sciforge_benchmark_per_task}
    \end{subfigure}
    \caption{Comparison of efficiency and accuracy across different LLM systems against our proposed coding agent.}
    \label{fig:sciforge_benchmark_combined}
\end{figure}

As can be seen in Figure~\ref{fig:sciforge_benchmark_combined}, MatSciAgent averages 62.5\% across seven benchmarks, about +7\% over strong baselines (GPT-5-mini 53.5\%, Claude-Sonnet-4 55\%, Gemini-2.5-Pro 53.3\%, Grok-4 53.5\%). Performance varies by task: Grok-4 edges MatSciAgent by ~3\% on Interpretable Alloy Design, and GPT-5-mini/Claude-Sonnet-4 are ~10\% higher on Phonon Dynamics Reconstruction, highlighting optimization opportunities. Error analysis shows strengths in conceptual understanding and correct library identification, but weaknesses include occasional architecture mistakes, structural/organizational issues, framework inconsistencies, missing key features, and oversimplification/placeholders.

\subsection*{Future Work}

MatSciAgent shows promise for moving from benchmark-style code completion to end-to-end, runnable
materials-science workflows, but several directions are important for maturation.

\paragraph{(1) Scientific correctness and domain verification.}
Future versions should go beyond ``it runs'' by adding domain-aware validation (units/dimensions, physical
constraints, invariances, sanity checks on toy inputs) and making assumptions explicit before generation.

\paragraph{(2) Execution-grounded refinement and reproducibility.}
Integrating an execution harness (isolated environments, automated runs, structured error capture, and
regression tests) would strengthen iterative refinement and improve reproducibility (versions, seeds, configs).

\paragraph{(3) Stronger retrieval with provenance.}
Retrieval can be extended to structured and multimodal sources (papers, repos, docs, figures/tables), with
provenance tracking so key code decisions are traceable to supporting evidence.

\paragraph{(4) Customization and role-based multi-agent collaboration.}
A structured user preference/constraint profile (libraries, compute limits, codebase conventions) plus
specialized roles (domain expert, engineer, reviewer) can reduce generic outputs and better match real
scientific workflows.

\paragraph{(5) Benchmark and evaluation expansion.}
The benchmark should scale in coverage and graded difficulty, maintain leakage-resistant protocols, and
move toward hybrid evaluation: LLM-as-judge complemented by execution-based tests and targeted expert
review to measure true system-level generalization.





\section{Implementing Knowledge Graphs as Long-term Memory for Scientific Agents}
\label{sec:chemunity}

Research agents are now attracting attention from researchers in material science, as they have shown the ability to plan and execute steps to successfully solve tasks. For example, El Agent by Zou et al. has demonstrated a strong ability to solve university-level course exercises \cite{zou2025agente}. However, these agents rely on pre-training or fine-tuning as a source of knowledge, which exposes the agents to hallucinations \cite{zhang2024knowledge, mckenna2023sources}. Therefore, developing methods to reduce hallucination and improve the reliability and scientific grounding of LLM is critical to enable AI scientists. 

\subsection*{Results}

The ChemUnityQA team explores using GraphRAG to improve on LLM Q\&A capabilities by leveraging knowledge graphs as long-term memory for a question answering agent equipped with machine learning (ML) models as tools \cite{edge2025graphrag}. The knowledge graph used in this work is MOF-ChemUnity, a knowledge graph that unifies the metal-organic framework (MOF) data from computational sources and experimental results in literature \cite{pruyn2025chemunity}. Using the knowledge graph as long-term memory allows the agent to synthesize knowledge across the domain, identifying non-obvious trends and answering questions that require high-level reasoning. The goal is to expand the agents' ability beyond simply fact retrieval, and enable them to generate novel insights by leveraging the entire structure of the knowledge graph.

\begin{figure}[htbp]
    \centering
    \includegraphics[width=\linewidth]{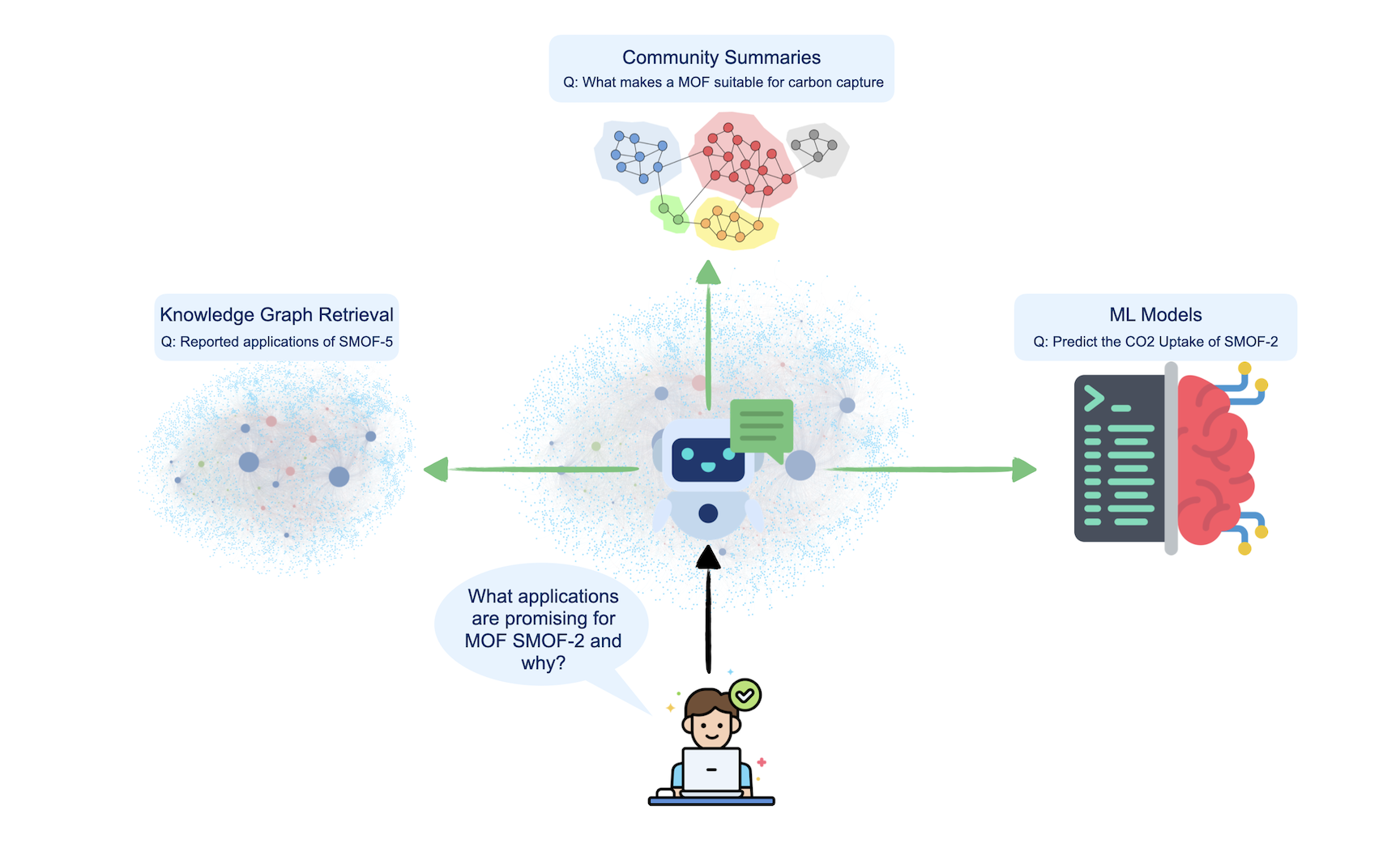}
    \caption{Illustration of using MOF-ChemUnity knowledge graph as long-term memory for AI agents.}
    \label{fig:chemunity}
\end{figure}

GraphRAG differs from traditional RAG in that it identifies multiple levels of communities within the knowledge graph \cite{edge2025graphrag}. For each community, the text is passed to an LLM to generate a summary of all text associated with its member nodes. This enables information retrieval from local information to global summaries and trends. Figure~\ref{fig:chemunity} demonstrates the GraphRAG methodology and its implementation with the MOF-ChemUnity knowledge graph using GPT-4o. The key result is that all explanations and justifications that the agent provides are grounded in the knowledge graph rather than the pre-trained weights. These explanations are generated from various community summaries that explain trends in the knowledge graphs. Therefore, the answers generated by the agent are less prone to hallucination and are grounded in scientific research.

\subsection*{Future Work}

This work demonstrates that providing agents with domain-specific knowledge enhances the agent's ability to use tools and answer research questions. This opens up the path to building domain-specific agents that use long-term knowledge to perform actions and draw conclusions. Additionally, using a knowledge graph enables a wide variety of techniques that improve the knowledge synthesis. The team envisions this AI agent can be implementing tools that determine whether the long-term knowledge available is sufficient to answer user queries. Later, it can discover new information through research tools to gather necessary information to address user questions.

\subsection*{Open-source Materials}

All code is available on GitHub: \href{https://github.com/AI4ChemS/ChemUnityQA}{\faGithub}

%




\section{AESOP: Accelerated Expert-in-loop Scientific Output Protocol}\label{sec:aesop}

In computational materials science and chemistry, researchers navigate a vast landscape of simulations spanning multiple size scales and theoretical frameworks. Selecting and implementing the most appropriate approach for a specific scientific question is often time-consuming, especially for incoming graduate students who must orient themselves within myriad tools, theoretical levels, and best practices. The AESOP team aims to streamline this process by introducing a flexible workflow powered by LLMs. The protocol employs two LLMs in tandem. First, literature-trained LLM suggests suitable scientific approaches based on a corpus of papers, institutional knowledge, and documentation. Second, Codebase-trained LLM instantiates these approaches as computational scripts, leveraging knowledge of open-source tools such as ASE\cite{larsen2017atomic} and GPAW.\cite{mortensen2024} This protocol helps users quickly utilize relevant literature, generate an initial workflow with suggested analyses, and avoids lengthy troubleshooting, which results in optimizing research time and accelerating scientific discovery.

The workflow begins with user interaction, where a researcher consults a general LLM, which was trained on institutional and domain-specific knowledge, to identify suitable methodological approaches for a given scientific task. Through an iterative refinement process, the user and model progressively sharpen the research question and clarify the desired outcomes. Once an approach is selected, a code-generation LLM, trained on scientific software repositories and documentation, produces executable scripts tailored to the problem. When the system detects low confidence or identifies a method as unusual or potentially error-prone, the generated scripts are routed to a human expert for verification, correction, and approval. This protocol supports the efficient production of robust and scientifically sound outputs while enabling senior scientists to mentor a larger number of users and thereby extending institutional research capacity.

\begin{figure}[h!]
    \centering
    \includegraphics[width=0.8\linewidth]{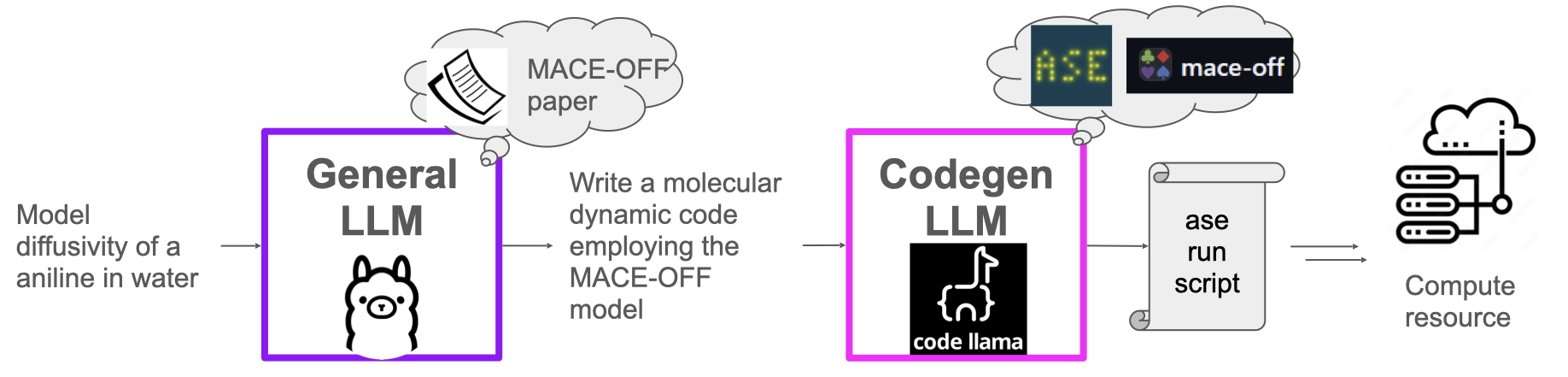}
    \caption{An example workflow for modeling diffusivity of an organic molecule in water.}
    \label{fig:AESOP-workflow}
\end{figure}

\subsection*{Results}

To demonstrate AESOP, the team considered a classic problem: modeling the diffusivity of a small organic molecule in water (Figure~\ref{fig:AESOP-workflow}). The literature-trained LLM evaluates available machine-learned interatomic potentials in the context of recent advances (e.g., MACE-OFF force field\cite{kovacs2025}) and recommends appropriate methodology. Then the code gen LLM draws on its repository knowledge to write executable Python scripts using appropriate libraries (e.g. ASE, GPAW, etc.), enabling automated or semi-automated deployment to compute clusters.
This initial deployment validates the protocol's potential to guide users from research question to computational result with minimal obstacles. 

\subsection*{Future Work}

Future expansions will focus on broadening the scope of inputs to both LLMs, improving autonomous operation on high-performance computing systems, and further structuring expert review and feedback. 

\subsection*{Open-Source Materials}

Code and data: \href{https://github.com/HallucinatingStrikeTeam/AESOP}{\faGithub}





\section{MATLAB-Integrated Generative Plugin for Text-Driven Chemical Compound Construction for General Chemical Engineering Workflows}\label{sec:MatGen}

Researchers in chemical engineering and materials science increasingly rely on computational tools for tasks such as molecular modeling, compound screening, and spectroscopic analysis. However, many of the most powerful toolkits—particularly those built around RDKit and other modern cheminformatics frameworks—require substantial scripting or programming expertise. This creates a barrier for experimentalists, limits rapid prototyping, and slows the pace of exploratory research, especially for users who are not proficient in Python or other programming languages \cite{BradBarrier}. To address this gap and enable more intuitive, rapid, and accessible experimentation, the MatGen team developed a MATLAB-integrated plugin that facilitates natural-language interaction with an agentic large language model (LLM) augmented with RDKit-based cheminformatics and materials-science analysis capabilities.

The system enables researchers to execute advanced molecular modeling, compound generation, and spectroscopic analysis workflows directly from the MATLAB environment without the need for extensive scripting or specialized programming expertise. By providing an intuitive text-driven interface to modern AI-driven scientific toolchains, the plugin enhances accessibility, accelerates experimental iteration, and establishes a seamless bridge between MATLAB and state-of-the-art computational chemistry and materials-informatics frameworks.

\subsection*{Results}

\begin{figure}[h!]
    \centering
    \includegraphics[width=0.8\linewidth]{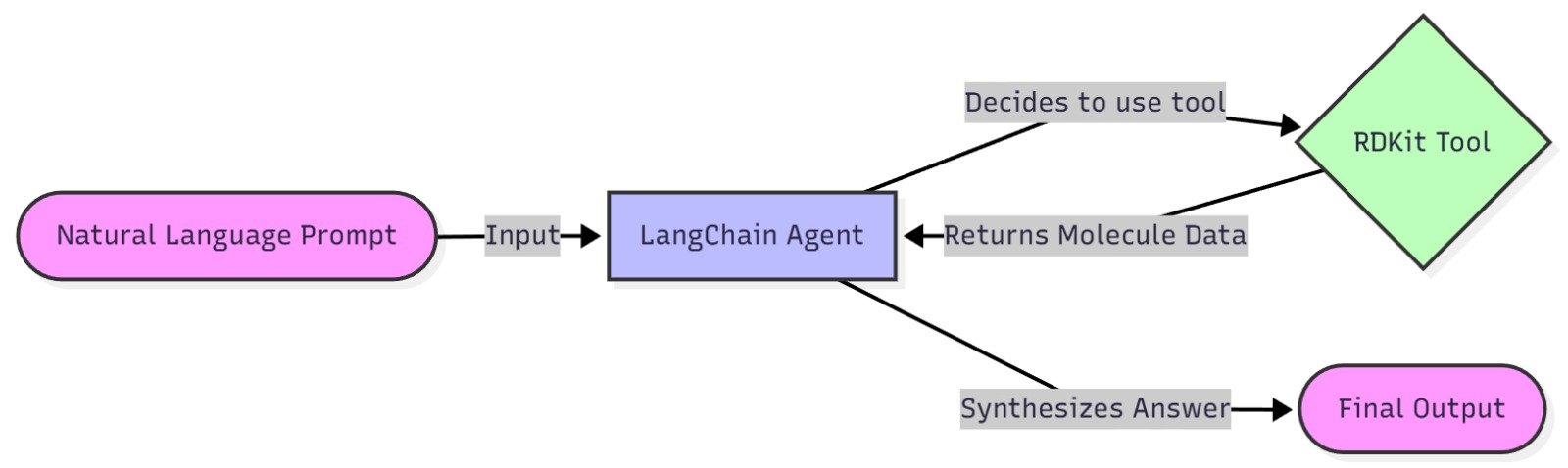}
    \caption{An example workflow for agent request from MATLAB CLI prompt}
    \label{fig:MatGen}
\end{figure}

As illustrated in Figure~\ref{fig:MatGen}, the plugin integrates MATLAB with an agentic LLM capable of executing RDKit-driven cheminformatics operations, including molecular property prediction, structural manipulation, and elementary reaction analysis. Users can issue natural-language instructions (e.g., "What is the charge of benzene? What is its SMILES representation and how can we visualize it") and receive computed outputs, molecular visualizations, and workflow results directly within MATLAB.

The system supports a complete end-to-end workflow consisting of natural-language input parsing, RDKit tool invocation via the LLM agent, post-processing and result packaging, and final rendering within MATLAB. This architecture enables robust AI-assisted compound analysis and visualization that integrates seamlessly into material-science and spectroscopy research pipelines.

\subsection*{Future Work}

Future development will focus on expanding the available toolset (e.g., reaction prediction, materials informatics functions), improving agent reliability, and potentially finetuning for added performance.

\subsection*{Open-source Materials}

The GitHub repository is still under development.
However, a video demo is available on YouTube: \href{https://www.youtube.com/watch?v=nIXMZN26804}{\faYoutube}.





\section{Explain that Automation}\label{sec:explain-automation}

As many scientific groups and institutions race towards building self-driving labs (SDLs), it has become clear that new tools are needed to inspect and utilize the large amount of semi-structured data produced by automated/autonomous instrument workflows. This data typically consists of instrument actions and, sometimes, related experimental data which are streamed to a database during a given session. Inspecting a particular action or extracting a specific experimental data from a session is tedious and time consuming. To address this, an LLM-agent was developed to interpret instrument interaction logs, thereby enabling AI-generated session summaries, conversational data inspection, and recommendations backed by experimental history. The agent uses multi-modal retrieval augmented generation (RAG) and query tools on the instrument action database for context injection related to the user's prompt. A simple web app was also developed around this agent for ease of use. A database of automated transition electron microscope experiments was used to demonstrate this method.

\subsection*{Results}

\begin{figure}[h]
    \centering
\includegraphics[width=1.0\linewidth]{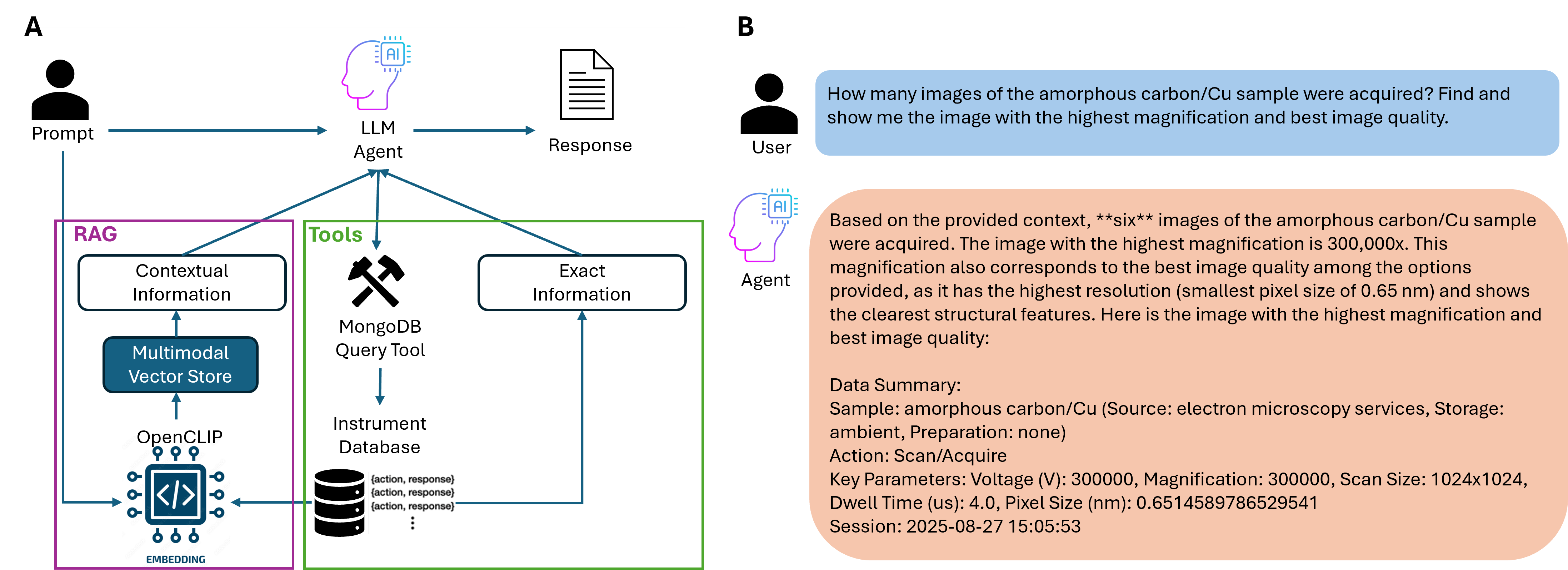}
    \caption{A) Workflow of instrument action database agent. B) Example prompt and agent response.}
    \label{fig:ETA-pipeline}
\end{figure}

Figure~\ref{fig:ETA-pipeline} shows the workflow of the instrument database agent. OpenCLIP was used to generate a vector store of joint, multimodal embeddings from text-based actions and 1D/2D experimental data (spectra/micrographs). The user's prompt is embedded by the same model for use in vector similarity search in the RAG context injection pipeline. When exact information is needed, the agent has access to a monogoDB query tool to make explicit queries based on the user's prompt and the RAG context. These three pieces of information (prompt+RAG+tool) are combined and used by the agent to produce the final response. Gemini 2.5 flash was used as the LLM.

\subsection*{Future Work}

The team plans to use this database agent as a sub-agent in a agentic experimental planning and instrument execution workflow.

\subsection*{Open-source Materials}

Code available on GitHub: \href{https://github.com/darianSmalley/explain-that-automation}{\faGithub}





\section{AIssistant: A Human-AI Collaborative Framework for Accelerating Scientific Discovery in Atomic Layer Deposition}\label{sec:AIssistant}

Scientific discovery in chemistry and materials science is an iterative process relying on expert intuition and experimental feedback. The TIB Assistant team introduces \texttt{AIssistant}\footnote{https://aissistant.tib.eu/}, a human-AI collaborative framework leveraging large language models (LLMs) to accelerate scientific discovery, specifically in Atomic Layer Deposition (ALD), by augmenting expert capabilities and preserving human oversight.

\begin{figure}[h]
    \centering
\includegraphics[width=1.0\linewidth]{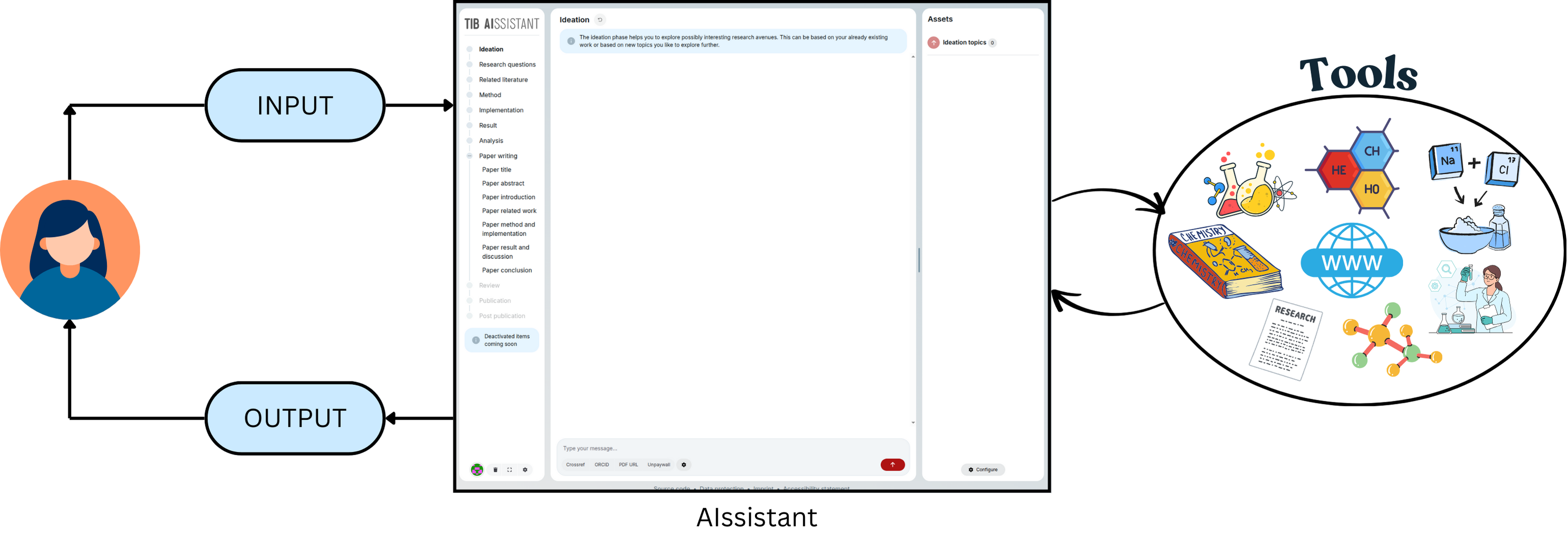}
    \caption{Workflow of \texttt{AIssistant} with MC-NEST and ChemCrow tools.}
    \label{fig:tib_ass_pipeline}
\end{figure}

\subsection*{Results}

The \texttt{AIssistant} framework integrates specialized tools like MC-NEST~\cite{rabby2024mc} for hypothesis generation and ChemCrow~\cite{bran2023chemcrow} for interactive refinement, enabling iterative cycles of AI-suggested hypotheses and human validation. Quantitative evaluation metrics were utilized to assess the alignment of AI-assisted outcomes with human reasoning. The high-level \textbf{average performance is summarized in Table~\ref{tab:average_performance}}, which details the average model score across various reference types for different papers. A more granular look at the performance of premium models (O3, O4-mini, and GPT-5) is presented in Table~\ref{tab:premium_model_performance}, comparing their scores against the Base LLM and Original Papers references. The workflow is summarized in Figure~\ref{fig:tib_ass_pipeline}.

\begin{table}[ht]
\centering
\caption{Average Model Performance by Paper and Reference Type.}
\label{tab:average_performance}
\resizebox{\columnwidth}{!}{%
\begin{tabular}{l|c|c|c|c|c}
\hline
\textbf{Paper Title (Abbreviated)} & \textbf{Base LLM} & \textbf{Original Papers} & \textbf{O3 Generated} & \textbf{O4-mini Generated} & \textbf{GPT-5 Generated} \\
\hline
Spatial ALD~\cite{hoye2025spatial} & 0.87 & 0.88 & 0.85 & \textbf{0.89} & 0.88 \\
Molecular Design ALD~\cite{lee2025molecular} & 0.81 & \textbf{0.96} & \textbf{0.82} & 0.80 & \textbf{0.82} \\
Topographically Selective ALD~\cite{janssen2025topographically} & 0.72 & \textbf{0.96} & 0.71 & \textbf{0.73} & \textbf{0.73} \\
Catalysts Design ALD~\cite{jung2025catalysts} & 0.81 & 0.90 & 0.84 & \textbf{0.85} & \textbf{0.85} \\
Subnano $Al_2O_3$ Coatings~\cite{li2025subnano} & 0.80 & 0.84 & \textbf{0.83} & 0.80 & \textbf{0.83} \\
\hline
\end{tabular}%
}
\end{table}

\begin{table}[ht]
\centering
\caption{Model Performance vs. Key References.}
\label{tab:premium_model_performance}
\resizebox{0.8\columnwidth}{!}{%
\begin{tabular}{l|l|c|c}
\hline
\textbf{Paper Title (Abbreviated)} & \textbf{Model} & \textbf{Base LLM Score} & \textbf{Original Papers Score} \\
\hline
\multirow{3}{*}{Spatial ALD~\cite{hoye2025spatial}} & O3 & 0.92 & 0.91 \\
& O4-mini & 0.90 & 0.93 \\
& GPT-5 & 0.92 & 0.93 \\
\hline
\multirow{3}{*}{Molecular Design ALD~\cite{lee2025molecular}} & O3 & 0.83 & 0.98 \\
& O4-mini & 0.86 & \textbf{0.99} \\
& GPT-5 & 0.88 & \textbf{0.99} \\
\hline
\multirow{3}{*}{Topographically Selective ALD~\cite{janssen2025topographically}} & O3 & 0.70 & 0.98 \\
& O4-mini & 0.74 & 0.98 \\
& GPT-5 & 0.80 & 0.98 \\
\hline
\multirow{3}{*}{Catalysts Design ALD~\cite{jung2025catalysts}} & O3 & 0.91 & 0.96 \\
& O4-mini & 0.91 & 0.96 \\
& GPT-5 & 0.89 & 0.96 \\
\hline
\multirow{3}{*}{Subnano $Al_2O_3$ Coatings~\cite{li2025subnano}} & O3 & 0.84 & 0.84 \\
& O4-mini & 0.84 & 0.84 \\
& GPT-5 & 0.83 & 0.93 \\
\hline
\end{tabular}%
}
\end{table}

\subsection*{Future Work}

Future work will focus on integrating reasoning and multimodal data (e.g., experimental measurements) to further improve hypothesis validity and explore integrating with laboratory hardware control systems.

\subsection*{Open-source Materials}

The code and paper materials are publicly available at: \href{https://github.com/DIYANAPV/llm-hackathon-MS-and-Chem-2025/tree/main}{\faGithub}





\section{MOF-Genie: A Multimodal AI-Driven Knowledge Graph Integrating ML, GNNs, and LLMs for Hydrogen Storage Discovery}
\label{sec:mof-genie}

Metal Organic Frameworks (MOFs) are promising candidates for solid-state hydrogen storage due to their tunable porosity and high surface areas. However, research progress is limited by fragmented datasets, labor-intensive modeling pipelines, and the lack of accessible tools for querying structure-property relationships. The H2-Guardians team introduces MOF-Genie, an end-to-end artificial intelligence platform that integrates machine learning, graph neural networks, and a Neo4j-based knowledge graph with a natural language interface to accelerate MOF discovery for hydrogen storage applications.

\subsection*{Results}

The platform processes 1,488 MOFs and constructs a heterogeneous knowledge graph comprising 5,479 nodes and 4,535 edges. A data processing module cleans and normalizes MOF structural and adsorption data, which is then used to train an ensemble of supervised learning models. These models achieve an R\textsuperscript{2} score of 99.6 percent for gravimetric hydrogen uptake prediction, enabling high-fidelity estimation of missing properties. The knowledge graph is further leveraged to train a relational graph convolutional network using PyTorch Geometric, achieving 65.4 percent accuracy on link prediction tasks.

A natural language interface powered by a local LLM (Ollama) translates user queries into Cypher, providing sub-second response times and achieving a 100 percent query success rate. Users can retrieve MOFs with specific structural attributes, metal compositions, or hydrogen uptake characteristics without requiring expertise in database languages. 

\subsection*{Future Work}

Future extensions include multi-modal data integration, synthesis pathway prediction, expansion to CO\textsubscript{2} capture and catalysis tasks.

\subsection*{Open-source Materials}

All source code, model definitions, and documentation are available in the public GitHub repository: \href{https://github.com/AntoRoyan/MOF-Genie}{\faGithub}





\section{SKY: An Agent for Materials Synthesis Planning}
\label{sec:sky}

Computational materials discovery has advanced rapidly, yet the transition from a predicted structure to an experimental reality remains a major bottleneck. This synthesis gap is driven by the multiplicity of viable pathways for a given target and by the fact that synthesis knowledge is distributed across heterogeneous sources and is often encoded in unstructured text.\cite{park2025closing} As a result, researchers largely depend on manual, heuristic-based reasoning to adapt existing protocols to novel targets. The SKY team introduces SKY, an agentic workflow that automates this reasoning loop by grounding Large Language Model synthesis proposals in similarity searches and recursive evidence retrieval.

\subsection*{Results}

SKY utilizes a modular retrieval-augmented architecture to bridge the gap between material prediction and experimental realization via an agentic workflow. The process begins with creating embeddings, using Magpie\cite{ward2016general} for compositional descriptors and MACE\cite{batatia2023mace} for structural embeddings to identify chemically and structurally analogous materials. It subsequently performs a nearest-neighbour search to identify chemically similar materials and retrieves synthesis procedures via the Synthesis Explorer API provided by the Materials Project. When no direct recipe is available for the closest neighbours, SKY expands the search recursively through the neighbour graph up to $k$ hops to increase coverage. Retrieved recipes and metadata are stored in an evidence cache and provided to the LLM, which combines this retrieved evidence with its prior chemical knowledge to generate structured synthesis route proposals with traceable provenance to the underlying sources.

\begin{figure}[h]
    \centering
    \includegraphics[width=1\linewidth]{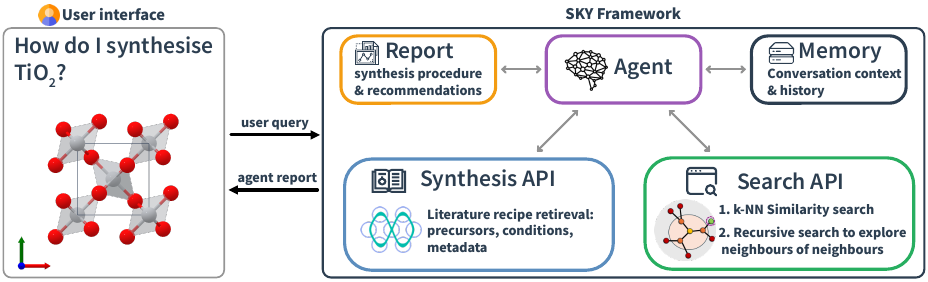}
    \caption{Overview of the SKY Workflow for Materials Synthesis Planning}
    \label{fig:sky}
\end{figure}

\subsection*{Future Work}

A key long-term goal is to build a persistent, searchable memory database that links retrieved synthesis evidence with experimental outcomes. By curating target materials, proposed procedures, lab conditions, and success/failure notes, it will continually improve retrieval, route ranking, and recommendation calibration using real-world data.

\subsection*{Open-source Materials}

Code available on GitHub: \href{https://github.com/hspark1212/sky}{\faGithub}

\section*{Overview of Hackathon Event}
The third LLM hackathon for Applications in Materials Science and Chemistry took place over two days (September 11-12 2025), across 16 in-person sites on four continents and one central online hub for worldwide access. The hackathon adopted a deliberately broad scope, inviting participants to harness open-source datasets and best-in-class multimodal models to address diverse challenges in materials science and chemistry. 
Details from the previous two hackathons held in 2023 and 2024 are available.~\cite{jablonka202314, llm2024, zimmermann202532}

\begin{figure}[h!]
    \centering
    \includegraphics[width=1\linewidth]{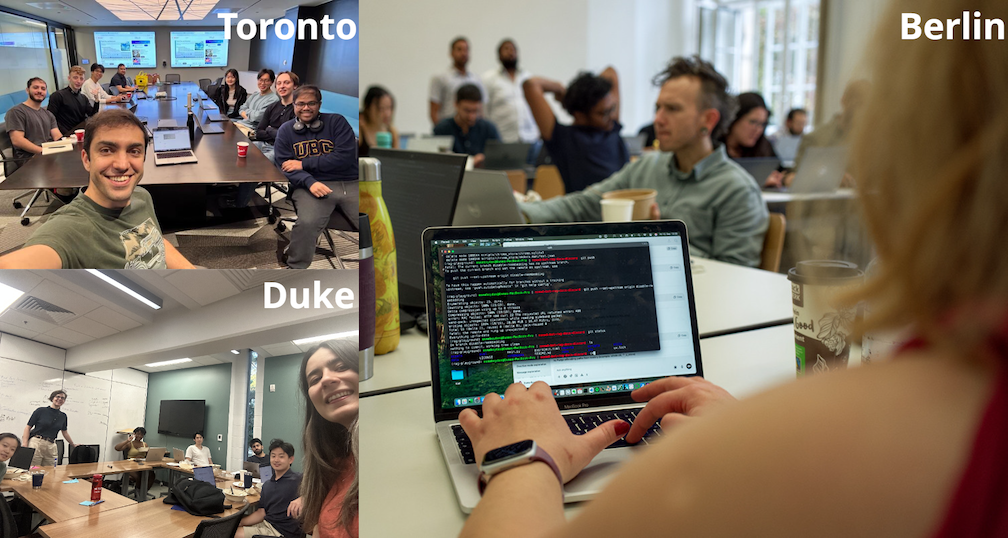}
    \caption{Participants collaborating at various physical hub locations during the 2025 LLM Hackathon for Applications in Materials Science and Chemistry.}
    \label{fig:on-site-images}
\end{figure}

\begin{figure}[h!]
    \centering
    \includegraphics[width=1\linewidth]{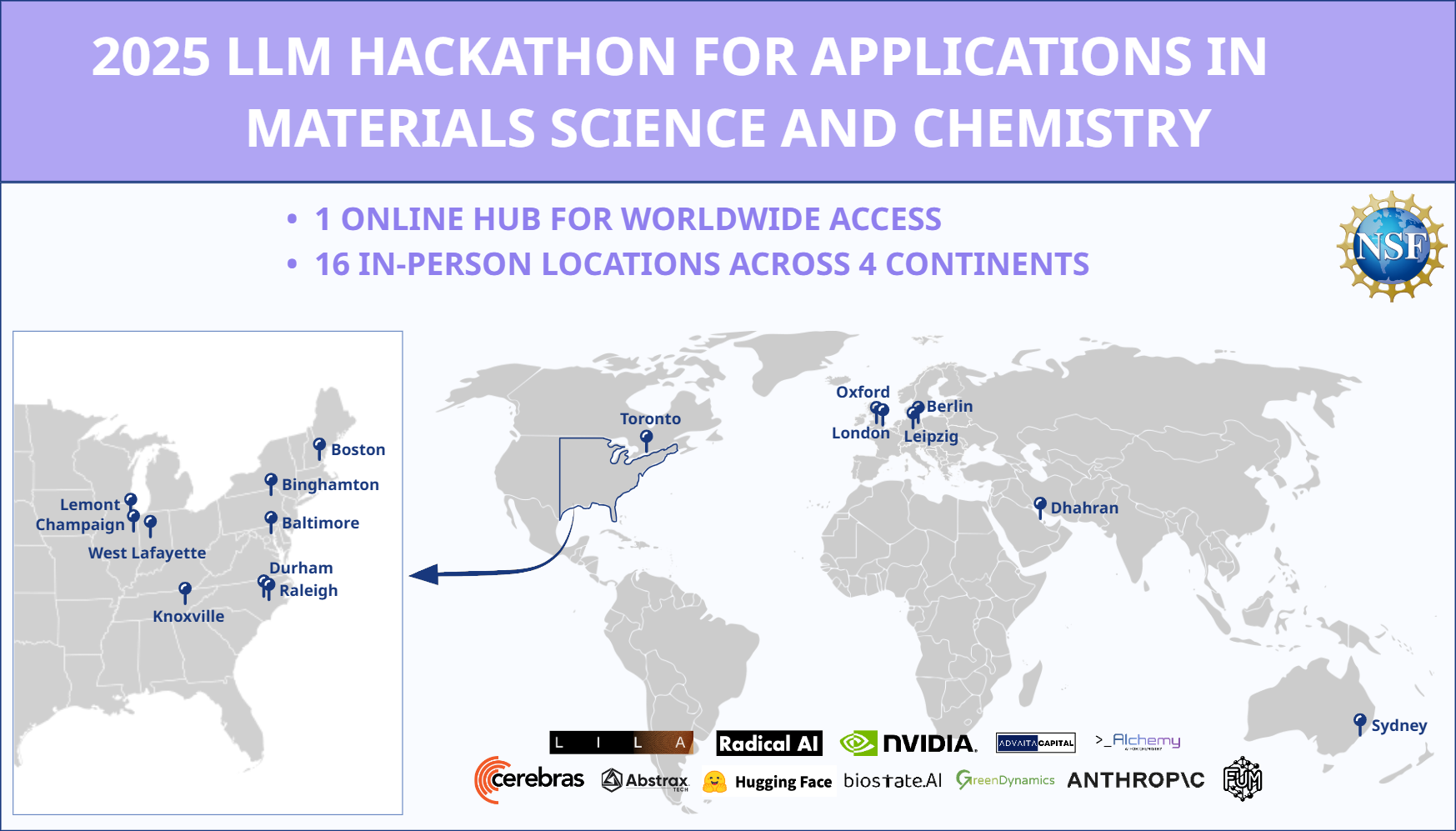}
    \caption{Hybrid nature and the sponsors of the 2025 LLM Hackathon for Applications in Materials Science and Chemistry. The markers in the map represent the on-site locations across four continents.}
    \label{fig:locations-and-sponsors}
\end{figure}

In 2025, 1,320 participants registered for the event (featured in Science News~\cite{savitsky2025researchers}). The hybrid format of the hackathon enabled researchers from geographically inaccessible areas to form teams online, connecting teams across the globe. The physical hubs facilitated networking through informal discussions and refreshments throughout the hackathon period (\autoref{fig:on-site-images}). Sixteen on-site locations were available across four continents including North America, Europe, Asia, and Australia, with North America hosting the most sites at 10 (Boston, Binghamton, Baltimore, Champaign, Durham, Knoxville, Lemont, Raleigh, West Lafayette, and Toronto). Europe hosted four locations (London, Oxford, Berlin, Leipzig), while Asia and Australia each had one site in Dhahran and Sydney, respectively. The hybrid nature created an accessible environment that transcended international and institutional boundaries, fostered diverse team collaborations, and was supported by multiple sponsors (\autoref{fig:locations-and-sponsors}). This approach yielded the projects documented in this paper and catalyzed the formation of a persistent online community of 1,483 researchers connected through Slack (see https://llmhackathon.github.io for more details or to join).

\section*{Conclusion and Future Directions}

With over 1,320 registered participants, 88 documented projects, and 16 physical hubs spanning four continents, the 2025 Large Language Models (LLM) Hackathon for Applications in Materials Science and Chemistry provided a large snapshot of researcher capability and how the community is deploying LLMs; especially as agentic interfaces and orchestrators of scientific work. The resulting prototypes span the full research lifecycle, from knowledge discovery and evidence synthesis to tool use, workflow automation, and early-stage design loops. A key takeaway is that the most compelling systems did not treat the LLM as a standalone predictor; instead, they combined LLM reasoning with structured retrieval, domain tools, and lightweight evaluation to create potentially reusable end-to-end workflows. These workflows can in turn become a resource to researchers operating under contraints including time, limited compute, data fragmentation, and heterogeneous modalities.

The broad division of projects into \emph{Knowledge Infrastructure} and \emph{Action Systems} further highlights a central pattern: i.e., progress accelerates when reliable scientific context (data, literature, provenance, and reproducibility) is tightly coupled with execution capabilities (planning, tool calling, simulation, and experiment coordination). Notably, simulation-oriented submissions frequently sat at the intersection of these two categories, underscoring simulations as both a knowledge substrate (generating mechanistic insight and labelled data) and an action substrate (enabling closed-loop optimization and design).
 
The cross-cutting themes identified in the Submissions Overview included the shift towards agentic orchestration and systems, the maturation of retrieval-augmented grounding, the importance of structured knowledge and data extraction, the expansion to multimodal and multilingual inputs, and the first steps toward physical laboratory integration. Each theme points to specific capability gaps that inform the future directions outlined below.

\paragraph{Future directions.} While the submissions highlight impressive momentum, they also point to what is needed to translate prototypes into reliable, field-deployable systems and to advance the longer-term vision of integrated, increasingly autonomous research workflows and self-driving laboratories:

\begin{enumerate}
 
    \item \textbf{From demos to defensible evaluation.}
    The prevalence of agentic architectures, present in over 40 of the 88 submissions, makes evaluation an urgent priority. The next generation of systems should be accompanied by standardized, domain-relevant benchmarks that measure not only fluency, but also scientific correctness, uncertainty calibration, robustness to distribution shift, and tool-execution reliability. In particular, agentic workflows need evaluation protocols that test multi-step planning, tool chaining, failure recovery, and reproducibility of outcomes.
 
    \item \textbf{Reproducibility-by-design for LLM-enabled science.}
    Many high-performing prototypes depend on rapidly evolving proprietary APIs. As retrieval-augmented generation matures from a technique into default infrastructure for scientific agents, reproducibility must be engineered purposefully. Future work should emphasize reproducible pipelines including explicit model/version logging, deterministic configuration (or testing) where possible, dataset and prompt provenance, containerized toolchains, and artifact packaging (e.g., structured run manifests and trace logs) so that scientific claims remain verifiable as models and platforms change.
 
    \item \textbf{Tool interoperability and scientific operating systems.}
    A recurring need is a unifying layer that standardizes how LLM agents interact with chemistry and materials tooling (e.g., databases, simulation engines, electronic lab notebooks, and synthesis/planning software). Common schemas for inputs/outputs, error reporting, and metadata could reduce brittle, one-off integrations and enable more modular, architectures.
 
    \item \textbf{Multimodal, experiment-in-the-loop autonomy with human oversight.}
    The growing number of submissions operating on non-textual inputs (e.g., S/TEM micrographs, NMR and XRD spectra, FTIR/UV-Vis data, nanoparticle images, audio laboratory logs, and multilingual patents) demonstrates that the community is already building towards multimodal input and grounding. Leveraging AI tools in self-driving laboratories will require extending this grounding to tight coupling between computational outputs and experimental execution. A practical near-term direction is human-in-the-loop autonomy, where systems propose actions, justify them with evidence, but still defer critical decisions to experts while learning from feedback.
 
    \item \textbf{Safety and governance for scientific agents.}
    As LLM agents gain execution power, from configuring HPC jobs to steering robotic liquid handlers, the community needs shared practices for safe tool use, access control, guardrails for laboratory operations, and clear responsibility boundaries. This includes transparent reporting of failure modes (hallucinations, invalid chemistry constraints, or unsafe actions) and mechanisms for auditing agent decisions. The human-in-the-loop designs that recur across this year's submissions represent a pragmatic starting point, but more formal frameworks will be needed as autonomy increases.
 
\end{enumerate}
 
Taken together, these submissions indicate that the field has entered a phase where the core question is no longer \emph{whether} LLMs can contribute to materials science and chemistry, but \emph{how} to engineer systems that are reliable, reproducible, and interoperable at scale, and further how to integrate them with the existing standards, practices, and norms of the scientific community. Future hackathons can accelerate this trajectory by pairing rapid prototyping with shared evaluation benchmarks, interoperability standards (such as common tool-calling schemas and structured run manifests), and explicit reproducibility requirements, thereby turning a growing portfolio of promising prototypes into a coherent, community-maintained ecosystem for trustworthy, AI-enabled scientific discovery.

\section*{Contributors and Acknowledgements}
This paper is the result of contributions from all enthusiastic participants in the 2025 LLM Hackathon for Applications in Materials Science and Chemistry. Below, we list all contributors alphabetically by last name, apart from lead and corresponding authors. Only author names and ORCID identifiers are listed for all participants except the corresponding author. Individual authors can be contacted using their ORCID details or by contacting the corresponding author.

Anthropic Claude 4.5 and 4.6 were used to help guide project tagging, and various summarization tasks in this work.
\vspace{5pt}


\begin{center}
Aritra Roy$^\dagger$\,\orcidlink{0000-0003-0243-9124},\;
Kevin Shen$^\dagger$\,\orcidlink{0000-0001-9715-7474},\;
Andrew MacBride$^\dagger$\,\orcidlink{0000-0001-9659-3776},\;
Awwal Oladipupo$^\dagger$\,\orcidlink{0009-0006-3979-2078},\;
Mudassra Taskeen$^\dagger$\,\orcidlink{0009-0009-6598-8359},\;
Wojtek Treyde$^\dagger$\,\orcidlink{0000-0003-1735-0640},\;
Ruaa A. E. A. Abakar\,\orcidlink{0009-0009-6972-2179},\;
Ahmad D. Abbas\,\orcidlink{0000-0003-0608-7074},\;
Elsayed Abdelfatah\,\orcidlink{0000-0003-1818-4605},\;
Abbas A. Abdullahi\,\orcidlink{0000-0002-3856-2709},\;
Seham S. Abyah,\;
Chahd Rahyl Adjmi,\;
Fariha Agbere\,\orcidlink{0009-0004-9090-8166},\;
Savyasanchi Aggarwal\,\orcidlink{0009-0007-7128-3465},\;
Muhammad Ahmed,\;
Tasnim Ahmed\,\orcidlink{0000-0002-0799-1180},\;
Motasem Ajlouni,\;
Mattias Akke\,\orcidlink{0000-0001-9207-0695},\;
Hussein AlAdwan\,\orcidlink{0009-0008-9940-7156},\;
Anwaar S. Alazani,\;
Zahra A. Alharbi,\;
Wajd A. Aljulyhi,\;
Mohammed A. AlKubaish,\;
Fatima A. Almahri,\;
Sayed A. Almohri\,\orcidlink{0000-0003-2104-0899},\;
David Obeh Alobo\,\orcidlink{0009-0007-7893-5366},\;
Mohammed Alouni,\;
Azizah S. Alqahtani\,\orcidlink{0009-0008-6031-9570},\;
Omar Alsaigh,\;
Husain Althagafi,\;
Md. Aqib Aman\,\orcidlink{0009-0007-3943-215X},\;
Lena Ara\,\orcidlink{0009-0007-0198-0871},\;
Arifin\,\orcidlink{0000-0001-7541-3326},\;
Ignacio Arretche\,\orcidlink{0000-0003-1443-0506},\;
Abdulaziz Ashy,\;
Syeda A. Asim,\;
Amro Aswad\,\orcidlink{0009-0000-2942-784X},\;
Adeel Atta,\;
Sören Auer,\;
Abdullah al Azmi\,\orcidlink{0009-0002-7168-6671},\;
Toheeb Balogun\orcidlink{0000-0006-8505-3196},\;
Suvo Banik\,\orcidlink{0000-0001-7239-8853},
Viktoriia Baibakova\,\orcidlink{0009-0004-5799-9510},\;
Shakira A. Baksh,\;
Neus G. Bastús\,\orcidlink{0000-0002-3144-7986},\;
Christina J. Bayard\,\orcidlink{0009-0002-5344-3713},\;
Adib Bazgir\,\orcidlink{0000-0001-6475-8505},\;
Louis Beal\,\orcidlink{0009-0000-6474-2722},\;
Lejla Biberić\,\orcidlink{0000-0003-0207-513X},\;
Wahid Billah\,\orcidlink{0009-0007-0214-0751},\;
Ankita Biswas\,\orcidlink{0009-0003-2627-6064},\;
Joshua Bocarsly\,\orcidlink{0000-0002-7523-152X},\;
Montassar T. Bouzidi\,\orcidlink{0009-0000-7404-9737},\;
Esma B. Boydas\,\orcidlink{0009-0006-2745-539X},\;
Youssef Briki\,\orcidlink{0009-0006-1083-5986},\;
Cailin Buchanan\,\orcidlink{0000-0001-9978-2687},\;
Mauricio Cafiero\,\orcidlink{0000-0002-4895-1783},\;
Damien Caliste\,\orcidlink{0000-0002-4967-9275},\;
Yi Cao\,\orcidlink{0009-0001-9151-0908},\;
Rafael E. Castañeda,\;
Sruthy K. Chandy\,\orcidlink{0000-0002-1061-647X},\;
Benjamin Charmes\,\orcidlink{0009-0007-9474-8632},\;
Shayantan Chaudhuri\,\orcidlink{0000-0003-4299-1837},\;
Yiming Chen\,\orcidlink{0000-0002-1501-5550},\;
Alexander Chen\,\orcidlink{0009-0001-7075-0152},\;
Jieneng Chen\,\orcidlink{0009-0008-8219-8909},\;
Min-Hsueh Chiu\,\orcidlink{0000-0003-0637-7856},\;
Defne Circi\,\orcidlink{0000-0002-5761-0198},\;
Cinthya H. Contreras\,\orcidlink{0000-0001-6405-0785},\;
Yoann Cur\'e,\;
Nathan Daelman\,\orcidlink{0000-0002-7647-1816},\;
Roshini Dantuluri\,\orcidlink{0000-0002-5101-8804},\;
Thomas Davy,\;
William Dawson\,\orcidlink{0000-0003-4480-8565},\;
Leonid Didukh\,\orcidlink{0000-0003-4900-5227},\;
Rui Ding\,\orcidlink{0009-0004-5909-8759},\;
Aminu R. Doguwa\,\orcidlink{0009-0001-4957-7850},\;
Claudia Draxl\,\orcidlink{0000-0003-3523-6657},\;
Sathya Edamadaka,\;
Oulaya Elargab,\;
Christina Ertural\,\orcidlink{0000-0002-7696-5824},\;
Matthew L. Evans\,\orcidlink{0000-0002-1182-9098},\;
Edvin Fako\,\orcidlink{0000-0002-0043-5907},\;
Hossam Farag\,\orcidlink{0000-0002-9314-3785},\;
Nur A. Fathurrahman\,\orcidlink{0000-0002-0386-8222},\;
Merve Fedai\,\orcidlink{0000-0002-6471-2408},\;
Rodrigo P. Ferreira\,\orcidlink{0009-0003-4030-9037},\;
Giuseppe Fisicaro\,\orcidlink{0000-0003-4502-3882},\;
Thomas Frank,\;
Sasi K. Gaddipati\,\orcidlink{0000-0003-3098-4592},\;
Abhijeet Gangan\,\orcidlink{0000-0002-8937-7984},\;
Jennifer Garland\,\orcidlink{0000-0003-2842-5951},\;
James Garrick,\;
Luigi Genovese\,\orcidlink{0000-0003-1747-0247},\;
Maryam Ghadrdran,\;
Sandip Giri\,\orcidlink{0000-0001-9015-5441},\;
Maxime Goulet\,\orcidlink{0000-0002-7505-1107},\;
Jeremy Goumaz\,\orcidlink{0009-0004-0637-2881},\;
Sara U. Gracia,\;
Jacob Graham, \;
Gabriel Graves,\;
Kevin P. Greenman\,\orcidlink{0000-0002-6466-1401},\;
Tim Greitemeier\,\orcidlink{0009-0005-2004-3088},\;
Cameron Gruich\,\orcidlink{0000-0002-3801-1296},\;
Sophie Gu,\;
Salomé Guilbert\,\orcidlink{0009-0000-6531-886X},\;
Hans Gundlach\,\orcidlink{0000-0001-5499-5072},\;
Muriel F. Gusta\,\orcidlink{0000-0001-9872-3079},\;
Mourad El Haddaoui,\;
Alexander J. Haibel\,\orcidlink{0009-0001-6249-5219},\;
Anubhab Haldar\,\orcidlink{0000-0002-2308-7415},\;
Vehaan Handa\,\orcidlink{0009-0000-4067-6660},\;
Hassan Harb\,\orcidlink{0000-0002-6016-3122},\;
Nathan D. Harms\,\orcidlink{0000-0003-2680-371X},\;
Abdullah Al Hasan\,\orcidlink{0009-0008-4820-9280},\;
Abir Hassan\,\orcidlink{0009-0004-7973-2497},\;
Qiyao He\,\orcidlink{0009-0005-7496-9126},\;
Andrés Henao-Aristizábal\,\orcidlink{0000-0002-2320-4693},\;
Bram Hoex\,\orcidlink{0000-0002-2723-5286},\;
Sungil Hong\,\orcidlink{0000-0001-8729-0861},\;
Alexander J. Horvath\,\orcidlink{0000-0002-8863-7158},\;
Md. Shaib Hossain\,\orcidlink{0009-0004-6663-5890},\;
Yanqi Huang\,\orcidlink{0009-0004-7958-1801},\;
Yuqing Huang,\;
Kostiantyn Hubaiev,\;
Donald Intal\,\orcidlink{0000-0003-3528-4894},\;
Katherine Inzani\,\orcidlink{0000-0002-3117-3188},\;
Kevin Ishimwe\,\orcidlink{0009-0008-4551-4570},\;
Tugba Isik\,\orcidlink{0000-0002-7298-1552},\;
Gopal R. Iyer\,\orcidlink{0000-0001-8117-6507},\;
Katharina Jager,\;
Jan Janssen\,\orcidlink{0000-0001-9948-7119},\;
Hyewon Jeong\,\orcidlink{0009-0003-1558-1848},\;
Michael Jirasek\,\orcidlink{0000-0002-4630-6457},\;
Tyler R. Josephson\,\orcidlink{0000-0002-0100-0227},\;
Nisarg Joshi\,\orcidlink{0000-0001-8497-7588},\;
Yassir Ben Kacem,\;
Remya A. M. Kalapurakal\,\orcidlink{0000-0002-9196-3448},\;
Rakesh R. Kamath\,\orcidlink{0000-0002-3291-7396},\;
Sugan Kanagasenthinathan\,\orcidlink{0009-0004-6821-0378},\;
Dohun Kang\,\orcidlink{0000-0001-5965-6432},\;
Jason Kantorow\,\orcidlink{0000-0003-1384-6112},\;
Kübra Kaygisiz\,\orcidlink{0000-0003-0077-7867},\;
Murat Ke\c{c}eli\,\orcidlink{0000-0001-8588-9272},\;
Farhana Keya\,\orcidlink{0000-0002-3782-8069},\;
Muhammad U. Khan\,\orcidlink{0009-0002-1004-7526},\;
Sartaaj Takrim Khan\,\orcidlink{0009-0009-2131-9700},\;
Hyungjun Kim\,\orcidlink{0000-0003-0879-0871},\;
Alexander Kister,\;
Sascha Klawohn\,\orcidlink{0000-0003-4850-776X},\;
Collin Kovacs,\;
Pranav Krishnan\,\orcidlink{0000-0002-5884-2521},\;
Maurycy Kryzanowski\,\orcidlink{0000-0002-9570-1955},\;
Ritesh Kumar\,\orcidlink{0000-0001-6345-6791},\;
Suman Kumari\,\orcidlink{0000-0002-3109-2847},\;
Gourav Kumbhojkar\, \orcidlink{0000-0003-4253-7247},\;
Ryo Kuroki,\;
Shashank Kushwaha\,\orcidlink{0000-0001-6295-1328},\;
Magdalena Lederbauer\,\orcidlink{0009-0008-0665-1839},\;
Jaejun Lee\,\orcidlink{0009-0003-8878-1752},\;
Seunghan Lee\,\orcidlink{0009-0005-3327-554X},\;
Jeonghwan Lee\,\orcidlink{0000-0002-3223-2573},\;
Bingcan Li\,\orcidlink{0000-0002-1744-6397},\;
Calvin Li,\;
Zhanzhao Li\,\orcidlink{0000-0001-7674-7424},\;
Shi Li\,\orcidlink{0000-0003-0505-6751},\;
Shicheng Li,\;
Chengyan Liu,\;
Hao Liu\,\orcidlink{0000-0001-8477-2558},\;
Tung Yan Liu\,\orcidlink{0000-0001-7188-6986},\;
Yutong Liu,\;
Lucia Vina-Lopez,\;
Chayaphol Lortaraparsert,\;
Andre K.Y. Low\,\orcidlink{0000-0002-0985-5123},\;
Saffron Luxford\,\orcidlink{0009-0008-5064-9568},\;
Carlos Madariaga\,\orcidlink{0009-0008-5252-5622},\;
Rishikesh Magar\,\orcidlink{0000-0001-6216-0518},\;
Piyush R. Maharana\,\orcidlink{0009-0004-4069-7102},\;
Rahul Mallela,\;
Shoaib Mahmud\,\orcidlink{0009-0002-3601-8467},\;
Natesan Mani\,\orcidlink{0000-0003-0635-8401},\;
Umair Mansoor,\;
Omar B. Mansour,\;
Cassandra Masschelein\,\orcidlink{0009-0001-7216-6320},\;
Kinga O. Mastej\,\orcidlink{0009-0006-5656-6646},\;
Ankit Mathanker\,\orcidlink{0000-0002-6738-2301},\;
Jeffrey Meng\,\orcidlink{0009-0007-8404-3153},\;
Omran Mezghani,\;
Yidong Ming\,\orcidlink{0009-0005-1417-1040},\;
Rishav Mitra\,\orcidlink{0009-0008-5244-8137},\;
Michail Mitsakis\,\orcidlink{0009-0003-3770-0000},\;
Matthew Miyagishima,\;
Ravikumar Mohan\,\orcidlink{0009-0003-0277-603X},\;
Naveen R. Mohanraj,\;
Trupti Mohanty\,\orcidlink{0000-0003-4270-1430},\;
Bernadette Mohr\,\orcidlink{0000-0003-0903-0073},\;
Francisco A. Molina-Bakhos,\;
Jeremy Monat\,\orcidlink{0009-0004-1360-0918},\;
Seyed Mohamad Moosavi\,\orcidlink{0000-0002-0357-5729},\;
Shayan Mousavi,\;
Arman Moussavi,\;
Rubel Mozumber\,\orcidlink{0009-0007-5926-6646},\;
Muhammad J. Mufti,\;
Diyana Muhammed\,\orcidlink{0009-0006-1178-4535},\;
Ram Munde\,\orcidlink{0009-0001-5597-1213},\;
Mrigi Munjal\,\orcidlink{0000-0003-2412-4764},\;
José A. Márquez\,\orcidlink{0000-0002-8173-2566},\;
Shankha Nag\,\orcidlink{0000-0001-9309-8366},\;
Giacomo Nagaro\,\orcidlink{0009-0001-8544-836X},\;
Juno Nam\,\orcidlink{0000-0002-9506-2938},\;
Jos\'e M. N\'apoles-Duarte\,\orcidlink{0000-0001-6823-4733},\;
Ry Nduma\,\orcidlink{0009-0008-1428-4637},\;
Xuan-Vu Nguyen\,\orcidlink{0000-0002-6078-2599},\;
Ebrahim Norouzi\,\orcidlink{0000-0002-2691-6995},\;
Oluwatosin Ohiro\,\orcidlink{0000-0001-7001-3518},\;
Ryotaro Okabe\,\orcidlink{0000-0002-5095-5951},\;
Viejay Ordillo,\orcidlink{0009-0004-7027-3052},\;
Shuichiro Ozawa\,\orcidlink{0009-0005-5420-8055},\;
Sebastian Pagel\,\orcidlink{0009-0003-6327-9803},\;
Daniel Palmer\,\orcidlink{0000-0002-2504-4195},\;
Angela Pan,\;
Akash Pandey,\;
Vivek Pandit\,\orcidlink{0009-0000-9121-6066},\;
Prakul Pandit\,\orcidlink{0000-0002-1343-3046},\;
Chiku Parida\,\orcidlink{0009-0005-9938-5696},\;
Jaehee Park\,\orcidlink{0000-0001-6956-7441},\;
Hyunsoo Park\,\orcidlink{0000-0001-9388-173X},\;
Hemangi Patel,\;
Shakul Pathak\,\orcidlink{0009-0006-3244-1670},\;
Taradutt Pattnaik\,\orcidlink{0009-0000-6673-9092},\;
Elena Patyukova\,\orcidlink{0000-0003-1641-5107},\;
Noah Paulson,\;
Deepak S. Pendyala\,\orcidlink{0009-0006-6691-1941},\;
Erick S. Pepek,\;
Martin H. Petersen\,\orcidlink{0000-0001-5840-1796},\;
Thang D. Pham\,\orcidlink{0000-0002-5232-4619},\;
Aniket Phutane\,\orcidlink{0009-0007-4043-6908},\;
Sabila K. Pinky,\;
Étienne Polack\,\orcidlink{0000-0003-0543-2008},\;
Alison Polasik,\;
Maria Politi\,\orcidlink{0000-0002-5815-3371},\;
Tim Pongratz\,\orcidlink{0009-0008-8217-3285},\;
Akhila Ponugoti\,\orcidlink{0009-0007-8947-7794},\;
Fabio Priante,\;
Thomas Michael Pruyn\,\orcidlink{0009-0006-7123-0454},\;
Sai S. Puppala\,\orcidlink{0009-0009-2572-3559},\;
Mohammad A. Qazi\,\orcidlink{0000-0002-0989-881X},\;
Heike Quosdorf\,\orcidlink{0000-0001-8348-7405},\;
Gollam Rabby,\;
Mohammad J. Raei\,\orcidlink{0009-0006-7464-154X},\;
Md. Habibur Rahman\,\orcidlink{0000-0002-7705-984X},\;
A.B.M. Ashikur Rahman\,\orcidlink{0000-0003-1375-4330},\;
Subhashree Rajasekaran,\;
Tawfiqur Rakib\,\orcidlink{0000-0001-6903-6667},\;
Hemanth N. Ramesh,\;
Vrushali Ranadive\,\orcidlink{0009-0000-0607-0956},\;
Karnamohit Ranka\,\orcidlink{0000-0002-2810-726X},\;
Bojana Rankovic\,\orcidlink{0000-0002-1476-6686},\;
Adwaith Ravichandran\,\orcidlink{0000-0002-0271-8994},\;
Ilija Rašović\,\orcidlink{0000-0001-9466-6281},\;
Sergei Rigin\,\orcidlink{0000-0003-4805-8060},\;
Tatem Rios,\;
Varun Rishi\,\orcidlink{0000-0003-0317-3815},\;
Victor Naden Robinson\,\orcidlink{0000-0002-9198-9154},\;
Lucas S. Rodrigues\,\orcidlink{0000-0002-3026-6403},\;
Oswaldo Rodriguez\,;
Mahule Roy\,\orcidlink{0009-0005-7259-771X},\;
Diptendu Roy\,\orcidlink{0000-0001-7718-3018},\;
Subhas Roy,\;
Arokia Anto Royan M,\;
Joseph F. Rudzinski\,\orcidlink{0000-0003-3403-640X},\;
Muhammad Sabih,\;
Subramanyam Sahoo,\;
Srusti Bheem Sain,\;
Thahira Saliya,\;
Vignesh Sampath,\;
Jesus Diaz Sanchez,\;
Arthur S. S. Santos\,\orcidlink{0009-0006-8803-6818},\;
Muliady Satria,\;
Hasan M. Sayeed\,\orcidlink{0000-0002-6583-7755},\;
Jörg Schaarschmidt\,\orcidlink{0000-0002-4389-2366},\;
Philippe Schwaller\,\orcidlink{0000-0003-3046-6576},\;
Nofit Segal\, \orcidlink{0000-0002-8891-8590},\;
Abhishec Senthilvel\,\orcidlink{0009-0004-4932-0627},\;
Sherjeel Shabih\,\orcidlink{0009-0008-6635-4465},\;
Devanshu Shah\,\orcidlink{0009-0009-8025-8397},\;
Faezeh Shahmoradi,\;
Samiha Sharlin\,\orcidlink{0000-0002-6379-9206},\;
Killian Sheriff\,\orcidlink{0000-0003-3613-2948},\;
Qiuyu Shi\,\orcidlink{0000-0002-0255-8257},\;
Abubakar D. Shuaibu\,\orcidlink{0000-0002-7674-9806},\;
Ayesha Siddiqua,\;
M.A. Shadab Siddiqui\,\orcidlink{0009-0008-8525-3133},\;
Darian Smalley\,\orcidlink{0000-0002-8980-0180},\;
Benjamin Smith\,\orcidlink{0000-0001-9673-2449},\;
Taylor D. Sparks\,\orcidlink{0000-0001-8020-7711},\;
Daniel T. Speckhard\,\orcidlink{0000-0002-9849-0022},\;
Elena Stojanovska\,\orcidlink{0000-0002-4350-3295},\;
Akshay Subramanian\,\orcidlink{0000-0002-9184-9979},\;
Jiwon Sun\,\orcidlink{0009-0007-0433-463X},\;
Yunkai Sun\,\orcidlink{0000-0001-8044-4936},\;
Abdul W. Syed,\;
Souvik Ta\,\orcidlink{0009-0009-5747-7639},\;
Izumi Takahara\,\orcidlink{0009-0004-6022-1945},\;
Kelly Tallau\,\orcidlink{0009-0006-9098-3080},\;
Guannan Tang\,\orcidlink{0000-0002-0269-2114},\;
Ans B. Tariq,\;
Sui X. Tay\,\orcidlink{0009-0006-1182-3610},\;
Nurlybek Temirbay\,\orcidlink{0000-0002-7037-1398},\;
Surya P. Tiwari,\;
Febin Tom,\;
Tajah Trapier\, \orcidlink{0009-0005-7609-364X},\;
Kasidet J. Trerayapiwat\,\orcidlink{0000-0002-2263-1343},\;
Samanvya Tripathi,\;
Hawra H. Tuhaifa,\;
Mustafa Unal\,\orcidlink{0000-0001-5829-4724},\;
Mohammad Uzair\,\orcidlink{0000-0003-2039-0870},\;
Vallabh Vasudevan\,\orcidlink{0000-0001-7933-4924},\;
Estefania Vazquez,\;
Victor Venturi,\;
Rahul Verma\,\orcidlink{0000-0003-4991-8376},\;
Ashwini Verma\,\orcidlink{0000-0001-9837-9597},\;
\'{A}lvaro V\'{a}zquez-Mayagoitia\,\orcidlink{0000-0002-1415-6300},\;
Nicholas Wagner,\;
Araki Wakiuchi\,\orcidlink{0000-0001-7607-2922},\;
Hao Wan,\;
Liaoyaqi Wang\,\orcidlink{0009-0006-5633-1417},\;
Wolfgang Wenzel\,\orcidlink{0000-0001-9487-4689},\;
Alexander Wieczorek\,\orcidlink{0000-0002-1025-128X},\;
Sze H. Wong\,\orcidlink{0009-0008-6629-3391},\;
Yue Wu\,\orcidlink{0000-0003-2874-8267},\;
Tong Xie\,\orcidlink{0000-0002-1659-4865},\;
Andrew Yi\,\orcidlink{0009-0002-4631-5588},\;
Ziqi Yin\,\orcidlink{0000-0001-8708-358X},\;
Jodie A. Yuwono\,\orcidlink{0000-0002-0915-0756},\;
Nahed A. Zaid\,\orcidlink{0000-0002-7830-6443},\;
Mohd Zaki\,\orcidlink{0000-0002-4551-3470},\;
Shehtab Zaman,\;
Maimuna U. Zarewa,\;
Mahtab Zehtab\,\orcidlink{0009-0000-4572-8941},\;
Baosen Zhang,\;
Wenyu Zhang\,\orcidlink{0000-0002-1924-6591},\;
Melody Zhang\,\orcidlink{0000-0001-9788-9958},\;
Yangfan Zhang\,\orcidlink{0009-0007-9824-2516},\;
Yuwen Zhang\,\orcidlink{0000-0001-8915-1769},\;
Runze Zhang,\;
Zongmin Zhang,\;
Huanhuan Zhao,\;
Yuanlong Bill Zheng,\;
Ramzi Zidani\,\orcidlink{0009-0008-1005-7242},\;
Xue Zong,\;
Ian Foster,\;
and Ben Blaiszik$^{*\textit{1,2}}$\,\orcidlink{0000-0002-5326-4902}

\end{center}

\begin{center}
\small
$^1$ Data Science \& Learning Division, Argonne National Laboratory, Lemont, IL, USA.\\
$^2$ University of Chicago, Chicago, IL, USA\\
\vspace{5pt}
$^*$Corresponding author: Ben Blaiszik (\href{blaiszik@uchicago.edu}{blaiszik@uchicago.edu}).\\$^\dagger$These authors contributed substantially to compiling team results and other paper writing tasks.
\end{center}

\FloatBarrier
\bibliographystyle{ieeetr}
\bibliography{ref}

\end{document}